\documentclass[dissertation, letterpaper, 10pt]{utthesis}

\title{Stability of Quantum Computers
}
\author{Samudra~Dasgupta}
\copyrightYear{2024}
\graduationMonth{May}
\majorProfessor{Dr. Travis Humble}	    
\viceProvost{Dixie L. Thompson} 
\major{Data Science and Engineering}	
\degree{Doctor of Philosophy}	    
\college{Engineering}           
\dept{Bredesen Center for Interdisciplinary Research}	
\university{The University  of Tennessee, Knoxville}	
\numberOfCommitteeMembers{4} 
\committeeMemberA {Dr. Travis Humble}	
\committeeMemberB {Dr. Russell Lee Zaretzki}	
\committeeMemberC {Dr. Himanshu Thapliyal}	
\committeeMemberD {Dr. Rebekah Herrman}	

\usepackage{xspace}
\usepackage{parskip, indentfirst}
\usepackage{pdflscape, rotating}
\usepackage{longtable}
\usepackage{etoc} 
\usepackage{amsmath,amssymb,amsfonts,amsthm, algorithm, algorithmic}
\usepackage{mathtools, braket, dsfont, graphicx, lipsum, enumitem}
\usepackage{textcomp, float, etoc, bibentry, multirow, tabularx}
\usepackage{nicefrac}       
\usepackage{microtype}      
\usepackage{nomencl}                    
\usepackage{float}                      
\usepackage{fancyhdr}                   
\usepackage{url}                        
\usepackage{relsize}                    
\usepackage{booktabs}                   
\usepackage{mathrsfs}                   
\usepackage{pdflscape}			
\usepackage{tikz}
\usetikzlibrary{quantikz}
\usepackage[numbers]{natbib}
\usepackage{xstring}
\usepackage[inactive]{srcltx}		 	
\usepackage[config, labelfont={bf}]{caption,subfig} 
\usepackage[titletoc]{appendix}			
\usepackage[normalem]{ulem}
\usepackage[english]{babel}
\usepackage[utf8]{inputenc} 
\usepackage[T1]{fontenc}    
\usepackage{hyperref}
\hypersetup{
  colorlinks=true, 
  allcolors=black,
  pdfauthor={...},
  pdftitle={...},
  pdfsubject={...},
}

\usepackage{xstring}
\usepackage{shorttoc}

\defcitealias{dasgupta2023reliability}{Reliability of Noisy Quantum Computing Devices}
\defcitealias{steed2023qvis_sc}{QVis: Visual Analysis of Quantum Computing Noise}
\defcitealias{steed2023qvis_ieee}{QVis: A Visual Analytics Tool for Exploring Noise and Errors in Quantum Computing Systems}
\defcitealias{dasgupta2023adaptive}{Adaptive mitigation of time-varying quantum noise}
\defcitealias{dasgupta2023reliable}{Reliable Devices Yield Stable Quantum Computations}
\defcitealias{dasgupta2022adaptive}{Adaptive stabilization of quantum circuits executed on unstable devices}
\defcitealias{dasgupta2022assessing}{Assessing the stability of noisy quantum computation}
\defcitealias{dasgupta2022characterizing}{Characterizing the reproducibility of noisy quantum circuits}
\defcitealias{dasgupta2021stability}{Stability of noisy quantum computing devices}
\defcitealias{dasgupta2021reproducibility}{Reproducibility in quantum computing}
\defcitealias{dasgupta2020characterizing}{Characterizing the stability of nisq devices}
\defcitealias{dasgupta2022characterizing_memory}{Characterizing the memory capacity of transmon qubit reservoirs}
\defcitealias{xu2023dynamic}{Dynamic Asset Allocation with Expected Shortfall via Quantum Annealing}
\defcitealias{dasgupta2019quantum}{Quantum Annealing Algorithm for Expected Shortfall based Dynamic Asset Allocation}

\newcolumntype{P}[1]{>{\raggedright\arraybackslash}p{#1}}
\newcommand{\figurewidth}{0.84\columnwidth} 

\newcommand{\unnumberedcaption}[1]{\captionsetup{labelformat=empty}#1\captionsetup{labelformat=default}}

\newcommand{\added}[1]{{\color{black}#1}}

\newcommand{\xrm}{\textrm{x}}
\newcommand{\lowk}{{{}_k}}
\newcommand{\low}[1]{{{}_{\text{#1}}}}

\newcommand{\BV}{Bernstein-Vazirani\xspace}
\newcommand{\pec}{probabilistic error cancellation\xspace}
\newcommand{\nsn}{non-stationary noise\xspace}
\newcommand{\ns}{non-stationary\xspace}
\newcommand{\SPAM}{state preparation and measurement\xspace}

\newcommand{\Tr}{\textrm{Tr}}
\newcommand{\Prob}{\textrm{Pr}} 

\newcommand{\sop}{\textrm{super-operator\xspace}}
\newcommand{\superc}{\textrm{superconducting}}

\newcommand{\belem}{ibm\_belem\xspace}
\newcommand{\mumbai}{ibm\_mumbai\xspace}
\newcommand{\kolkata}{ibm\_kolkata\xspace}
\newcommand{\washington}{ibm\_washington\xspace}
\newcommand{\toronto}{ibm\_toronto\xspace}
\newcommand{\yorktown}{ibm\_yorktown\xspace}

\def\BibTeX{{\rm B\kern-.05em{\sc i\kern-.025em b}\kern-.08em
    T\kern-.1667em\lower.7ex\hbox{E}\kern-.125emX}}

\graphicspath{
{figures/}
}

\begin{document}
\pagenumbering{roman}



\pagenumbering{alph}
\makeTitlePage
\pagenumbering{roman}
\setcounter{page}{2}
\makeCopyrightPage
\chapter*{}
\begin{center}
{\centering \it Dedicated to my parents.}
\end{center} 

\chapter*{Acknowledgments}
I am deeply thankful to Travis for his invaluable guidance, support, and expertise during my entire PhD journey. His steadfast dedication and insightful mentorship has significantly influenced my research and personal development. I also want to express my gratitude for the opportunity to utilize the resources of the Oak Ridge Leadership Computing Facility (a United States Department of Energy, Office of Science User Facility).

\chapter*{Abstract}\label{ch:abstract}

Quantum computing's potential is immense, promising super-polynomial reductions in execution time, energy use, and memory requirements compared to classical computers. This technology has the power to revolutionize scientific applications such as simulating many-body quantum systems for molecular structure understanding, factorization of large integers, enhance machine learning, and in the process, disrupt industries like telecommunications, material science, pharmaceuticals and artificial intelligence. However, quantum computing's potential is curtailed by noise, further complicated by non-stationary noise parameter distributions across time and qubits. This dissertation focuses on the persistent issue of noise in quantum computing, particularly non-stationarity of noise parameters in transmon processors. It establishes a framework comprising computational accuracy, device reliability, outcome stability, and result reproducibility for assessing noisy outcomes amidst time-varying quantum noise. It further aims to determine the upper and lower bounds for this framework using available noise characterization data, in terms of the distance between time-varying noise densities. Using real data from a transmon processor, it validates the bounds on a test quantum circuit. It also demonstrates that if the physical platform's noise stays within the bounds determined by the analysis, experimental reproducibility can be guaranteed with a high degree of certainty. Furthermore, it develops a Bayesian algorithm to enhance outcome stability and accuracy for probabilistic error cancellation (PEC) in presence of time-varying quantum noise. The results obtained from experiments using a 5-qubit implementation of the Bernstein-Vazirani algorithm conducted on the \kolkata device, underscore the effectiveness of the adaptive algorithm, showing a 42\% improvement in accuracy over non-adaptive PEC and a 60\% improvement in stability. Considering the time-varying stochastic nature of quantum noise, integrating adaptive estimation in error mitigation is crucial. In summary, by delving into the complexities of non-stationary noise in quantum computing, this dissertation provides valuable insights into quantifying and enhancing stability of outcomes from noisy quantum computers.

\renewcommand{\contentsname}{Table of Contents}
\makeatletter
\renewcommand\tableofcontents{%
  \chapter*{\contentsname
    \@mkboth{%
       \MakeUppercase\contentsname}{\MakeUppercase\contentsname}}%
 \@starttoc{toc}%
}
\makeatother 
\setcounter{tocdepth}{1} 
\tableofcontents

\clearpage\newpage 
\listoftables\listoffigures
\clearpage\newpage 
\chapter*{List of abbreviations}

\newcolumntype{P}[1]{>{\centering\arraybackslash}p{#1}}

\begin{longtable}{p{2cm} p{12cm}}
\caption*{}
\label{tab:abbreviations} \vspace{4mm}\\
NSN & Non-stationary noise \vspace{4mm}\\
PEC & probabilistic error cancellation \vspace{4mm}\\
BC & Bhattacharyya coefficient \vspace{4mm}\\ 
BV & Bernstein-Vazirani \vspace{4mm}\\ 
SPAM & State preparation and measurement \vspace{4mm}\\
Tr & Trace \vspace{4mm}\\ 
WSS & Wide-sense stationary \vspace{4mm}\\ 
MCMC & Markov chain Monte-Carlo \vspace{4mm}\\ 
\end{longtable}
\clearpage\newpage 
\chapter*{Notations}

\newcolumntype{P}[1]{>{\centering\arraybackslash}p{#1}}

\vspace{-12mm}
\begin{longtable}{p{4cm} p{10cm}}
\caption*{}
\label{tab:Notations}\vspace{4mm}\\
$\ket{\psi}$ & a pure quantum state\vspace{4mm}\\
$\otimes$   & tensor operator\vspace{4mm}\\ 
Tr$(\cdot)$, Pr$(\cdot)$& Trace and probability operators respectively\vspace{4mm}\\ 
$\rho$ & density matrix\vspace{4mm}\\ 
$\Omega, \lambda$ & Hamiltonian and its eigenvalue\vspace{4mm}\\ 
$T_1, T_2$ & Qubit relaxation and dephasing time respectively\vspace{4mm}\\ 
$N_s$    & Number of shots per circuit execution\vspace{4mm}\\ 
$L$    & Number of times a circuit is executed for collecting statistics\vspace{4mm}\\
$n$ & quantum register size\vspace{4mm}\\ 
$\ket{v}, \ket{v_{n-1}\cdots v_0}, v_i, v$ & $\ket{v}$ is the short form for $\ket{v_{n-1} \cdots v_0}$. The latter denotes an n-bit string with each $v_i \in \{0, 1\}$. Lastly, $v$ denotes the decimal form for the bit-string e.g. if $\ket{v} = \ket{0011}$, then $v=3$. \vspace{4mm}\\ 
$\mathcal{U}, \tilde{\mathcal{U}}$ & a unitary matrix and its noisy version (could be non-unitary) \vspace{4mm}\\ 
$R_Y(\theta)$ & operator for rotation by an angle ($\theta$) about the Y-axis on the Bloch sphere \vspace{4mm}\\
$\mathds{I}, \mathds{X}, \mathds{Y}, \mathds{Z}$ & 2x2 identity and the Pauli X, Y, Z matrices\vspace{4mm}\\ 
$\mathds{H}$, CNOT, SX, RZ & Various quantum gates (Hadamard gate, Controlled-NOT, Square root of NOT and Unitary gate for rotation about Z-axis\vspace{4mm}\\ 
$T_D$ & delay gate duration\vspace{4mm}\\ 
$\tilde{\mathds{H}}$ & noisy Hadamard gate\vspace{4mm}\\ 
$\mathcal{I}(X, Y)$ & Mutual Information between two random variables $X$ and $Y$\vspace{4mm}\\ 
$\mathcal{H}(X)$ & Entropy of $X$\vspace{4mm}\\ 
$\mathcal{O}$& A quantum observable\vspace{4mm}\\ 
$\mathcal{E}_{\xrm}, \mathcal{E}_{\xrm_i}(\rho)$& Multi-qubit quantum noise channel parameterized by the vector x; $\mathcal{E}_{\xrm_i}(\rho)$ is a single-qubit quantum noise channel\vspace{4mm}\\ 
$M_k$ & Kraus operators characterizing the error channel $\mathcal{E}(\cdot)$\vspace{4mm}\\ 
$F_{\text{SPAM}}, \epsilon_\text{SPAM}$ & SPAM Fidelity and SPAM error respectively\vspace{4mm}\\ 
$F_{\text{G}}, e_G$ & Gate Fidelity and error per Clifford gate respectively\vspace{4mm}\\ 
$T_G$ & CNOT Gate Length\vspace{4mm}\\ 
$\tau_G$ & Duty Cycle\vspace{4mm}\\ 
$\text{Beta}(\alpha, \beta)$& Beta function $(= \int_{0}^1 t^{\alpha-1}(1-t)^{\beta - 1}dt)$\vspace{4mm}\\ 
$\Gamma(z)$& Gamma function $(= \int_{0}^{\infty} t^{z-1}\exp^{-t}dt)$\vspace{4mm}\\ 
$\mathds{R}, \mathds{C}$& Set of real and complex numbers respectively\vspace{4mm}\\ 
$P_2(\mathds{R})$& Set of real polynomials of degree not more than 2\vspace{4mm}\\ 
$\xrm, \xrm_i$ & $x=(\xrm_1, \cdots, \xrm_{d})$ is a vector of parameters characterizing the noise of the quantum circuit \vspace{4mm}\\ 
$X, X_i$& $X$ is a vector of random variables (stochastic device parameters) corresponding to x above; $X_i$ is a scalar random variable\vspace{4mm}\\ 
$f_{X}(\xrm;t)$& multi-variate continuous probability distribution function for x; same as $f_{X}(\xrm;t)$\vspace{4mm}\\ 
$F_{X}(\xrm;t)$& multi-variate continuous cumulative distribution function for x; same as $F_{X}(\xrm;t)$\vspace{4mm}\\ 
$\eta(X, Y)$ & Normalized Mutual Information between two random variables $X$ and $Y$\vspace{4mm}\\ 
$F_{\text{A}}$ & Addressability\vspace{4mm}\\ 
$BC$ & Bhattacharyya coefficient\vspace{4mm}\\
$H_{X}(t_1, t_2)$& Hellinger distance between the multi-variate joint distributions of $X$ at times $t_1$ and $t_2$; if $t_1=0$ then we sometimes simply write it as $H_{X}(t)$\vspace{4mm}\\ 
$H_{\text{avg}}(t_1, t_2)$ & average of the Hellinger distances over the $d$ univariate marginal distributions $\left(=\frac{1}{d} \sum\limits_{k=1}^d H_{X_k} (t_1, t_2)\right)$\vspace{4mm}\\ 
$H_\text{normalized}(t_1, t_2)$ & Hellinger distance normalized with respect to the dimension $(=\sqrt{1 - BC^{1/d}})$\vspace{4mm}\\ 
$\Theta(\cdot)$ & copula function\vspace{4mm}\\ 
$f^{\text{BV}}(\cdot)$ & Oracle function in the Bernstein-Vazirani problem\vspace{4mm}\\ 
$r$ & secret $n$-bit string in the Bernstein-Vazirani problem\vspace{4mm}\\ 
$U_f$ & Unitary for the Oracle function in the Bernstein-Vazirani problem\vspace{4mm}\\ 
$P_a, P_C, P_T$ & Pauli gates \vspace{4mm}\\
$\mathcal{Q}$ & quasi probability distribution\vspace{4mm}\\ 
$\{ \eta_\lowk \}$ & generally refers to a PEC linear combination coefficient\vspace{4mm}\\ 
$\tilde{\mathcal{B}}_i$ & noisy basis superoperators in PEC\vspace{4mm}\\ 
\end{longtable}

\newpage\pagenumbering{arabic}
\setcounter{page}{1}

\newcommand{\removefigs}{}
%
\removefigs
\chapter{Introduction}\label{ch:introduction}
\section{Background}
\subsection{Promise of quantum computing}
Error-resilient quantum computing holds great promise, offering significant advancements over conventional computing. Once realized, it is expected to super-polynomially reduce execution time, energy consumption, and memory storage needs compared to conventional state-of-the-art computers~\cite{humble2019quantum}. The potential impact of error-resilient quantum computing includes revolutionizing scientific applications such as simulating many-body quantum systems~\cite{browaeys2020many}, solving large-scale optimization problems~\cite{montanaro2016quantum}, efficiently sampling high-dimensional probability distributions~\cite{shang2015monte}, factorizing large integers, and enhancing the security of communication networks~\cite{espitia2021role}. Consequently, this technology is expected to be disruptive to sectors such as telecommunications, cyber-security, pharmaceuticals, logistics, supply chain management, artificial intelligence, and materials science~\cite{how2023business}.
\subsection{Quantum computing vs classical computing}
Classical mechanics, rooted in the laws of Newtonian physics, has served as a successful framework for understanding the macroscopic world for centuries. However, when examining the behavior of particles at the atomic and subatomic scales, classical mechanics began to exhibit limitations and inconsistencies. For instance, classical mechanics predicted absurd outcomes, like suggesting that a blackbody emits an infinite amount of energy across all wavelengths.

As scientists delved deeper into the microscopic realm, counter-intuitive phenomena such as wave-particle duality, quantized energy levels, and non-locality emerged, challenging the classical paradigm. These challenges necessitated the development of quantum mechanics, which offered a novel and revolutionary approach to describe the behavior of particles at the quantum level~\cite{feynman1965feynman}. 

Quantum mechanics introduced probabilistic interpretations, superposition states, and entanglement, providing a more accurate and comprehensive understanding of the intricate workings of nature~\cite{cohen1977quantum}. For example, consider a particle moving in one dimension under the influence of a conservative force, such as a harmonic oscillator. In classical mechanics, we can describe the particle's motion using Newton's second law and the equation of motion for a harmonic oscillator, which yields a sinusoidal trajectory and continuous energy levels. In contrast, in quantum mechanics, we describe the particle using the Schr\"{o}dinger equation for a harmonic oscillator, which results in quantized energy levels and wave functions corresponding to discrete energy states. However, as the value of the Planck's constant (h), representing the fundamental scale of quantum mechanics, approaches zero ($h \rightarrow 0$), the quantum system converges towards the classical limit. In this limit, the quantized energy levels of the quantum harmonic oscillator become densely spaced and form a continuous energy spectrum, matching the classical behavior. The quantum wave function also converges to the classical trajectory, and the classical and quantum results become indistinguishable. 

This convergence phenomenon is known as the correspondence principle~\cite{sakurai2020modern}, where classical mechanics emerges as the limiting case of quantum mechanics at large scales or when the quantum effects become negligible. Understanding the system's scale is crucial in selecting the appropriate framework for a given physical problem, with classical mechanics suitable for macroscopic objects with well-defined trajectories and negligible quantum effects, while quantum mechanics is employed for microscopic particles, providing a more accurate description of phenomena like wave-particle duality, quantization of energy levels, and quantum entanglement. 

Quantum mechanics is built upon several fundamental postulates~\cite{mermin2007quantum, nielsen2002quantum} that were formulated to address experimental observations in the early 20th century. These postulates provide the framework for understanding the behavior of quantum systems. The first postulate states that every quantum system is associated with a complex Hilbert space~\cite{bhatia1997matrix}. The quantum state of a system is described by a density operator~\cite{kraus1983states, bhatia1997matrix}, often denoted by $\rho$, which belongs to the set of density operators defined on the Hilbert space. The second postulate deals with measurements~\cite{kraus1983states}. When a measurement is performed on a quantum system, it can have random outcomes with finite probabilities. 
The third postulate connects the quantum state with the measurement outcomes. When a measurement $M$ is performed on a quantum system in the state $\rho$, then the observed outcome is a realization of the random variable $M\rho$~\cite{nielsen2002quantum}. 
The fourth postulate addresses composite quantum systems. When we have two quantum systems, each associated with its own Hilbert space, say $\mathcal{H}_1$ and $\mathcal{H}_2$, the combined Hilbert space of the composite system is given by the tensor product $\mathcal{H}_1 \otimes \mathcal{H}_2$. This tensor product construction allows us to represent the joint states of the individual systems~\cite{sakurai2020modern}. 

These fundamental postulates provide the tools to analyze quantum systems. The concepts of quantum state, measurement, and composite systems are the key building blocks for understanding the intriguing phenomena that occur in the quantum world.

Classical computing is the conventional form of computing that relies on classical bits, represented by the binary numbers 0 and 1. Classical computers crucially rely on components like transistors, which function based on quantum mechanical principles. However, despite the quantum nature of the transistors, the interactions between these components within a classical computer follow a classical framework. This distinction can sometimes lead to confusion, as it may seem unsatisfactory to say that classical computers, which are built using components based on quantum principles, operate according to classical laws. But while specific underlying components leverage quantum phenomena, the interactions between these components, and the analysis of the data produced by them, are modeled satisfactorily using classical physics and classical information theory. 

Analogously, the link between quantum computing~\cite{zeguendry2023quantum} and quantum mechanics is fundamental, as the fundamental interaction between the components of quantum computers depend on quantum mechanical phenomenon~\cite{nielsen2002quantum} such as superposition and entanglement. Quantum computations use quantum bits or qubits~\cite{schumacher1995quantum} as the basic unit of information. Superposition enables qubits to exist in multiple states simultaneously, allowing quantum computers to perform parallel calculations. Entanglement~\cite{bennett1998quantum} creates strong correlations between qubits, even when they are physically separated, potentially leading to increased computational power. Different qubit technologies, such as superconducting qubits, trapped ions, and topological qubits, each utilize distinct quantum phenomena. Researchers continuously draw upon quantum mechanical principles to optimize performance, addressing challenges like quantum decoherence~\cite{bejanin2021interacting, burnett2019decoherence, carroll2021dynamics, wrightgroup} resulting from interactions with the external environment.

Classical computing remains the practical and efficient choice for many computing needs. In fact, the computing power of classical computers has been doubling every two years, as per Moore's law. But still classical computing faces limitations with computationally challenging problems that grow exponentially with problem size such as unstructured large-scale optimization, factorization of large integers, and simulation of many-body systems. The quantum equivalent of Moore's law~\cite{nielsen2002quantum} states that adding just one perfect qubit to a quantum computer doubles its computational capability. Therefore, to match the progress of classical computers, a single error-resilient qubit needs to be integrated into quantum computers every two years. 

However, quantum algorithms~\cite{mermin2007quantum} are more efficient for tackling only a sub-set of classically computationally challenging problems (not universally). Example of problems that have an efficient quantum algorithm include prime factorization and discrete logarithm, both of which were developed by Shor. An efficient algorithm~\cite{cormen1990introduction} operates within a time frame that corresponds to a polynomial function of the problem size, whereas an inefficient algorithm takes time that corresponds to a super-polynomial function of the problem size.

The study of algorithm efficiency~\cite{cormen1990introduction} is a fundamental aspect of complexity theory~\cite{cook2023complexity}, a branch of computer science. Problems are categorized based on their resource requirements, such as time and memory. For instance, problems solvable in polynomial time by classical computers fall into the class P, whereas those with solutions verifiable in polynomial time belong to class NP. While it's evident that P is a subset of NP, the question of whether there are problems in NP not in P remains unresolved. 

In the realm of quantum computing, problems solvable by quantum algorithms within polynomial time (with bounded error probability) are classified as belonging to class BQP~\cite{nielsen2002quantum}. It has not been formally established whether BQP contains P. So, we are not certain that quantum computing contains classical computing as a special case, but evidence supports this assertion. 

BQP of course contains QP which represents problems that a quantum computer can solve with a 100\% probability of success in polynomial time. Examples that belong to the BQP class include Deutsch-Jozsa, Bernstein-Vazirani, and Simon's algorithm.

Why can we not simulate quantum computation using classical computers? The reason lies in the exponential space and time complexity involved in storing quantum gates as classical matrices and tracking entangled qubits after logical operations. For instance, even a system with just 500 atoms would need $2^{500}$ complex coefficients for perfect description. Attempting computations with such requirements would overwhelm classical computers. However, quantum computers excel in simulating such scenarios by storing, representing, and evolving states as native quantum states on qubit registers, bypassing the need for managing $2^{500}$ complex floating-point numbers with limited precision 

Note that, even if we restricted ourselves to a small scale register, then also a perfect quantum computer can never be built using classical computers because quantum measurement cannot be perfectly simulated as there is no perfect random number generator.

In classical computing, logic gates are basic building blocks that manipulate classical bits (0s and 1s) to perform logical operations. Two examples of classical logic gates are the AND gate and the NOT gate. The AND gate takes two input bits, and its output is 1 (true) only when both input bits are 1; otherwise, the output is 0 (false). The NOT gate takes a single input bit and produces the opposite value as output.

\begin{table*}[h]
\centering
\begin{tabular}{|c|c|c|}
\hline
Input 1 & Input 2 & AND Gate Output \\
\hline
0 & 0 & 0 \\
\hline
0 & 1 & 0 \\
\hline
1 & 0 & 0 \\
\hline
1 & 1 & 1 \\
\hline
\end{tabular}
\hspace{2cm}
\begin{tabular}{|c|c|}
\hline
Input & NOT Gate Output \\
\hline
0 & 1 \\
\hline
1 & 0 \\
\hline
\end{tabular}
\end{table*}

In quantum computing, instead of bits, we have qubits, which are represented as vectors in a complex two-dimensional Hilbert space. The basis vectors spanning this space are commonly expressed in three ways: the $\mathds{Z}$-Basis (also called computational basis or standard basis), the $\mathds{X}$-basis, and the $\mathds{Y}$-basis. Each of these bases is made up of orthogonal vectors in two dimensions, ideal for representing a two-level quantum system in a two-dimensional vector space.

The need for these bases becomes clear when studying the underlying physics of quantum systems. Take, for example, the spin of an electron in a magnetic field. The electron's spin resembles a minuscule magnetic moment. Its orientation relative to the magnetic field influences the electron's energy.

For the $\mathds{Z}$-Basis, consider a magnetic field applied vertically. An electron's spin might align with this field, represented by the lower energy state $\ket{0}$, or it could oppose the field, corresponding to the higher energy state $\ket{1}$. This basis provides an intuitive way to think about qubits, likening the lower and higher energy states to the classical binary values of $0$ and $1$ respectively. A qubit's state can be in a superposition of both $\ket{0}$ and $\ket{1}$, expressed as $c_1 \ket{0} + c_2 \ket{1}$, where $c_1$ and $c_2$ are complex amplitudes satisfying: 
\begin{equation}
|c_1|^2 + |c_2|^2 = 1.
\end{equation}

However, if we change our perspective and measure the spin horizontally, along the $\mathds{X}$-axis, the electron's spin might point left or right. These orientations, when related back to the $\mathds{Z}$-Basis, are actually superpositions of the $\ket{0}$ and $\ket{1}$ states. These superpositions, 
\begin{equation}
\ket{+} = \frac{\ket{0}}{\sqrt{2}}+\frac{\ket{1}}{\sqrt{2}},
\end{equation}
and 
\begin{equation}
\ket{-} = \frac{\ket{0}}{\sqrt{2}}-\frac{\ket{1}}{\sqrt{2}},
\end{equation}
define the $\mathds{X}$-basis.

The $\mathds{Y}$-basis offers yet another viewpoint. Measuring perpendicular to both the $\mathds{X}$ and $\mathds{Z}$ axes gives states that are complex superpositions of the $\mathds{Z}$-Basis: 
\begin{equation}
\ket{i} = \frac{\ket{0}}{\sqrt{2}}+i\frac{\ket{1}}{\sqrt{2}},
\end{equation}
and 
\begin{equation}
\ket{-i} = \frac{\ket{0}}{\sqrt{2}}-i\frac{\ket{1}}{\sqrt{2}},
\end{equation}


Quantum operations can be visualized as rotations around these different basis axes, offering valuable insights for designing quantum algorithms. Furthermore, using different bases for measurements can be instrumental in pinpointing various types of errors in quantum systems.

Quantum gates, represented by unitary matrices, manipulate qubits, with essential examples including the Pauli gates ($\mathds{I}$, $\mathds{X}$, $\mathds{Y}$, and $\mathds{Z}$), the Hadamard gate ($\mathds{H}$), and the two-qubit entangling CNOT gate ($\mathds{U}_\text{CNOT}$). The Pauli-$\mathds{X}$ gate acts as the quantum analog of the NOT gate, flipping the qubit's state between $\ket{0}$ and $\ket{1}$. The Pauli-$\mathds{Y}$ gate transforms $\ket{0}$ to $i\ket{1}$ and $\ket{1}$ to $-i\ket{0}$. The Pauli-$\mathds{Z}$ gate introduces a relative phase shift between the basis states of a qubit (it leaves $\ket{0}$ unchanged and flips the sign of $\ket{1}$). The Hadamard gate puts a qubit in an equal superposition of $\ket{0}$ and $\ket{1}$. The CNOT gate is a two-qubit gate that flips the target qubit if and only if the control qubit is in state $\ket{1}$. 

\begin{table*}[h]
\centering
\begin{minipage}[t]{0.45\textwidth}
\centering
\unnumberedcaption{Pauli-$\mathds{X}$ Gate}\\
\begin{tabular}{|c|c|}
\hline
Input & Output \\
\hline
$\ket{0}$ & $\ket{1}$ \\
\hline
$\ket{1}$ & $\ket{0}$ \\
\hline
\end{tabular}
\end{minipage}\hfill
\begin{minipage}[t]{0.45\textwidth}
\centering
\unnumberedcaption{Pauli-$\mathds{Y}$ Gate}\\
\begin{tabular}{|c|c|}
\hline
Input & Output \\
\hline
$\ket{0}$ & $i\ket{1}$ \\
\hline
$\ket{1}$ & $-i\ket{0}$ \\
\hline
\end{tabular}
\end{minipage}

\vspace{1cm}

\begin{minipage}[t]{0.45\textwidth}
\centering
\unnumberedcaption{Pauli-$\mathds{Z}$ Gate}\\
\begin{tabular}{|c|c|}
\hline
Input & Output \\
\hline
$\ket{0}$ & $\ket{0}$ \\
\hline
$\ket{1}$ & -$\ket{1}$ \\
\hline
\end{tabular}
\end{minipage}\hfill
\begin{minipage}[t]{0.45\textwidth}
\centering
\unnumberedcaption{Hadamard Gate}\\
\begin{tabular}{|c|c|}
\hline
Input & Output \\
\hline
$\ket{0}$ & $(\ket{0} + \ket{1})/\sqrt{2}$ \\
\hline
$\ket{1}$ & $(\ket{0} - \ket{1})/\sqrt{2}$ \\
\hline
\end{tabular}
\end{minipage}
\end{table*}

\begin{table*}[h]
\centering
\unnumberedcaption{CNOT gate}\\
\label{tab:truth_table_cnot}
\begin{tabular}{|c|c|c|}
\hline
Control qubit & Target qubit & Output (state of target qubit) \\
\hline
$\ket{0}$ & $\ket{0}$ & $\ket{0}$\\
\hline
$\ket{0}$ & $\ket{1}$ & $\ket{1}$\\
\hline
$\ket{1}$ & $\ket{0}$ & $\ket{1}$\\
\hline
$\ket{1}$ & $\ket{1}$ & $\ket{0}$\\
\hline
\end{tabular}
\end{table*}

In classical circuits, information flows through the movement of electrons from one transistor to another in a well-defined spatial layout. Classical logic gates, such as AND, OR, and NOT, manipulate classical bits (0 or 1) and perform logical operations. In contrast, quantum circuits process information using qubits. 
Quantum circuits evolve the quantum state in situ, which modify the quantum amplitudes and phases of the state of the quantum register. Quantum gates, like Pauli-$\mathds{X}$, CNOT, and Hadamard, implement quantum algorithms by performing quantum operations on the qubits. 

Designing quantum algorithms~\cite{mermin2007quantum} is significantly more challenging than classical algorithms~\cite{cormen1990introduction} for several reasons. Firstly, quantum computing requires a departure from classical intuition, as quantum phenomena behave differently from classical physics. For instance, computer scientists experienced in conventional parallel programming understand the challenges associated with designing algorithms that can effectively harness GPU parallelism. They can thus empathize with the complexity of leveraging computing power through superposition, a form of parallel computing utilized in quantum systems. However, phenomena such as entanglement and quantum interference present unique opportunities that lack analogues in classical algorithm development. Secondly, to demonstrate the utility of a quantum algorithm, it must be more efficient than the best-known classical algorithm for a specific problem. The competitiveness of the latter introduces a moving target for quantum algorithm developers, where the best classical algorithms keep evolving, demanding continuous advancements in quantum algorithms to maintain claims of utility. Lastly, mapping a real-world (often classical) use-case into a quantum representation is a non-trivial task.

A quantum program is a series of instructions that can be executed by a quantum device in a specific sequence to perform a specific task. These instructions are typically written in a high-level programming language like Qiskit~\cite{alexander2020qiskit}, designed to be readable and writable by humans. 

A classical computer architecture consists of several key components that work together to perform various computational tasks. At its core, a classical computer contains a central processing unit (CPU). The CPU executes instructions, and coordinates data movement. It is supported by memory units, including random-access memory (RAM) and cache memory, where data and instructions are temporarily stored for faster access. The architecture also includes input and output (I/O) devices, such as keyboards, mice, monitors, and storage devices like hard drives or solid-state drives. These allow users to interact with the computer and store data for future use. The CPU communicates with other components via buses, which are pathways that transfer data and control signals between different parts of the computer. The system clock generates regular pulses that synchronize the activities of various components, ensuring smooth coordination of operations. Moreover, a classical computer architecture often involves a graphics processing unit (GPU) dedicated to handling graphics-intensive tasks, such as rendering images and videos.

The development and elaboration of a quantum computer architecture~\cite{ball2021quantum, 8123664, 8638598} are still in their early stages, mainly because we have not achieved fault-tolerant quantum computing yet. The full stack architecture will need focus beyond physical layer and must include error correction, feedback stabilization, hardware-aware compilation, logical level compilation, circuit optimization, application layer, and user interface. While progress is being made towards this vision, it remains a distant goal. 

\subsection{The problem of noise}The behavior of an ideal quantum computer can be modeled as follows. An $n$-qubit register spans a complex Hilbert Space denoted by $(\mathbb{C}^2)^{\otimes n}$. The initial state of the register can be represented as a tensor:
\begin{equation}
\ket{\psi} = \ket{0}^{\otimes n}.
\end{equation}
A logical operation on the register state (also called a quantum gate) can be represented by a linear, unitary operation \textit{U}:
\begin{equation}
\ket{\psi} \rightarrow U\ket{\psi}.
\end{equation}
A measurement reads out a $n$-bit string \textit{v} in the computational basis:
\begin{equation}
Pr(v) = |\braket{v|\psi}|^2,
\end{equation}
where $v \in \{0, 1\}^{\otimes n}$

Unlike modern classical computers with extremely low failure rates (e.g., $10^{-17}$ or less), \superc~quantum computers exhibit higher gate-level failure rates (nearly $0.01$)~\cite{van2023probabilistic, bravyi2021mitigating}. Thus, it is imperative to investigate the causes of noise and mitigate and correct them so that quantum computers can provide correct results. 

Noise, in the context of this dissertation, refers to deviations from the ideal description of a quantum computer. Practical efforts to build quantum computers introduce noise, which affects technologies like superconducting qubits, trapped ions, and silicon quantum dots. Our focus is primarily on superconducting quantum computers~\cite{krantz2019quantum, roth2021introduction}. The underlying noisy processes that impact such a computer can be classified into three groups: noise affecting the the quantum register (such as leakage~\cite{vittal2023eraser}, undesired coupling, decay processes, non-uniformity, and cross-talk~\cite{ben2011approximate}), noise affecting the quantum operations (such as pulse distortion, attenuation, drift, and mis-calibration), and noise in the thermodynamic isolation system (due to issues with dilution refrigerators, vacuum chambers, shields, and vibration suppression mechanisms~\cite{martinis2015qubit}).

Quantum computers today are referred to as existing in the NISQ era~\cite{preskill2019quantum}, which stands for noisy intermediate-scale quantum. The noise threshold for NISQ is defined by a single-qubit error rate being worse than $10^{-4}$. The intermediate-scale label is often associated with having fewer than a hundred thousand qubits. Computing done with NISQ devices is called NISQ computing. The bare minimum requirements for NISQ computing includes: quantum registers for storing data, quantum gates to execute logical operations on the registers, and a measurement interface for extracting the computation outcomes. The field has witnessed rapid advancements, with NISQ devices now operating as systems with hundreds of interacting qubits. Remarkably, the field is already witnessing a transition towards a phase where NISQ devices are performing scientific computations at a scale that rivals classical supercomputers in terms of computational power~\cite{kim2023evidence}. These experiments apply error mitigation techniques to the outcomes of the noisy computations performed on the NISQ devices. 

Note that error mitigation and error correction are distinct strategies to tackle challenges arising from noise: mitigation employs statistical techniques focused on minimizing noise effects rather than eliminating errors entirely, such as zero noise extrapolation and probabilistic error cancellation, while error correction seeks to actively detect and rectify errors during computations with an objective of fault tolerance, but requires the device noise to be below a threshold, which has not been achieved yet. One example of the unique challenges in the area of quantum error-correction is the no-cloning theorem states that it's impossible to create an exact duplicate of an unknown quantum state. This makes it challenging to incorporate redundancy into quantum computing systems to protect against information corruption.

\section{Research focus}
The fact that noise exhibits non-stationarity, underscores the core motivation of this dissertation. A quantum noise channel is often used to model how quantum information becomes distorted during its passage through a physical system. It describes how interactions with the environment can modify the quantum state of a system. Examples include depolarizing channel, Pauli noise channel, amplitude damping channel, phase damping channel, and the SPAM noise channel. 

\begin{figure}[H]
  \centering
  \includegraphics[width=\figurewidth]{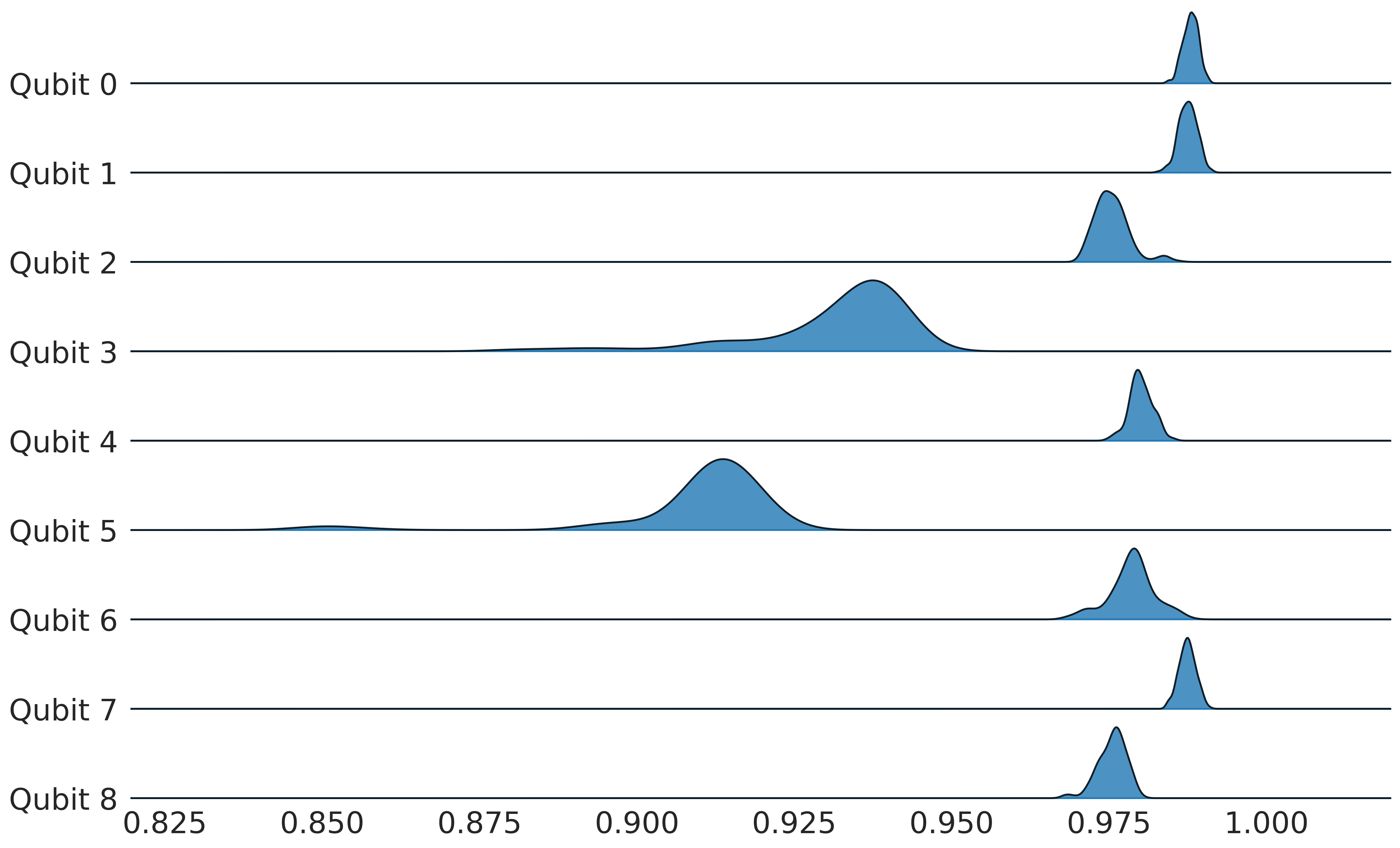}%
  \caption{\SPAM (SPAM) fidelity distributions on \toronto for qubits $0-8$ as measured on 8 April 8 2021, between 8:00-10:00pm (UTC-05:00).}
  \label{fig:f0f1_toronto_qubit_0_onwards_spruce_2021}
\end{figure}

\begin{figure}[H]
\centering
\includegraphics[width=\figurewidth]{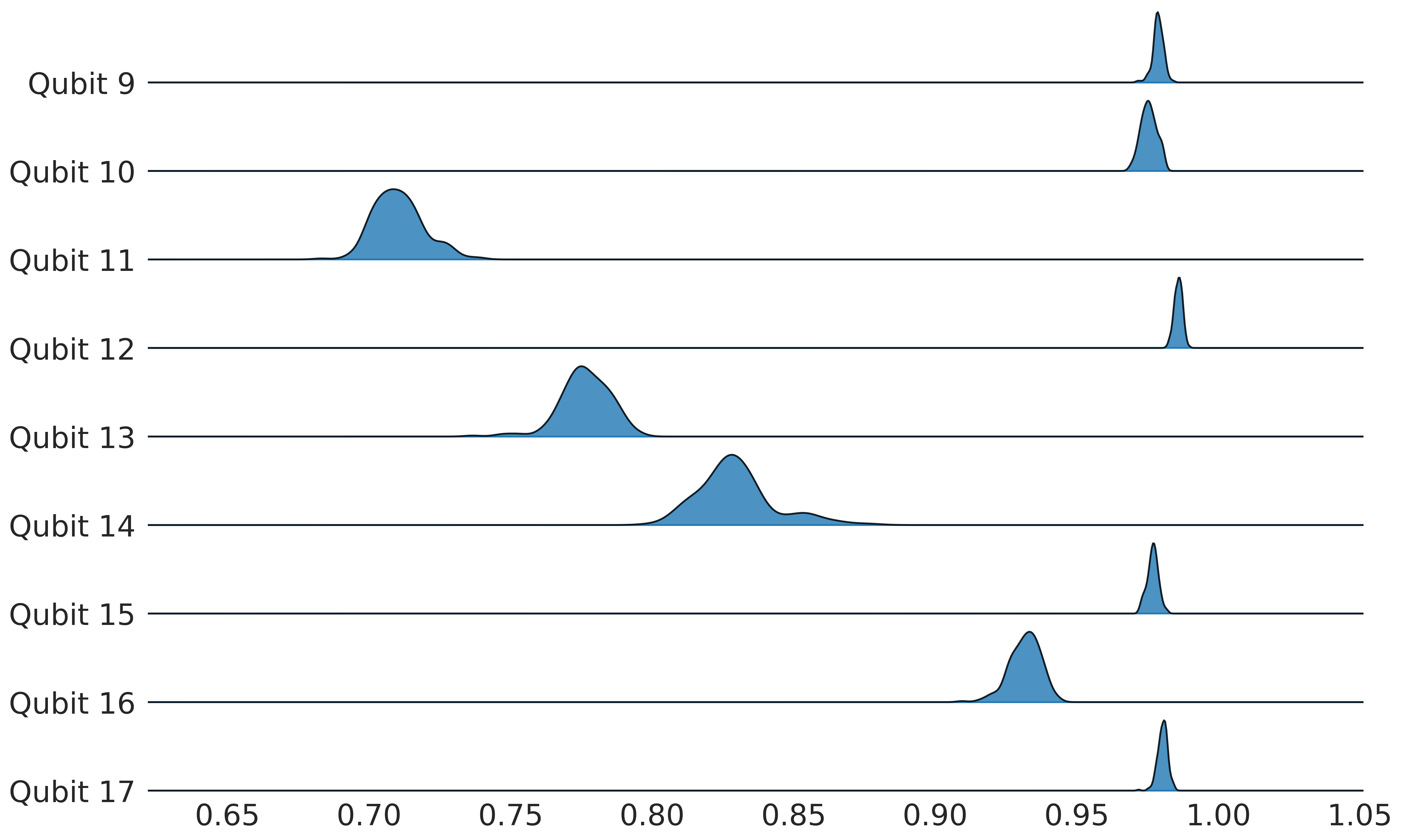}%
\caption{
\SPAM (SPAM) fidelity distributions on \toronto for qubits $9-17$ as measured on 8 April 8 2021, between 8:00-10:00pm (UTC-05:00).}
\label{fig:f0f1_toronto_qubit_9_onwards_spruce_2021}
\end{figure}

\begin{figure}[H]
\centering
\includegraphics[width=\figurewidth]{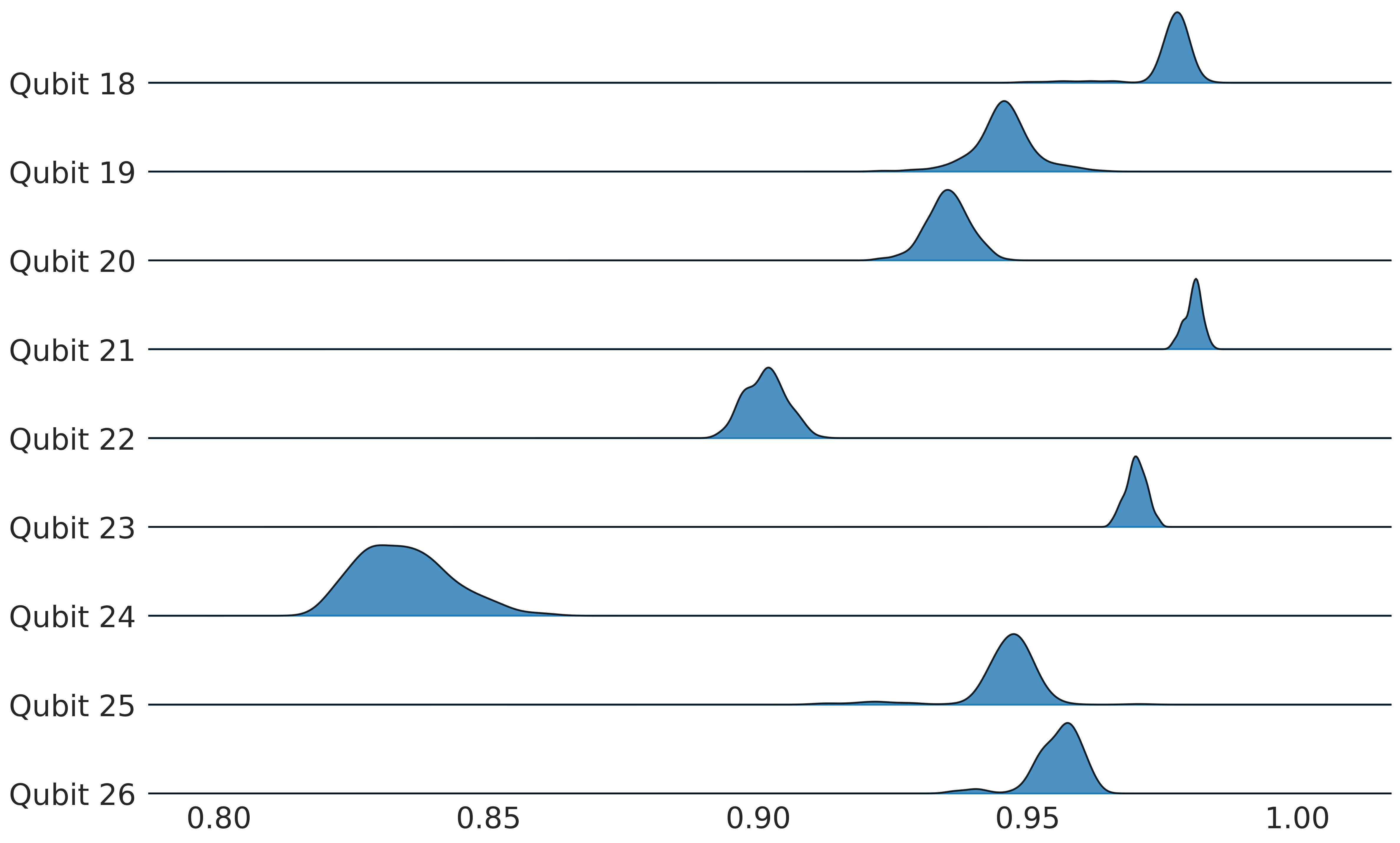}%
\caption{
\SPAM (SPAM) fidelity distributions on \toronto for qubits $18-26$ as measured on 8 April 8 2021, between 8:00-10:00pm (UTC-05:00).}
\label{fig:f0f1_toronto_qubit_18_onwards_spruce_2021}
\end{figure}

The term SPAM denotes state preparation and measurement. The statistics of the SPAM error channel are commonly quantified using the SPAM fidelity, a metric that evaluates a device's ability to prepare and measure a qubit. Specifically, SPAM fidelity quantifies the likelihood that the device readies the qubit (or a set of qubits) in the desired state and subsequently measures it in that same state. Fig.~\ref{fig:f0f1_toronto_qubit_0_onwards_spruce_2021}~-~\ref{fig:f0f1_toronto_qubit_18_onwards_spruce_2021} depicts the time-varying noise densities of SPAM fidelity.

\begin{figure}
\centering
\includegraphics[width=0.5\columnwidth]{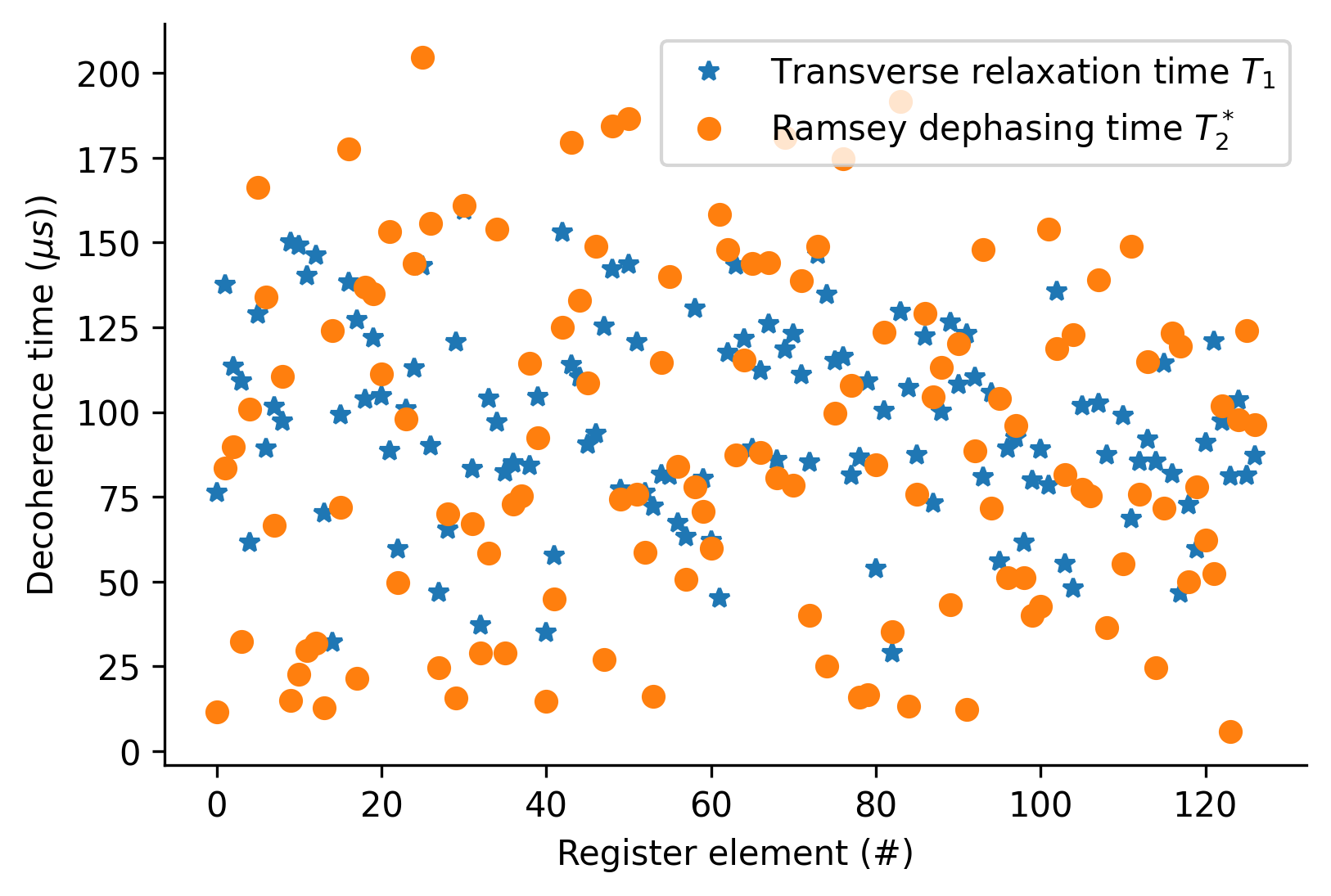}
\caption{Evidence of spatial non-stationarity in the mean values for quibt decoherence times $T_1$ and $T_2$ for the 127-qubits of the \washington device, generated on 14 Jan 2023 10:20 PM UTC. 
}
\label{fig:T1T2_spatial}
\end{figure}

Concurrent experimental studies~\cite{bejanin2021interacting, burnett2019decoherence, carroll2021dynamics, klimov2018fluctuations, mcrae2021reproducible} support our concern that the assumption of fixed and invariant parameters for quantum processors is flawed. 
In fact, these studies show that the noise parameters can exhibit time variations of up to 50\% of their mean value within an hour. 
Spatially varying noise in quantum devices have also been extensively studied, including the role of circuit geometry~\cite{klesse2005quantum, gupta2020integration} and cross-talk between neighbouring qubits~\cite{parrado2021crosstalk, fang2022crosstalk}. 
Unlike temporal variations in noise, the effects of which are magnified by the complexity of a quantum circuit, the spatial variations, such as seen in Fig.~\ref{fig:T1T2_spatial}, are dependent on the geometry and scale of the circuit implementation. 

\begin{figure}[htbp]
\centering
\includegraphics[width=\figurewidth]{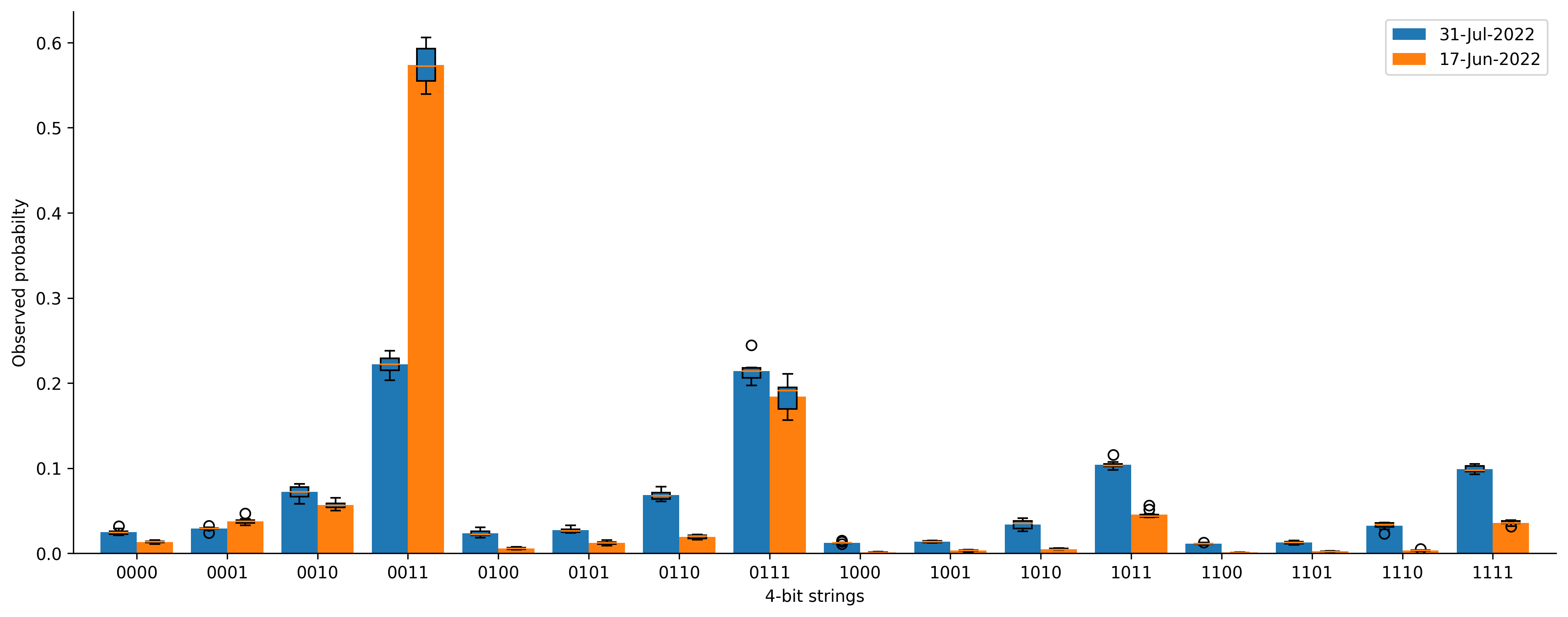}
\caption{
Impact of the non-stationary noise, leading to irreproducible outcomes. The results show histograms (after \SPAM (SPAM) noise mitigation) upon executing the Bernstein-Vazirani circuit on \mumbai. 
}
\label{fig:BV_2022_accuracy_reproducibility}
\end{figure}

To understand the impact of such non-stationary noise on program outcomes, consider Fig.~\ref{fig:BV_2022_accuracy_reproducibility} which depicts the time-varying histogram obtained from IBM's \mumbai device for the \BV circuit. It is evident that the associated error bars on a particular day do not provide insights into results from a different date, highlighting the problem of reproducibility of results in quantum computing today. 

The causes of non-stationary noise are not fully understood but are believed to stem from  TLS defects in transmon registers which might be arising from deviations in crystalline order. The current consensus attributes these defects to the presence of certain oxides on the superconductors' surface~\cite{muller2015interacting, klimov2018fluctuations}. 
Thus, static quantum channel models do not accurately capture the dynamics in realistic quantum computations, particularly in superconducting qubits. Cosmic rays~\cite{xu2022distributed, mcewen2022resolving} also contribute by ionizing the substrate upon impact, leading to the emission of high-energy phonons, which in turn triggers a burst of quasi-particles. These quasi-particles disrupt qubit coherence across the device. It has been shown that quantum computers can experience catastrophic errors in multi-qubit registers approximately every 10 seconds due to cosmic rays originating from outer space~\cite{mcewen2022resolving}. Studies that address non-stationary noise in \superc~quantum computers include 
investigations on output reproducibility~\cite{proctor2020detecting},  
noise modeling~\cite{etxezarreta2021time}, 
tracking the \ns~profile of quantum noise~\cite{danageozian2022noisy}, and 
quantum error mitigation using continuous control~\cite{majumder2020real}. 
Non-stationary quantum channel models have been proposed~\cite{etxezarreta2021time, martinez2020approximating, demarti2023decoding, demarti2022performance} that use stochastic processes. 
Our dissertation focuses on understanding the effect of non-stationary noise on program outcomes, as well as devising strategies to address it. 
Specifically, we model the noise channel as a random variable and implement adaptive methods to manage it.
%

Before we can begin, we need to fix the precise language for performance assessment. 
This task is not trivial due to the complexity inherent in quantum technology, which both distinguishes it from classical computing and hinders its rigorous checking~\cite{vapnik1999nature, carrasco2021theoretical}. Challenges include the inherent randomness in quantum measurements, error accumulation without clear source attribution, the curse of dimensionality, and the inability to step-through program execution in quantum circuits. 

In fact, the performance evaluation~\cite{kliesch2021theory} of noisy quantum computations is a vast topic that is crucial for several additional reasons, apart from our motivation of studying the impact of non-stationary noise. Firstly, as quantum computing is still in its early stages~\cite{roadmap, acin2018quantum, hughes2004quantum}, understanding the sources of errors and noise is vital. Through rigorous evaluation, researchers can model~\cite{khatri2020information}, identify, model and quantify sources of noise, such as decoherence, gate errors, and readout errors. Secondly, this understanding is essential for developing error mitigation techniques~\cite{bharti2022noisy}. Thirdly, reproducibility of results is critical, and rigorous performance evaluation ensures experiments can be replicated by other researchers, contributing to the validation and verification of quantum algorithms. An additional challenge is the diverse range of terms encountered in quantum computing today which can blur distinctions between them, making it challenging to appreciate their nuanced differences. Examples include verification~\cite{harper2020efficient, gheorghiu2019verification, PhysRevA.101.042315} (ensuring correct transpilation), validation~\cite{10.1371/journal.pone.0206704, 10.1109/qcs54837.2021.00013} (validating correctness of output or the quantum nature of a device), benchmarking~\cite{mccaskey2019quantum, BlumeKohout2020volumetricframework, 10.1038/s41534-022-00628-x} (assigning a performance measure to a processor), accreditation~\cite{ferracin2021experimental}, and certification~\cite{kliesch2021theory}. Thus out first task is to precisely define computational accuracy, result reproducibility, device reliability, and observable stability in the presence of non-stationary noise.

Our next objective is to experimentally assess hardware reliability, with a particular emphasis on analyzing spatial and temporal variations in noise statistics. IBM~\cite{roadmap} has introduced a range of processors in recent years, each with an expanding register size. These include the Canary processors with 2-16 qubits, Falcon processors with 27 qubits, Egret processors with 33 qubits, Hummingbird processors with 65 qubits, Eagle processors with 127 qubits, and Osprey processors with 433 qubits. Quantifying the spatial and temporal reliability of these quantum computers is crucial to understand system-wide performance changes over time. This evaluation should encompass both component-level metrics, such as individual gates and qubits, and composite-level metrics, such as circuits, to assess the degree of non-stationarity in noise and its implications on program outcomes. Holistically measuring reliability at the circuit level is essential, as examining thousands or millions of qubits and gates may not provide conclusive insights at the application level~\cite{divincenzo2000physical}.

Our third objective in this dissertation is to establish stability bounds for error-mitigated outcomes affected. We aim to determine the upper and lower bounds for our performance evaluation metrics. Our inquiries include determining the minimum sample size necessary to ensure histogram reproducibility with a confidence level of $ 1 - \delta $, bounding outcome stability based on the variation in time-varying noise densities, and establishing reliability bounds to achieve stable outcomes.

Numerous studies on noise modeling in quantum computing systems have highlighted the challenges associated with noise estimation~\cite{blume2020modeling, blume2010optimal}. 
A natural question arises regarding how can we effectively counteract the detrimental impacts of non-stationary noise using adaptive algorithms? 
In the final chapter of our dissertation, we consolidate the various elements of our investigation in the context of adaptive probabilistic error cancellation (PEC)~\cite{temme2017error}. 
Our focus is on enhancing the stability of PEC outcomes, using a Bayesian~\cite{lukens2020practical, zheng2020bayesian, gordon1993novel, kotecha2003gaussian} updating of the quasi-probability distributions, in the presence of non-stationary noise.

This research focuses on a limited scope. 
Firstly, the experiments exclusively uses the superconducting platforms provided by IBM. 
Other platforms such as trapped ion, neutral atom, photonic, or quantum dot are not considered. 
Secondly, our performance evaluation framework mainly concerns with the output measured in computational basis and how it is impacted by device noise. 
Thirdly, not all the superconducting devices provided by IBM have been characterized; only a sub-set of the devices, mainly \kolkata, \mumbai, \washington, \toronto, and \yorktown, are used in this study. 
Lastly, we do not focus on the problem of optimal selection of a statistical model for a given noisy device. Instead, a generic error channel formalism is employed whenever possible. For verifying the theory, quantum noise channel models like the Pauli noise~\cite{gottesman1998heisenberg} channel is chosen, with a specific focus on parameter estimation. The research does not explore the question of identifying the best noise model for a given device. 

For our research, we have made use of the daily characterization data stored on IBM's servers as-is. For data at time-scales of minutes and below, we collected the data ourselves and offer all the associated collection and preparation software. For the latter case, the data collection frequency was limited by network time lags and constraints in the qiskit software, such as the maximum number of circuits and shots allowed. These limitations have been gradually improving over time
\section{Organization and notation}
The document is structured as follows. Chapter~\ref{ch:decoherence_characterization} provides background on noise in quantum computing, quantum channel modeling, and experimental characterization of quantum decoherence, emphasizing the non-stationary statistics of noise. Chapter~\ref{ch:statistical_taxonomy} establishes a systematic framework for assessing noisy quantum computer performance. Chapter~\ref{ch:exp_characterization} focuses on the testing of reliability. The evaluation encompasses both component-level metrics (such as individual gates and qubits), and composite-level metrics (such as circuits). Chapter~\ref{ch:analytical_bounds} seeks to determine the bounds on the assessment framework developed in Chapter~\ref{ch:statistical_taxonomy}, using available noise characterization data. Specifically, it discusses how to bound outcome stability in terms of the distance between time-varying noise densities~\cite{spehner2017geometric}. Chapter~\ref{ch:adaptive_algorithms} explores methods to improve accuracy in the presence of non-stationary noise. Chapter~\ref{ch:adaptPEC} brings together the various concepts discussed till date in the context of adaptive probabilistic error cancellation. Chapter~\ref{sec:conclusion} provides concluding remarks.

Notations in this dissertation vary in meaning depending on font, although it should be clear from the context (see Table~\ref{tab:Notations} ). This was needed because the work draws upon concepts from physics, information theory, computer science, and statistics, each of which has established conventions. $\ket{\psi}$ represents a pure quantum state, and $\rho$ represents density matrices. The symbol $\otimes$ signifies a tensor operator, and Tr$(\cdot)$ is an abbreviation for the trace operator. The system Hamiltonian is denoted as $\Omega$, with its eigenvalues represented by $\lambda$ (however $\omega$ signifies angular frequency). Quantum observables are typically denoted by $\hat{O}$, and the uppercase letter $U$ typically stands for a unitary matrix. A single-qubit rotation by an angle $\theta$ on the Bloch sphere is often denoted as $R(\theta)$. However, the uppercase Greek letter $\Theta(\cdot)$ denotes the copula~\cite{nelsen2007introduction} function from statistics. $\mathcal{E}_{\xrm}(\cdot)$ represents a quantum noise channel, while $\Lambda$ symbolizes the SPAM noise channel, a classical channel operating on probabilities. The canonical Pauli matrices are denoted as $\mathds{X}, \mathds{Y}, \mathds{Z}, \mathds{I}$. Note that we do not use the small Greek letter $\sigma$ for Pauli matrices, reserving it for standard deviation instead. The Pearson correlation matrix is denoted by the capital Greek letter $\Sigma$. Note that the small z signifies a standard normal variable. The identity matrix is represented as $\mathds{I}$, while $\mathcal{I}(X, Y)$ in calibrated font signifies the mutual information between random variables X and Y. The letter $\xrm$ typically signifies noise parameter(s), whereas an uppercase X corresponds to a specific realization of $\xrm$. If $\xrm$ is not deterministic, then $f(\xrm)$ denotes the probability distribution of $\xrm$. This distribution can exhibit temporal fluctuations, denoted as $f(\xrm; t)$, with its cumulative distribution function indicated by $F(\xrm; t)$. Additionally, the curly capital $\mathcal{F}$ stands for Fisher Information, while $f_s$ with the subscript $s$ represents the data sampling frequency. The symbol $\Pi_r$ with a subscript is reserved for the projector operator onto the eigenstate $\ket{\lambda_r}$ (however, when presented without a subscript, $\Pi$ signifies the normal product operator). The measurements are conducted in the computational basis (or $\mathds{Z}$ basis), resulting in qubits yielding classical bits. We employ the notation $b_i(t)$ to represent the observed classical bit value on qubit $i$ at time $t$. The state of an $n$-qubit quantum register is denoted by $\ket{v} = \ket{v_{n-1} \cdots v_0}$, with $n$ generally denoting the quantum register size. Upon measurement, this state yields an $n$-bit string, with each $v_i \in {0, 1}$. We utilize the symbol $W$ to denote the dataset consisting of collected bit-strings from repeated circuit executions. The total number of samples collected is typically denoted as $L$, where $l$ denotes the l-th circuit execution (however, note that the curly $\mathcal{L}$ represents the likelihood function). The Hellinger distance between probability distributions is denoted as $H$, while the curly capital $\mathcal{H}$ is exclusively reserved for entropy. The symbol $\mathds{H}$ represents the Hadamard gate. A quantum circuit is represented by the capital $C$, and while, the small c usually signifies the control qubit in a CNOT gate. Additionally, $c$ serves as a constant in certain information theoretic results. The symbols $\alpha$ and $\beta$ may assume different meanings depending on the context, referring either to quantum amplitudes for $\ket{0}$ and $\ket{1}$ or the parameters of the beta distribution, both of which find common usage in this dissertation. The small greek $\gamma$ is used as a proxy parameter encapsulating various device noise parameters, while the capital $\Gamma$ is reserved for the gamma function. The symbol $\eta$ typically represents a PEC linear combination coefficient. Capital $D$ denotes a noisy quantum device, while the small letter $d$ typically dimensionality. In the context of the Bernstein-Vazirani problem, $r$ is used to denote the secret $n$-bit string, with the latter problem being extensively employed as an illustrative quantum circuit. The small letter $s$ is used to denote the stability metric. Absolute time is typically represented by the small letter $t$, while time intervals are denoted as $\tau$. For instance, $\tau_c$ signifies the circuit execution time, and $\tau_N$ denotes network delay. Time duration is represented as $\delta t$. However, the capital letter $T$ is primarily reserved for parameters related to decoherence characterization, such as $T_1$ and $T_2$ (representing qubit relaxation and dephasing time, respectively). In some instances, $T$ (without a subscript) is used to denote the target qubit in a CNOT gate, which is generally clear from the context. Finally, the non-standard abbreviations that have been used in this dissertation are: NSN (non-stationary noise), PEC (probabilistic error cancellation), BC (Bhattacharyya coefficient), BV (Bernstein-Vazirani), SPAM (State preparation and measurement) and WSS (Wide-sense stationary).

\clearpage
\vspace{0.5in}
\clearpage\vspace{0.5in}
\vspace{0.5in}
\vspace{0.5in}

\removefigs
\chapter{Noise in quantum computing}\label{ch:decoherence_characterization}
Practical efforts to realize a quantum computer (e.g. transmons, trapped ions, silicon quantum dots \cite{burnett2019decoherence, wan2019quantum, PhysRevApplied.10.044017, humble2019quantum}) introduce various physical processes, referred to as noise, which deviate from the ideal description of a quantum computer. 
Unlike modern classical computers, which boast device components with extremely low failure rates (e.g., $10^{-17}$ or less), the current state-of-the-art quantum computers exhibit higher gate-level failure rates (e.g., $10^{-2}$). In this dissertation, we mainly use experimental data from transmon\cite{roth2021introduction, martinis2015qubit} based realizations of a quantum computer. Transmon qubits are a variant of superconducting charge qubits designed to reduce sensitivity to charge noise\cite{krantz2019quantum}. 
\section{Physical sources of noise}
The various noise processes \cite{bejanin2021interacting, burnett2019decoherence, carroll2021dynamics} can be classified into three groups:
\subsection{Quantum register}
One of the pathways for noise is the implementation of the quantum register, which encounter phenomena such as:
(i) Leakage i.e. unintended energy states outside the computational subspace, 
(ii) Undesired coupling to the external environment (such as spurious charge, magnetic fields, stray photons, lattice vibrations (phonons), nuclear spins) leading to loss of coherence, 
(iii) Spontaneous decay processes that transition a qubit from an excited state to a lower energy state, 
(iv) Non-uniformity in qubit's coupling strength to the control field, results in errors in quantum gates, 
(v) Inter-qubit cross-talk arising from shared control lines or capacitive coupling between neighboring qubits. In trapped ion systems, cross-talk could arises from motional coupling between ions, affecting the states of neighboring ions.
\subsection{Control system}
The quantum register undergoes four fundamental control operations: initialization or reset, measurement, single-qubit rotation gates, and 2-qubit entangling gates. These operations require the application of precisely calibrated control pulses on the qubits. Imperfections in the control system used for logic implementation can arise from several sources. Firstly, pulse distortion occurs when the desired shape and duration of pulses encoding quantum information are altered due to the finite time resolution and frequency response limitations, as well as pulse timing errors. Secondly, control pulses may experience attenuation caused by electromagnetic interference and material imperfections in the quantum system's vicinity. Thirdly, qubits can drift either physically (in the case of trapped ions) or in parameter space. Fourthly, the noise may be an effect mis-calibration. 
\subsection{Thermodynamic isolation system}
Transmon qubits, a specific kind of superconducting qubit, require cooling to approximately 10 milli-kelvin in order to mitigate the presence of thermal noise. 
The thermodynamic isolation system \cite{jin2015thermal} helps achieve this using a system of dilution refrigerators, vacuum chambers, electromagnetic shields, and vibration suppression mechanisms. 
The dilution refrigerators employ a multi-stage cooling process that gradually reaches colder temperatures, using substances like liquid helium to progressively lower the temperature. 
The vacuum chambers effectively eliminate gas molecules and particles that could potentially couple with the qubits.
The electromagnetic shields are responsible for blocking external radiation and fields from disturbing the quantum state. Inadequate electromagnetic shielding could allow disruptive external radiation to interfere with the qubits. 
The vibration suppression systems minimize mechanical vibrations and movements that could potentially jeopardize the quantum states. 
Noise from imperfect thermodynamic control systems can be non-Markovian in nature, which are difficult to rectify using quantum error correction tools. 

\section{Cause of non-stationarity}
Non-stationary noise refers to noise in a quantum system that exhibits time-varying statistical properties. 
The temporal fluctuations of the mean and variance of the energy relaxation times ($T_1$), dephasing times ($T_2$), and qubit frequencies, are well-studied topics\cite{burnett2019decoherence, carroll2021dynamics, klimov2018fluctuations, mcrae2021reproducible} that suggest suggest that noise in NISQ\cite{preskill2019quantum} devices can fluctuate unpredictably. 
For example, $T_1$ times have been found to fluctuate by approximately 50 percent within an hour \cite{burnett2019decoherence}. 
Similarly, many advances have also been made for spatially varying noise in quantum devices \cite{klesse2005quantum}, its effect on the choice of circuit geometry \cite{gupta2020integration}, as well as the interplay with cross-talk between qubits for single and two-qubit gates \cite{parrado2021crosstalk, fang2022crosstalk}.

The causes and mechanisms behind quantum noise non-stationarity are poorly understood. 
In transmon registers, potential sources of fluctuations include TLS (two-level system) defects, quasi-particles, parasitic microwave modes, phonons, nuclear spins, paramagnetic impurities, spurious resonances, critical current noise, background charges, gate voltage fluctuations, and the electromagnetic environment \cite{hughes2004quantum}. 
Among these, TLS defects have been identified as the primary cause of decoherence \cite{bejanin2021interacting, burnett2019decoherence, carroll2021dynamics}. 
These defects arise from deviations from crystalline order in the naturally occurring oxide layers of transmons, resulting in trapped charges, dangling bonds, tunneling atoms, or collective motion of molecules.

The findings not only highlight the necessity for frequent re-calibration in qubit setups but also question the reproducibility of device characterizations, and their use in error mitigation. 
Consequently, modeling time-varying quantum noise has become an active area of interest \cite{etxezarreta2021time}, such as through the inclusion of $T_1$ and $T_2$ fluctuations in quantum channel models to investigate the concept of time-varying quantum channels (TVQC).
\section{Decoherence characterization}\label{sec:decoherence_data}
Decoherence refers to loss of unitarity in state evolution. 
The traditional definition of decoherence, which describes the decay of off-diagonal terms in the density matrix, is now referred to as dephasing and considered one kind of decoherence \cite{hughes2004quantum}. Decoherence studies typically focus on three metrics: transverse relaxation time $(T_1)$, longitudinal relaxation time $(T_2)$, and dephasing time $(T_\phi)$. 

$T_1$, also known as the transverse relaxation time or relaxation time, measures the attenuation of amplitude in a quantum system. It represents the probability that an excited state $\ket{1}$ will decay to the ground state $\ket{0}$ after time $t$, and is modeled by the function:
\begin{equation}
\textrm{Pr}(\ket{1} \rightarrow \ket{0}) = 1-\exp(-t/T_1) \; .
\end{equation}
The decay-time probability density $f_T(t)$ can be described by the exponential function:
\begin{equation}
f_T(t) = T_1 \exp^{-t/T_1} \; ,
\end{equation}
whose mean is the density parameter $\mathds{E}(T) = T_1$. 

$T_2$ is a measure of how long it takes for a qubit in the superposition state to decay. Specifically, it measures the decay of the off-diagonal elements of the density matrix and is modeled by an exponential decay function. Therefore, it captures the loss of synchronization between the basis states of an arbitrary quantum ensemble. There are two types of $T_2$ time often quoted in literature \cite{wrightgroup}:
\begin{itemize}
\item Ramsey dephasing time $T_2^*$: measures the time-scale at which a quantum register experiences de-phasing effects when left to evolve freely
\item Hahn-echo dephasing time $T_2^{\textrm{echo}}$: uses intermediate $\pi$ pulses for re-focusing to increase relaxation time.
\end{itemize}
When simulating noisy circuits, the appropriate $T_2$ value to use depends on whether the physical implementation of the circuit uses Hahn-echo for noise suppression or not. We will specifically focus on the Hahn-echo with one echo $T_2$ time.

Finally, the pure dephasing time $(T_\phi)$ is an upper bound on the decoherence time for a qubit, since thermal fluctuations in the environment inevitably cause a loss of phase coherence. In practice, the dominant relaxation time is usually $T_2$ (or sometimes $T_1$), rather than $T_\phi$ \cite{wrightgroup}. 

The three decoherence benchmarks are related by:
\begin{equation}
\frac{1}{T_2} = \frac{1}{2T_1} + \frac{1}{T_\phi} \; .
\label{eq:Bloch-Redfield}
\end{equation} 

\subsection{Experimental characterization}
We analyzed decoherence (i.e. $T_1$ and $T_2$ times) in the transmon processor \kolkata. We had 24-hour access on Tuesday, September 12, to September 13 (from 12 noon to 12 noon) through OLCF. We chose this time-frame as it is typical for user program queues for execution on the IBM platform. We measured fluctuations in $T_1$ and $T_2$ times for all 27 qubits on the device during this 24-hour period. We validated our software through numerical simulations (detailed in the end of this section).

\begin{figure}[htbp]
\centering
\includegraphics[width=\figurewidth]{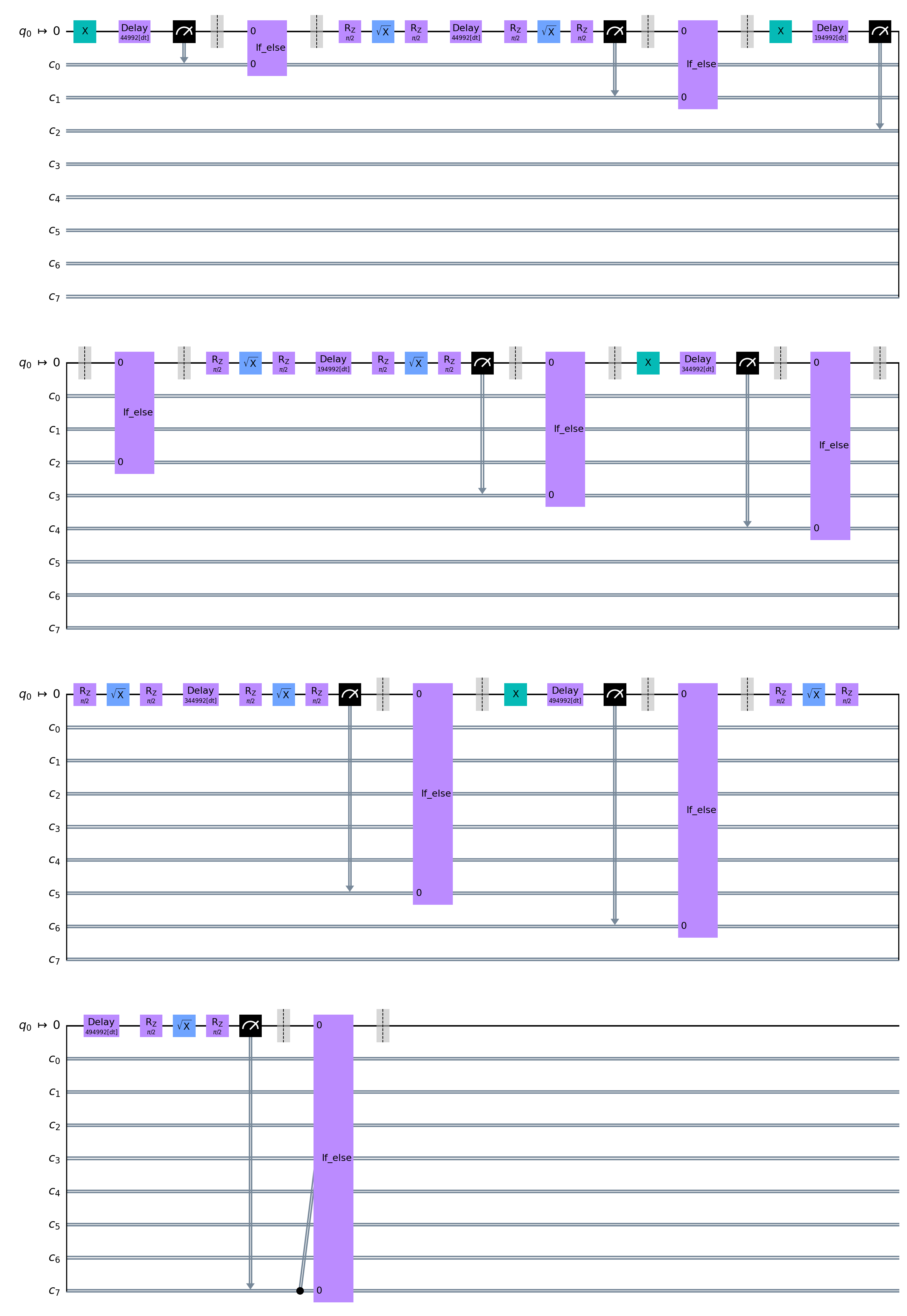}
\caption{Interleaved sub-circuit post transpilation.}
\label{fig:qc_t1t2_transpiled}
\end{figure}

The complete quantum circuit used to gather the decoherence parameters $T_1$ and $T_2$ is too large to display in its entirety. However, in Fig.~\ref{fig:qc_t1t2_transpiled}, we provide a concise representation of a section of the circuit, specifically for qubit $\ket{0}$. It's important to clarify that this circuit structure is replicated for all $27$ qubits in \kolkata, and the entire circuit is executed multiple times to obtain statistical data.

In the sub-circuit presented in Fig.~\ref{fig:qc_t1t2_transpiled}, we illustrate only one mid-circuit reset for the sake of clarity. In reality, the full circuit employs three conditional resets to ensure a high probability of mid-circuit reset success. This choice of three resets aligns with qiskit guidelines to optimize the likelihood of successful resets.

It's worth noting that mid-circuit measurement allows for the simultaneous collection of the decoherence parameters $T_1$ and $T_2$ with a time interval of just a few hundred microseconds. This simultaneous data collection facilitates the empirical calculation of temporal correlations between these parameters.

The basic $T_1$ measurement circuit begins by initializing a qubit to the $\ket{0}$ state and then applying an X gate to transition it to the $\ket{1}$ state. Subsequently, a phase gate is introduced, during which the qubit is affected by noise. Following this, a measurement is performed in the Z basis. In the absence of noise, the measurement would yield the $\ket{1}$ state with complete certainty. However, in the presence of noise, the probability of obtaining the $\ket{1}$ state is less than 100\%, and this probability depends solely on $T_1$. Therefore, by analyzing the observed probability of measuring $\ket{1}$, we can deduce an estimate for the $T_1$ time. 

The basic $T_2$ circuit starts by setting a qubit to the $\ket{0}$ state. It then uses the Hadamard gate to create an equal superposition of $\ket{0}$ and $\ket{1}$. After a brief phase-shift delay, another H gate is applied, followed by a measurement in the Z basis. In the absence of noise, the circuit guarantees a 100 percent chance of retrieving the $\ket{0}$ state. However, if there's dephasing noise, the probability decreases, and this can be used to estimate the $T_2$ time.

It's important to note that these calculations also consider SPAM noise, which will be discussed in detail later.

To calculate $T_1$ and $T_2$, data is fitted to an exponential decay plot using four different evolution times: 10 $\mu$s, 50 $\mu$s, 100 $\mu$s, and 160 $\mu$s. For each qubit, the process involves the following sequence:
\begin{enumerate}
\item basic $T_1$ circuit for a 10 $\mu$s, followed by measurement and reset
\item basic $T_2$ circuit for a 10 $\mu$s, followed by measurement and reset
\item basic $T_1$ circuit for a 50 $\mu$s, followed by measurement and reset
\item basic $T_2$ circuit for a 50 $\mu$s, followed by measurement and reset
\item basic $T_1$ circuit for a 100 $\mu$s, followed by measurement and reset
\item basic $T_2$ circuit for a 100 $\mu$s, followed by measurement and reset
\item basic $T_1$ circuit for a 160 $\mu$s, followed by measurement and reset
\item basic $T_2$ circuit for a 160 $\mu$s, followed by measurement
\end{enumerate}
The statistical estimation of $T_1$ and $T_2$ are impacted by SPAM noise. 
We model the SPAM noise using a binary asymmetric model with the parameters:
\begin{equation}
\begin{aligned}
u_{00} + u_{01} &= 1\\
u_{10} + u_{11} &= 1
\end{aligned}
\end{equation}
where, \( u_{00} \) denotes the probability of getting 0 given an input of 0, \( u_{01} \) denotes the probability of getting 1 given an input of 0, \( u_{10} \) denotes the probability of getting 0 given an input of 1, and \( u_{11} \) denotes the probability of getting 1 given an input of 1. 
We define \( p \) as the survival probability for the \( T_1 \) circuit, indicating the probability of an excited state enduring beyond time \( t \) in \( T_1 \) measurement. Similarly, \( q \) denotes the survival probability for the \( T_2 \) circuit, reflecting the probability of observing a ground state after time \( t \) in \( T_2 \) measurement. 
\begin{equation}
\begin{aligned}
\tilde{p}_1 &= \frac{\# 1^{\prime}s \text{ observed post-measurement of $T_1$ circuit}}{{\# \text{Circuit repetitions}}}\\
\tilde{q}_0 &= \frac{\# 0^{\prime}s \text{ observed post-measurement of $T_2$ circuit}}{{\# \text{Circuit repetitions}}}\\
\end{aligned}
\end{equation}
The survival probabilities in absence of SPAM error are given by:
\begin{equation}
\begin{aligned}
p_1 & = e^{-\tau / T_1} \\
q_0 & = \frac{1}{2}( 1 + e^{-\tau / T_2} ) \\
\end{aligned}
\end{equation}

While the formula for $p_1$ (for $T_1$) is straightforward, the derivation for $q_0$ (for Hahn-echo $T_2$ with one echo) is a little more involved. 
The steps are as follows. 
First we initialize the qubit in the ground state:
\begin{equation}
\rho_0 = \left(\begin{array}{ll}
1 & 0 \\
0 & 0
\end{array}\right)
\end{equation}
Then, we subject it to a Hadamard gate:
\begin{equation}
\rho_1 = 
H \rho_0 H^{\dagger}=\frac{1}{2}\left(\begin{array}{ll}
1 & 1 \\ 1 & 1
\end{array}\right)
\end{equation}
Then, we evolve the density matrix for time $\tau/2$ using the Hamiltonian $\Omega=\lambda \ket{1}\bra{1}$. The unitary operator for this phase gate $D$ is:
\begin{equation}
D = \exp (-i \Omega \frac{\tau}{2}) = \exp (-i \lambda \frac{\tau}{2})\ket{1}\bra{1}+\ket{0}\bra{0}\\
\equiv\left(\begin{array}{ll}
1 & 0 \\
0 & e^{-i \lambda \frac{\tau}{2}}
\end{array}\right)
\end{equation}
Thus, the state becomes:
\begin{equation}
\rho_2 = D \rho_1 D^{\dagger}=\frac{1}{2}\left(\begin{array}{cc}
1 & e^{-i \lambda \frac{\tau}{2}} \\ 
e^{i \lambda \frac{\tau}{2}} & 1
\end{array}\right)
\end{equation}
However, during this time-evolution, it is acted upon by an amplitude and phase damping (APD) channel. 
The state after taking APD noise into account is:
\begin{equation}
\rho_3 =\mathcal{E}_\text{APD}\left(\rho_2\right)=\left(\begin{array}{cc}
1-\frac{1}{2} e^{-\frac{\tau}{2} / T_1} & \frac{1}{2} e^{-i \lambda t} e^{-\frac{\tau}{2} / T_2} \\ 
\frac{1}{2} e^{i \lambda t} e^{-\frac{\tau}{2} / T_2} & \frac{1}{2} e^{-\frac{\tau}{2} / T_1}
\end{array}\right)
\end{equation}
After this comes a deliberate bit-flip through a $\mathds{X}$ gate (assumed noiseless):
\begin{equation}
\rho_4 =X \rho_3 X=\left(\begin{array}{cc}
\frac{1}{2} e^{-\frac{\frac{\tau}{2}}{T_1}} & \frac{1}{2} e^{i \lambda \frac{\tau}{2}} e^{-\frac{\tau}{2} / T_2} \\
\frac{1}{2} e^{-i \lambda \frac{\tau}{2}} e^{-\frac{\tau}{2} / T_2} & 1-\frac{1}{2} e^{-\frac{\tau}{2} / T_1}
\end{array}\right)
\end{equation}
This is followed up with another phase gate $D$ subject to APD noise:
\begin{equation}
\rho_5 =D \rho_4 D^{\dagger}=\left(\begin{array}{cc}
\frac{1}{2} e^{-\frac{\tau}{2} / T_1} & \frac{1}{2} e^{-\frac{\frac{\tau}{2}}{T_2}} \\ 
\frac{1}{2} e^{-\frac{\tau}{2} / T_2} & 1-\frac{1}{2} e^{-\frac{\frac{\tau}{2}}{T_1}}
\end{array}\right)
\end{equation}
\begin{equation}
\rho_6 = \mathcal{E}_\text{APD}(\rho_5) = \left(\begin{array}{cc}
1-e^{-\frac{\tau}{2} / T_1}\left(1-\frac{1}{2} e^{-\frac{\tau}{2} / T_1}\right) & \frac{1}{2} e^{-\frac{2\frac{\tau}{2}}{T_2}} \\ 
\frac{1}{2} e^{-2\frac{\tau}{2} / T_2} & e^{-\frac{\tau}{2} / T_1}\left(1-\frac{1}{2} e^{-\frac{\tau}{2} / T_1}\right)
\end{array}\right)
\end{equation}
Then another Hadamard is applied:
\begin{equation}
\rho_7 = \mathds{H} \rho \mathds{H}^\dagger
\end{equation}
Finally, we measure the probability of the qubit being in the ground state. The probability of observing $\ket{0}$ in the final measurement is:
\begin{equation}
\begin{aligned}
q_0 = \text{Pr}(0) &= 
\frac{1}{2}\left(1+e^{-\tau / T_2}\right)
\end{aligned}
\end{equation}

The survival probabilities in presence of SPAM error are given by:
\begin{equation}
\begin{aligned}
\tilde{p}_1 & =p_{1} u_{11} + p_{0} u_{01} = e^{-\tau / T_1}\left[1-u_{10}-u_{01}\right]+u_{01}\\
\tilde{q}_0 & =q_{0} u_{00} + q_{1} u_{10} = e^{-\tau / T_2} \frac{1-u_{10}-u_{01}}{2} + \frac{1+u_{10}-u_{01}}{2}\\
\label{eq:roem_agnostic_t1t2}
\end{aligned}
\end{equation}
\subsection{$T_1$ estimation}
For the $T_1$ circuit, the observed data $y_l$ measured in the Z-basis follows a Bernoulli distribution:
\begin{equation}
\begin{aligned}
y_l &=
\begin{cases}
    1, \;\;\;\; \text{ with probability } \tilde{p}_1\\
    0, \;\;\;\; \text{ with probability } 1-\tilde{p}_1\\
\end{cases}\\
\end{aligned}
\end{equation}
where $\tilde{p}_1 = e^{-\tau / T_1}\left[1-u_{10}-u_{01}\right]+u_{01}$. The Likelihood function is given by:
\begin{equation}
\begin{aligned}
\mathcal{L} =& \prod_{l=1}^L \text{Pr}\left(Y_l=y_l\right) = \prod_{l=1}^L \tilde{p}_1^{y_l} (1-\tilde{p}_1)^{1-y_l} \\\end{aligned}
\end{equation}

Setting:
\begin{equation}
\frac{\partial \log \mathcal{L}}{\partial \tilde{p}_1} =0,
\end{equation}
we get,
\begin{equation}
\begin{aligned}
\tilde{p}_1^* = \sum \frac{y_l}{L}
\end{aligned}
\end{equation}

Thus we have obtained the estimator (denoted by the * sign) for $\tilde{p}_1$. Note that
\begin{equation}
\mathds{E}( \tilde{p}_1^* ) = \sum \frac{\mathds{E}(y_l)}{L} = \sum \frac{\tilde{p}_1}{L} = \tilde{p}_1 
\end{equation}
and hence it is an unbiased estimator.

The variance of this estimate is given by:
\begin{equation}
\begin{aligned}
\text{Var}( \tilde{p}_1^* ) = \sigma^2(\tilde{p}_1^*) = \sum \frac{\text{Var}(y_l)}{L^2} = \frac{ \tilde{p}_1(1-\tilde{p}_1) }{L} \approx \frac{ \tilde{p}_1^*(1-\tilde{p}_1^*) }{L}
\end{aligned}
\end{equation}

Using the above, we can get K different equations, one for each phase gate with evolution time $\tau_k$ and whose post-measurement results are denoted by $\{ y_{l,k} \}$:
\begin{equation}
\begin{aligned}
e^{-\tau_1 / T_1}\left[1-u_{10}-u_{01}\right]+u_{01} =& \frac{\sum y_{l,1}}{L}\\
e^{-\tau_2 / T_1}\left[1-u_{10}-u_{01}\right]+u_{01} =& \frac{\sum y_{l,2}}{L}\\
&\cdots\\
e^{-\tau_K / T_1}\left[1-u_{10}-u_{01}\right]+u_{01} =& \frac{\sum y_{l,K}}{L}\\
\end{aligned}
\end{equation}
Each of these K equations has three unknowns: $T_1, u_{10}$ and $u_{01}$. From these K equations, we find the best fit value for $T_1$ using the scipy.optimize.minimize module in python. K has to be at least 3 so that the problem is not underspecified (overspecified is okay).

\subsection{$T_2$ estimation}
In an analogous manner, for the $T_2$ circuit, the observed data $y_l$ measured in the Z-basis follows a Bernoulli distribution:
\begin{equation}
\begin{aligned}
y_l &=
\begin{cases}
    1, \;\;\;\; \text{ with probability } 1-\tilde{q}_0\\
    0, \;\;\;\; \text{ with probability } \tilde{q}_0\\
\end{cases}\\
\end{aligned}
\end{equation}
where $\tilde{q}_0 = e^{-\tau / T_2} \frac{\left[1-u_{10}-u_{01}\right]}{2} + \frac{\left[1+u_{10}-u_{01}\right]}{2}$. 

The Likelihood function is given by:
\begin{equation}
\begin{aligned}
\mathcal{L} =& \prod_{l=1}^L \text{Pr}\left(Y_l=y_l\right) = \prod_{l=1}^L \tilde{q}_0^{1-y_l} (1-\tilde{q}_0)^{y_l}\\
\end{aligned}
\end{equation}

Setting:
\begin{equation}
\frac{\partial \log \mathcal{L}}{\partial \tilde{q}_0} =0,
\end{equation}
we get,
\begin{equation}
\begin{aligned}
\tilde{q}_0^* = 1-\sum \frac{y_l}{L}
\end{aligned}
\end{equation}

Thus we have obtained the estimator (denoted by the * sign) for $\tilde{q}_0$. Note that
\begin{equation}
\mathds{E}( \tilde{q}_0^* ) = \sum \frac{\mathds{E}(y_l)}{L} = \sum \frac{\tilde{q}_0}{L} = \tilde{q}_0 
\end{equation}
and hence it is an unbiased estimator.

The variance of this estimate is given by:
\begin{equation}
\begin{aligned}
\text{Var}( \tilde{q}_0^* ) = \sigma^2(\tilde{q}_0^*) = \sum \frac{\text{Var}(y_l)}{L^2} = \frac{ \tilde{q}_0(1-\tilde{q}_0) }{L} \approx \frac{ \tilde{q}_0^*(1-\tilde{q}_0^*) }{L}
\end{aligned}
\end{equation}

Using the above, we can get K different equations, one for each phase gate with evolution time $\tau_k$ and whose post-measurement results are denoted by $\{ y_{l,k} \}$:
\begin{equation}
\begin{aligned}
e^{-\tau_1 / T_2} \frac{1-u_{10}-u_{01}}{2} + \frac{1+u_{10}-u_{01}}{2} =& \frac{\sum y_{l,1}}{L}\\
e^{-\tau_2 / T_2} \frac{1-u_{10}-u_{01}}{2} + \frac{1+u_{10}-u_{01}}{2} =& \frac{\sum y_{l,2}}{L}\\
&\cdots\\
e^{-\tau_K / T_2} \frac{1-u_{10}-u_{01}}{2} + \frac{1+u_{10}-u_{01}}{2} =& \frac{\sum y_{l,K}}{L}\\
\end{aligned}
\end{equation}
Each of these K equations has three unknowns: $T_2, u_{10}$ and $u_{01}$. From these K equations, we find the best fit value for $T_2$ using the scipy.optimize.minimize module in python. K has to be at least 3 so that the problem is not underspecified (overspecified is okay).

\subsection{Error bars on decoherence estimates}
For the $T_1$ circuit:
\begin{equation}
\begin{aligned}
p_1 =& e^{-\tau / T_1}\\
\Rightarrow \log p_1 =& -\frac{\tau}{T_1}\\
\Rightarrow \frac{\delta p_1}{p_1} =& \frac{\tau}{T_1^2}\delta T_1 \\
\Rightarrow \sigma^2_{T_1} =& \frac{T_1^4 \left( e^{\tau/T_1}-1\right)}{L\tau^2}\\ 
\end{aligned}
\end{equation}
Now suppose that we use $K$ different evolution times in the circuit: $\tau_1, \tau_2, \cdots \tau_K$.
Let the desired standard deviation of the $T_1$ estimate be $\sigma_\text{desired}$.
Since the data underlying the estimation obtained at different times are independent:
\begin{equation}
\begin{aligned}
\sigma_\text{desired}^2 = \frac{1}{K} \sum \sigma_i^2
\end{aligned}
\end{equation}
where $\sigma_i^2$ is the variance obtained when $T_1$ is estimated using a delay gate with delay time $\tau_i$. This gives:
\begin{equation}
\begin{aligned}
L_\text{min} = \frac{T_1^4}{K \sigma^2_\text{desired}} \sum \frac{e^{\tau_i/T_1}-1}{\tau_i^2}
\end{aligned}
\end{equation}

For the $T_2$ circuit:
\begin{equation}
\begin{aligned}
q_0 =& \frac{1}{2}(1+e^{-\tau / T_2})\\
\Rightarrow \log (2q_0 -1) =& -\frac{\tau}{T_2}\\
\Rightarrow \frac{2\delta q_0}{2q_0-1} =& \frac{\tau}{T_2^2}\delta T_2 \\
\Rightarrow \sigma^2_{T_2} =& \frac{T_2^4 \left( e^{2\tau/T_2}-1\right)}{L\tau^2}\\ 
\end{aligned}
\end{equation}
Now suppose that we use $K$ different evolution times in the circuit: $\tau_1, \tau_2, \cdots \tau_K$.
Let the desired standard deviation of the $T_2$ estimate be $\sigma_\text{desired}$.
Since the data underlying the estimation obtained at different times are independent:
\begin{equation}
\begin{aligned}
\sigma_\text{desired}^2 = \frac{1}{K} \sum \sigma_i^2
\end{aligned}
\end{equation}
where $\sigma_i^2$ is the variance obtained when $T_2$ is estimated using phase-shift time $\tau_i$. This gives:
\begin{equation}
\begin{aligned}
L_\text{min} = \frac{T_2^4}{K \sigma^2_\text{desired}} \sum \frac{e^{2\tau_i/T_2}-1}{\tau_i^2}
\end{aligned}
\end{equation}
Since we have a circuit that measures $T_1$ and $T_2$ in one go, we have to take the max of the $L_\text{min}$ for each of the two cases (i.e. max of $L_\text{min}$ for $T_1$ and $L_\text{min}$ for $T_2$).

Using a noisy simulation with known error estimates (detailed in the program validation section next), we arrived at $L_\text{min}$ at each time to be $20,000$ samples. There were four delay gates (aka four separate decay experiments), so a total of $80,000$ samples went into the computation of each $T_1$ and $T_2$ value for each time-stamp. The standard-deviation we aimed for is $1 \mu s$.

\subsection{Program validation}
Program validation is crucial to ensure that, within the assumptions of our theoretical noise model, our simulations of quantum circuits, utilizing known noise parameters, yield precise results. Specifically, we aimed for our statistical analysis to accurately recover the expected values. Additionally, program validation helps us determine the required number of circuit repetitions for achieving a specified level of outcome precision. While minor deviations in the final standard deviation are expected due to various factors, they should generally align with our target precision.

Since we individually measure the $T_1$ and $T_2$ times for each of the 27 qubits of \kolkata, validating the program for a single qubit suffices. We found that a sample size of 10,000 is sufficient to attain the desired precision of 1 microsecond, but for safety, we opted for 20,000 samples in the final run on real device. Given the use of four delay gates, each $T_1$ or $T_2$ data point estimation relies on a total of $80,000$ samples. For consistency with the \kolkata processor's specifications, we set the readout error at 0.028 $T_1$ time at 134 microseconds and $T_2$ time at 93 microseconds in our noise simulation. 

Our program validation yielded an estimated T1 time of 134.52 microseconds, which falls within the 1-microsecond precision target. The estimated T2 time was 94.05 microseconds, slightly exceeding the 1-microsecond precision by a difference of 1.03 microseconds which we deemed acceptable.

\subsection{Summary of results}
\begin{figure}
\centering
\includegraphics[width=\figurewidth]{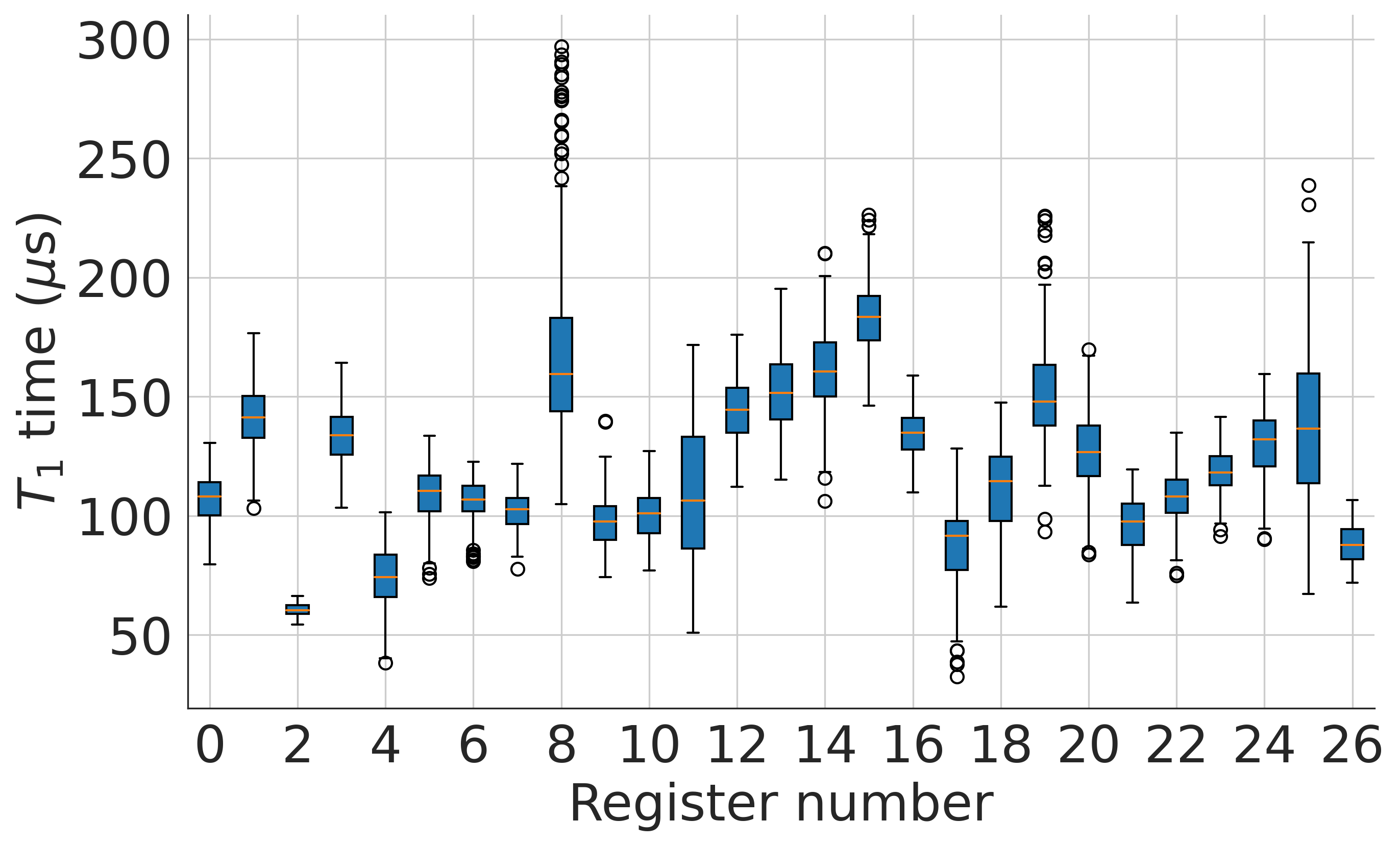}
\caption*{(a)}
\vspace{0.5in}
\includegraphics[width=\figurewidth]{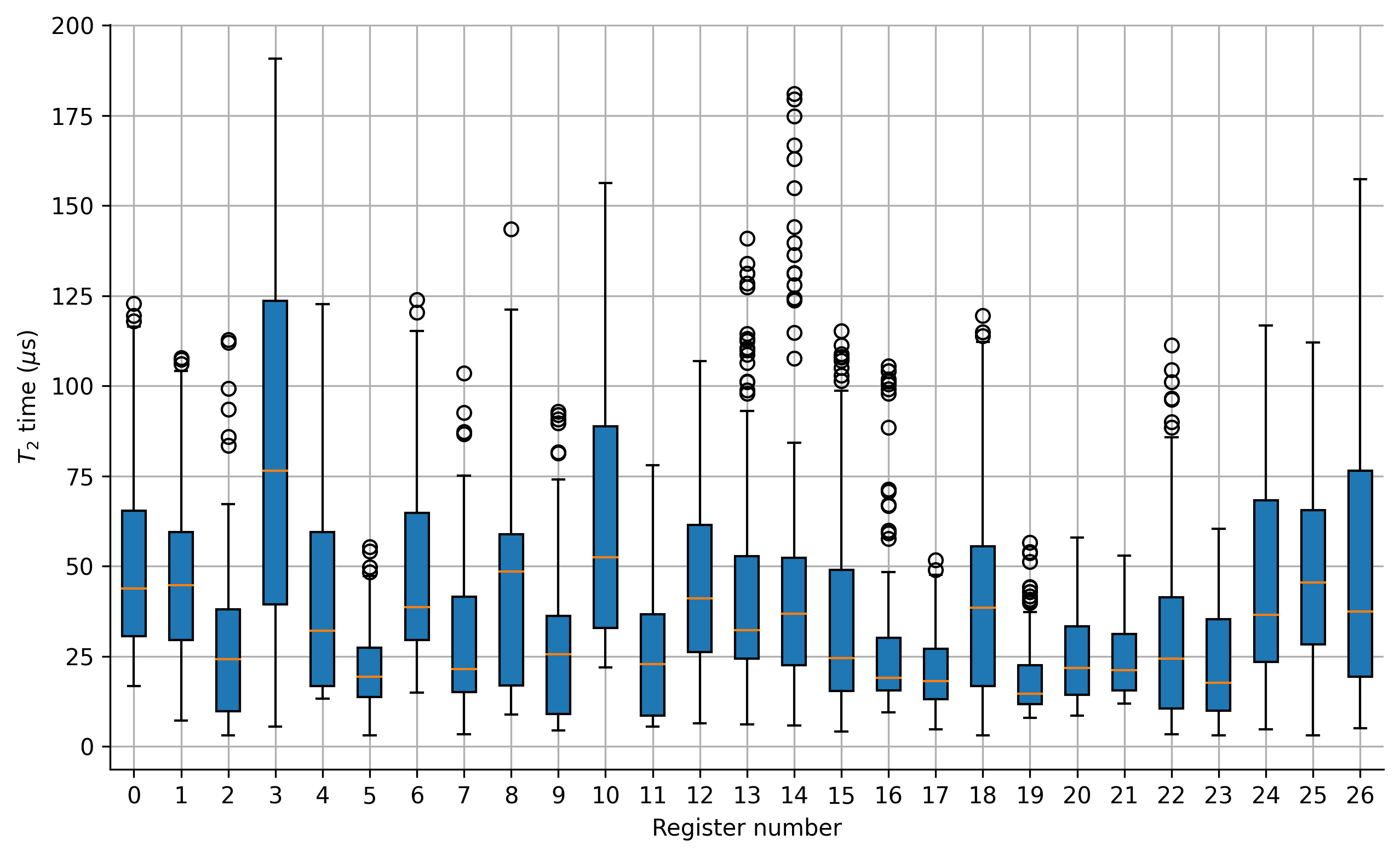}
\caption*{(b)}
\caption{
Spatial non-stationarity of decoherence times. Indivudal error-bars do not capture the variation across qubits. (a) In terms of the $T_1$ parameter, qubit 15 performed the best, exhibiting a median $T_1$ time of 184 microseconds, while qubit 4 performed the worst with a median $T_1$ time of 74 microseconds. Across all 27 qubits, the median $T_1$ time was 116 microseconds, with a standard deviation of 1.5 microseconds, and the range of $T_1$ times spanned from 32 to 297 microseconds. (b) For the $T_2$ parameter, qubit 3 demonstrated the highest performance, displaying a median $T_2$ time of 76 microseconds, while qubit 19 had the poorest performance with a median $T_2$ time of 15 microseconds. Across all 27 qubits, the median $T_2$ time averaged 30 microseconds, with a standard deviation of 1.1 microseconds, and the range of $T_2$ times varied from 3 to 191 microseconds.
}
\label{fig:T1T2_boxplot}
\end{figure}

\begin{figure}
\centering
\includegraphics[width=\figurewidth]{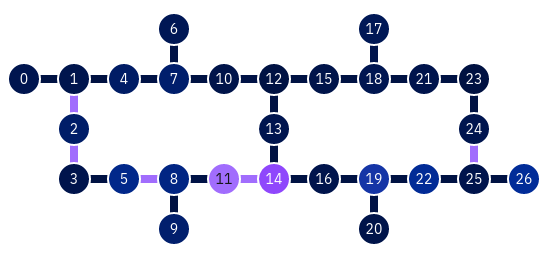}
\caption*{(a)}
\vspace{0.5in}
\includegraphics[width=\figurewidth]{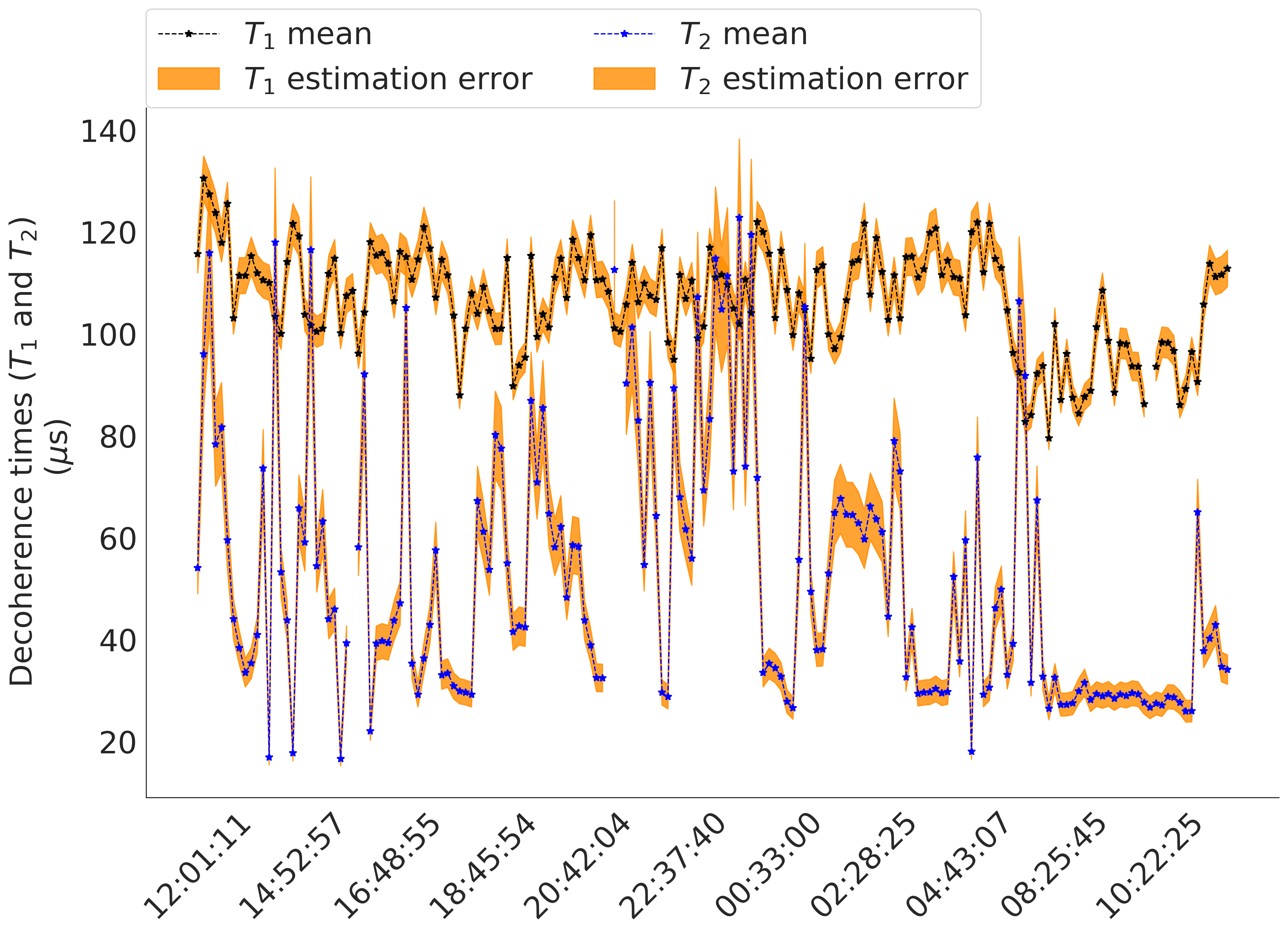}
\caption*{(b)}
\caption{
(a) Schematic of the 27-qubit device \kolkata. 
(b) Estimated $T_1$ and $T_2$ time-series for qubit 0 as collected between 12:00 P.M. ET on Sep 12, 2023 and 12:00 P.M. ET on Sep 13, 2023. 
}
\label{fig:estimated_T1T2_timeseries_qubit_first}
\end{figure}

\begin{figure}
\centering
\includegraphics[width=\figurewidth]{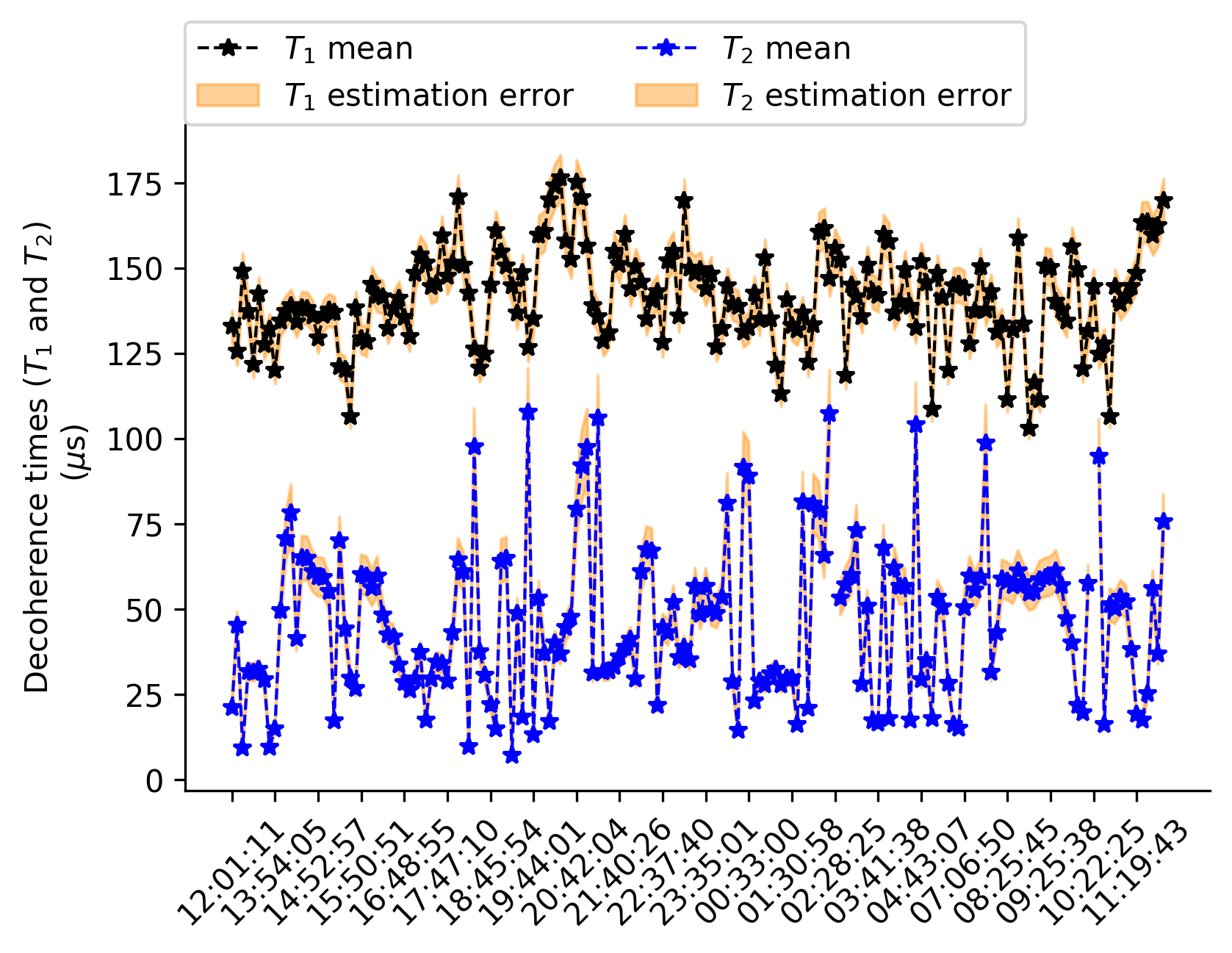}
\caption*{(a)}
\vspace{0.5in}
\includegraphics[width=\figurewidth]{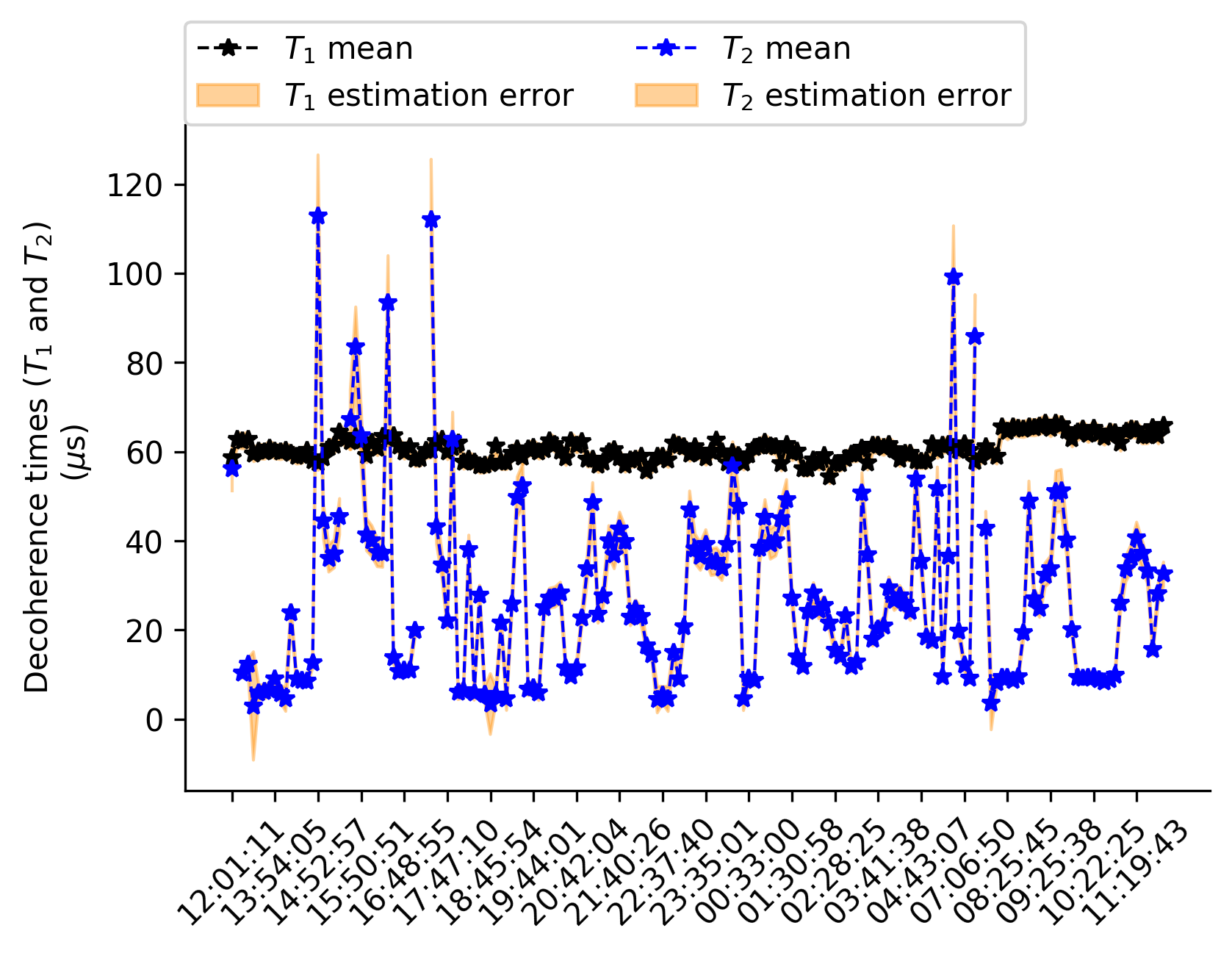}
\caption*{(b)}
\caption{
Estimated $T_1$ and $T_2$ time-series as collected between 12:00 P.M. ET on Sep 12, 2023 and 12:00 P.M. ET on Sep 13, 2023 for (a) qubit 1 and (b) qubit 2.
}
\label{fig:estimated_T1T2_timeseries_qubit_1_2}
\end{figure}

\begin{figure}
\centering
\includegraphics[width=\figurewidth]{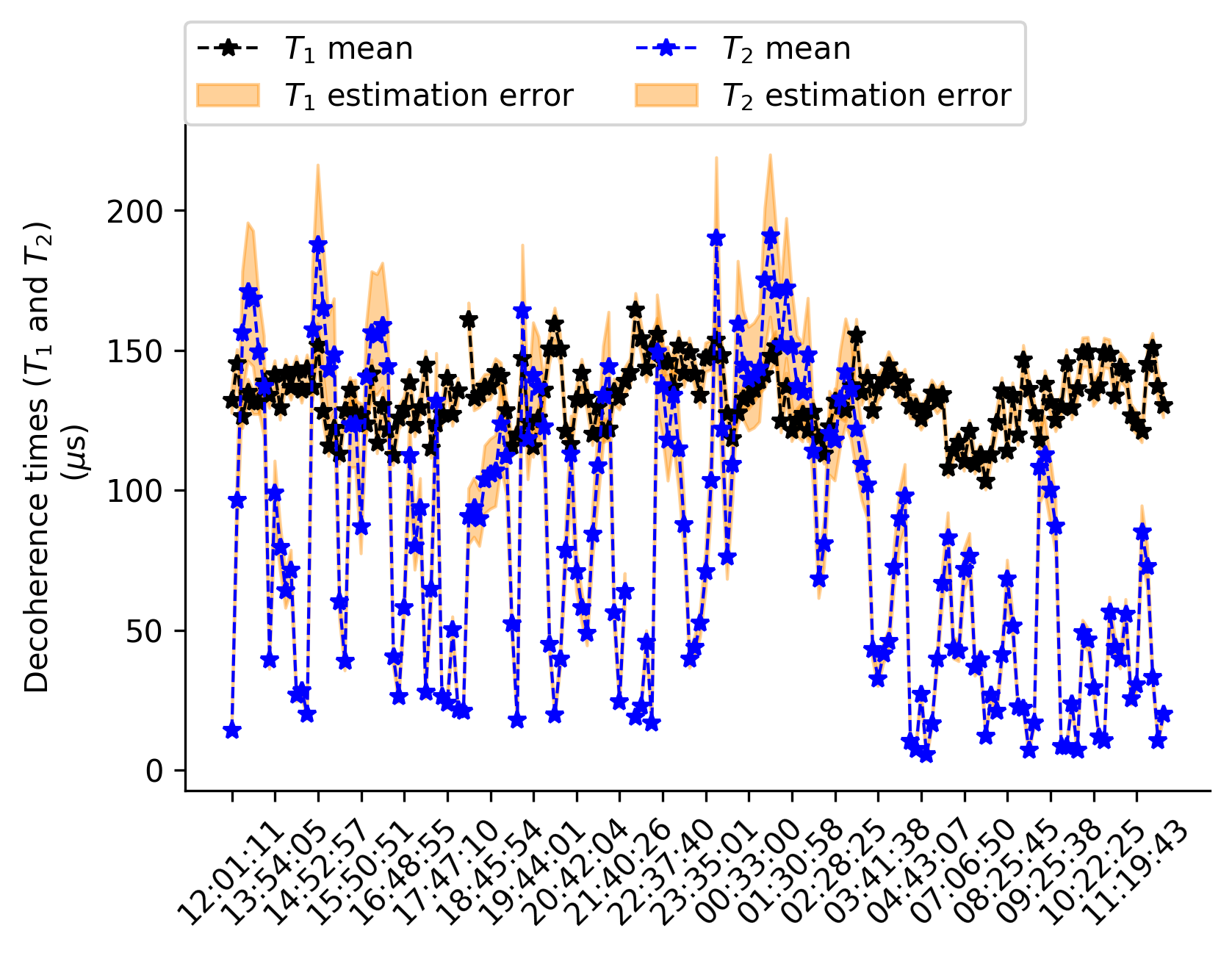}
\caption*{(a)}
\vspace{0.5in}
\includegraphics[width=\figurewidth]{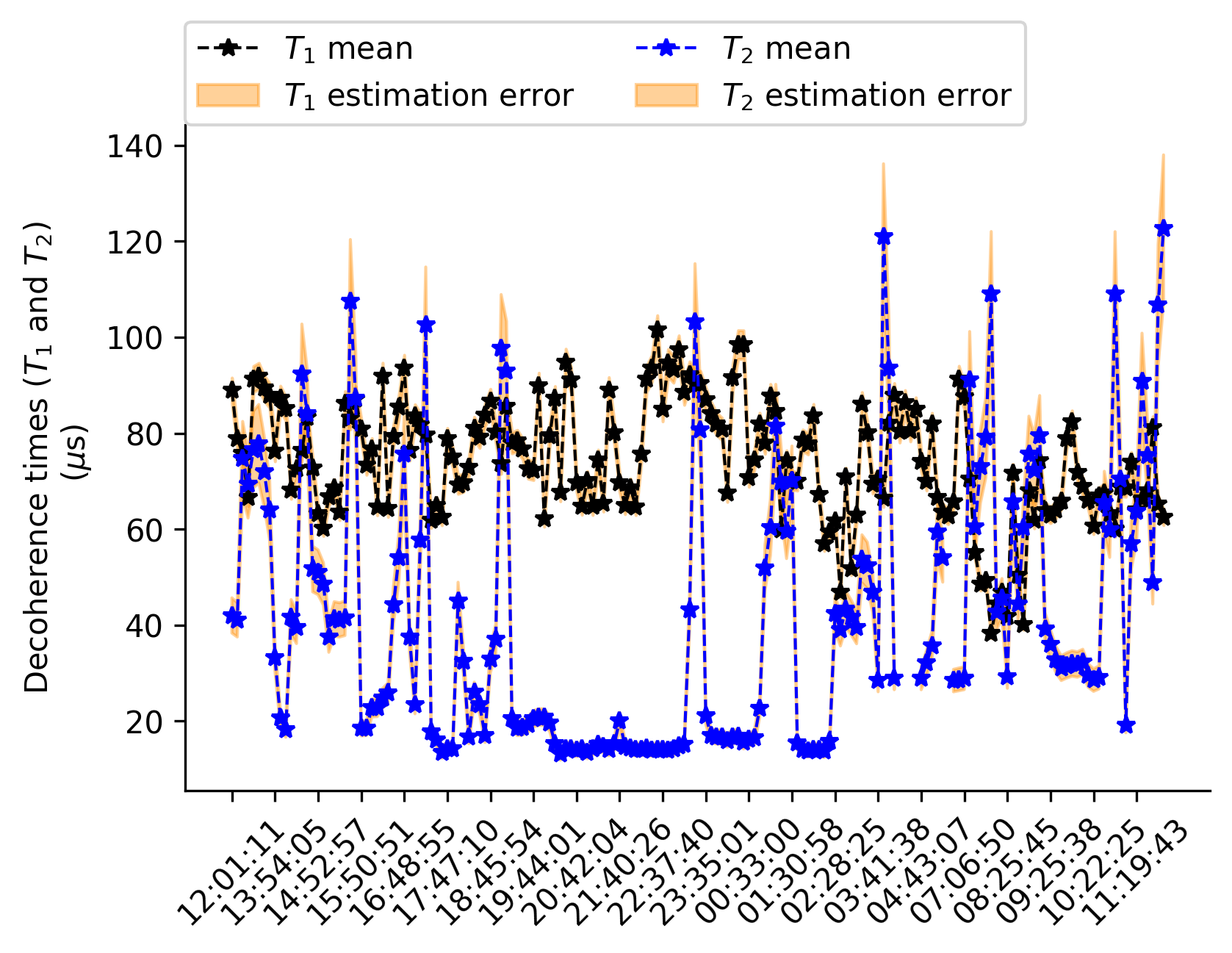}
\caption*{(b)}
\caption{
Estimated $T_1$ and $T_2$ time-series as collected between 12:00 P.M. ET on Sep 12, 2023 and 12:00 P.M. ET on Sep 13, 2023 for (a) qubit 3 and (b) qubit 4.
}
\label{fig:estimated_T1T2_timeseries_qubit_3_4}
\end{figure}

\begin{figure}
\centering
\includegraphics[width=\figurewidth]{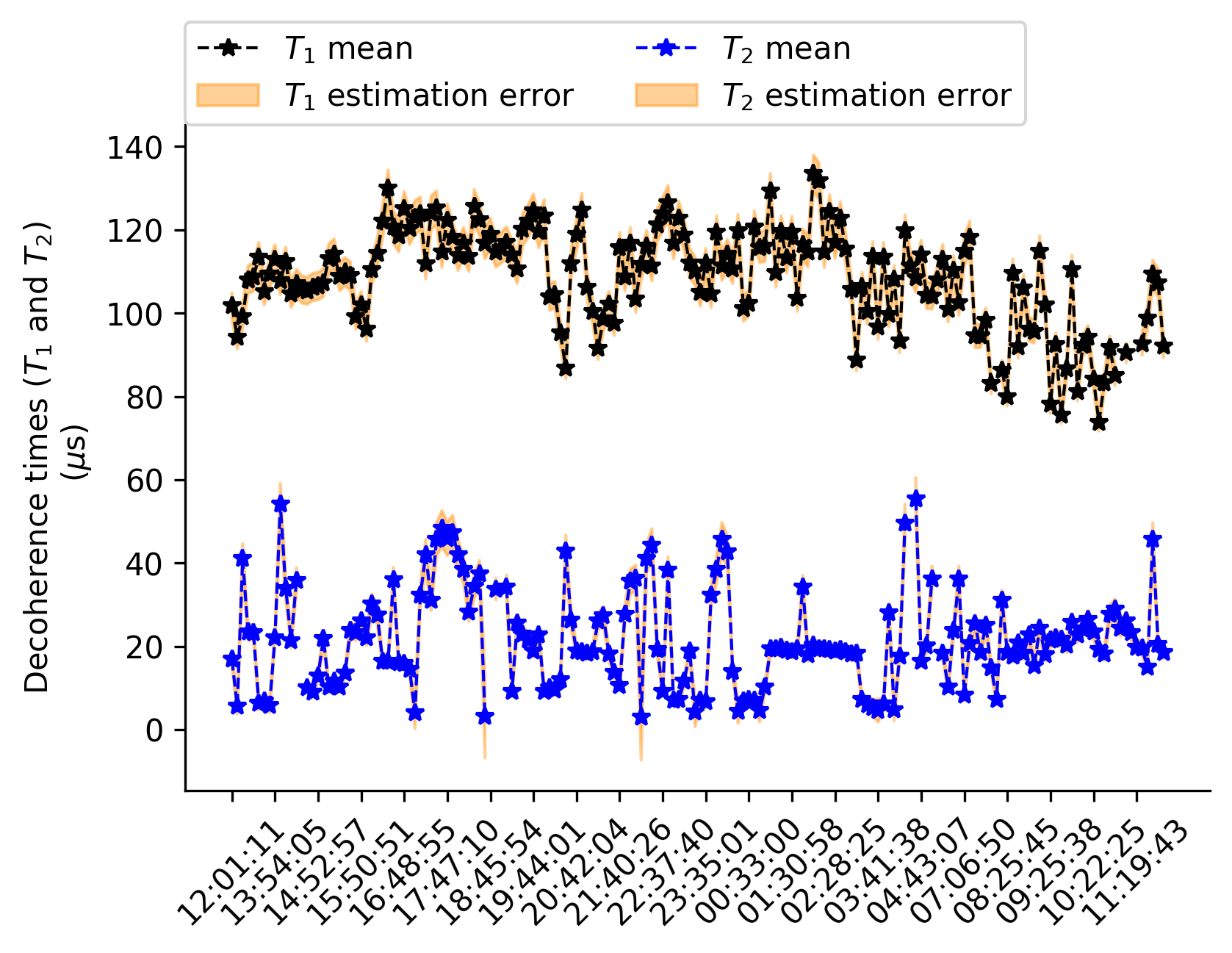}
\caption*{(a)}
\vspace{0.5in}
\includegraphics[width=\figurewidth]{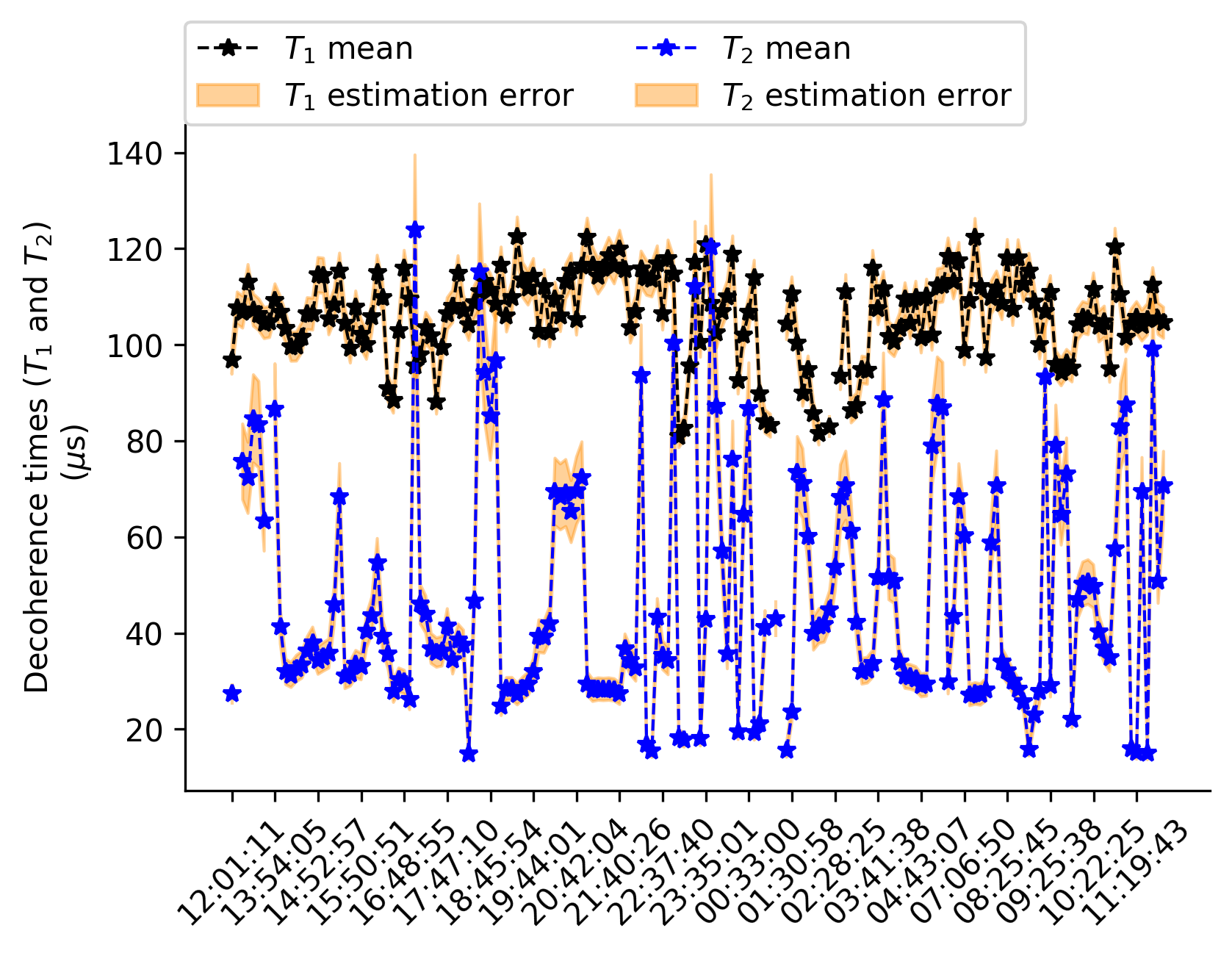}
\caption*{(b)}
\caption{
Estimated $T_1$ and $T_2$ time-series as collected between 12:00 P.M. ET on Sep 12, 2023 and 12:00 P.M. ET on Sep 13, 2023 for (a) qubit 5 and (b) qubit 6.
}
\label{fig:estimated_T1T2_timeseries_qubit_5_6}
\end{figure}

\begin{figure}
\centering
\includegraphics[width=\figurewidth]{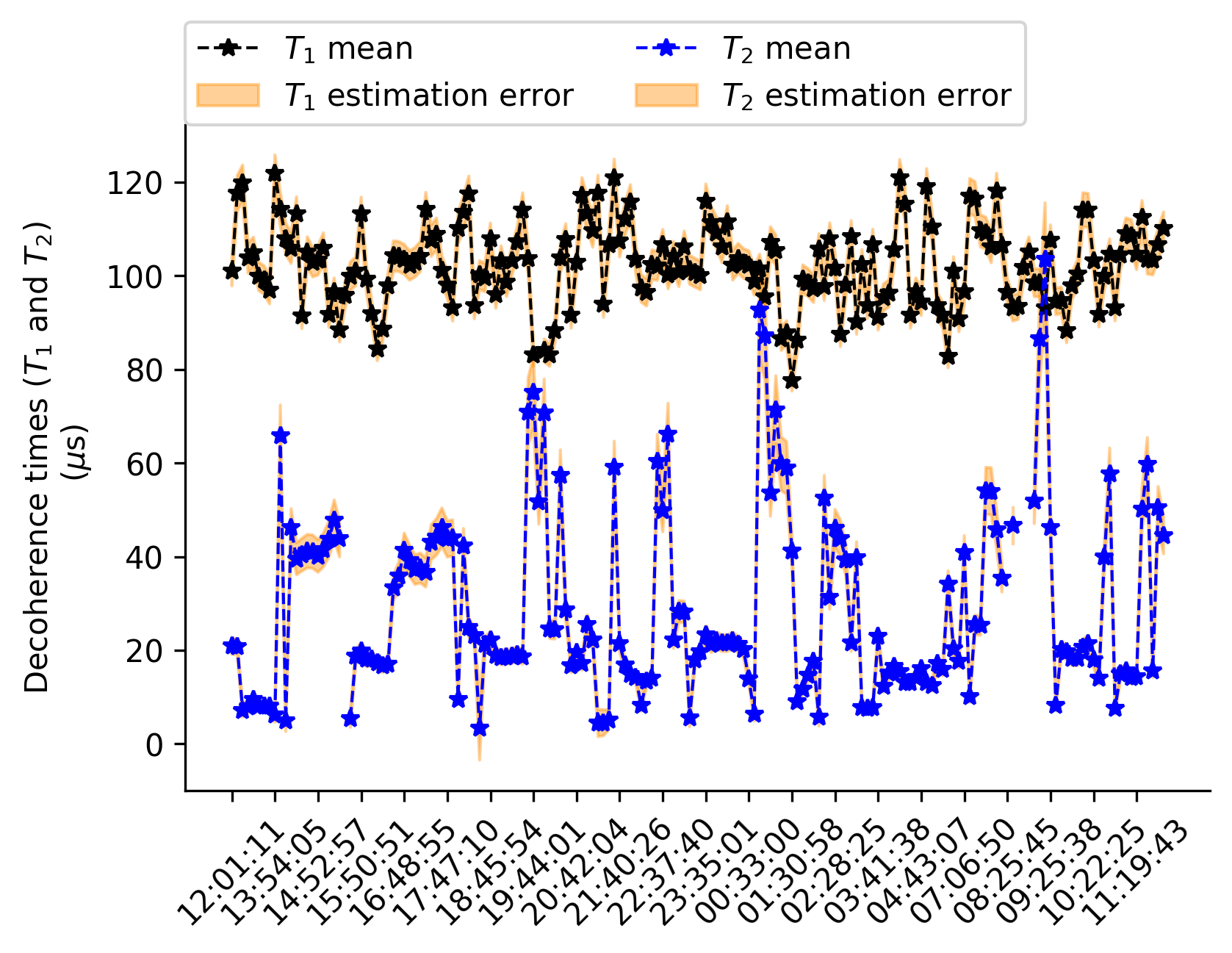}
\caption*{(a)}
\vspace{0.5in}
\includegraphics[width=\figurewidth]{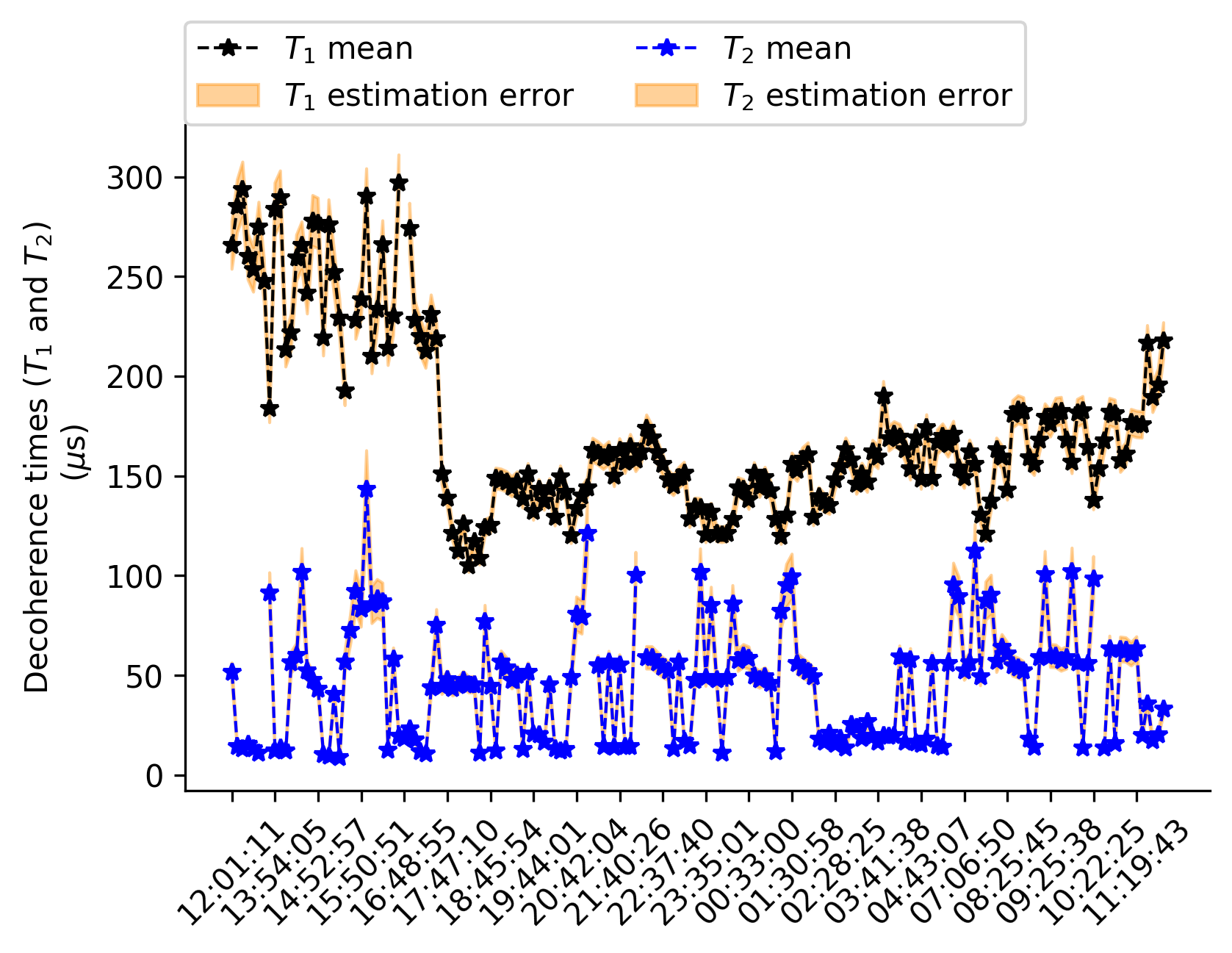}
\caption*{(b)}
\caption{
Estimated $T_1$ and $T_2$ time-series as collected between 12:00 P.M. ET on Sep 12, 2023 and 12:00 P.M. ET on Sep 13, 2023 for (a) qubit 7 and (b) qubit 8.
}
\label{fig:estimated_T1T2_timeseries_qubit_7_8}
\end{figure}

\begin{figure}
\centering
\includegraphics[width=\figurewidth]{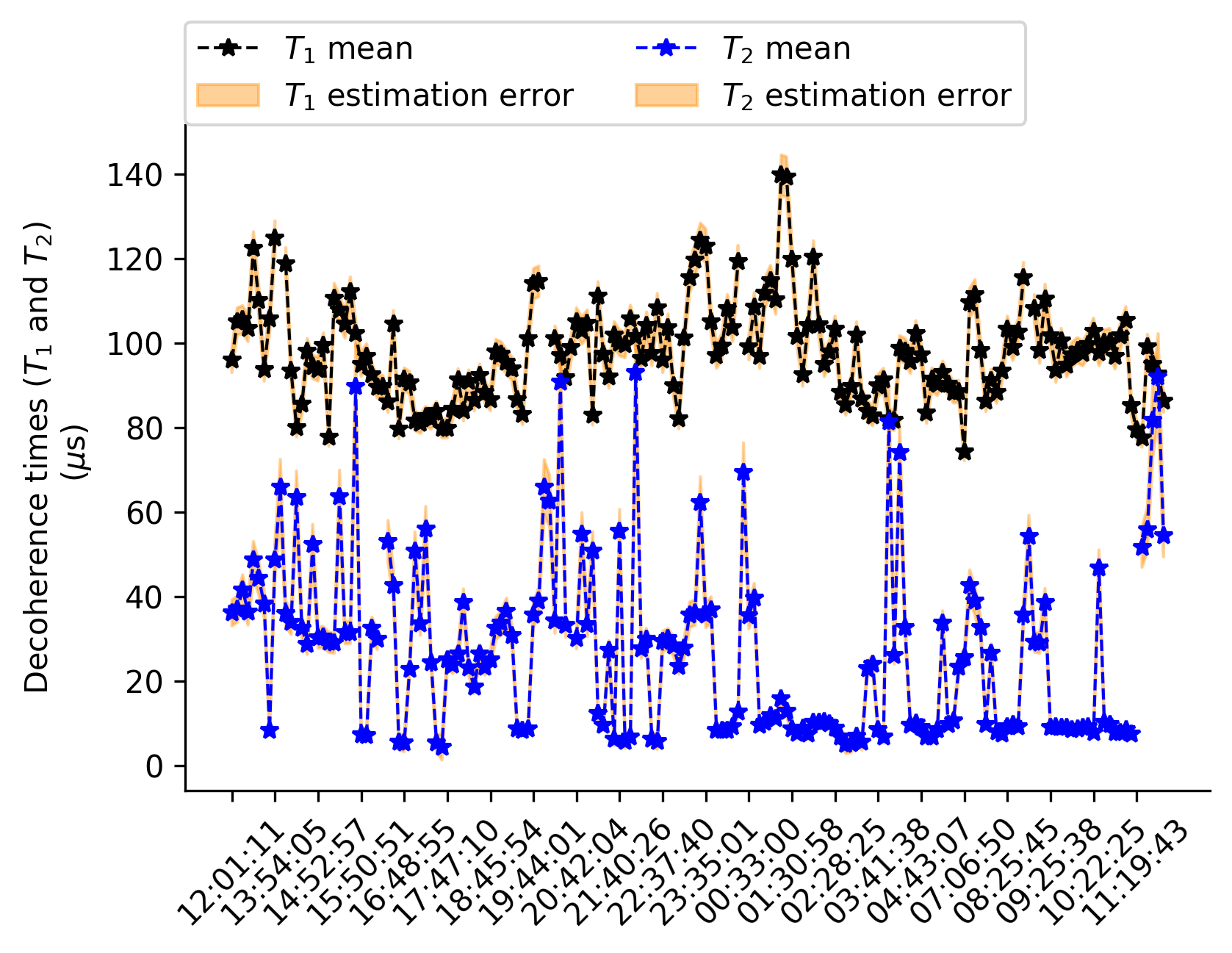}
\caption*{(a)}
\vspace{0.5in}
\includegraphics[width=\figurewidth]{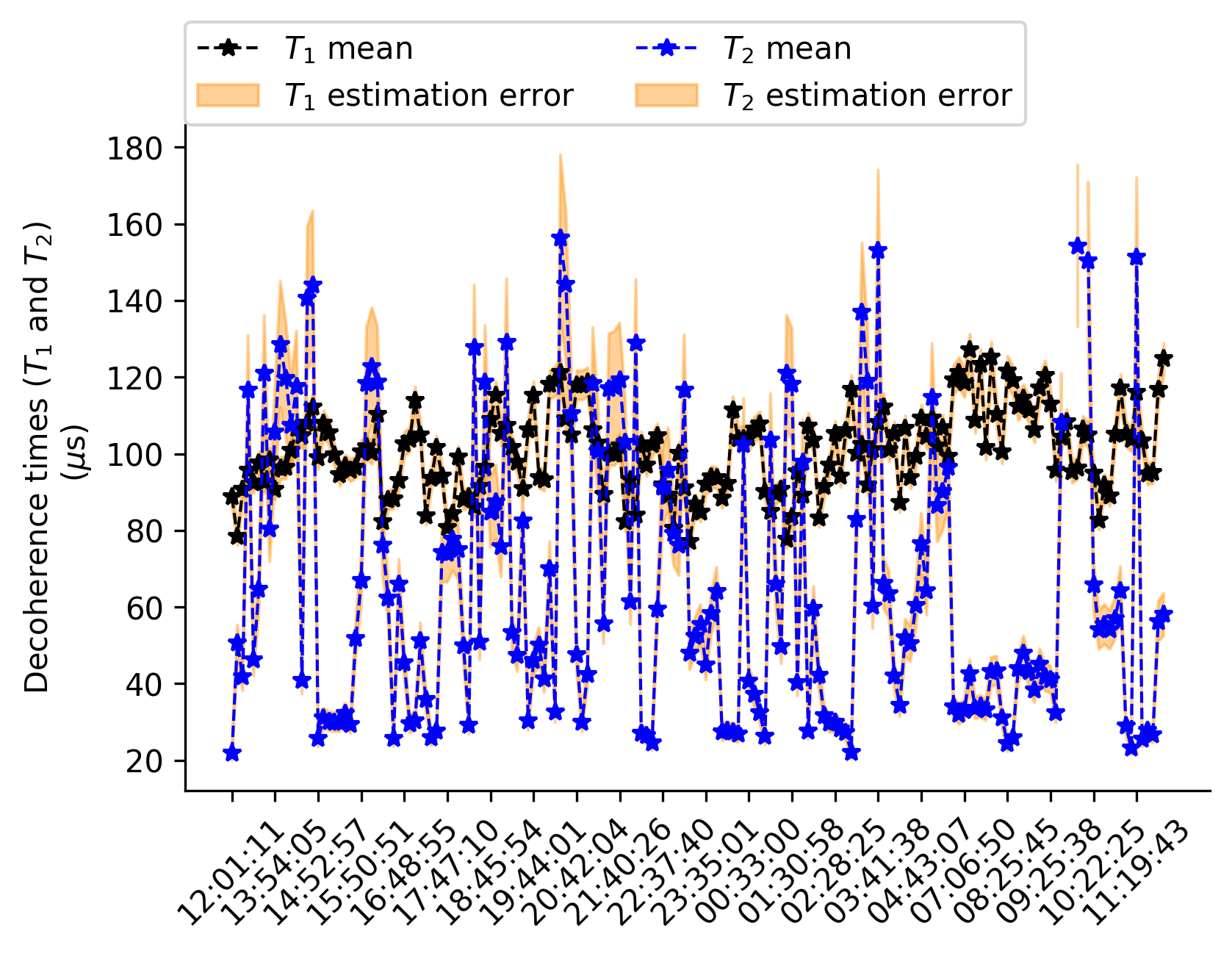}
\caption*{(b)}
\caption{
Estimated $T_1$ and $T_2$ time-series as collected between 12:00 P.M. ET on Sep 12, 2023 and 12:00 P.M. ET on Sep 13, 2023 for (a) qubit 9 and (b) qubit 10.
}
\label{fig:estimated_T1T2_timeseries_qubit_9_10}
\end{figure}

\begin{figure}
\centering
\includegraphics[width=\figurewidth]{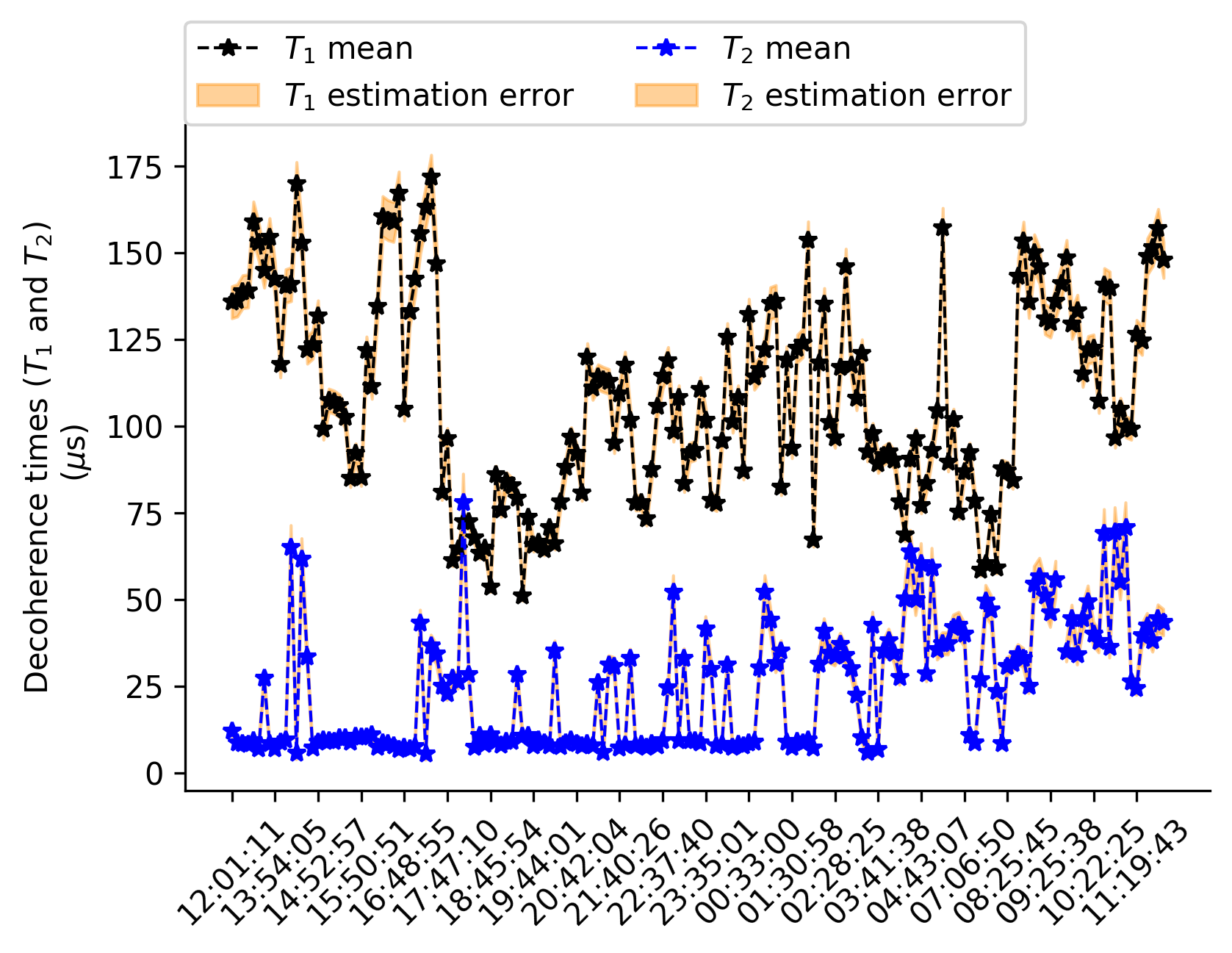}
\caption*{(a)}
\vspace{0.5in}
\includegraphics[width=\figurewidth]{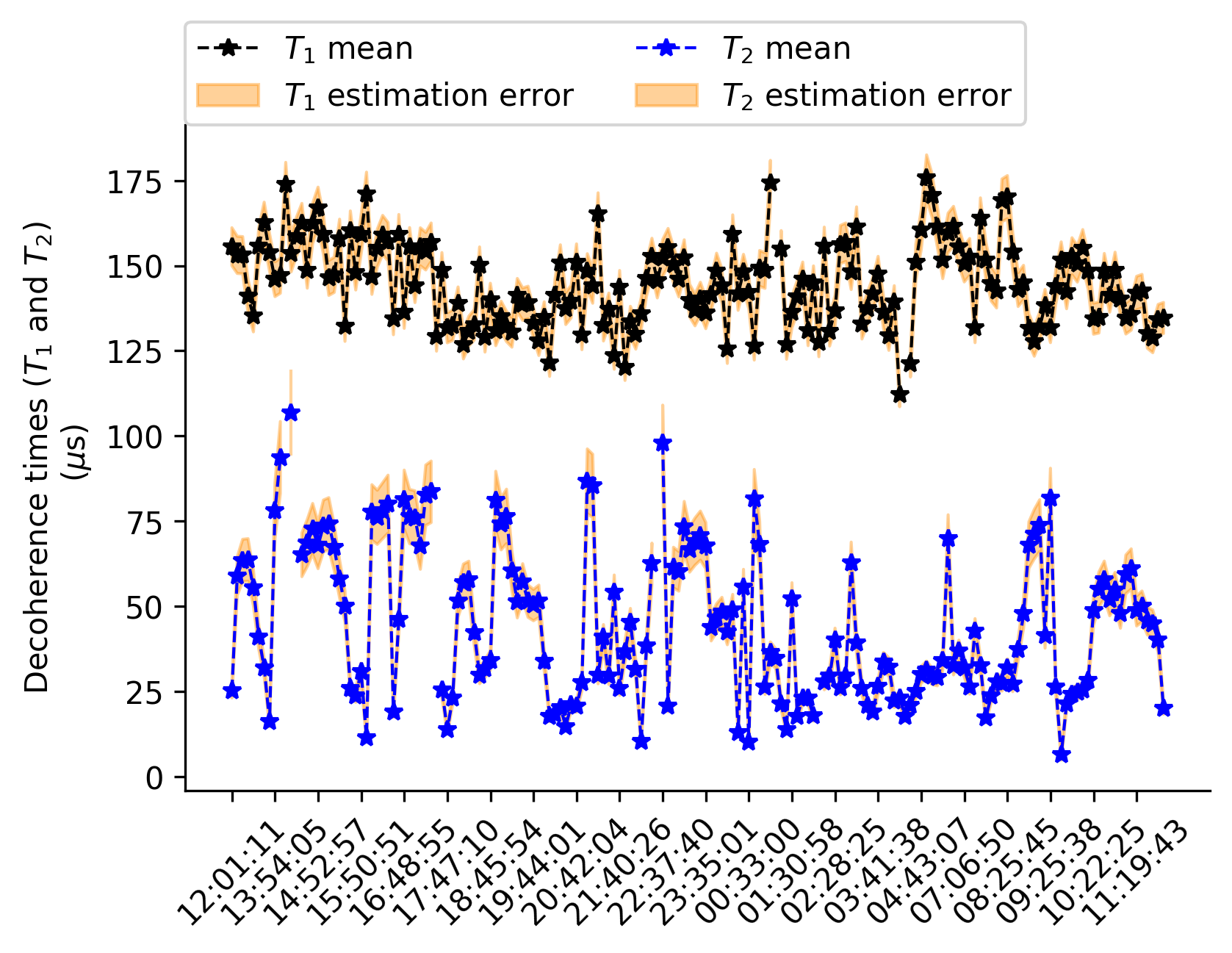}
\caption*{(b)}
\caption{
Estimated $T_1$ and $T_2$ time-series as collected between 12:00 P.M. ET on Sep 12, 2023 and 12:00 P.M. ET on Sep 13, 2023 for (a) qubit 11 and (b) qubit 12.
}
\label{fig:estimated_T1T2_timeseries_qubit_11_12}
\end{figure}

\begin{figure}
\centering
\includegraphics[width=\figurewidth]{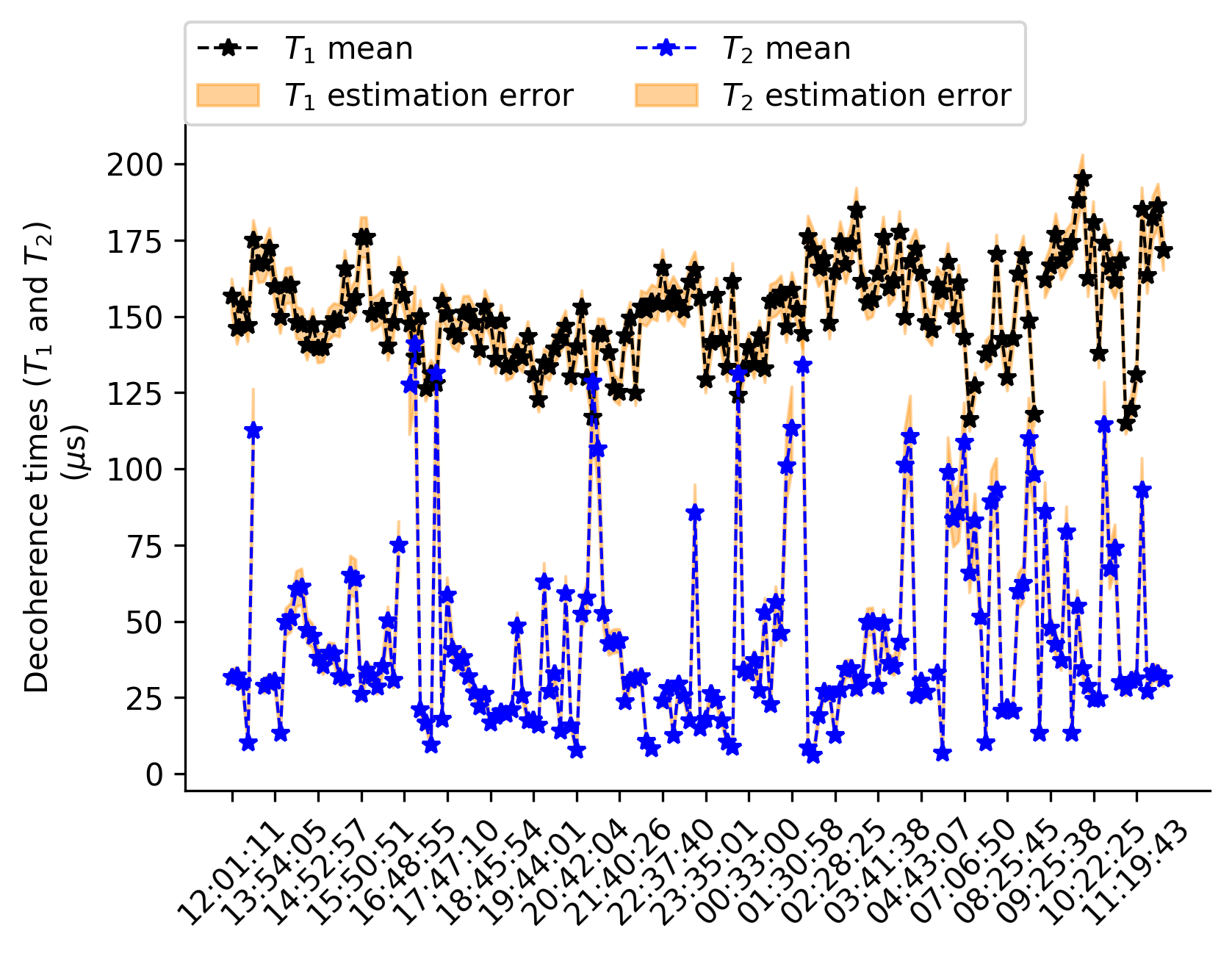}
\caption*{(a)}
\vspace{0.5in}
\includegraphics[width=\figurewidth]{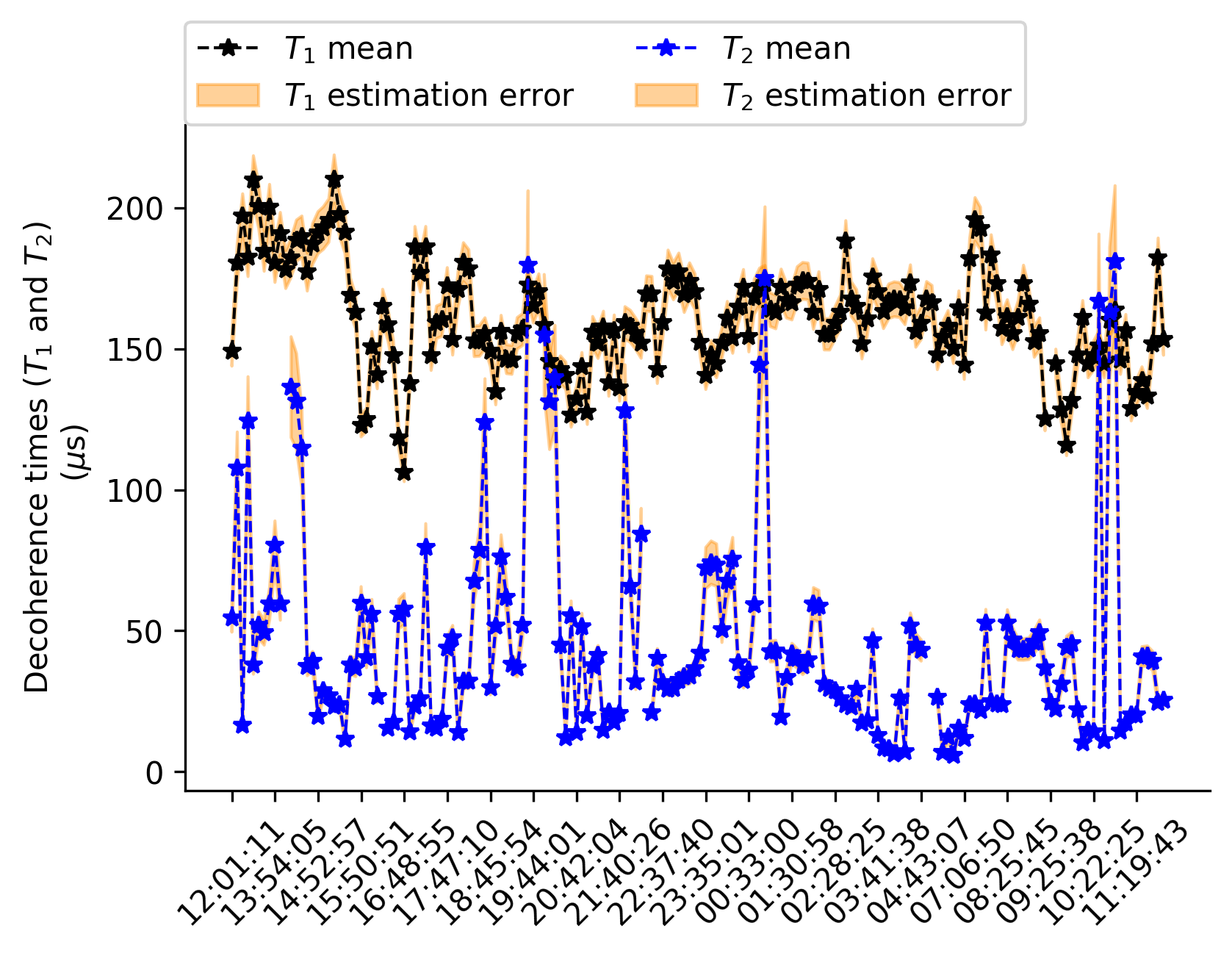}
\caption*{(b)}
\caption{
Estimated $T_1$ and $T_2$ time-series as collected between 12:00 P.M. ET on Sep 12, 2023 and 12:00 P.M. ET on Sep 13, 2023 for (a) qubit 13 and (b) qubit 14.
}
\label{fig:estimated_T1T2_timeseries_qubit_13_14}
\end{figure}

\begin{figure}
\centering
\includegraphics[width=\figurewidth]{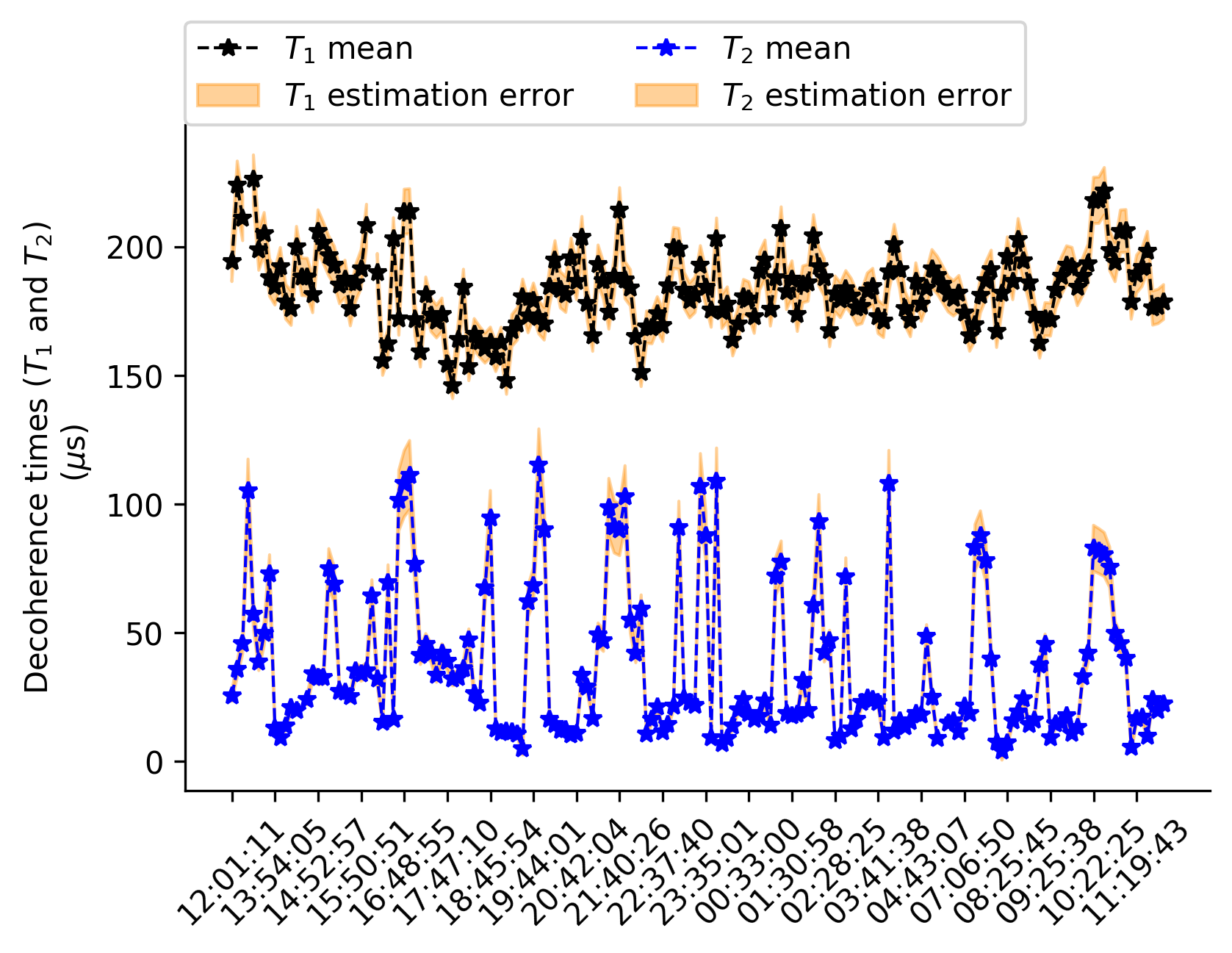}
\caption*{(a)}
\vspace{0.5in}
\includegraphics[width=\figurewidth]{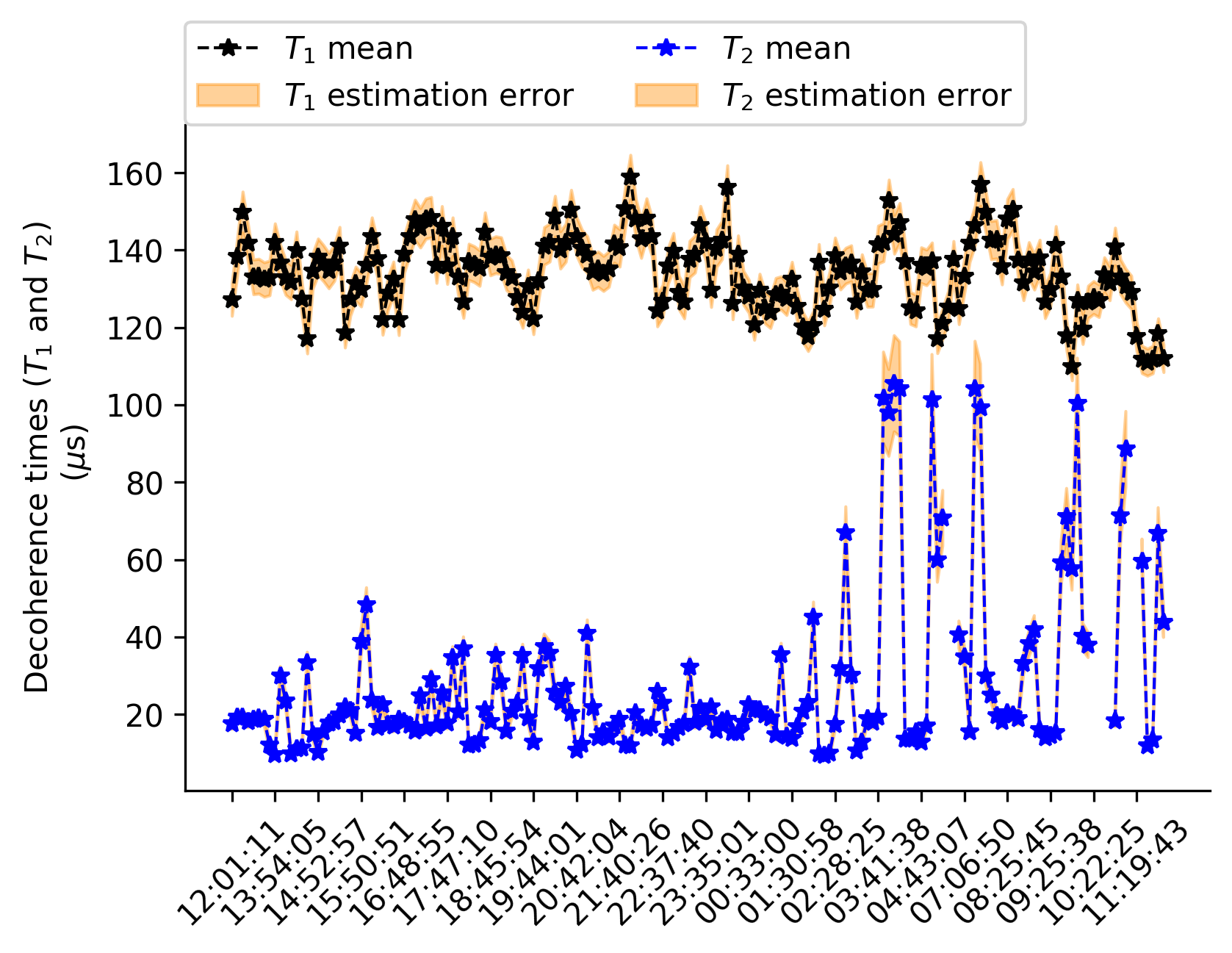}
\caption*{(b)}
\caption{
Estimated $T_1$ and $T_2$ time-series as collected between 12:00 P.M. ET on Sep 12, 2023 and 12:00 P.M. ET on Sep 13, 2023 for (a) qubit 15 and (b) qubit 16.
}
\label{fig:estimated_T1T2_timeseries_qubit_15_16}
\end{figure}

\begin{figure}
\centering
\includegraphics[width=\figurewidth]{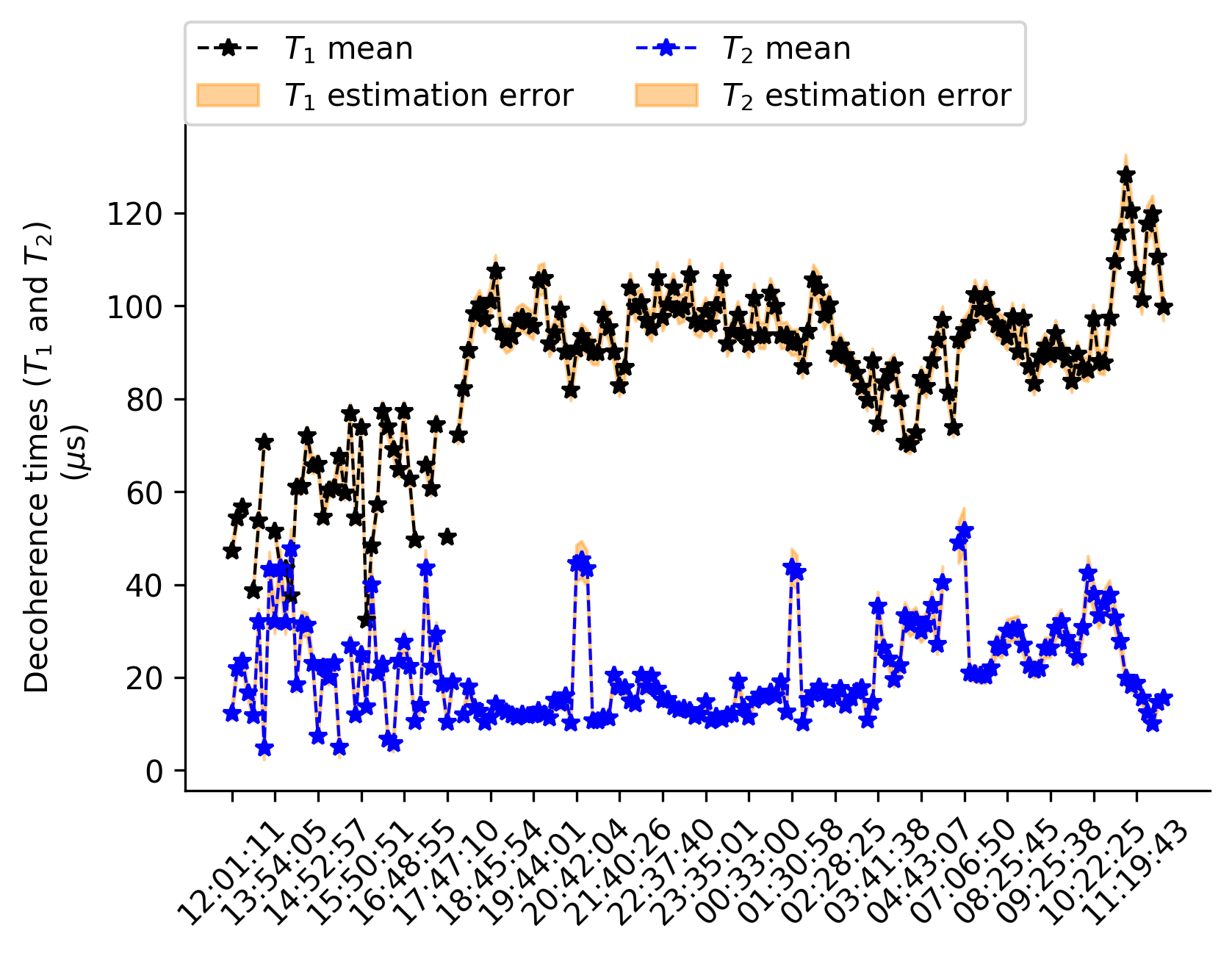}
\caption*{(a)}
\vspace{0.5in}
\includegraphics[width=\figurewidth]{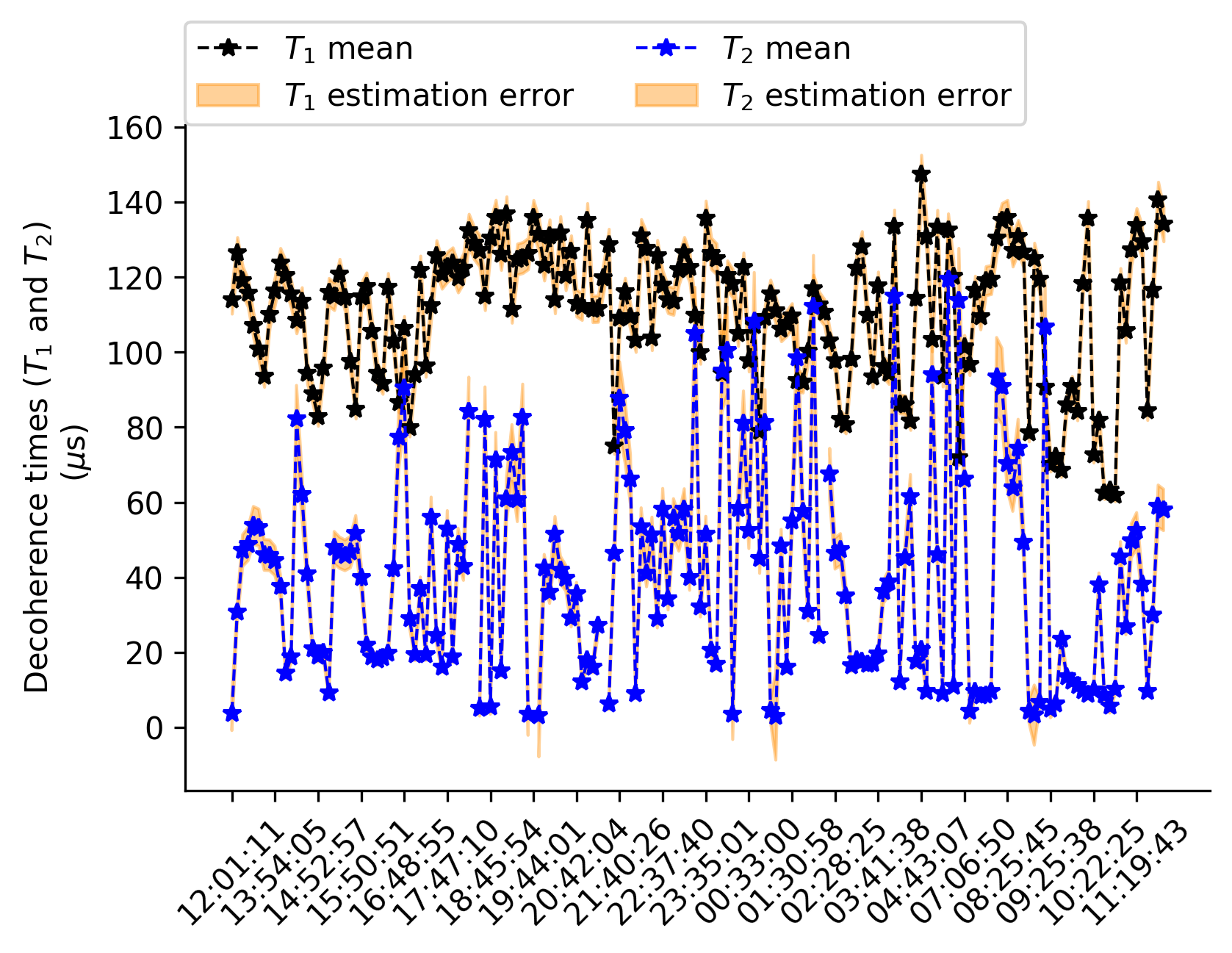}
\caption*{(b)}
\caption{
Estimated $T_1$ and $T_2$ time-series as collected between 12:00 P.M. ET on Sep 12, 2023 and 12:00 P.M. ET on Sep 13, 2023 for (a) qubit 17 and (b) qubit 18.
}
\label{fig:estimated_T1T2_timeseries_qubit_17_18}
\end{figure}

\begin{figure}
\centering
\includegraphics[width=\figurewidth]{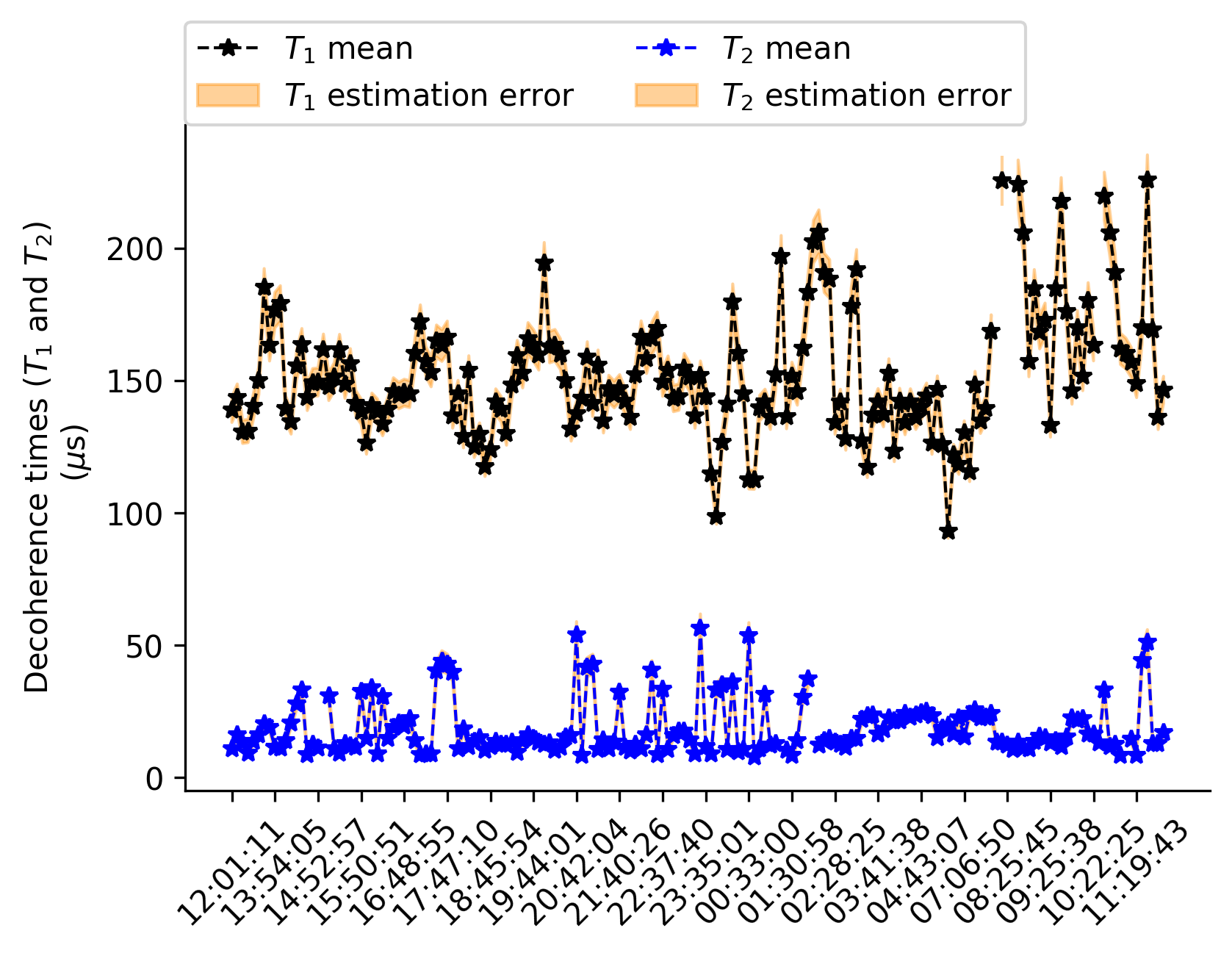}
\caption*{(a)}
\vspace{0.5in}
\includegraphics[width=\figurewidth]{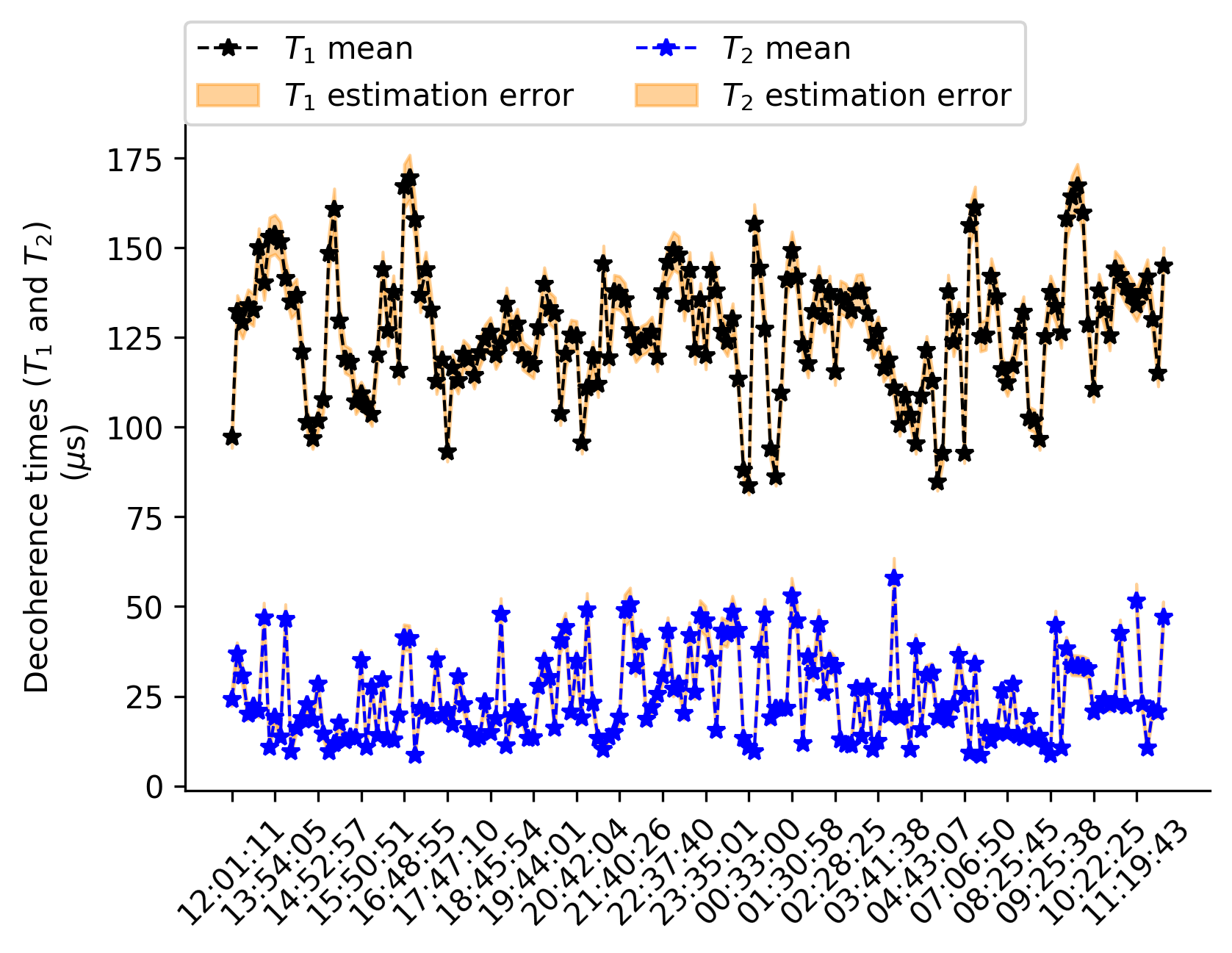}
\caption*{(b)}
\caption{
Estimated $T_1$ and $T_2$ time-series as collected between 12:00 P.M. ET on Sep 12, 2023 and 12:00 P.M. ET on Sep 13, 2023 for (a) qubit 19 and (b) qubit 20.
}
\label{fig:estimated_T1T2_timeseries_qubit_19_20}
\end{figure}

\begin{figure}
\centering
\includegraphics[width=\figurewidth]{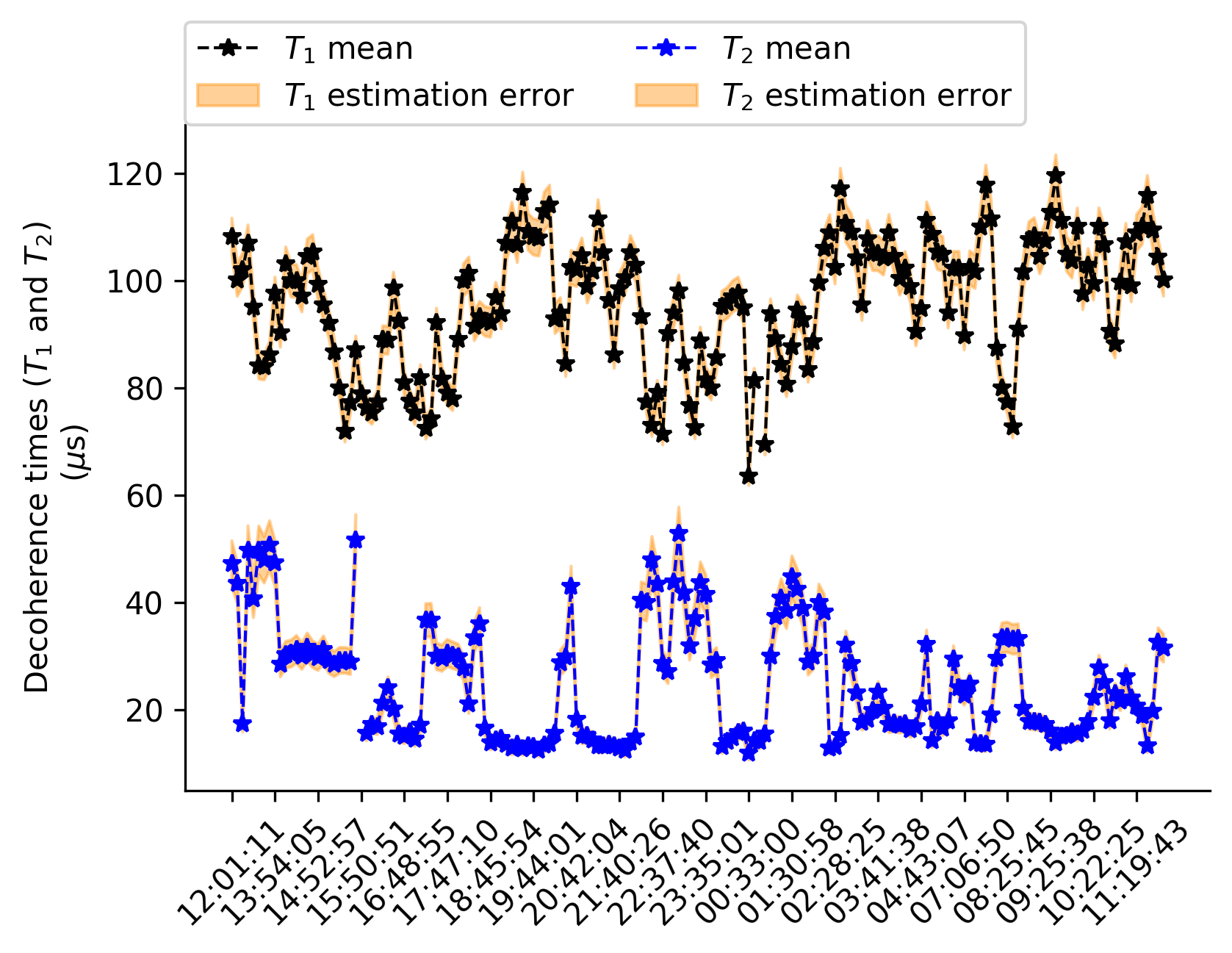}
\caption*{(a)}
\vspace{0.5in}
\includegraphics[width=\figurewidth]{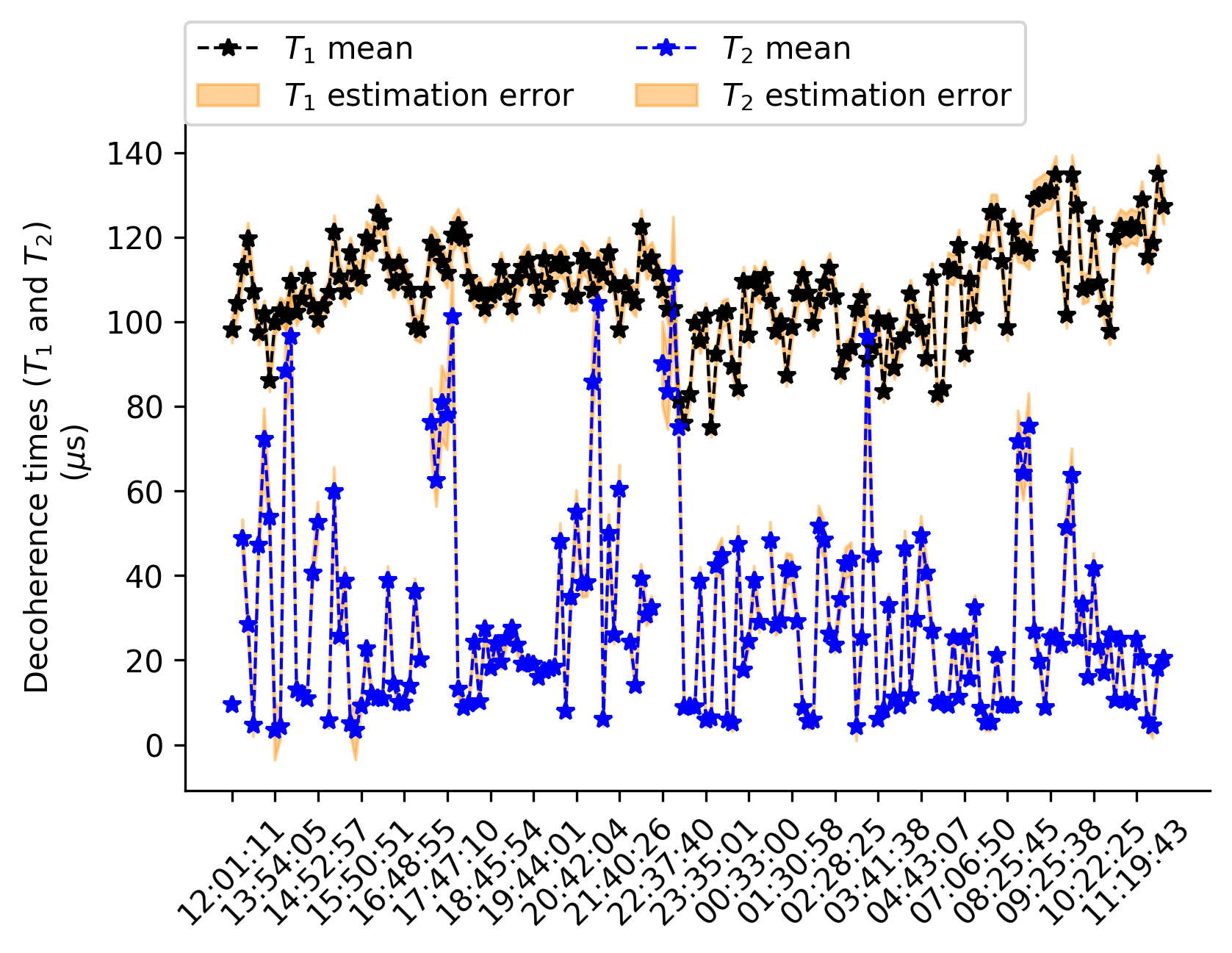}
\caption*{(b)}
\caption{
Estimated $T_1$ and $T_2$ time-series as collected between 12:00 P.M. ET on Sep 12, 2023 and 12:00 P.M. ET on Sep 13, 2023 for (a) qubit 21 and (b) qubit 22.
}
\label{fig:estimated_T1T2_timeseries_qubit_21_22}
\end{figure}

\begin{figure}
\centering
\includegraphics[width=\figurewidth]{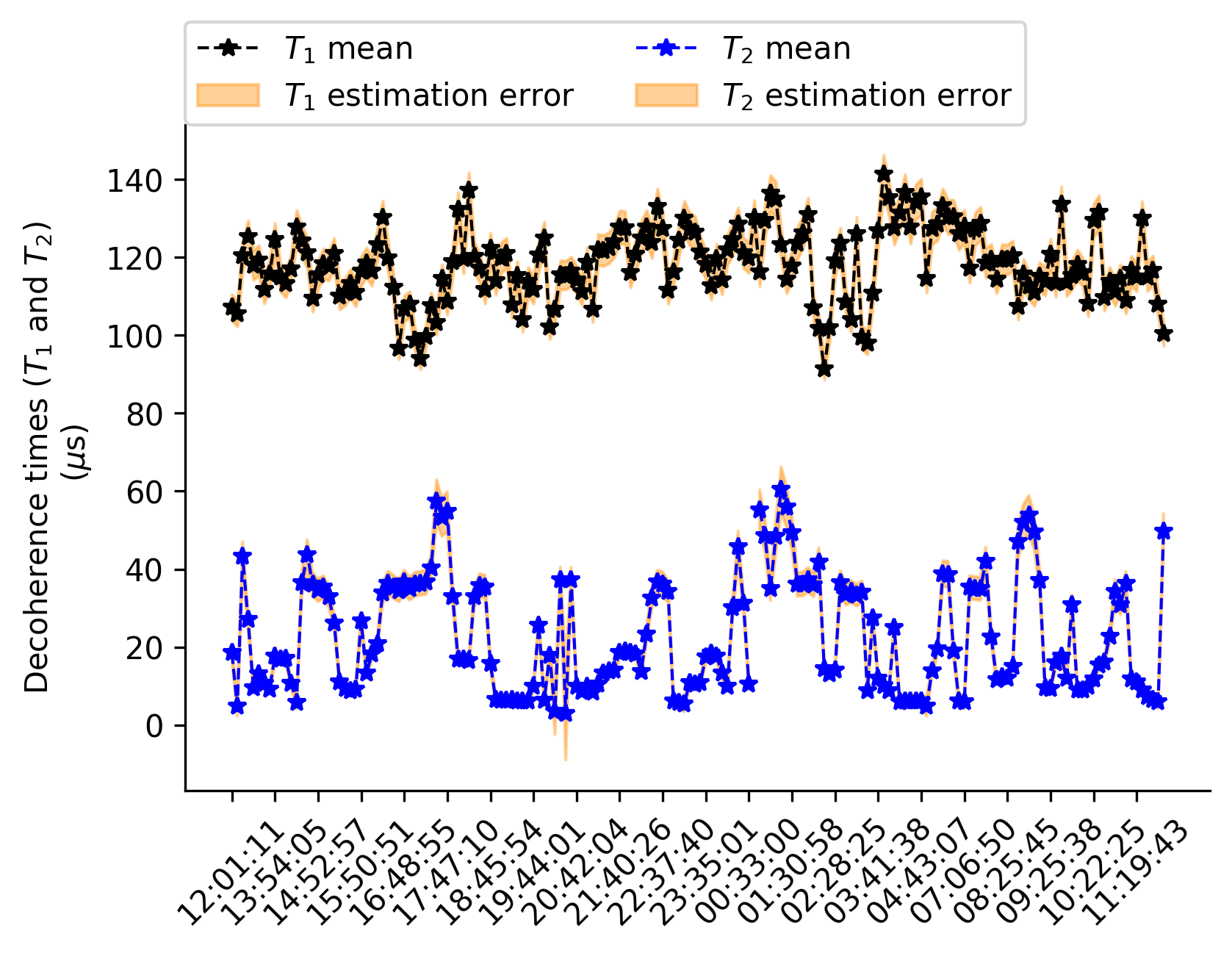}
\caption*{(a)}
\vspace{0.5in}
\includegraphics[width=\figurewidth]{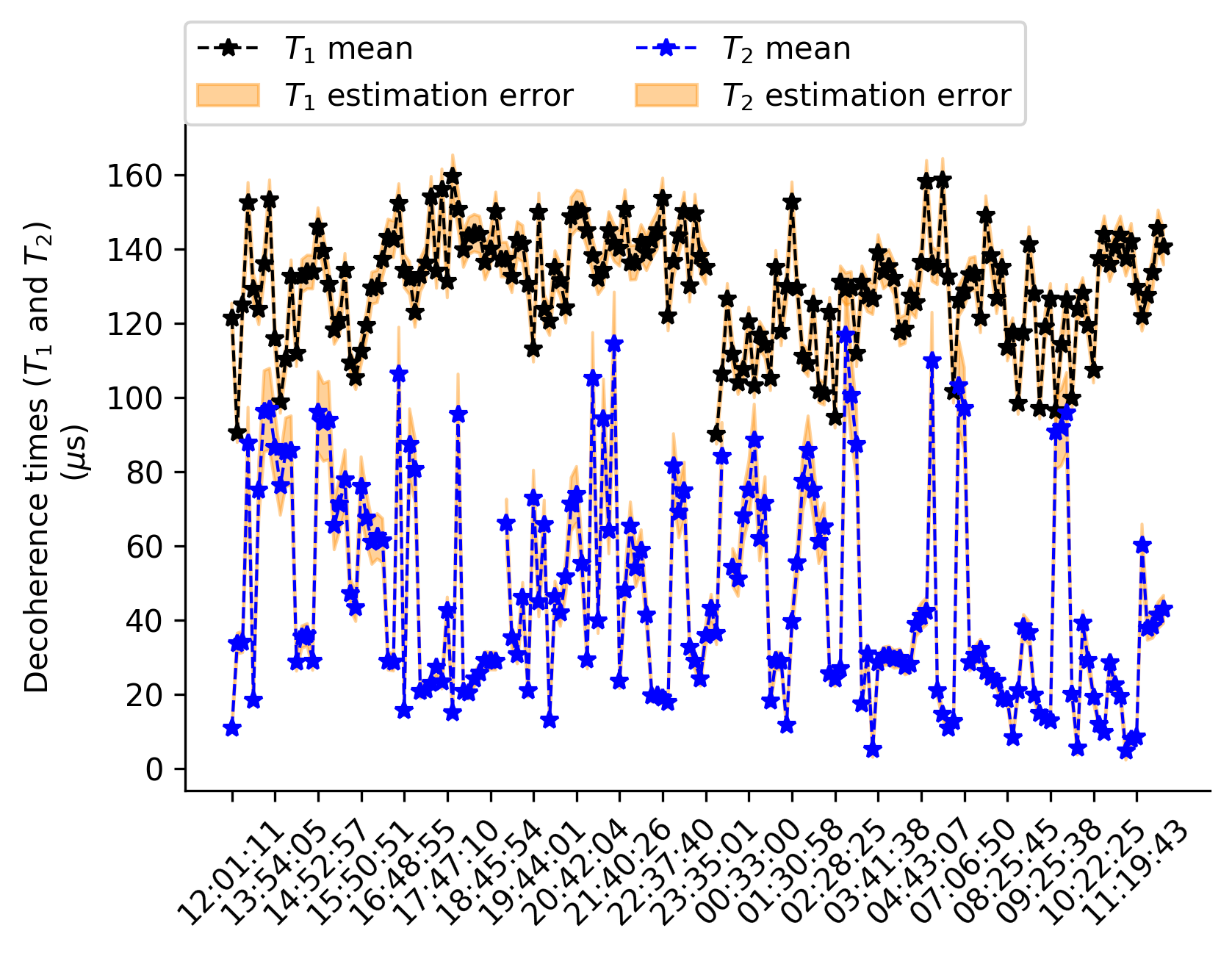}
\caption*{(b)}
\caption{
Estimated $T_1$ and $T_2$ time-series as collected between 12:00 P.M. ET on Sep 12, 2023 and 12:00 P.M. ET on Sep 13, 2023 for (a) qubit 23 and (b) qubit 24.
}
\label{fig:estimated_T1T2_timeseries_qubit_23_24}
\end{figure}

\begin{figure}
\centering
\includegraphics[width=\figurewidth]{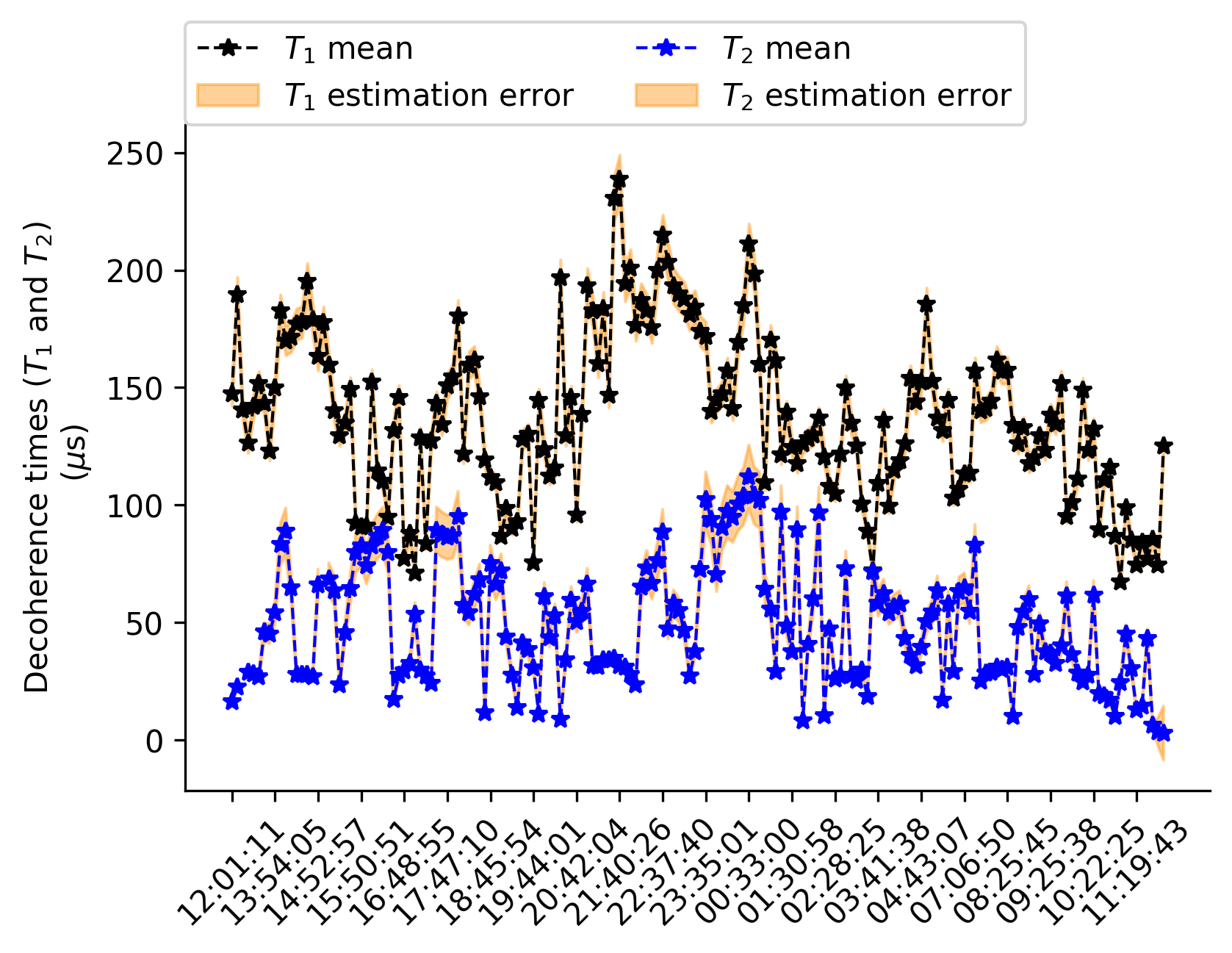}
\caption*{(a)}
\vspace{0.5in}
\includegraphics[width=\figurewidth]{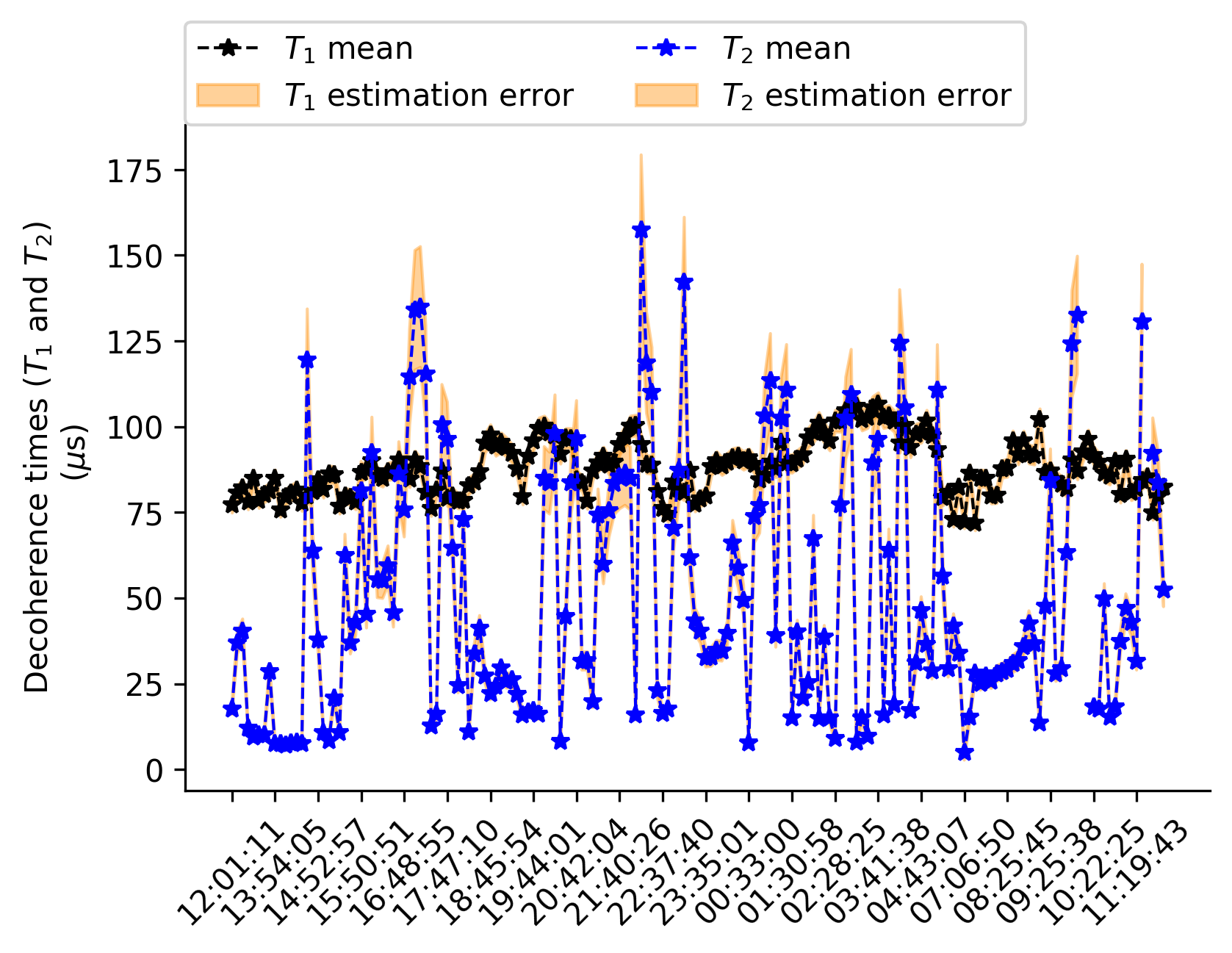}
\caption*{(b)}
\caption{
Estimated $T_1$ and $T_2$ time-series as collected between 12:00 P.M. ET on Sep 12, 2023 and 12:00 P.M. ET on Sep 13, 2023 for (a) qubit 25 and (b) qubit 26.
}
\label{fig:estimated_T1T2_timeseries_qubit_last}
\end{figure}

In terms of the $T_1$ parameter, qubit 15 performed the best, exhibiting a median $T_1$ time of 184 microseconds, while qubit 4 performed the worst with a median $T_1$ time of 74 microseconds. Across all 27 qubits, the median $T_1$ time was 116 microseconds, with a standard deviation of 1.5 microseconds, and the range of $T_1$ times spanned from 32 to 297 microseconds (see Fig.~\ref{fig:T1T2_boxplot}~(a)). 
As for the $T_2$ parameter, qubit 3 demonstrated the highest performance, displaying a median $T_2$ time of 76 microseconds, while qubit 19 had the poorest performance with a median $T_2$ time of 15 microseconds. Across all 27 qubits, the median $T_2$ time averaged 30 microseconds, with a standard deviation of 1.1 microseconds, and the range of $T_2$ times varied from 3 to 191 microseconds (see Fig.~\ref{fig:T1T2_boxplot}~(b)). 
The time-series for all the 27 qubits over the 24 hour period is shown in Figs.~\ref{fig:estimated_T1T2_timeseries_qubit_first}~(b)-~\ref{fig:estimated_T1T2_timeseries_qubit_last}.

\section{Modeling quantum noise channels}
\subsection{Amplitude and Phase damping channel}
The amplitude damping channel $\mathcal{E}^\textrm{AD}(\cdot)$ and de-phasing channel $\mathcal{E}^\textrm{PD}(\cdot)$ are two fundamental sources of quantum de-coherence and information loss in transmons \cite{burnett2019decoherence, klimov2018fluctuations, nielsen2002quantum}. A realistic model for this noise channel, denoted as APD, involves a combination of amplitude damping and de-phasing. Amplitude damping can be described by the Kraus operators $E_0^{\textrm{AD}}$ and $E_1^{\textrm{AD}}$, while phase damping can be described by $E_0^{\textrm{PD}}$ and $E_1^{\textrm{PD}}$, as follows \cite{nielsen2002quantum}:
\begin{equation}
\mathcal{E}^{\textrm{AD}}(\rho) = \sum\limits_{k=0}^1 E_k^{\textrm{AD}}\rho E_k^{\textrm{AD}\dagger} \; ,
\end{equation}
\begin{equation}
E_0^{\textrm{AD}} = 
\begin{pmatrix}
1 & 0\\
0 & \sqrt{1-\gamma}
\end{pmatrix} \; ,
\end{equation}
\begin{equation}
E_1^{\textrm{AD}} = 
\begin{pmatrix}
0 & \sqrt{\gamma}\\
0 & 0
\end{pmatrix} \; ,
\end{equation}
\begin{equation}
\mathcal{E}^{\textrm{PD}}(\rho) = \sum\limits_{k=0}^1 E_k^{\textrm{PD}} \rho E_k^{\textrm{PD}\dagger} \; ,
\end{equation}
\begin{equation}
E_0^{\textrm{PD}} = 
\begin{pmatrix}
1 & 0\\
0 & \sqrt{1-\lambda}
\end{pmatrix} \; ,
\end{equation}
\begin{equation}
E_1^{\textrm{PD}} = 
\begin{pmatrix}
0 & 0\\
0 & \sqrt{\lambda}
\end{pmatrix} \; .
\end{equation}
Here, $\gamma = 1-\exp(-t/T_1)$ and $\lambda = 1-\exp(-t/T_2)$, where $t$ is the time scale of the decoherence process. The relation between $T_1$, $T_\phi$, and $T_2$ was previously discussed in Eq.~(\ref{eq:Bloch-Redfield}). The Kraus decomposition of the combined amplitude and phase damping channel $\mathcal{E}^\textrm{APD}(\cdot)$, valid for a single qubit, can be expressed as $E_0^{\textrm{APD}}$, $E_1^{\textrm{APD}}$, and $E_2^{\textrm{APD}}$. 
\begin{equation}
\mathcal{E}^{\textrm{APD}}(\rho) \coloneqq \mathcal{E}^{\textrm{PD}}\circ \mathcal{E}^{\textrm{AD}} = \sum\limits_{k=0}^2 E_k^{\textrm{APD}} \rho E_k^{\textrm{APD}^\dagger} \; ,
\end{equation}
where, 
\begin{equation}
E_0^{\textrm{APD}} = E_0^{\textrm{PD}} E_0^{\textrm{AD}} =
\begin{pmatrix}
1 & 0\\
0 & \sqrt{[1-\gamma][1-\lambda]}
\end{pmatrix} \; ,
\end{equation}
\begin{equation}
E_1^{\textrm{APD}} = E_0^{\textrm{PD}} E_1^{\textrm{AD}} = 
\begin{pmatrix}
0 & \sqrt{\gamma}\\
0 & 0
\end{pmatrix} \; ,
\end{equation}
\begin{equation}
E_2^{\textrm{APD}} = E_1^{\textrm{PD}} E_0^{\textrm{AD}} =
\begin{pmatrix}
0 & 0\\
0 & \sqrt{[1-\gamma]\lambda}
\end{pmatrix} \; .
\end{equation}
Using the fact that:
\begin{align}
    & E_0^{\textrm{APD}} = \frac{1+\sqrt{1-\lambda-\gamma+\lambda\gamma}}{2}I + \frac{1-\sqrt{1-\lambda-\gamma+\lambda\gamma}}{2}Z \; , \\
    & E_1^{\textrm{APD}} = \frac{\sqrt{\gamma}}{2}X+  \frac{\sqrt{\gamma}}{2}iY \; , \\
    & E_2^{\textrm{APD}} = \frac{\sqrt{\lambda-\lambda\gamma}}{2}I - \frac{\sqrt{\lambda-\lambda\gamma}}{2}Z \; .
\end{align}
the APD channel can be expressed as:
\begin{equation}
\begin{split}
\mathcal{E}^{\textrm{APD}}(\rho) =& \frac{2-\gamma+2\sqrt{1-\lambda-\gamma+\lambda\gamma}}{4}\rho
+\frac{\gamma}{4} X\rho X
-\frac{\gamma}{4} Y\rho Y\\
&+\frac{2-\gamma-2\sqrt{1-\lambda-\gamma+\lambda\gamma}}{4} Z \rho Z\\
&+\frac{\gamma}{4} I \rho Z
+\frac{\gamma}{4} Z \rho I
-\frac{\gamma}{4i} X \rho Y
-\frac{\gamma}{4i} Y \rho X \; ,
\end{split}
\end{equation}
where $\lambda$ and $\gamma$ are the APD parameters, and $I$, $X$, $Y$, and $Z$ are the Pauli matrices.

\subsection{Depolarizing channel}
Next, the depolarizing channel is a common type of quantum noise channel. 
It works by randomly applying one of the Pauli operators $(\mathds{X}, \mathds{Y}, \mathds{Z})$ to the quantum state with a certain (but equal) probability, causing a loss of information about the state. For a qubit:
\begin{equation}
\mathcal{E}_D(\rho) = (1 - \xrm) \rho + \xrm \frac{\mathds{I}}{d},
\end{equation}
where $\rho$ is the input quantum state, $\xrm$ is the probability of noise occurring, $\mathds{I}$ is the identity operator, and $d$ is the dimension of the Hilbert space ($d=2$ for a qubit).

\subsection{Pauli noise channel}
The Pauli noise channel is a generalization of the depolarizing channel. 
It encompasses the effect of the bit-flip ($\mathds{X}$), phase-flip ($\mathds{Z}$), and bit-phase-flip ($\mathds{Y}$) errors with unequal probabilities. 

The impact of Pauli noise on quantum information encoded in an $n$-qubit register is shown below:
\begin{equation}
\mathcal{E}_\xrm(\rho) = \sum\limits_{i=0}^{N_p-1} \xrm_i P_i(n) \rho P_i(n)^\dagger
\end{equation}
where $N_p$ denotes the total number of Pauli coefficients and $P_i(n)$ represents $n$-qubit Pauli operators. The coefficients contribute to a simplex:
\begin{equation}
\sum\limits_{i=0}^{N_p-1}\xrm_i=1, \;\;\;\; \xrm_i \geq 0
\end{equation}

Pauli noise channel is widely used in quantum error correction because it is a simple and natural model for random quantum noise \cite{emerson2007symmetrized, silva2008scalable, martinez2020approximating}. It is a well-understood and easily implementable noise model that can simulate a variety of realistic physical processes that lead to quantum errors, such as dephasing, amplitude damping, and phase-flip errors. Additionally, the Pauli noise channel is mathematically tractable and can be efficiently simulated, making it a useful tool for developing and testing quantum error correction protocols. It can be used to estimate the average fidelity of a quantum gate subject to the original APD channel and identify codes that work for the APD channel \cite{silva2008scalable}. The Pauli noise channel, although not a completely general noise model, still manages to model many practical situations. It is widely used because of two reasons: (a) it is efficiently simulatable on a classical computer (per the Gottesman-Knill theorem) and (b) when used as a proxy for physically accurate noise models (such as the amplitude and phase damping noise) which are not efficiently simulatable on a classical computer, it still manages to preserve important properties like entanglement fidelity \cite{horodecki1999general}.

Remarkably, Pauli twirling can map\cite{eggeling2001separability, dankert2009exact, magesan2008gaining} a more complex quantum noise channel (e.g. APD) to a simple Pauli channel while preserving certain features such as the average channel fidelity and the entanglement fidelity \cite{horodecki1999general}. 
Consider a single-qubit amplitude and phase damping channel (APD) \cite{khatri2020information}. 
Upon Pauli twirling \cite{cai2020mitigating}:
\begin{align}
\mathcal{E}_\textrm{twirl}(\rho) =& \frac{1}{4} \sum\limits_{A \in \{I, X, Y, Z\}} A^\dagger \mathcal{E}_\text{APD} \left( A\rho A^\dagger \right)A \\
=& \sum\limits_{k=0}^3 c_k \sigma_k \rho \sigma_k
\end{align}
an APD channel becomes a Pauli noise channel. 
Here, $\{ \sigma_k \}_{k=0}^3 = \{I, X, Y, Z\}$ are the Pauli matrices. 
Thus, the coefficients of the Pauli noise channel are functions of the coefficients of the original APD channel, which in turn are functions of the decoherence times $T_1$ and $T_2$ \cite{etxezarreta2021time}: 
\begin{align}
c_1 = c_2 = & \frac{1}{4}\left[1-\exp\left(-t/T_1\right)\right]\\
c_3 =& \frac{1}{4}\left[1-\exp\left(-t/T_2\right)\right]\\
c_0 =& 1- (c_1 + c_2 + c_3)
\label{eq:depol_coeffs}
\end{align}
This directly links the estimation of Pauli channels to the decoherence data collected in Sec.~\ref{sec:decoherence_data}.

\clearpage
\vspace{0.5in}
\vspace{0.5in}
\vspace{0.5in}
\begin{figure}
\centering
\includegraphics[width=\figurewidth]{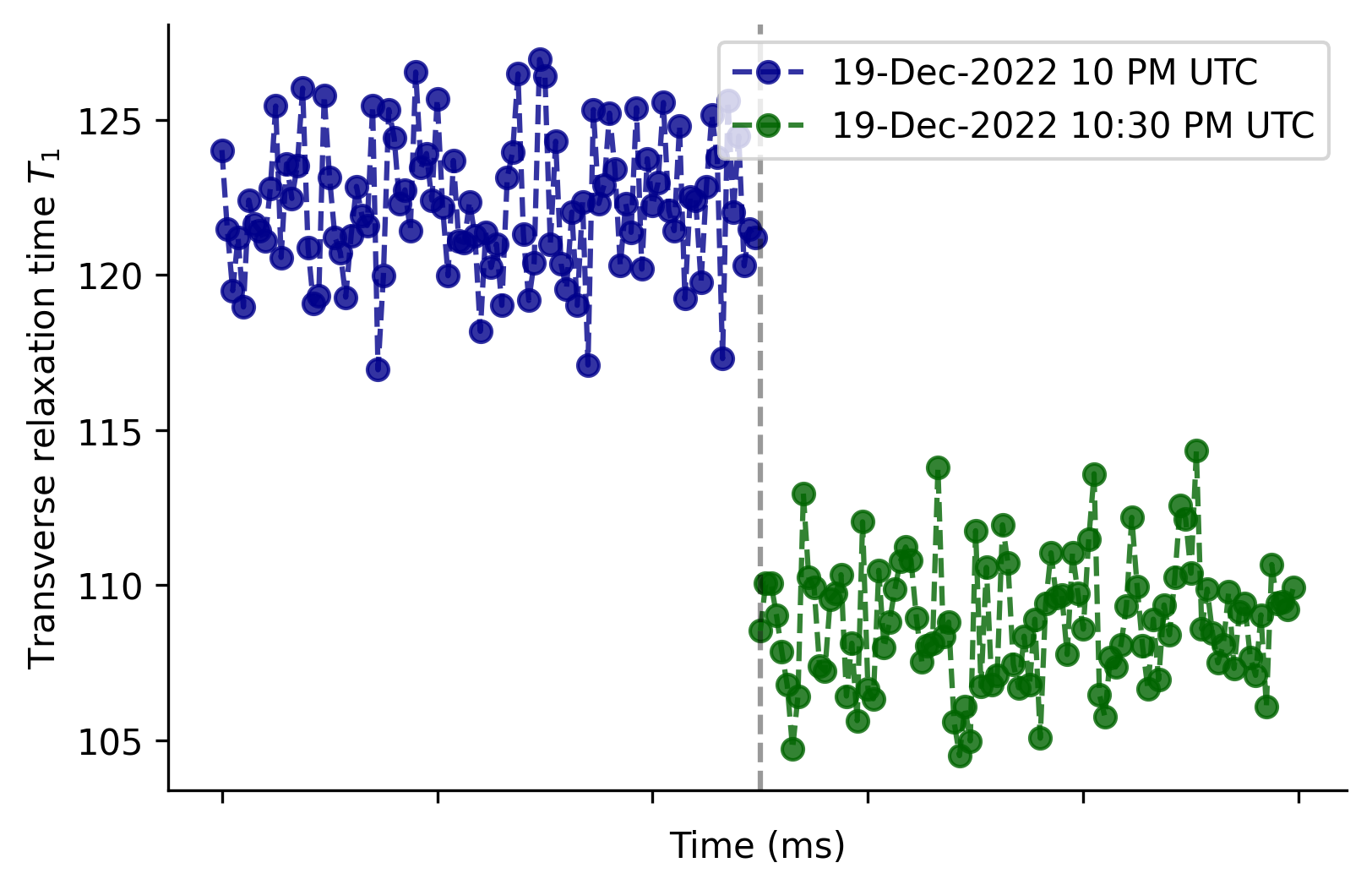}\\ \vspace{0.5in}
\includegraphics[width=\figurewidth]{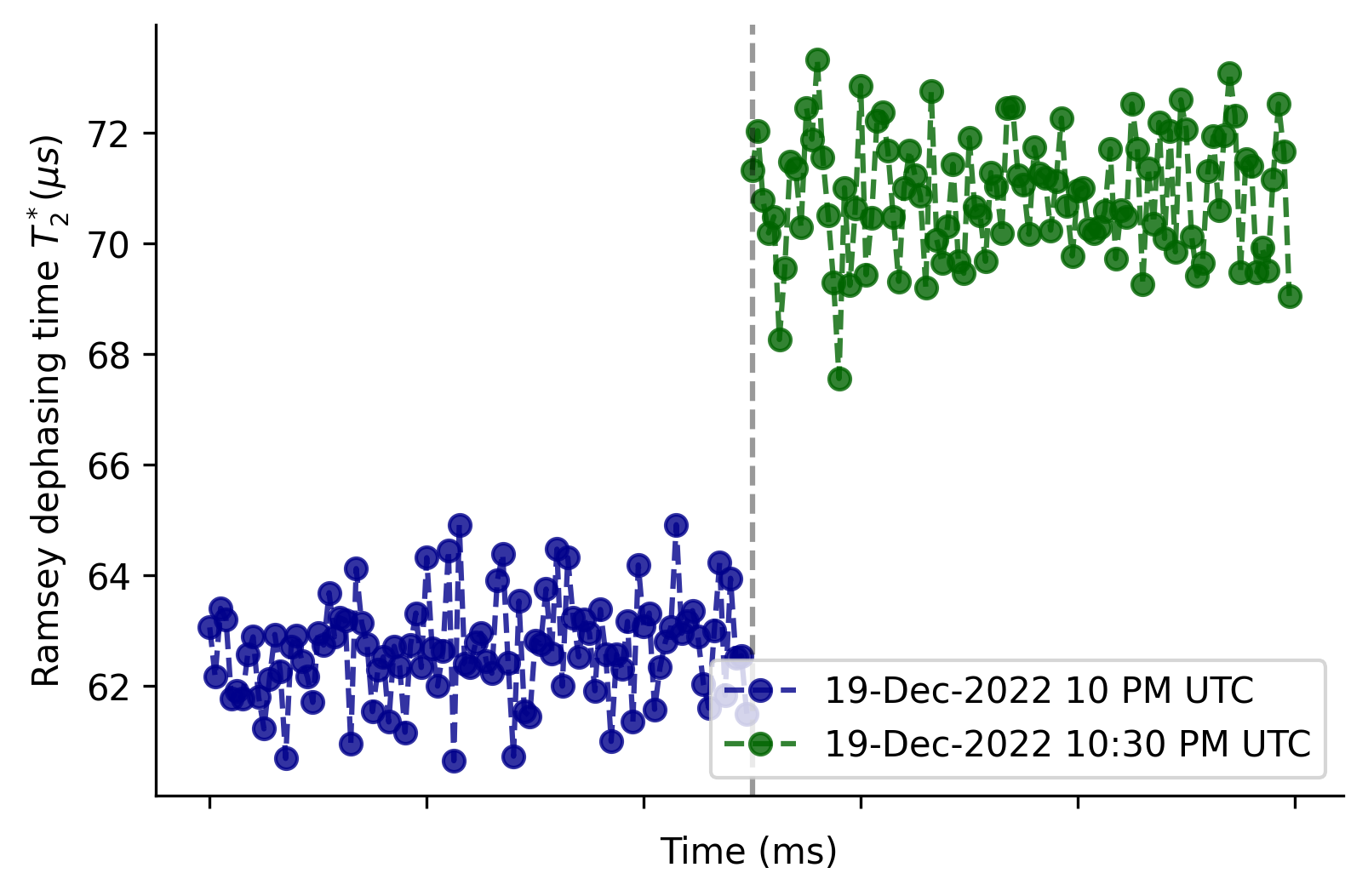}
\caption{This figure presents non-stationary temporal dynamics of decoherence for a qubit on IBM's belem device. The top figure shows $T_1$ relaxation time series for qubit 0, where two datasets were collected for 5 ms each on the same day, separated by a vertical line. The blue dataset varies between 116-126$\mu$s with a mean of 122 $\mu$s and a standard deviation of 2 $\mu$s, while the green dataset varies between 104-114$\mu$s with a mean of 108 $\mu$s and a standard deviation of 2 $\mu$s. The bottom figure displays Ramsey dephasing time ($T_2$) series for qubit 0, where two datasets were collected for 5 ms each on the same day, separated by a vertical line. The blue dataset varies between 67-73$\mu$s with a mean of 70 $\mu$s and a standard deviation of 1 $\mu$s, while the green dataset varies between 60-64$\mu$s with a mean of 62 $\mu$s and a standard deviation of 1 $\mu$s, collected around different times on the same day. The data shows significant non-stationarity in decoherence values over a 30-minute interval.}
\label{fig:T1T2_temporal}
\end{figure}
\vspace{0.5in}
\vspace{0.5in}
\vspace{0.5in}
\vspace{0.5in}
\vspace{0.5in}
\vspace{0.5in}
\vspace{0.5in}
\vspace{0.5in}
\vspace{0.5in}
\vspace{0.5in}
\vspace{0.5in}
\vspace{0.5in}
\vspace{0.5in}
\vspace{0.5in}
\vspace{0.5in}

\removefigs
\chapter{Performance evaluation framework}\label{ch:statistical_taxonomy}
Performance evaluation of noisy quantum computing in crucial for several reasons. 
Firstly, quantum computing is still in its early stages of development \cite{roadmap, acin2018quantum, hughes2004quantum}, and understanding the sources of errors and noise\cite{khatri2020information} is vital to improve the performance of quantum computers. 
By conducting rigorous performance evaluation, researchers can identify and quantify the various sources of noise, such as decoherence, gate errors, and readout errors. 
Secondly, this understanding is essential for developing error mitigation techniques \cite{bharti2022noisy}, which are necessary for scaling up small-scale quantum computations. 
Thirdly, reproducibility of results from quantum computing is critical. Rigorous performance evaluation ensure that experiments can be replicated by other researchers, contributing to validation and verification of quantum algorithms.

However, the task is not simple as the same complexity that gives quantum technology an advantage over classical computing also hinders its rigorous checking \cite{kliesch2021theory}. 
Reasons include the inherent randomness in quantum results due to the Born rule, error accumulation without clear source attribution, the curse of dimensionality \cite{vapnik1999nature}, and the inability to step-through program execution in quantum circuits \cite{carrasco2021theoretical}.

Also, the diverse range of terms encountered in quantum computing today can blur the distinctions between them, making it challenging to appreciate their nuanced differences. 
Examples include verification\cite{harper2020efficient, gheorghiu2019verification, PhysRevA.101.042315} (which pertains to ensuring correct transpilation), validation\cite{10.1371/journal.pone.0206704, 10.1109/qcs54837.2021.00013} (which can have two connotations: (a) validating correctness of output by comparison to theory or concurrent classical simulation (akin to accuracy), or validating the quantum nature of a device), benchmarking\cite{mccaskey2019quantum, BlumeKohout2020volumetricframework, 10.1038/s41534-022-00628-x} (which involves assigning a performance measure, often a simple scalar number, to a quantum processor/ subsystem/ subroutine, with reproducibility as a key defining characteristic), accreditation \cite{ferracin2019accrediting, ferracin2021experimental} and certification \cite{kliesch2021theory}.

\section{Distance measures}
In this chapter, our focus is on the development of a systematic performance evaluation framework for the outcomes from noisy quantum computers. To accomplish this, it is necessary to compare probability distributions. Various options exist for quantifying the distance between high-dimensional densities, including Hellinger distance\cite{beran1977minimum}, total variation distance\cite{rahimian2019identifying}, Kolmogorov-Smirnov statistic\cite{lilliefors1967kolmogorov}, Mahalanobis distance\cite{xiang2008learning}, Wasserstein metric\cite{cuesta1989notes}, Levy-Prokhorov metric\cite{fisher1969distinguishability}, and non-metric divergence measures like Kullback-Leibler divergence\cite{van2014renyi}, Jensen-Shannon divergence\cite{nielsen2020generalization}, Renyi's divergence\cite{van2014renyi} and Tsallis divergence\cite{vigelis2019properties} for quantum states. Metric measures are advantageous because they obey the triangle inequality, enabling rigorous comparisons, while non-metric measures (often called divergence) are useful for obtaining performance bounds in specific problem settings. However, all distance measures suffer from the curse of dimensionality\cite{vapnik1999nature}, which results in exponentially increasing resource requirements to accurately represent information as the number of qubits in the quantum system grows \cite{10.1103/physreva.84.032120, 10.3934/fods.2019007}.

\subsection{Hellinger distance}
In our work, we primarily employ the Hellinger distance. The Hellinger distance between two probability distributions $f_X(\xrm)$ and $f_Y(\xrm)$ for the random variables $X$ and $Y$ is defined by:
\begin{equation}
H( f_X, f_Y ) = \sqrt{1-BC(f_X, f_Y)},
\label{eq:hellinger_unmodified}
\end{equation} 
where the Bhattacharyya coefficient BC is
\begin{equation}
BC(f_X, f_Y) = \int\limits_{\xrm} \sqrt{f_X(\xrm) f_Y(\xrm)} d\xrm.
\end{equation}
The Hellinger distance provides a practical and meaningful approach to measuring the similarity of distributions. Firstly, it operates directly on observed data, eliminating the need to compute intermediate abstractions like entropy. Secondly, the Hellinger distance is easy to interpret and apply in practical scenarios. Thirdly, it can be easily extended to quantum states through Tsallis divergence \cite{vigelis2019properties}. Lastly, the Hellinger distance is proportional to the Fisher information\cite{toth2014quantum}, which quantifies the partial knowledge a density carries about some unknown. 

In particular, suppose $\xrm$ is a realization of the random noise parameter $X$ drawn from the time-varying distribution $f_X(\xrm; t)$, where $t$  denotes time. Let, 
\begin{equation}
H_{X}(t_1, t_2) = H(f_X(t_1), f_X(t_2)),
\label{eq:hxy}
\end{equation}
measure the distance between the densities of $X$ at time $t_1$ and $t_2$. 

For example, the Hellinger distance between two beta distributions (used for SPAM and CNOT fidelity characterizations) is given by:
\begin{equation}
\begin{split}
f(\xrm; \alpha_1, \beta_1)  &= \frac{\xrm^{\alpha_1-1}(1-\xrm)^{\beta_1-1}}{\text{Beta}(\alpha_1, \beta_1)}, \;\;\;\;
g(\xrm; \alpha_2, \beta_2) = \frac{\xrm^{\alpha_2-1}(1-\xrm)^{\beta_2-1}}{\text{Beta}(\alpha_2, \beta_2)}\\
d_H &= \sqrt{1 - \int\limits_0^1 \sqrt{f(\xrm; \alpha_1, \beta_1) g(\xrm; \alpha_2, \beta_2)} d\xrm} 
= \sqrt{1-\frac{\text{Beta}(\alpha_1+\alpha_2-1, \beta_1+\beta_2-1)}
{\text{Beta}(\alpha_1, \beta_1)\text{Beta}(\alpha_2, \beta_2)}}
\end{split}
\end{equation}
while that between two gamma distributions (used for duty cycle characterization) is given by:
\begin{equation}
\begin{split}
f(\xrm; m_1, \alpha_1)  &=  \frac{1}{\Gamma(m_1)} \xrm^{m_1-1}\alpha_1^{m_1} e^{-\alpha_1 \xrm}, \;\;\;\;
g(\xrm; m_2, \alpha_2) = \frac{1}{\Gamma(m_2)} \xrm^{m_2-1} \alpha_2^{m_2} e^{-\alpha_2 \xrm}\\
d_H &= \sqrt{ 1 - \int\limits_0^1 \sqrt{f(\xrm; n, \alpha) g(\xrm; m, \beta)} d\xrm } 
= \sqrt{1-\frac{\alpha_1^{m_2}\alpha_2^{m_1}}{(\alpha_1+\alpha_2)^{m_1+m_2-1}}\frac{\Gamma(m_1+m_2-1)}{\Gamma(m_1)\Gamma(m_2)}}
\end{split}
\end{equation}
For non-standard distributions $f(x)$ and $g(x)$ where an analytical closed form solution is not available, we numerically integrate the empirical distributions using the standard trapezoidal method:
\begin{equation}
\begin{split}
d_H^2-1 &= \int\limits_a^b \sqrt{f(x)g(x)}dx = \lim_{n\rightarrow\infty} \sum\limits_{i=0}^{n-1}
\left[\frac{1}{2}\left[f(a-ih/2)g(a-ih/2)+ f(a+ih/2)g(a+ih/2)\right]\right]^{1/2}h\\
\end{split}
\end{equation}
where $h = \frac{b-a}{n}$.

Despite its ease of interpretation, the Hellinger distance scales exponentially in the number of noise parameters. 
This has the effect that even small changes in a distribution yield large changes in the distance value. 
This is called the curse of dimensionality. To see this, consider $d$ independent and identically distributed noise parameters $\{\xrm_1, \cdots, \xrm_d\}$, whose marginals are given by $f_{X_i}(\xrm;t)$. 
Let $h$ be the Hellinger distance between the marginals at time $t_1$ and $t_2$. Thus, 
\begin{equation}
H_{X_i}(t_1, t_2)=h \;\;\;\; \forall i
\end{equation}
Since the parameters are independent:
\begin{equation}
\begin{split}
\log \left( 1 - H_{X}^2 \right) &= -d |\log (1-h^2)|\\
\Rightarrow H_{X} &= \sqrt{1-\exp\left[ -d |\log (1-h^2)|\right]}.
\end{split}
\label{eq:dimcurse}
\end{equation}
Thus the distance approaches $1$ quickly as the number of dimensions increases.

A more sensitive measure can be defined using $H_{\textrm{avg}}$, defined as the average over the distances for the $d$ univariate $(X_k)$ marginal distributions:
\begin{equation}
H_{\textrm{avg}}(t_1, t_2) = \frac{1}{d} \sum\limits_{k=1}^d H_{X_k} (t_1, t_2).
\label{eq:hellinger_avg}
\end{equation}
When the joint distributions are time-invariant, then the marginals must also be time-invariant, resulting in a small average value for $H_{\textrm{avg}}$. This test is more sensitive as it mitigates the curse of dimensionality and offers higher dispersion for improved calibration. 

Another sensitive approach is to normalize the distance relative to the dimensionality $d$ of the distribution:
\begin{equation}
H_\textrm{normalized}(t_1, t_2) = \sqrt{ 1 - BC^{1/d} }.
\label{eq:hellinger_normalized}
\end{equation}
We refer to this statistic as the normalized Hellinger distance (note that although we call it distance, this statistic is not technically a metric as it does not satisfy the triangle inequality). 
%
\subsection{Moment-Based Distance}
We also developed a new disance measure as part of our research which we call Moment-Based Distance (MBD). The key advantage of MBD is its ability to incorporate the geometric shape of the underlying noise distribution while still being a metric. Thus, it takes into account higher order effects like kurtosis and skewness. Specifically, we define the moment-based metric ($d$) between two histograms ($f$ and $g$) based on the equality of their moments.
\begin{equation}
d(f,g) = \sum\limits_{m=0}^{\infty} S_m(f, g)
\end{equation}
where
\begin{equation}
S_{m}(f,g) = \frac{1}{(m)!}\int\limits_a^{b} \left| \left( \frac{x}{\gamma} \right)^{m}(f(x) - g(x))\right|dx
\end{equation}
and,
\begin{equation}
\gamma = \max( |a|,|b|)\\
\end{equation}
for a bounded real variable $x$. Here, $a$ and $b$ are the minimum and maximum values of $x$ and can be derived from theoretical considerations (e.g., when the random variable is a probability then $\gamma=1$) or from empirical histogram data.

The moment-based-distance $d(f,g)$ satisfies the following properties:
\begin{enumerate}
\item $d(f,g) \geq 0$ follows from the definition of $d$.
\item $d(f, g) = d(g, f)$ follows from the definition of $d$.
\item $d(f, g) = 0$ iff $f(x) = g(x)$.\\
Proof: If $f(x) = g(x)$, then $d = 0$ because $S_m = 0$ for every $m$. Conversely, if $d = 0$, then $S_{m} = 0$ for all $m$.If $S_{m} = 0$, then 
for all $x$, the integrand  must satisfy
\begin{equation*}
\left| \left(\frac{x}{\gamma}\right)^m (f-g)\right| = 0
\end{equation*}
As $(x)^{m} \neq 0$ for all $x$, it must be that $|f(x)-g(x)| = 0$ for all $x$ and, hence, $f(x) = g(x)$.
\hfill $\tiny\blacksquare$
\item $d(f,g) \leq d(f, h) + d(h, g)$ \\
Proof: For every $m$,
\begin{equation*}
\begin{split}
S_m(f,g) =& \int\limits_a^{b} \left| \left( \frac{x}{\gamma} \right)^{m}\frac{1}{m!}(f(x) - g(x))\right|dx\\
=& \int\limits_a^{b} \bigg| \left( \frac{x}{\gamma} \right)^{m}\frac{1}{m!}(f(x) - h(x) \\
& \hspace{2.0cm}  + h(x) - g(x))\bigg|dx\\
\leq & \int\limits_a^{b} \left| \left( \frac{x}{\gamma} \right)^{m}\frac{f(x) - h(x)}{m!}\right|dx\\
& +\int\limits_a^{b} \left| \left( \frac{x}{\gamma} \right)^{m}\frac{h(x) - g(x)}{m!}\right|dx\\
\leq& S_m(f,h) + S_m(h,g)\\
\end{split}
\end{equation*}
and whence the sum satisfies the inequality as well.\hfill $\blacksquare$
\item The series $d = S_0 + S_1 + S_2 + \cdots$ converges.\\
Proof: The distance $d$ converges if, after some fixed term, the ratio of each term to the preceding term is less than some quantity $r$, which is itself numerically less than unity. If $S_{m+1} < S_m$ for all $m \geq 1$, then
\begin{equation*}
\begin{split}
d =& S_0 + S_1 + S_1 \cdot \frac{S_2}{S_1} + S_1\cdot \frac{S_2}{S_1}\cdot \frac{S_3}{S_2} + \cdots\\
<& S_0 + S_1(1+r+r^2 +r^3 + \cdots)\\
=& S_0 + \frac{S_1}{1-r}  \textrm{ since } r<1\\
\end{split}
\end{equation*}
To prove that $S_{m+1} < r S_m$ for $m\geq 1$, we proceed as follows:
\begin{equation*}
\begin{split}
S_{m+1} &= \int\limits_a^{b} \left| \left( \frac{x}{\gamma} \right)^{m+1}\frac{1}{(m+1)!}(f(x) - g(x))\right|dx \\
&= \int\limits_a^{b} \left| \frac{x}{\gamma(m+1)}  \right|\left|\left( \frac{x}{\gamma} \right)^{m}\frac{f(x) - g(x)}{m!}\right|dx\\
&\leq \left| \frac{x}{\gamma} \right|_{max} \frac{1}{m+1} S_m\\
&\leq \frac{1}{m+1}S_m \textrm{ since }  \left| \frac{x}{\gamma} \right|_{max} = 1 \\
&\leq \frac{1}{2}S_m \textrm{ since } m \geq 1
\end{split} 
\end{equation*}
where $|x|_{max}$ is the maximum of $x$. \hfill $\blacksquare$
\end{enumerate}

An important consequence of the latter convergence property is that the moment-based distance satisfies the practical requirement that lower-order moments contribute more than higher-order moments to the distance (for $m>1$). This proves essential to our subsequent use of the moment-based distance below, as we rely on the approximate distance defined to order $n$ as
\begin{equation}
d_{n} = \sum_{m=0}^{n}{S_m}
\end{equation}

We next present a series of simulation studies to develop intuition for how the moment-based distance behaves in the presence of both stable and unstable distributions. In particular, we will show that moment-based distance is small but non-zero for distributions that are similar but not identical, while such deviations grow with dissimilarity. For our studies, we computed the distance of 10 different distributions with respect to a reference distribution. Table~\ref{table:dis_distance} summaries the list of tests as well as their moment-based distance from the reference normal distribution $\mathcal{N}(\mu, \sigma)$. 
\begin{table}[htbp]
\renewcommand{\arraystretch}{1.3}
\caption{Moment-based distance by Distribution}
\label{table:dis_distance}
\centering
\begin{tabular}{|l|c|c|c|}
\hline
Distribution & $d_4$ & $d_{20}$ & Error(\%)\\
\hline
$N(\mu, \sigma)$ & 0.00000 & 0.00000 & NA\\
\hline
$N(\mu+\Delta, \sigma)$ & 2.70868 & 2.70876 & -0.00289\\
\hline
$N(\mu, 2\sigma)$ & 0.83252 & 0.83253 & -0.00104\\
\hline
$N(\mu, 4\sigma)$ & 1.47301 & 1.47304 & -0.00180\\
\hline
$N(2\mu, \sigma)$ & 2.93489 & 2.93520 & -0.01033\\
\hline
$N(\mu, 1.5\sigma)$ & 0.49215 & 0.49216 & -0.00091\\
\hline
$N(1.01\mu, \sigma)$ & 0.11739 & 0.11740 & -0.00079\\
\hline
$Skewed Normal(\mu, 2\sigma)$ & 0.80887 & 0.80888 & -0.00140\\
\hline
$Gumbel(\mu, 2\sigma)$ & 0.95131 & 0.95134 & -0.00246\\
\hline
\end{tabular}
\end{table}
For testing purpose, the parameters are $\mu = 0.4, \Delta = 0.2$ and $\sigma = 0.04$. We note that, as expected, the distribution `closest' to $\mathcal{N}(\mu, \sigma)$ is $\mathcal{N}(1.01\mu, \sigma)$ and the `farthest' are $\mathcal{N}(2\mu, \sigma)$ and $\mathcal{N}(\mu+2\Delta, 2\sigma)$.

\begin{figure}[htbp]
\centering
\includegraphics[width=\figurewidth]{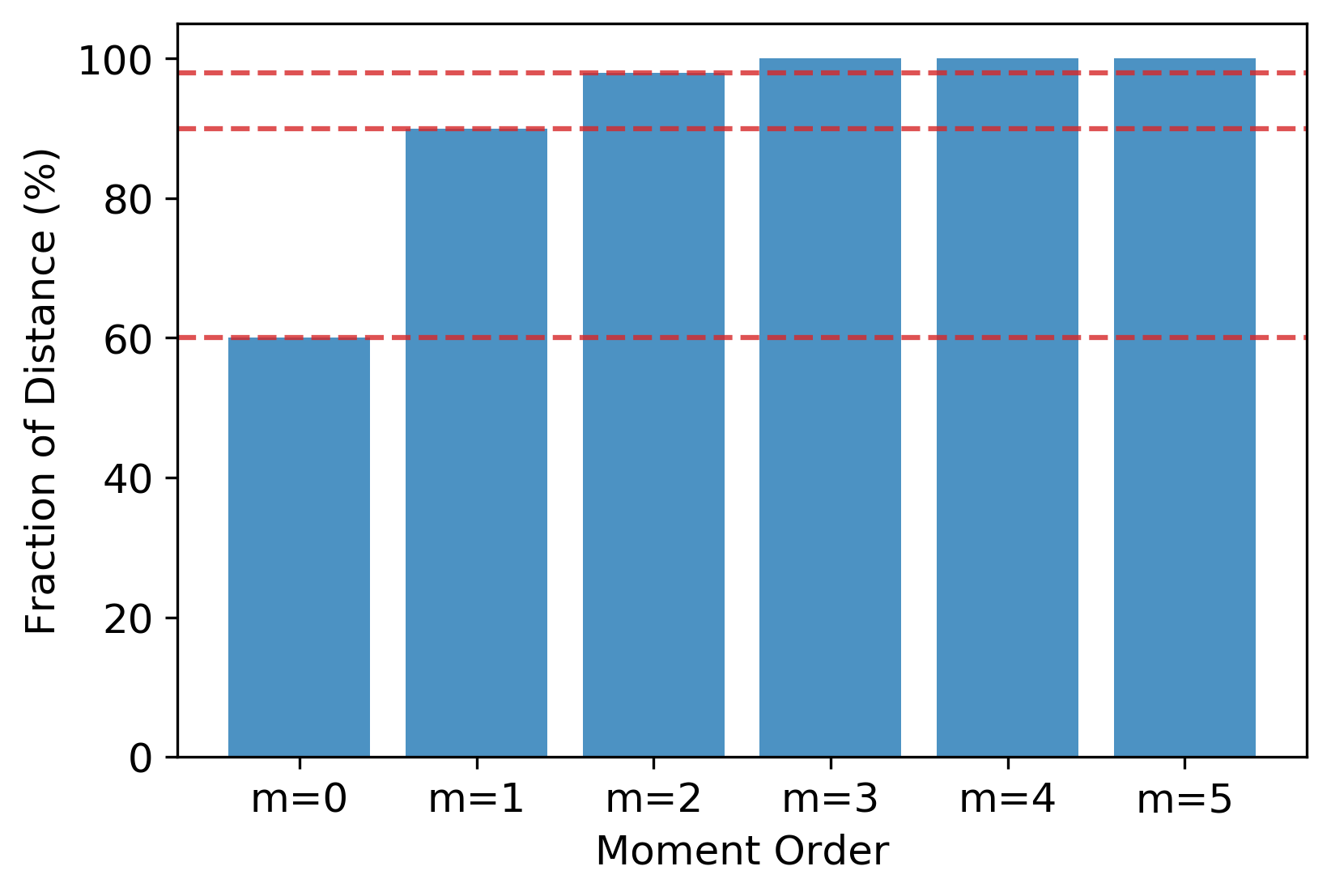}
\caption{Contribution to moment based distance ($d$) from increasing moment orders. The graph shows the results of comparing two normal distributions: $\mathcal{N}_1(\mu = \mu_0, \sigma=\sigma_0)$ and $\mathcal{N}_2(\mu = 2\mu_0, \sigma=2\sigma_0)$ where $\mu_0 = 40$ and $\sigma_0=4$.}
\label{fig:dnN_unequal_relative}
\end{figure}

We next study how the order of the series $d_n$ increases the accuracy of the distance measured. In our simulation studies of well-defined distributions, we find that $d_n$ converges for $n = 4$ when the distributions are sufficiently dissimilar. As shown in Fig. ~\ref{fig:dnN_unequal_relative}, the relative contributions of each $S_m$ to $d_n$ decreases with increasing $m$ as expected from the convergence property. 
Thus, $m=0$ accounts for about $60\%$ of the total distance while $m=1$ accounts for $90\%$ and $m=2$ reaches $98\%$. For $m=4$, $d_{m}$ is nearly $100\%$ of the $d_\infty$. Consequently, we will consider $m=4$ sufficient to accurately characterize the moment-based distance for the remainder of our analysis. This is certainly an approximation in the sense that two histograms which start to differ only after the fourth order moment will be erroneously classified as same. Is $d_4$ still a valid \textit{distance metric}? Yes. A glance at the proofs will reveal that properties (1) to (4) are still satisfied when we truncate the d series at a finite m (say m=4). Moreover, it converges too (i.e. Property (5) is satisfied too) because a finite number of terms (in this case 5 terms) is by definition convergent when the individual terms are finite. The latter is true because each $S_m$ is bounded between finite $a$ and $b$ as per Equation (2).

\begin{figure}[htbp]
\centering
\includegraphics[width=\figurewidth]{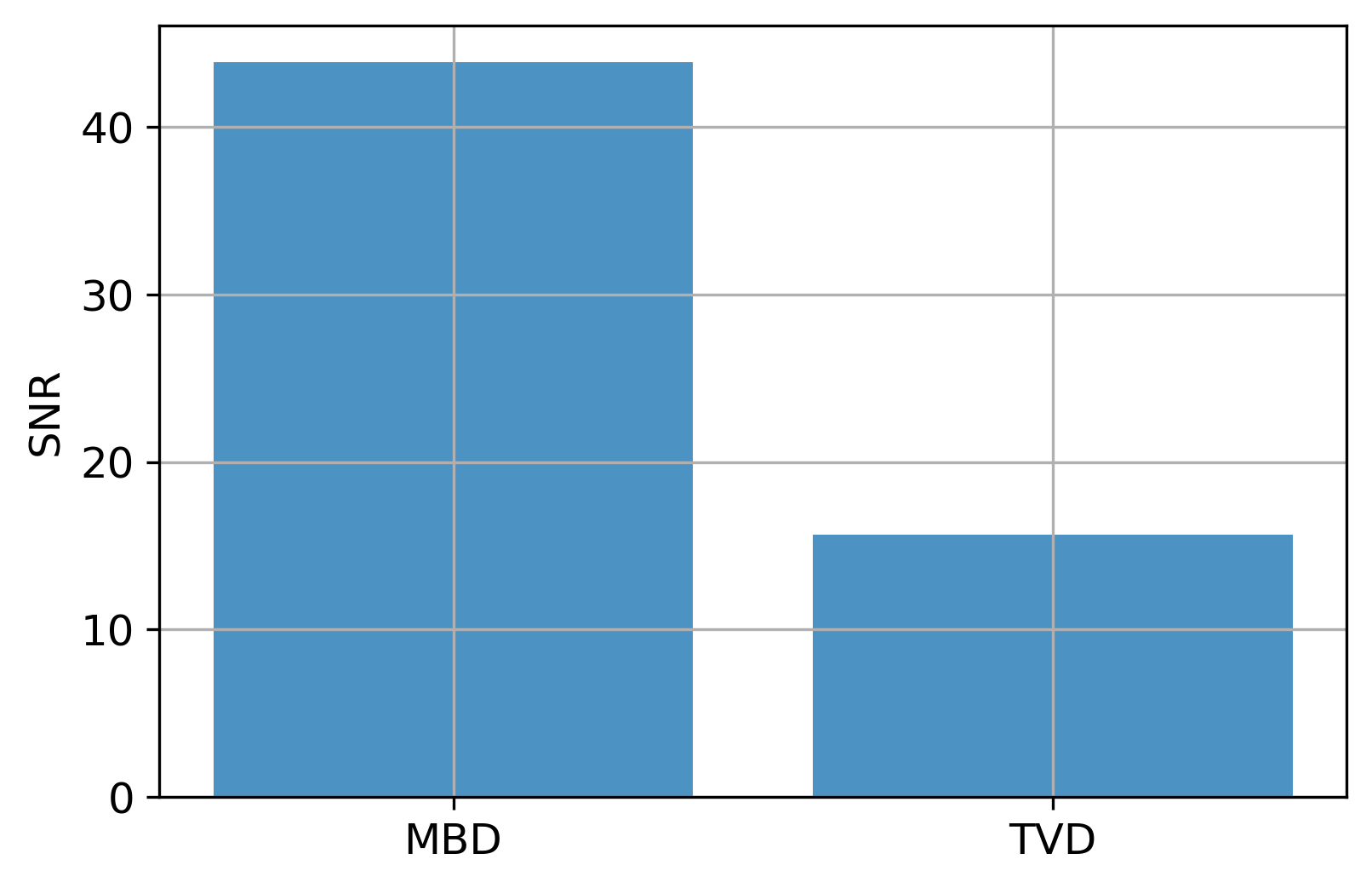}
\caption{Simulated signal-to-noise ratios of the moment-based distance and total variation distance for two normal distributions of varying width.}
\label{fig:MBDstability}
\end{figure}

\begin{figure}[htbp]
\centering
\includegraphics[width=\figurewidth]{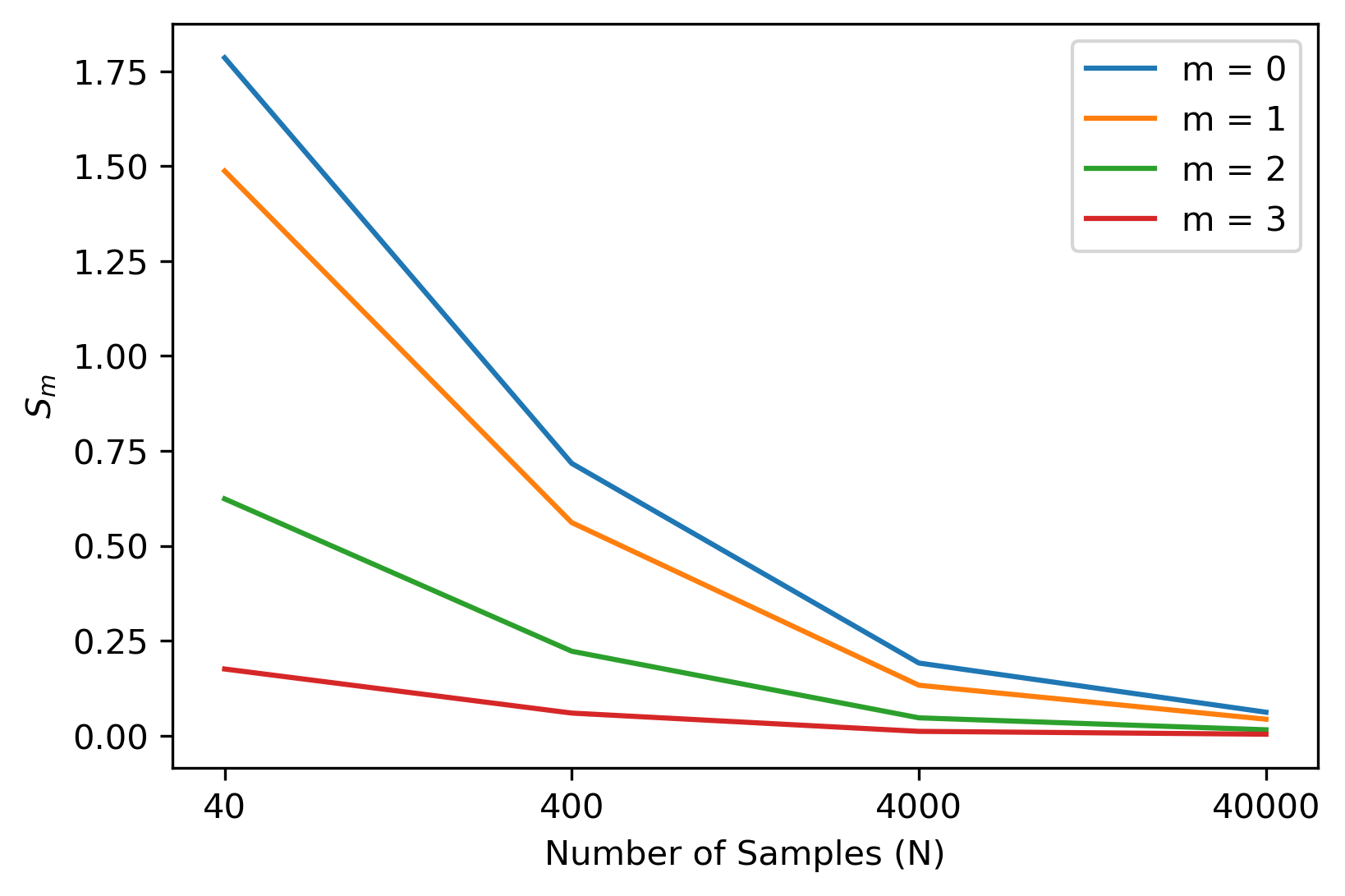}
\caption{When the two distributions are similar, then we expect each $S_m$
to be zero. Empirically, that happens as we increase the sample size. The
lower order moments take longer to go to zero.}
\label{fig:SnN_equal}
\end{figure}

As a point of comparison, we contrast the moment-based distance to total variation distance (TVD), a state-of-the-art metric which has proven useful in earlier experimental investigations \cite{buadescu2019quantum,rudinger2019probing}. We note that the magnitudes of the moment-based distance and total variation distance are not directly comparable as they follow very different methodologies but one can compare the corresponding signal-to-noise ratio (SNR) of the two metrics as the inverse of the coefficient of variation. For our numerical studies, we generated two time series, each of length 8192, by sampling two different probability distributions. The first was a normal distribution with mean 10 and standard deviation 1, and the second a normal distribution with mean 10 and standard deviation 4. We calculate the moment-based distance and total variation distance between these two time series, and then we repeated this numerical experiment 400 times to generate a distribution of the TVD and MBD distances. Using the average $\mu$ and standard deviation $\sigma$ of these distributions, we calculated the respective SNR as
\begin{equation}
\textrm{SNR} = \frac{\mu}{\sigma}
\end{equation}
As shown in Fig.~\ref{fig:MBDstability}, our results indicate that the moment-based distance has more statistical power as indicated by a higher SNR. As an aside, a practical concern is the dependence of precision of the moment-based distance on sampling size. Although each $S_m$ should vanish when two distributions are similar, finite sampling lead to approximations and ultimately a non-zero distance. As shown in Fig.~\ref{fig:SnN_equal}, increasing sampling may be used to reduce the relative error in each moment to a desired relative precision. Since MBD lacks direct comparability with measures such as Fisher information (which have deep physical interpretations), we primarily use the Hellinger distance for the rest of the document. 
\section{Evaluation framework}
Let $\rho$ be a density matrix representing the state of an $n$-qubit quantum register. 
Suppose $\rho$ undergoes a unitary transformation $U$, which can be decomposed into $K$ unitaries:
\begin{equation}
U = U_K \cdots U_1.
\end{equation}
The noiseless output state of the quantum register is:
\begin{equation}
\rho^\text{ideal}_\text{out} = U \rho U^\dagger.
\end{equation}
The projection operators $\{\Pi_i = \ket{i}\bra{i}\}$ project the output state into one of the $2^n$ computational basis states $\{\ket{0}, \cdots, \ket{2^n-1}\}$. 
The probability distribution for the results generated by a noiseless quantum computer is denoted by:
\begin{equation}
\mathds{P}^\text{ideal} = \{p_i^\text{ideal}\} \;\;\;\;\; i \in \{0, 1, \cdots, 2^n-1\}
\end{equation}
where $p_i^{ideal} = \textrm{Tr}[\Pi_i \rho^\text{ideal}_\text{out} ]$. In general, it is not efficient to construct the set $\mathds{P}^\text{ideal}$ using classical computing as the resource needs scale exponentially with $n$. 
However, such demanding calculations are feasible if either $n < 50$ or if the circuit has exhibits high-degree of structure (such as the quantum search). 

In the presence of noise, the evolution of the quantum register no longer adheres to a unitary evolution\cite{nielsen2002quantum}. This leads to mixed states in the output. 
Let $\mathcal{E}$ denote the super-operator that characterizes a noisy quantum channel. 
It can be defined by a set of Kraus operators $\{M_k\}$. 
In particular, 
\begin{equation}
\rho_{\textrm{out}}^\textrm{noisy} = 
\mathcal{E}_K\left(\cdots
\mathcal{E}_2\left(
U_2
\mathcal{E}_1\left(
U_1 \rho_\text{in} U_1^\dagger
\right)
U_2^\dagger
\right)
\cdots\right)
\end{equation}
where the action of each $\mathcal{E}_k$ is given by:
\begin{equation}
\mathcal{E}_k \left( \rho \right) = \sum\limits_{k} M_k \rho M_k^\dagger.
\end{equation}
We will sometimes use the notation $\mathcal{E}_{\xrm}$ to emphasize the dependence of the error channel on a vector of noise parameters $(\xrm_1, \cdots, \xrm_d)$.

The corresponding probability distribution for a noisy computer is:
\begin{equation}
\mathds{P}^{\text{noisy}} = \{p_i^{\text{noisy}}\} \text{ for } i \in \{0, 1, \cdots, 2^n-1\}
\end{equation}
where $p_i^{\text{noisy}} = \text{Tr}[M_i^\dagger M_i \rho^\text{ideal}_\text{out}]$, and $M_i$ is the measurement operator for a noisy readout channel \cite{smith2021qubit}:
\begin{equation}
\begin{split}
M_0 =& \sqrt{f_0}\ket{0}\bra{0}+\sqrt{1-f_1}\ket{1}\bra{1}\\
M_1 =& \sqrt{1-f_0}\ket{0}\bra{0}+\sqrt{f_1}\ket{1}\bra{1}\\
\end{split}
\end{equation}
The Hellinger distance between $\mathds{P}^{ideal}$ and $\mathds{P}^{noisy}$ is:
\begin{equation}
H(\mathds{P}^{\textrm{ideal}}, \mathds{P}^{\textrm{noisy}}) = 
\sqrt{1-BC(\mathds{P}^{\textrm{ideal}},\mathds{P}^{\textrm{noisy}})}
\label{eq:hellinger}
\end{equation}
with the Bhattacharyya coefficient $BC(\mathds{P}^{\textrm{ideal}}, \mathds{P}^{\textrm{noisy}}) \in [0,1]$ defined as:
\begin{equation}
BC(\mathds{P}^{\textrm{ideal}}, \mathds{P}^{\textrm{noisy}}) = 
\sum\limits_{i=0}^{2^n-1} \sqrt{p_i^{\textrm{ideal}} p_i^{\textrm{noisy}}}.
\end{equation}

Next, we turn our attention to the notation for the mean of a quantum observable as an outcome of a noisy quantum computer. Let $\hat{O}$ symbolize the operator associated with an observable computed from the results of the quantum circuit. The operator can be broken down into its spectral decomposition:
\begin{equation}
O = \sum_m \lambda_m \ket{\lambda_m}\bra{\lambda_m},
\end{equation}
where $\lambda_m$ represents the real eigenvalues of $O$ and $\ket{\lambda_m}$ denotes the corresponding eigen-states. The expectation of the observable $O$, relative to the noisy quantum state described by the density matrix $\rho_{\textrm{out}}^\textrm{noisy}$, is given by:
\begin{equation}
\braket{O_{\xrm}} = \textrm{Tr}\left( \hat{O} \rho_{\textrm{out}}^\textrm{noisy}\right)
= \textrm{Tr}\left( \hat{O} \mathcal{E}_{\xrm} ( \rho_{\textrm{out}}^{\textrm{ideal}} ) \right)
= \sum_m \lambda_m \textrm{Tr}\left( \Pi_m\mathcal{E}_\xrm( \rho_{\textrm{out}}^{\textrm{ideal}} )\right),
\end{equation}
where $\Pi_m = \ket{\lambda_m}\bra{\lambda_m}$ stands as the projective operator.

For example, consider the case of a register with $n=1$ qubits in the presence of depolarizing noise. The latter channel operator is characterized by a noise parameter $\xrm$ for which the Kraus operators $M_{k} \in \{\sqrt{1-\xrm} \mathds{I}, \sqrt{\xrm}\mathds{X}, \sqrt{\xrm} \mathds{Y}, \sqrt{\xrm} \mathds{Z} \}$ yield  
\begin{align}
\rho^\text{noisy}_\text{out} = \mathcal{E}_\xrm(\rho) =& (1-\xrm) \rho + 
\frac{\xrm}{3} \mathds{X} \rho \mathds{X} + 
\frac{\xrm}{3} \mathds{Y} \rho \mathds{Y} + 
\frac{\xrm}{3} \mathds{Z} \rho \mathds{Z}
\end{align}
Assuming $\ket{\psi} = \alpha \ket{0} + \beta \ket{1}$, the state-dependent noisy observable $Z$ is: 
\begin{align}
\braket{Z_\xrm} = (2|\beta|^2-1)\left(1-\frac{4}{3}\xrm \right).
\end{align}
With the notations out of the way, now we can focus on assessing the quality of the digital histograms in the presence of time-varying quantum noise\cite{etxezarreta2021time}. 
We reduce the complexity in assessment by developing an intuitive performance evaluation framework. 
Specifically, we differentiate between computational accuracy, result reproducibility, program stability, and device reliability. 
These notions are related yet still distinct. 
\subsection{Computational accuracy}
We begin by defining computational accuracy. 
We say that a quantum computation is $\epsilon-$accurate if the Hellinger distance between  $\mathds{P}^{\text{noisy}}$ and $\mathds{P}^{\text{ideal}}$ is upper bounded by $\epsilon$:
\begin{equation}
H(\mathds{P}^{\textrm{ideal}}, \mathds{P}^{\textrm{noisy}}) \leq \epsilon
\label{eq:accuracy_def}
\end{equation}
The above definition requires a-priori knowledge of the noiseless reference distribution $\mathds{P}^{\text{ideal}}$. This may be an impractical requirement when testing the accuracy for large problem sizes. In such cases, instead of looking at histogram accuracy, we might choose to look at the accuracy of the mean of an observable. The accuracy condition is then described as:
\begin{align}
&|\braket{O_{noisy}} - \braket{O_{noiseless}}| \leq \epsilon\\
\Rightarrow &|\sum\limits_m \lambda_m \left( p^\text{noisy}(m) - p^\text{noiseless}(m)\right)| \leq \epsilon\\
\Rightarrow &| \sum\limits_m \lambda_m \textrm{Tr}\left[\Pi_m \left(\mathcal{E}_\xrm(\rho_\text{out}) -\rho_\text{out}\right)\right]| \leq \epsilon 
\end{align}
where $\mathcal{E}_\xrm(\cdot)$ denotes the effective noise channel, $\lambda_m$ is an eigenvalue of the observable $\hat{O}$, and $\Pi_m$ is the projection operator corresponding to the $m$-th eigenstate.

Consider the single-qubit example in the presence of depolarizing noise. The accuracy metric (in terms of the Hellinger distance) for this case is state-dependent and is given as:
\begin{align}
    H = \left(1-|\alpha|^2\sqrt{1-\frac{2\xrm}{3}\left(1-\left|\frac{\beta}{\alpha}\right|^2\right)}-|\beta|^2\sqrt{1-\frac{2\xrm}{3}\left(1-\left|\frac{\alpha}{\beta}\right|^2\right)}\right)^{1/2}
\end{align}
The accuracy metric in terms of the $Z$ observable is:
\begin{equation}
|\braket{Z}_\text{noisy}-\braket{Z}_\text{noiseless}| = \left|\frac{4\xrm ( 1-2 |\beta|^{2} ) }{3}\right|
\end{equation}
Requiring $\epsilon$-accuracy places an upper bound on the depolarizing channel parameter as:
\begin{align}
\xrm \leq \frac{3\epsilon}{4}.
\end{align}
\subsection{Distribution reproducibility}
Next, consider the problem of reproducibility in quantum computing. 
We will call our empirical histogram $\delta$-reproducible if:
\begin{equation}
\textrm{Pr}(H \leq \epsilon) \geq 1 - \delta,
\label{eq:hist_repr}
\end{equation}
where $1-\delta$ is the statistical confidence level. 
This analysis requires an ensemble of histograms to be created through multiple executions on a noisy quantum computer. 

With respect to the mean of a quantum observable, we may similarly pose the reproducibility condition as: 
\begin{align}
\textrm{Pr}(\left|\braket{O^{ideal}}-\braket{O^{noisy}}\right| \leq \epsilon) &\geq 1 - \delta
\end{align}
where $\braket{O^{noisy}_\xrm}$ is a random variable due to the presence of both shot noise as well as the non-stationarity of $\xrm$. This reproducibility condition may be used to derive a stronger bound on the device noise. For example, consider again the single-qubit example in the presence of depolarizing noise. 
Suppose the depolarizing parameter $\xrm$ follows an exponential distribution:
\begin{align}
f_{X}(\xrm) = \nu\exp^{-\nu\xrm}.
\end{align}
Then the $\delta$-reproducibility condition requires that the mean of the depolarizing parameter $\xrm$ should be bounded as:
\begin{equation}
\mathds{E}(\xrm) = \frac{1}{\nu} \leq \frac{3\epsilon}{4|\log \delta |}.
\end{equation}
\subsection{Hardware reliability}\label{sec:reliability_metrics}
While device characterization metrics can be technology specific, there is a subset of five abstractions that represent the fundamenatal criteria for achieving a functional quantum computer \cite{divincenzo2000physical}. These are: (1) Register size, $n$, a measure of the information capacity, (2) SPAM fidelity, $F_\text{SPAM}$, a measure of the noise in preparing a fiducial state and subsequently measuring it, (3) gate fidelity, $F_\text{G}$, a measure of the noise in implementing a quantum operation, (4) duty cycle, $\tau_G$, a measure of the number of operations feasible before a quantum state decoheres, and (5) addressability, $F_\text{A}$, a measure of unwanted inter-qubit cross-talk. We use this subset to characterize the reliability of a NISQ\cite{preskill2019quantum} computer. 

SPAM fidelity is defined as:
\begin{equation}
F_\text{SPAM} = 1 - \epsilon_\text{SPAM}.
\end{equation}
where 
$F_\text{SPAM}$ stands for the probability of preparing and measuring the $n$-qubit register in a fiducial state while $\epsilon_\text{SPAM}$ is the probability of observing any other erroneous outcome. 

Gate fidelity ($F_G$) is defined by the error per Clifford gate $\epsilon_G$, often measured using randomized benchmarking:
\begin{equation}
F_G = 1 - \epsilon_{G}.
\label{eqn:fg}
\end{equation}
We focus our attention specifically on characterizing the two-qubit CNOT gate, which often plays a decisive role in the performance limits of NISQ computing (second only to SPAM noise).

We define the duty cycle $\tau_G$ as the ratio of the duration $T_G$ of a given gate to the register de-coherence time:
\begin{equation}
    \tau_G = T_G/T_D.
\end{equation}
The composite metric $\tau_G$ measures the number of quantum operations that can be executed before the register de-coheres, providing a quantification of the circuit depth achievable.

Addressability characterizes how well each register element can be measured individually. 
We quantify this in terms of the intra-register correlations \cite{maciejewski2021modeling, perez2011quantum} that arise during quantum operations and measurement, due to either unwanted entanglement or classical cross-talk. 
Specifically, we define addressability $F_A$ as:
\begin{equation}
F_A = 1-\frac{2 \mathcal{I}(X, Y)}{\mathcal{H}(X)+\mathcal{H}(Y)}.
\label{eq:FM}
\end{equation}
where 
$\mathcal{I}(X, Y)$
is the mutual information between $X$ and $Y$ and $\mathcal{H}(\cdot)$ denotes the entropy.

\begin{figure}[!htbp]
\centering
\includegraphics[width=.5\columnwidth]{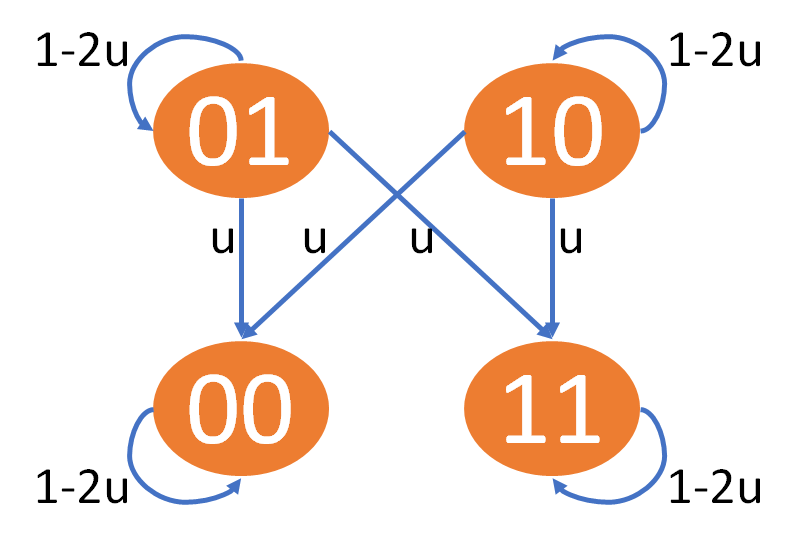}
\caption{A classical Markov model for the correlated error process.
}
\label{fig:markov}
\end{figure}

\begin{figure}[htbp]
\centering
\includegraphics[width=\figurewidth]{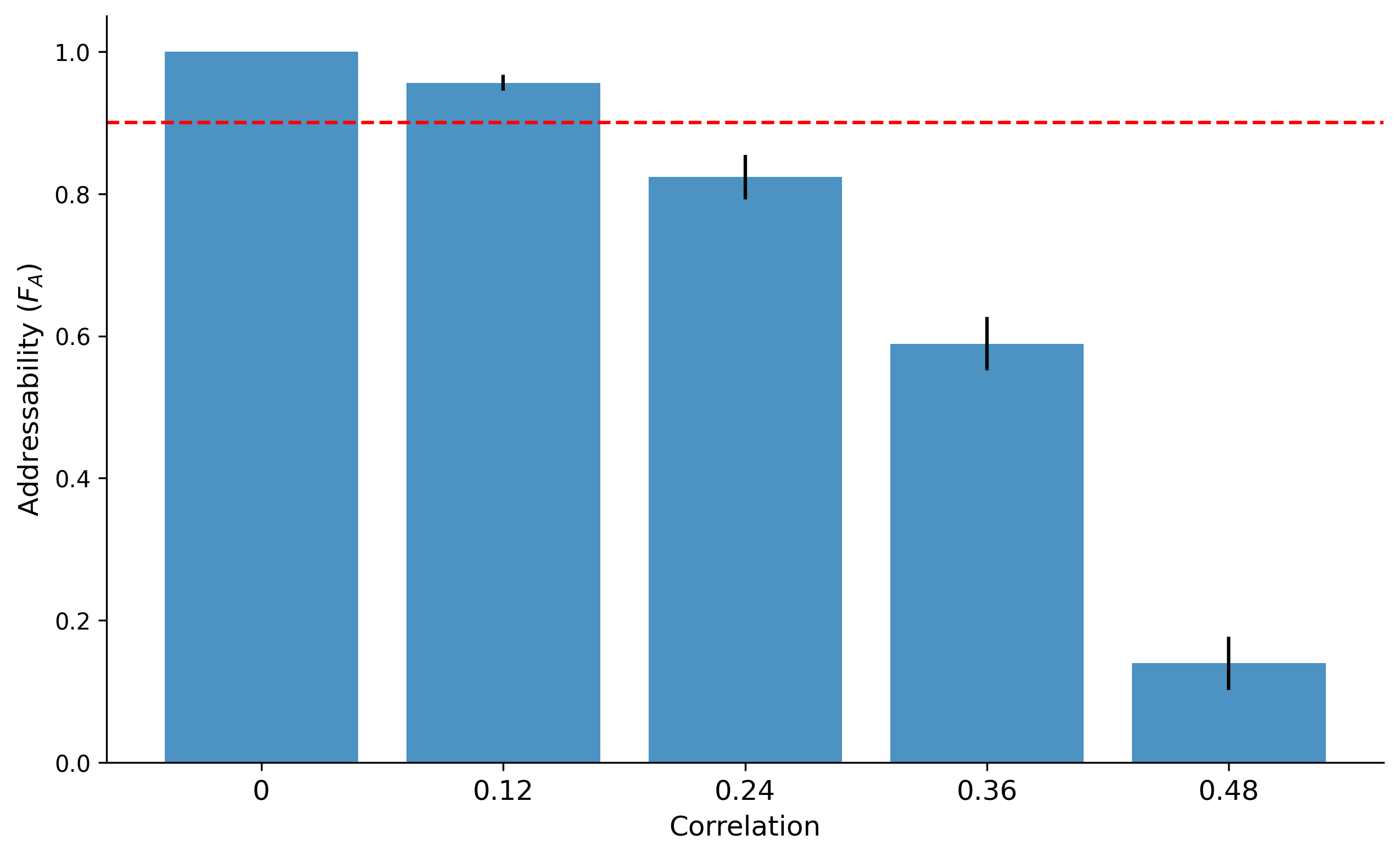}
\caption{Simulation result for addressability $F_A$ showing sensitivity to intra-register correlation.}
\label{fig:mf_vs_p}
\end{figure}

\begin{figure}[htbp]
\centering
\includegraphics[width=\figurewidth]{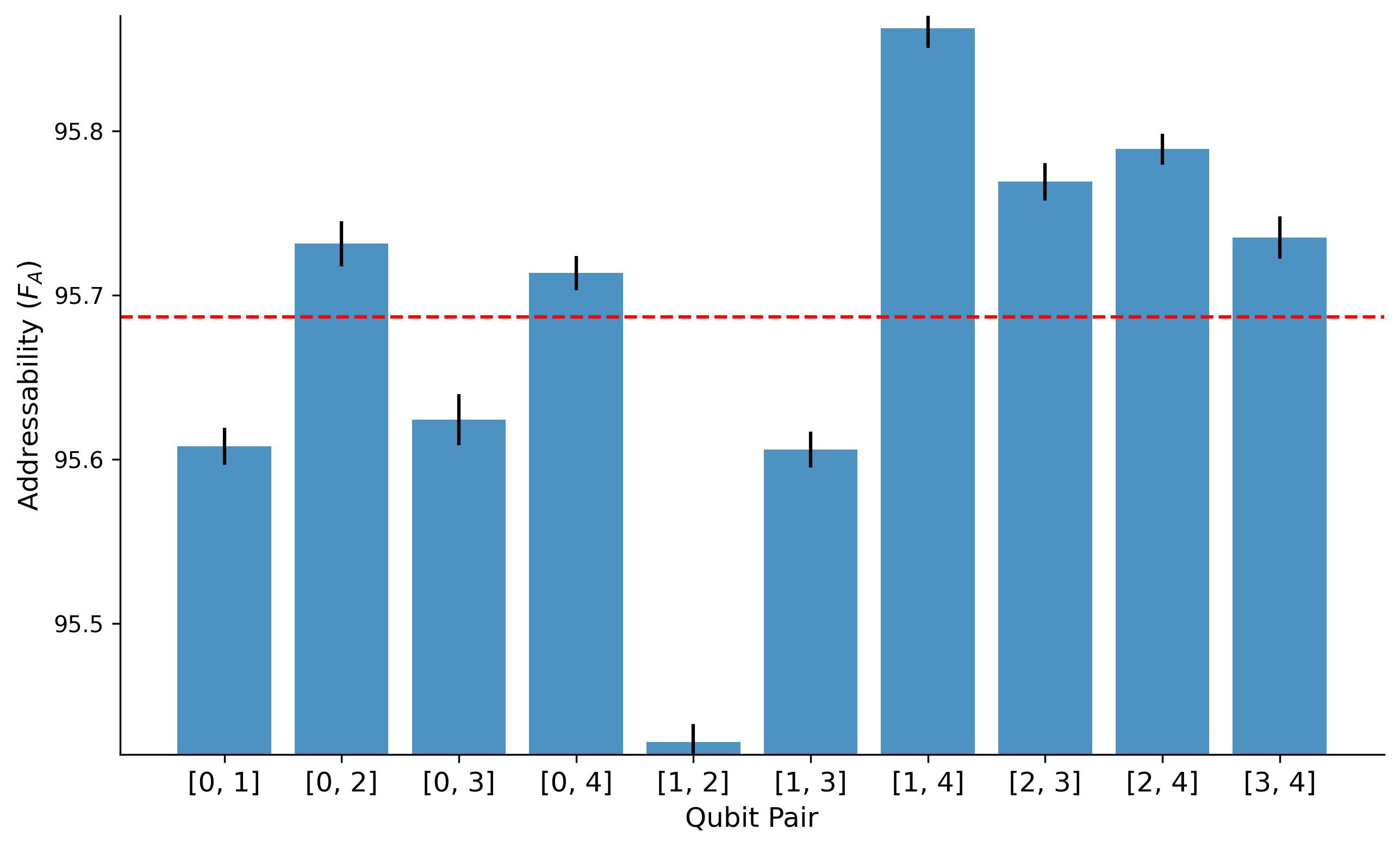}
\caption{Addressability: pairwise comparison when correlation parameter $u = 0.12$. Ideally, there should have been zero spatial variation (as idealized simulations do not differentiate between qubits at different physical locations). However, small fluctuations are seen about a mean value of 95.80 with a standard deviation of 0.02. This arises due to the readout error fluctuations in the \yorktown noise model.}
\label{fig:mf_spatial}
\end{figure}

We use a numerical simulation to illustrate addressability $F_A$. 
Let us characterize a two-qubit device where $q_0$ is the first qubit and $q_1$ is the second qubit. 
Assume that the state prior to measurement ($\ket{\alpha}$) is in one of four computational basis states: $\ket{00}$, $\ket{01}$, $\ket{10}$ and $\ket{11}$. 
Suppose this prior state is impacted by an uncorrelated binary noise process to become $\ket{\beta}$. 
The transition probability from $\ket{\alpha}$ to $\ket{\beta}$ is given by Table~\ref{table:p_table}. 
The input to this model is the intermediate state $\ket{\beta}$, which is subjected to a correlated noise process as shown in Fig.~\ref{fig:markov}. 
Let $P(X)$ denote the probability of state $X$. Using the noise models represented by Table~\ref{table:p_table}:
\begin{equation}
\begin{split}
P(\ket{s} = \ket{00}) =& \frac{1+2u}{4}\\
P(\ket{s} = \ket{01}) =& \frac{1-2u}{4}\\
P(\ket{s} = \ket{10}) =& \frac{1-2u}{4}\\
P(\ket{s} = \ket{11}) =& \frac{1+2u}{4}\\
\end{split}
\end{equation}
The probability of observing individual measurement outcomes $Q_0$ and $Q_1$ are:
\begin{equation}
\begin{split}
\textrm{Pr}(Q_0 = 0) =& 1-\textrm{Pr}(Q_0 = 1)= \textrm{Pr}(\ket{s}=\ket{00})+\textrm{Pr}(\ket{s}=\ket{01})\\
\textrm{Pr}(Q_1 = 0) =& 1-\textrm{Pr}(Q_1 = 1)= \textrm{Pr}(\ket{s}=\ket{00})+\textrm{Pr}(\ket{s}=\ket{10})\\
\end{split}
\end{equation}
which yields 
\begin{equation}
\begin{split}
\textrm{Pr}(Q_0 = 0) &= \textrm{Pr}(Q_0 = 1) = \frac{1}{2}\\
\textrm{Pr}(Q_1 = 0) &= \textrm{Pr}(Q_1 = 1) = \frac{1}{2}\\
\end{split}
\end{equation}
The binary entropy is therefore maximal, i.e., $H(Q_0) = H(Q_1) = 1$.
This leads to a final expression for the addressability as
\begin{align}
F_A &= 1 - \frac{1+2u}{2}\log(1+2u) - \frac{1-2u}{2}\log(1-2u)
\label{eq:MFtoy}
\end{align}
This model analyzed addressability for a simple Markov model as shown in Fig.~\ref{fig:markov}. We show in Fig.~\ref{fig:mf_vs_p} the variability of addressability using a numerical simulation of a correlated noise model. A similar spatial characterization (numerical simulation) is shown in Fig.~\ref{fig:mf_spatial}.
We quantify reliability using the distance between quantum noise densities at different times and register locations. For example, $H_{\text{SPAM}}(t_1, t_2)$ measures the similarity in distributions of SPAM fidelity at different times. A reliable, but not necessarily ideal, device maintains the characteristic density for the fidelity at both times. By similar considerations, spatially-varying noise processes can be subjected to reliability analysis using this definition.  
We will call a device $\varepsilon$-reliable if:
\begin{equation}
H_{X} < \varepsilon
\end{equation}
The normalized version in Eqn.~\ref{eq:hellinger_normalized} is more sensitive and useful for reliability testing.
\subsection{Observable stability}
Suppose that the distribution of quantum noise exhibits a time-dependence:
\begin{equation}
f_X(\xrm) = f(\xrm;t)
\end{equation}
Observable stability studies bounds on the output of a noisy quantum circuit in presence of time-varying quantum noise\cite{dasgupta2022assessing}. 

Specifically, the instability between the results obtained at two different times $t_1$ and $t_2$ can be quantified by:
\begin{equation}
s(t_1, t_2) = | \braket{O}_{t_1} - \braket{O}_{t_2} |
\label{eq:stability}
\end{equation}
where 
\begin{equation}
\braket{O}_t = \int\limits \braket{O_{ \xrm}}  f(\xrm; t) d\xrm.
\label{eq:O_td_def}
\end{equation}
Here, $\xrm$ denotes a specific realization of the random quantum noise parameter $X$, which has a density characterized by $f(\xrm;t)$. $\braket{O_{ \xrm}}$ is the mean of the quantum observable $\hat{O}$, computed from results obtained from a noisy quantum circuit subject to noise $\xrm$.

The presence of time-varying noise renders the observable (a random number) non-stationary stochastic process when observed as a time-series. Hence, the mean of the observable may exhibit symptoms characteristic of a non-stationary stochastic process such as drifting means or time-varying error bars. 

As side note, stationary noise may produce irreproducible results because of large static variance. Also, stationary (stable) noise with small variance may produce reproducible results and yet be inaccurate because of bias. The point is that accuracy, reproducibility and stability of outcomes are all conceptually distinct (albeit related) concepts.

The $\epsilon-$stability condition can be stated as:
\begin{equation}
s(t_1, t_2) \leq \epsilon
\label{eq:stability_condition}
\end{equation}

This framework enables the question of time-scale for re-calibration. For example, consider a depolarizing channel characterized by the depolarizing parameter $p$. Suppose, $p$ exhibits non-stationarity and its stochastic behavior can be modeled by an exponential distribution with a time-varying parameter $\lambda(t)$:
\begin{equation}\lambda(t) = \lambda_0 -\chi \delta t.\end{equation}
This models the situation when error bars on $p$ increases with time (as often happens in between calibrations) since the spread (or variance) of an exponential distribution $\propto \lambda^{-2}$. The time-scale at which the device transitions from stable to unstable (in the absence of re-calibration) can then be estimated from Eqn.~\ref{eq:stability_condition} as:
\begin{align}
\delta t_{\textrm{stable}} &\leq 2\sqrt{2} \frac{\lambda_0 }{\chi}\sqrt{1-\sqrt{2\phi^2-1}}
\end{align}
where $\phi = \frac{3\sqrt{3}\epsilon}{14\sqrt{2}}$. Thus, in presence of this variance-drift model, we expect a stable device to remain stable until $t=\delta t_{stable}$. This gives an estimate of the frequency of device calibration required to meet the stability condition.

We end this section with a result for a special case. Assume that the quantum noise channel can be assumed to separable. Thus, the effect of noise on the quantum state can be understood by examining the impact of noise on individual qubits. Additionally, suppose that the channel is a first-order polynomial in $\xrm$ (for instance, a depolarizing channel). Also, assume that $\xrm$ is wide-sense stationary (WSS). Thus, its mean remains constant over time, and its standard deviation depends on the time interval between observations. These are all reasonable assumptions that correspond to noise data collected from NISQ computers. 

Under these assumptions, it can be shown that the mean of the observable $\braket{O}$ is invariant with respect to time:
\begin{equation}
\begin{split}
\braket{O}_t &= \int \braket{O_{X}} f_{X}(\xrm;t) d\xrm\\
&= \prod_{i=1}^n \int \bra{r_i} \mathcal{E}_{\xrm_{i}} (\ket{r_i}\bra{r_i} )\ket{r_i} f_{X_i}(\xrm_{i};t) d\xrm_{i}\\
&= \prod_{i=1}^n \int \bra{r_i} \left[\ket{r_i}\bra{r_i} + \xrm_i g(\ket{r_i}\bra{r_i})\right] \ket{r_i} f_{X_i}(\xrm_{i};t) d\xrm_{i}\\
&\text{ (where $g(\cdot)$ is an arbitrary function that outputs a valid density matrix)}\\
&= \prod_{i=1}^n \left[1+\mu_{\xrm_i}\bra{r_i}g(\ket{r_i}\bra{r_i})\ket{r_i}\right]
\end{split}
\end{equation}
which is independent of time because of the wide-sense stationarity assumption. 
Thus, $s(t_1, t_2) = 0$ when the quantum noise channel is separable, WSS and has noise terms till first-order only.
\section{Test circuit used}\label{sec:BV}
We use the Bernstein-Vazirani \cite{bernstein1993quantum} circuit as a test circuit for performance evaluation because it is a well-known example of quantum advantage, requiring only a modest number of gates, and is commonly used as a benchmarking tool for quantum computers. It was conceived by Bernstein and Vazirani in 1992 as an extension of Simon's algorithm. Using a quantum algorithm, it transforms a problem of $O(n)$ complexity to one of $O(1)$.

The algorithm is tasked with deciphering a $n$-bit secret string $r$, embedded in a black-box oracle function. 
The oracle responds with a yes/ no answer to the question: Is the secret string $w$? The algorithm locates the secret via just one query, irrespective of the value of $n$. The classical algorithm requires $n$ queries (worst case).

Mathematically, the oracle function takes an $n$-bit string as input ($w$) and produces the following output:
\begin{equation}f(w) = w \cdot r \text{ mod } 2,\end{equation}
where $(\cdot)$ represents bitwise multiplication and mod $2$ ensures the output is either 0 or 1. 

\begin{figure}[!t]
  \centering
\includegraphics[width=\figurewidth]{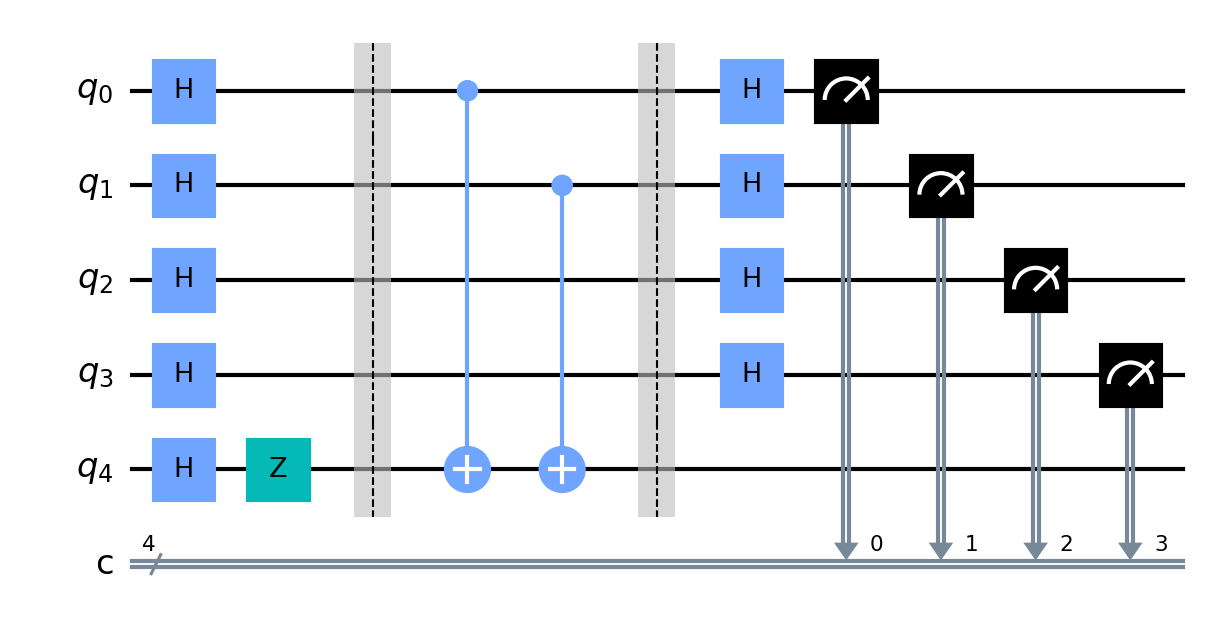}
\caption{The Bernstein-Vazirani circuit for a 4-bit secret string $r = 1000$. $H$ represents the Hadamard gate and $Z$ represents the $Z$-gate. The meter symbols are measurement operations that project to the computational basis states. The measurement results are recorded in a classical register $c$.}
\label{fig:bv_qiskit_ckt}
\end{figure}

The quantum circuit for the secret string $r=1000$ is shown in Fig.~\ref{fig:bv_qiskit_ckt}. 
In the first step, an equal superposition across all $2^n$ possible input bit strings is generated using a layer of Hadamard gates acting on $\ket{0}^{\otimes n}$. The second layer has an implementation  of the oracle function using a layer of CNOT gates. This is followed by another layer of Hadamard gates which yields the binary representation of the secret string $r$ at the output. 

Let us now understand precisely how this algorithm works. We will sometimes omit the normalization factors for clarity here. To begin, for 1-qubit, it is easily verified that:
\begin{equation}H\ket{\xrm} = \sum\limits_{w \in \{0, 1\}} (-1)^{\xrm w} \frac{\ket{w}}{\sqrt{2}}\end{equation}
Thus, 
\begin{equation} H\ket{0} =\frac{\ket{0}+\ket{1}}{\sqrt{2}} \end{equation}
\begin{equation} H\ket{1} =\frac{\ket{0}-\ket{1}}{\sqrt{2}} \end{equation}
When the input register is a $n$-bit state $\ket{w_0 \ldots w_{n-1}}$ and it is acted upon by a layer of Hadamard gates as shown in Fig.~\ref{fig:bv_qiskit_ckt}, then the output is:
\begin{equation}
\begin{split}
H^{\otimes n} \ket{w_0 \ldots w_{n-1}} &= H\ket{w_0} \otimes H\ket{w_2} \cdots \otimes H\ket{w_{n-1}}\\
&=\frac{1}{\sqrt{2^n}}\sum\limits_{\xrm}(-1)^{w\cdot x}\ket{\xrm}\\
\end{split}
\end{equation}
Thus,
\begin{equation}
H^{\otimes n} \ket{w} = \frac{1}{\sqrt{2^n}}\sum\limits_{\xrm}(-1)^{w\cdot \xrm}\ket{\xrm}
\label{eq:BV_Hw}
\end{equation}
For our 5-qubit circuit, the initial input is $\ket{0}^{\otimes 5}$. After the first unitary layer, the output becomes: 
\begin{equation}
\ket{\psi_1} = \ket{++++}\ket{-}.
\end{equation}

Next comes the unitary layer $U_f$ that encodes the oracle function $f(w)$. Specifically,
\begin{equation} 
U_f ( \ket{x y} ) = \ket{x} \ket{ y + f(x)} 
\end{equation}
When $f(x) =0$: 
\begin{equation}
\begin{split}
U_f \left( \ket{x}\ket{-} \right) &= U_f\left( \ket{x}(\ket{0}-\ket{1}\right) \\
&= U_f\left( \ket{x}\ket{0}\right) - U_f\left( \ket{x}\ket{1}\right)\\
&= \ket{x}\ket{0+f(x)} - \ket{x}\ket{1+f(x)}\\
&= \ket{x}\ket{0+0} - \ket{x}\ket{1+0}\\
&= \ket{x}\ket{-}\\
&= (-1)^{f(x)}\ket{x}\ket{-}\\
\end{split}
\end{equation}
When $f(x) = 1$: 
\begin{equation}
\begin{split}
U_f \left( \ket{x}\ket{-} \right) &= U_f\left( \ket{x}(\ket{0}-\ket{1}\right) \\
&= U_f\left( \ket{x}\ket{0}\right) - U_f\left( \ket{x}\ket{1}\right)\\
&= \ket{x}\ket{0+f(x)} - \ket{x}\ket{1+f(x)}\\
&= \ket{x}\ket{0+1} - \ket{x}\ket{1+1}\\
&= -\ket{x}\ket{-}\\
&= (-1)^{f(x)}\ket{x}\ket{-}\\
\end{split}
\end{equation}
We shorten this to write:
\begin{equation}
\begin{split}
U_f\ket{x} &= (-1)^{f(x)}\ket{x}\\
&= (-1)^{ r^\prime \cdot x}\ket{x}\\
\end{split}
\end{equation}
with the implicit assumption that the ancilla was set to $\ket{-}$. Here, $r^\prime$ is the 1-bit secret string. It follows then that,
\begin{equation}
\begin{split}
\sum\limits_{\xrm} U_f\ket{x} &= \sum\limits_{\xrm} (-1)^{r\cdot x}\ket{x} \\
&= H^{\otimes n} \ket{r} \text{ (from Eqn.~\ref{eq:BV_Hw})}
\end{split}
\end{equation}
where $r$ is the n-bit secret string. Thus, at the end of the second layer of the circuit, the output is of the form:
\begin{equation}
\ket{\psi_2} = H^{\otimes n} \ket{r}\ket{-}.
\end{equation}
So, to retrieve the secret string $r$, the third layer simply needs another layer of Hadamard gates. This yields the final output as:
\begin{equation}
\ket{\psi_3}=\ket{r}\ket{1}.
\end{equation}
\clearpage
\vspace{0.5in}
\vspace{0.5in}
\vspace{0.5in}
\vspace{0.5in}
\vspace{0.5in}
\begin{table}
\centering
\caption{Transition Probabilities for Uncorrelated Noise}
\vspace{1pt}
\begin{tabular}{l |c|c|c|c|}
      & $\ket{00}$ & $\ket{01}$ & $\ket{10}$ & $\ket{11}$ \\
\hline
$\ket{00}$ & $1-p$ & $p/3$ & $p/3$ & $p/3$ \\
\hline
$\ket{01}$ & $p/3$ & $1-p$ & $p/3$ & $p/3$ \\
\hline
$\ket{10}$ & $p/3$ & $p/3$ & $1-p$ & $p/3$ \\
\hline
$\ket{11}$ & $p/3$ & $p/3$ & $p/3$ & $1-p$ \\
\hline
\end{tabular}
\label{table:p_table}
\end{table}
\vspace{0.5in}
\vspace{0.5in}
\vspace{0.5in}

\removefigs
\chapter{Reliability of device characterization}\label{ch:exp_characterization}
In this chapter, we quantify and assess the reliability \cite{dasgupta2023reliability} of noisy quantum computers \cite{preskill2019quantum} using the performance evaluation framework introduced in the previous chapter. 
Our reliability testing will use experimental data from IBM \cite{ibm_quantum_experience_website} at various time-scales (monthly, daily, hourly and seconds level). 
Spatial reliability testing \cite{dasgupta2021stability} will explore variations across different parts of the system while the temporal reliability testing will study changes over time. 
Both component level metrics (such as individual gates and qubits) as well as composite level metrics (such as circuits) will be discussed. 
The chapter will highlight the non-stationary nature of noise in contemporary quantum computers. 
It will discuss the implications of such unreliable devices on program outcomes \cite{dasgupta2023reliability}. 
This will in turn motivate the need for rigorous stability analyses (in subsequent chapters) to ensure confidence in results from noisy quantum computers. It will provide the setting for dynamic mitigation \cite{zheng2020bayesian} to enhance reliability which will be discussed in later chapters.

\section{Experimental data}
\subsection{Device}
We used quantum computers provided by IBM, which are based on the transmon qubit architecture \cite{roth2021introduction}. 
IBM has unveiled a series of processors over the past few years, with steadily increasing register size \cite{roadmap}. 
Some of the earliest processors were categorized as Canary, featuring 2-16 qubits. 
After that came the Falcon processors with 27 qubits, Egret processors with 33 qubits, Hummingbird processors with 65 qubits, Eagle processors with 127 qubits and Osprey processors with 433 qubits. 
As an aside, the classification of processors goes beyond mere qubit count and can encompass details like the connectivity graph \cite{processor_types}. 
Our research used data from \yorktown (5 qubits), \toronto (27 qubits), and \washington (127 qubits), which belong to the Canary, Falcon and Eagle families respectively.

\subsection{Data}
We employed the Qiskit software library \cite{aleksandrowicz2019qiskit} to generate our data sets. 
The access to IBM quantum computers was provided by the Oak Ridge Leadership Computing Facility (OLCF) located at Oak Ridge National Laboratory, Tennessee \cite{olcf}. 

The device characterization data (at daily time-scale) for the \washington device spanned 16 months (starting from January 1, 2022, and ending on April 30, 2023). 
Examples of elements recorded include the date and time of the device calibration, the SPAM (State Preparation and Measurement) error rate for each individual qubit, the CNOT gate error rates (calculated via randomized benchmarking \cite{knill2008randomized} employing varying lengths of two-qubit Clifford gates \cite{bravyi2005universal}), the duration of the CNOT gates \cite{hurant2020asymmetry}, and the de-coherence times ($T_1$ and $T_2$) for each qubit \cite{burnett2019decoherence}. 
Similar daily data was gathered for the quantum computer called \yorktown (from March 1, 2019, to December 30, 2020).

A separate set of intra-day data was gathered from the 27-qubit \toronto device on December 11, 2020, at the following time intervals: 8:00-8:30 am, 11:00-11:30 am, 2:00-2:30 pm, 5:00-5:30 pm, 8:00-8:30 pm, and 11:00-11:30 pm (UTC-5). In this dataset, each individual qubit was sequentially sampled a total of $212,992$ times. It had a total of $5,750,784$ recorded outcomes.

\section{Component reliability}
\begin{figure}[H]
\center
\includegraphics[width=0.4\columnwidth]{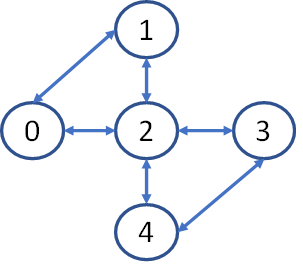}
\caption{Schematic layout of the \yorktown device produced by IBM. Circles denote register elements and edges denote connectivity of 2-qubit operations.}
\label{fig:yorktown}
\end{figure}

In this section, we evaluate the temporal and spatial reliability of the DiVincenzo metrics \cite{divincenzo2000physical} that we discussed in Sec.~\ref{sec:reliability_metrics}. We use data from \yorktown and \toronto whose layouts are shown in Fig.~\ref{fig:yorktown} and \ref{fig:toronto}.
\subsection{Reliability of SPAM noise characterization}
\begin{figure}[htbp]
\centering
\includegraphics[width=\figurewidth]{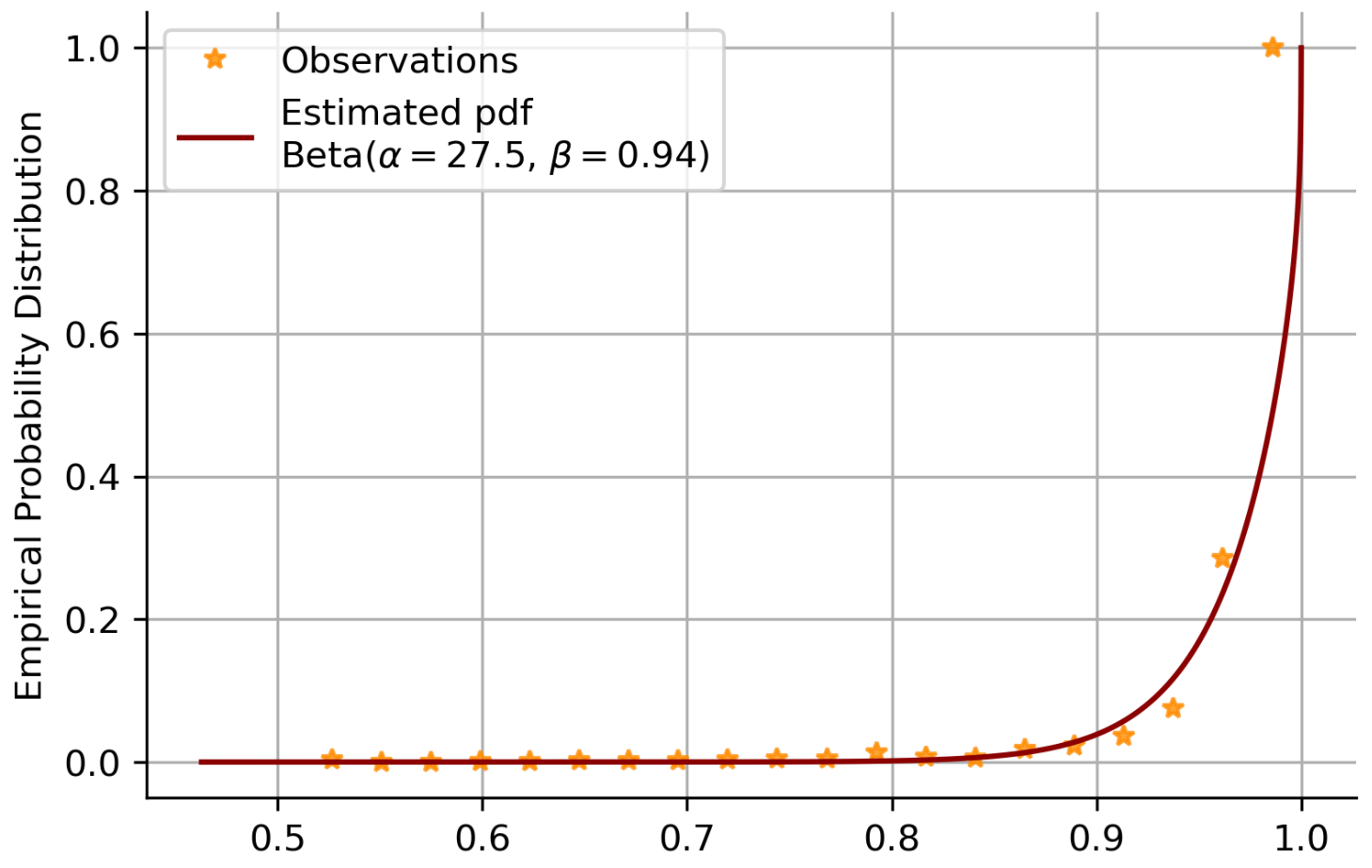}
\caption{Experimentally observed probability density for SPAM (state preparation and measurement) fidelity of one of the register elements of the IBM transmon device named \washington.}
\label{fig:washington_FI_beta_fit_example}
\end{figure}

\begin{figure*} 
\centering
\subfloat[\yorktown qubit 0]{\includegraphics[width=0.4\columnwidth]{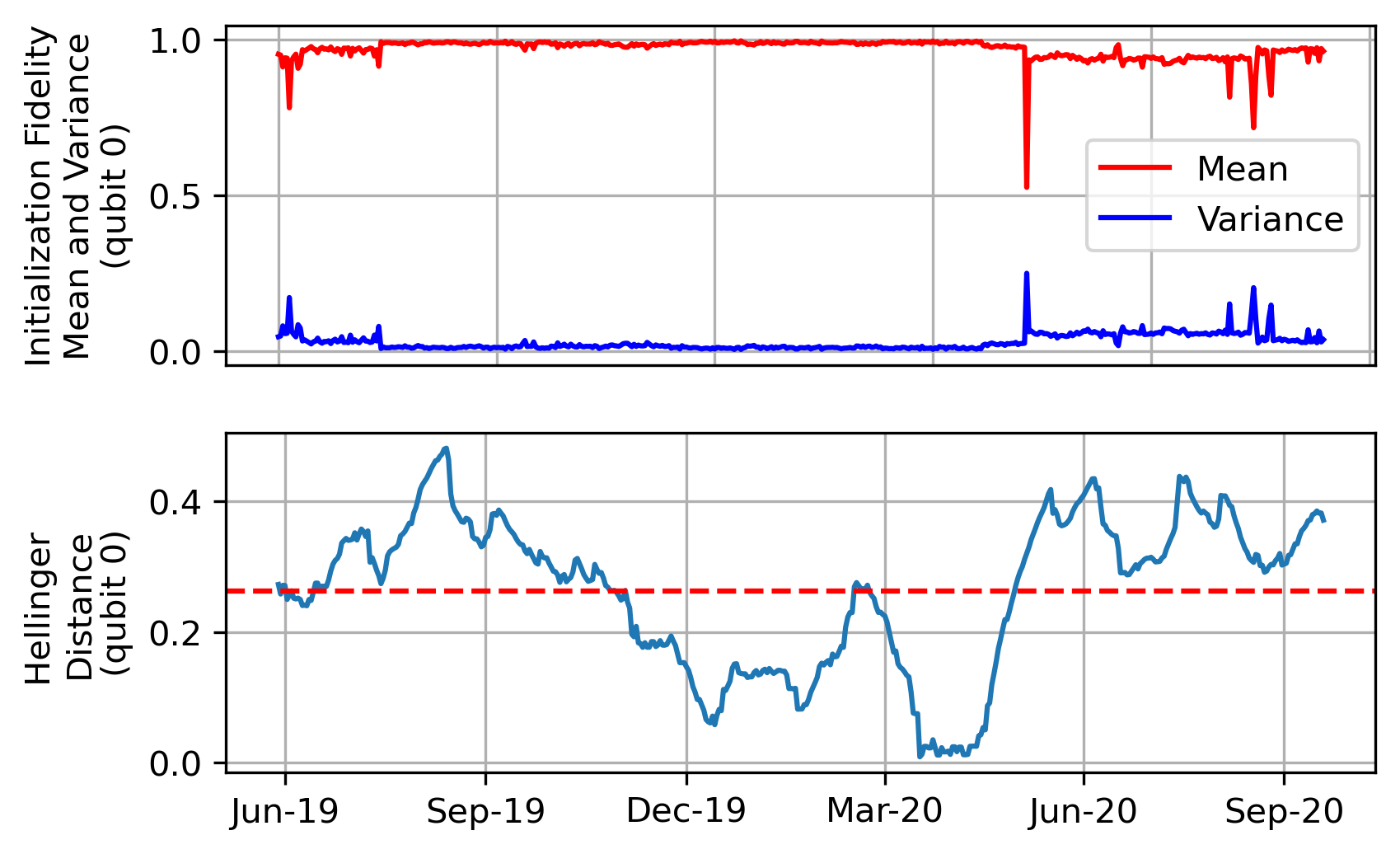}\label{fig:FI_temporal_MBD_willow_0}}
\hspace{0.5in}\subfloat[\yorktown qubit 1]{\includegraphics[width=0.4\columnwidth]{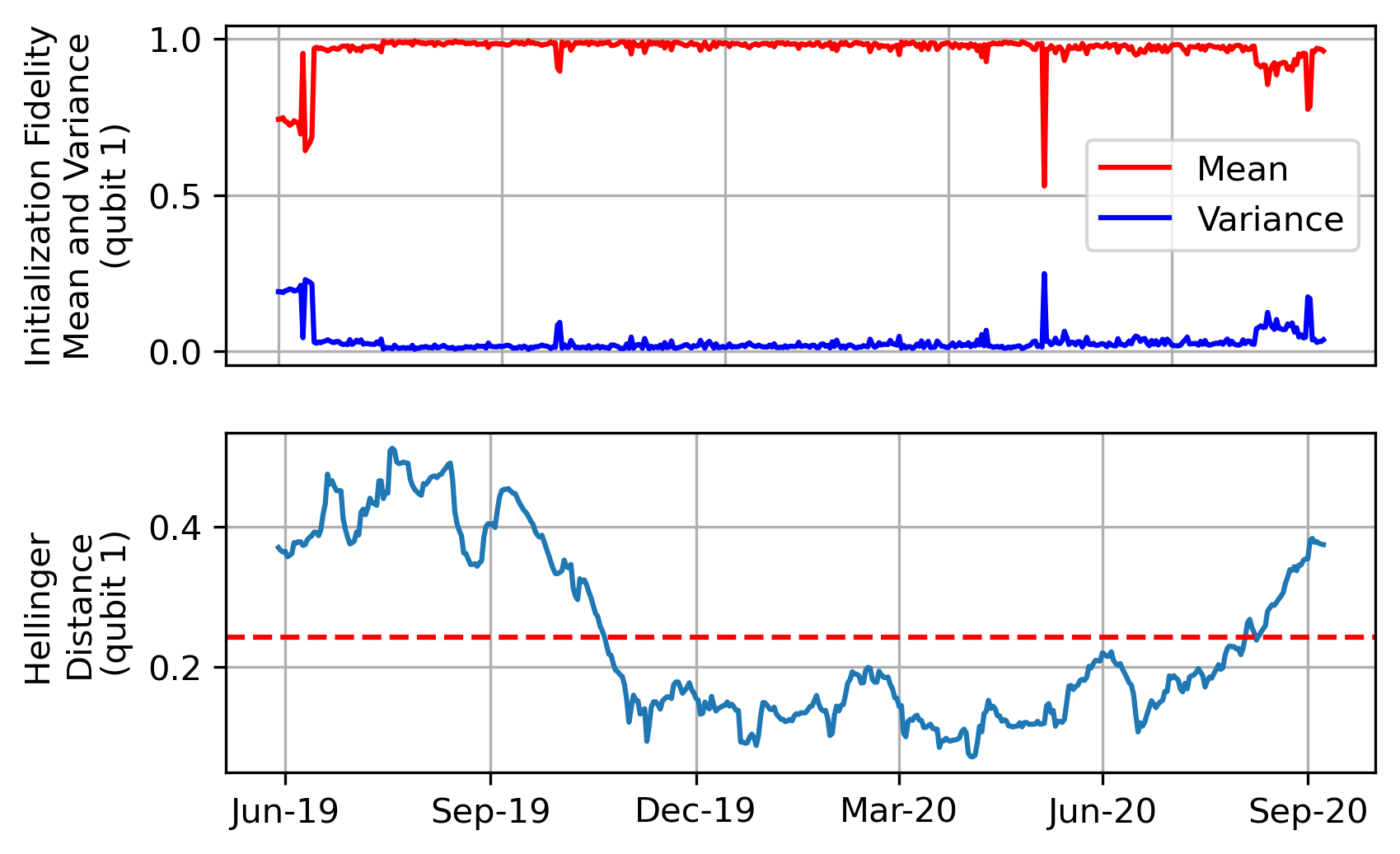}\label{fig:FI_temporal_MBD_willow_1}}
\hspace{0.5in}\subfloat[\yorktown qubit 2]{\includegraphics[width=0.4\columnwidth]{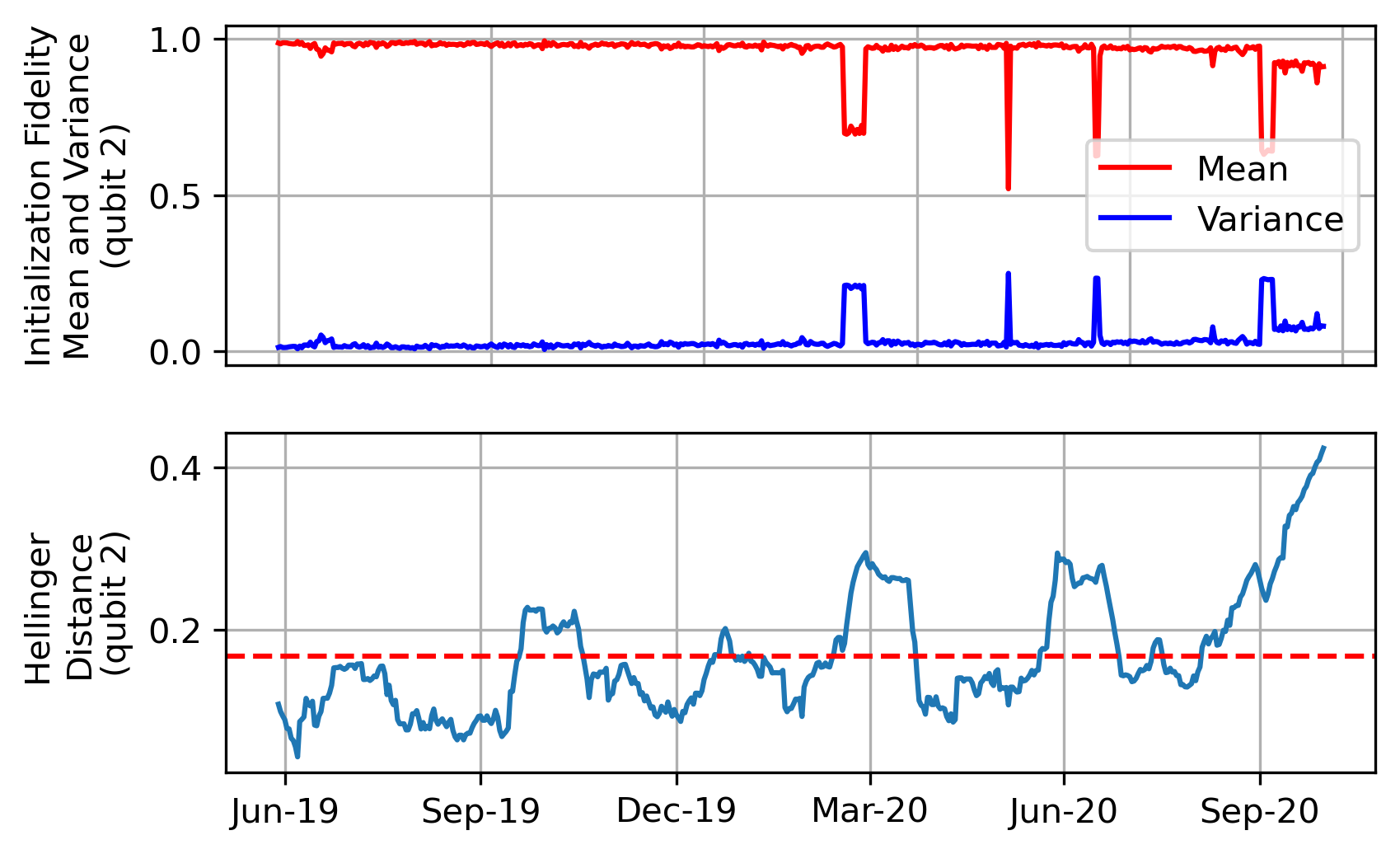}\label{fig:FI_temporal_MBD_willow_2}}
\hspace{0.5in}\subfloat[\yorktown qubit 3]{\includegraphics[width=0.4\columnwidth]{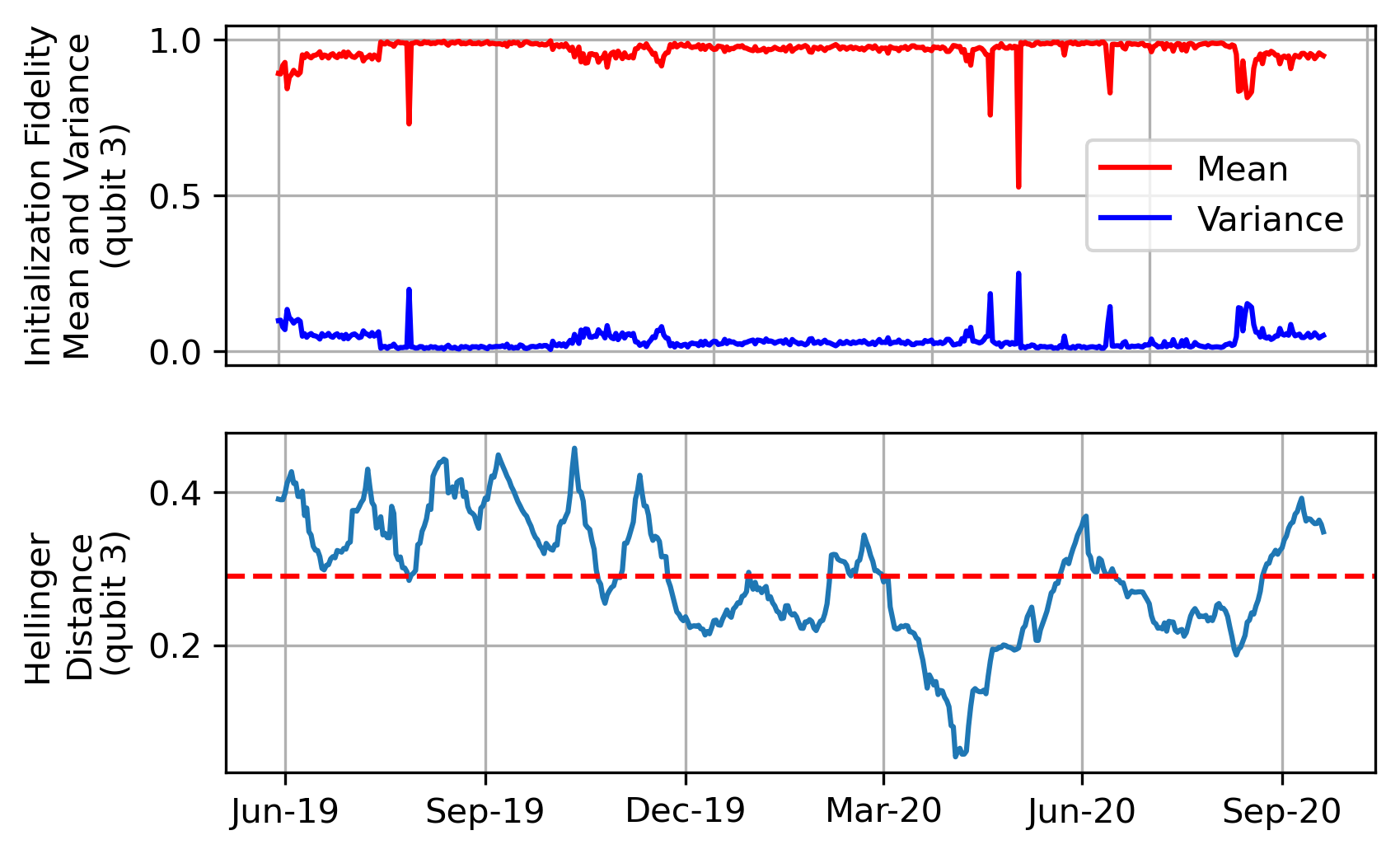}\label{fig:FI_temporal_MBD_willow_3}}
\hspace{0.5in}\subfloat[\yorktown qubit 4]{\includegraphics[width=0.4\columnwidth]{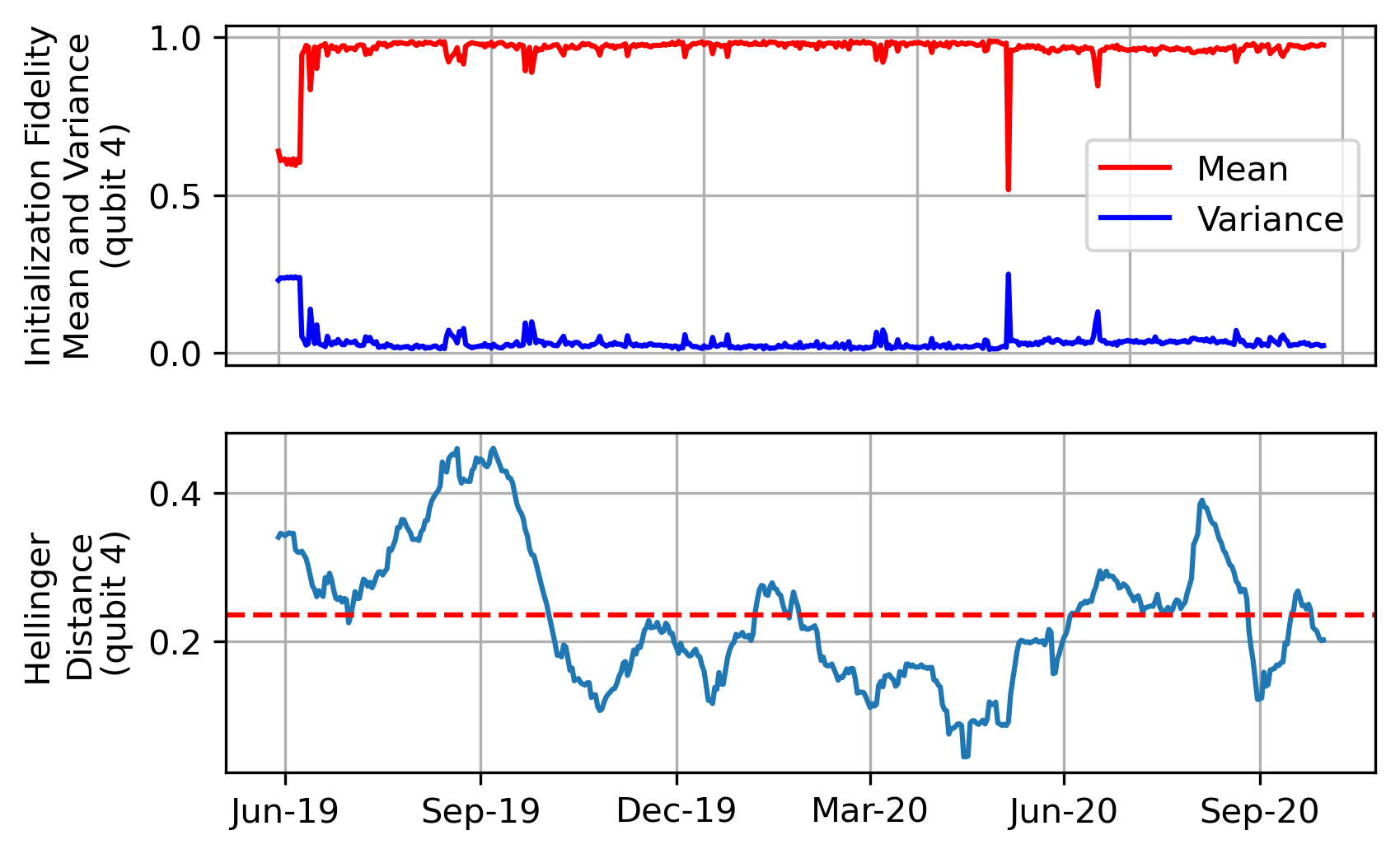}\label{fig:FI_temporal_MBD_willow_4}}
\caption{
(a)-(e) Temporal stability of the SPAM fidelity $F_\text{SPAM}$ of each register element in the \yorktown device. The top panel shows the average $F_\text{SPAM}$ of the register with associated variance, and the bottom panel shows a running calculation of the Hellinger distance using a one-month window. The dashed red line is the median value.
}
\label{fig:willow}
\end{figure*}

\begin{figure}[htbp]
\centering
\includegraphics[width=\textwidth]{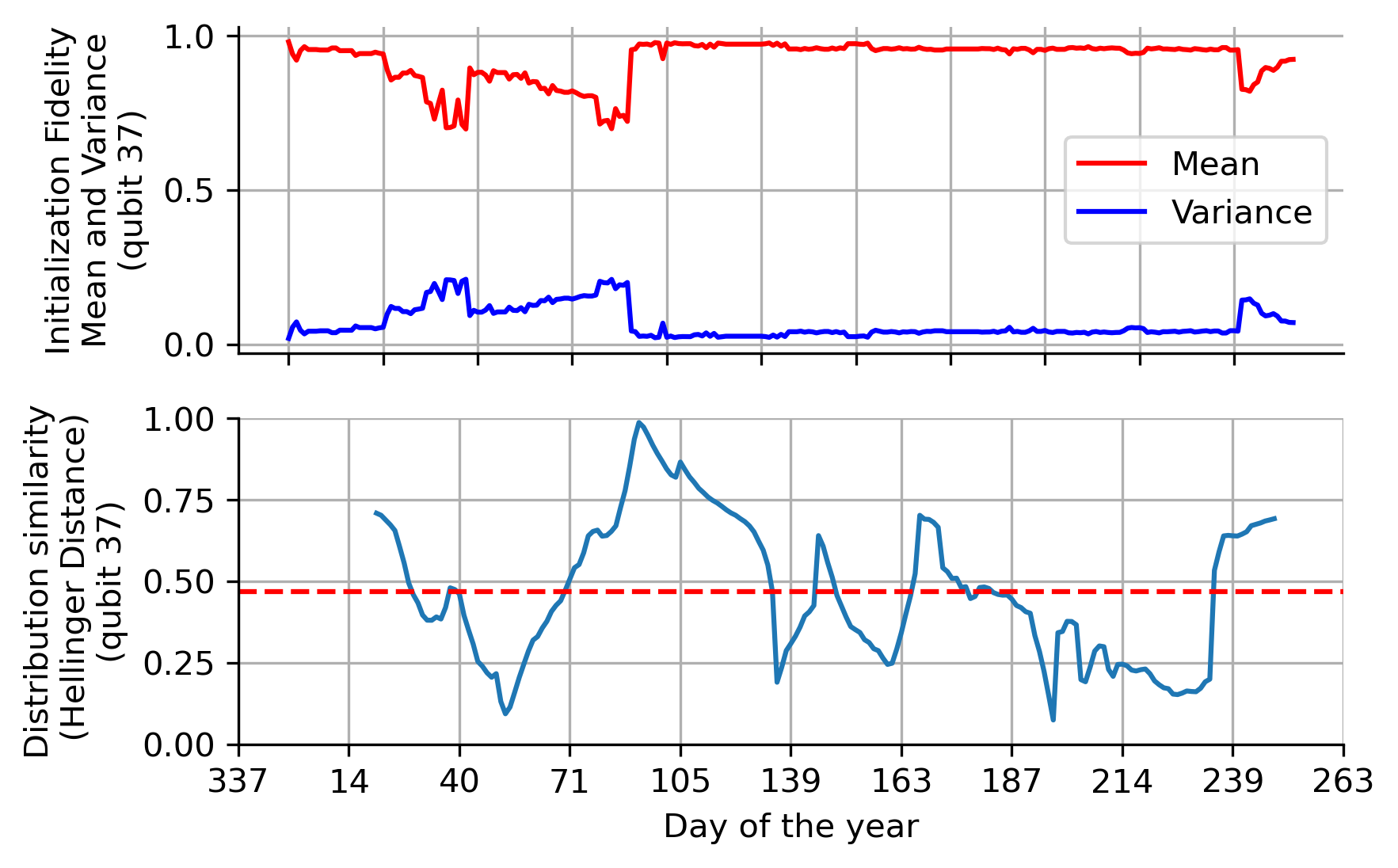}
\caption{Temporal reliability of the SPAM fidelity $F_\text{SPAM}$ of register element $q=37$}
\label{fig:washington_FI_hellinger_temporal_37}
\end{figure}

We find that the SPAM fidelity for all three IBM computers (\yorktown, \toronto, and \washington) fluctuates significantly over time, even when the average fidelity is tightly controlled. 
This is indicative of poor device reliability. 
Fig.~\ref{fig:washington_FI_beta_fit_example} shows an example of the SPAM noise density at a particular time.  
Fig.~\ref{fig:willow}(a)-(e) show the temporal fidelity of Yorktown.  
The top panel has the time series for the mean and variance, while the bottom panel plots the distance between the time-varying densities. 
The red line is the median. 
Fig.~\ref{fig:washington_FI_hellinger_temporal_37} shows the results for washington (qubit 37). 
The series starts in Dec-21 (when the device was commissioned) and ends in Oct-22. 
The red line is the median. 

All these plots show that there can be long periods when the fidelity is tightly controlled, but there are also times when it fluctuates significantly. 
These fluctuations are not reflected in the variance and could be due to changes in the underlying physics \cite{martin2007stationary}. 

\begin{figure}[htbp]
\centering
\includegraphics[width=\figurewidth]{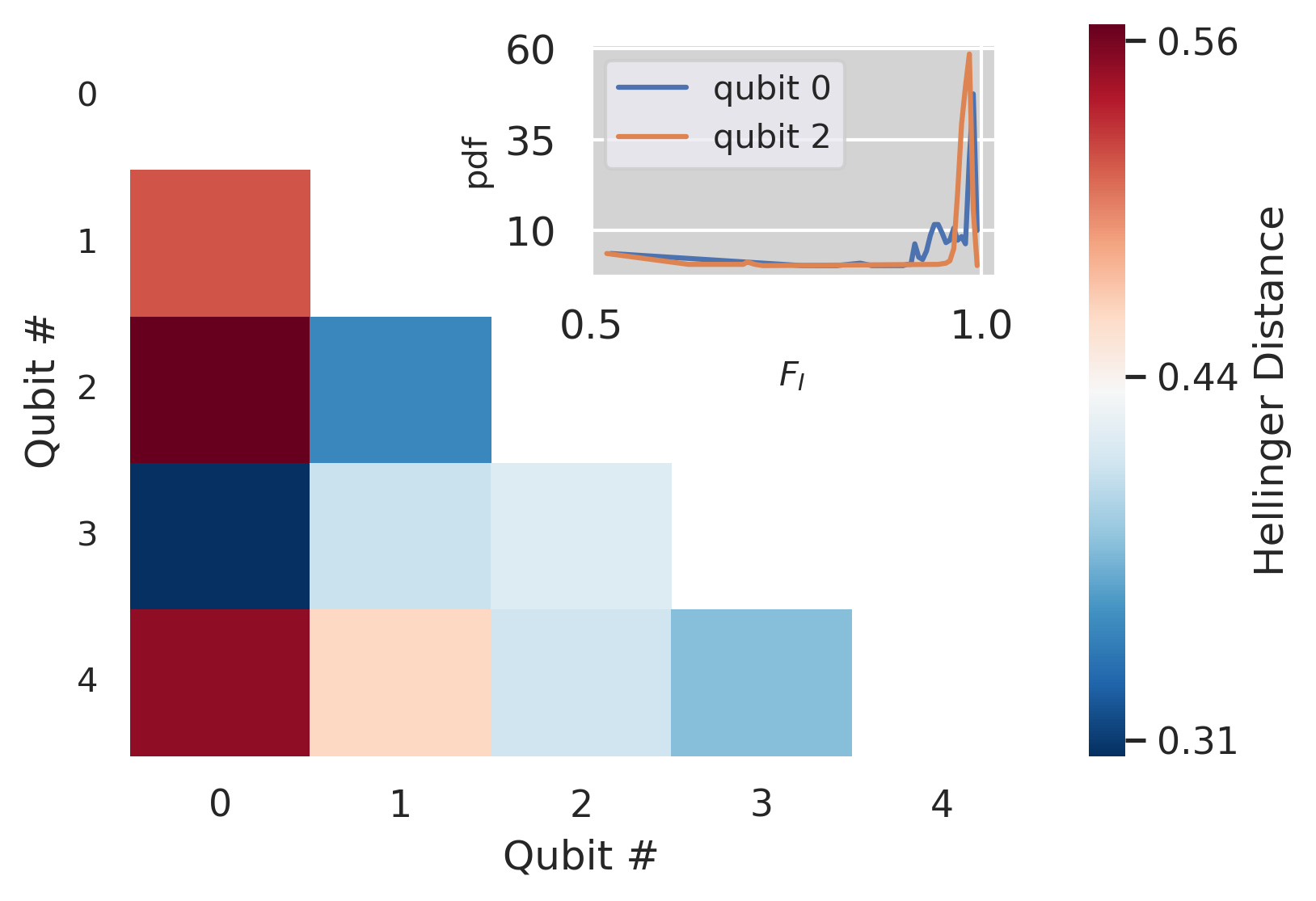}
\caption{Spatial stability of the SPAM fidelity for the \yorktown device from May 2019 to December 2020, where the inset highlights the registers with the maximum distance.}
\label{fig:yorktown_FI_spatial_birch}
\end{figure}

\begin{figure}[htbp]
\centerline{
\includegraphics[width=\figurewidth]{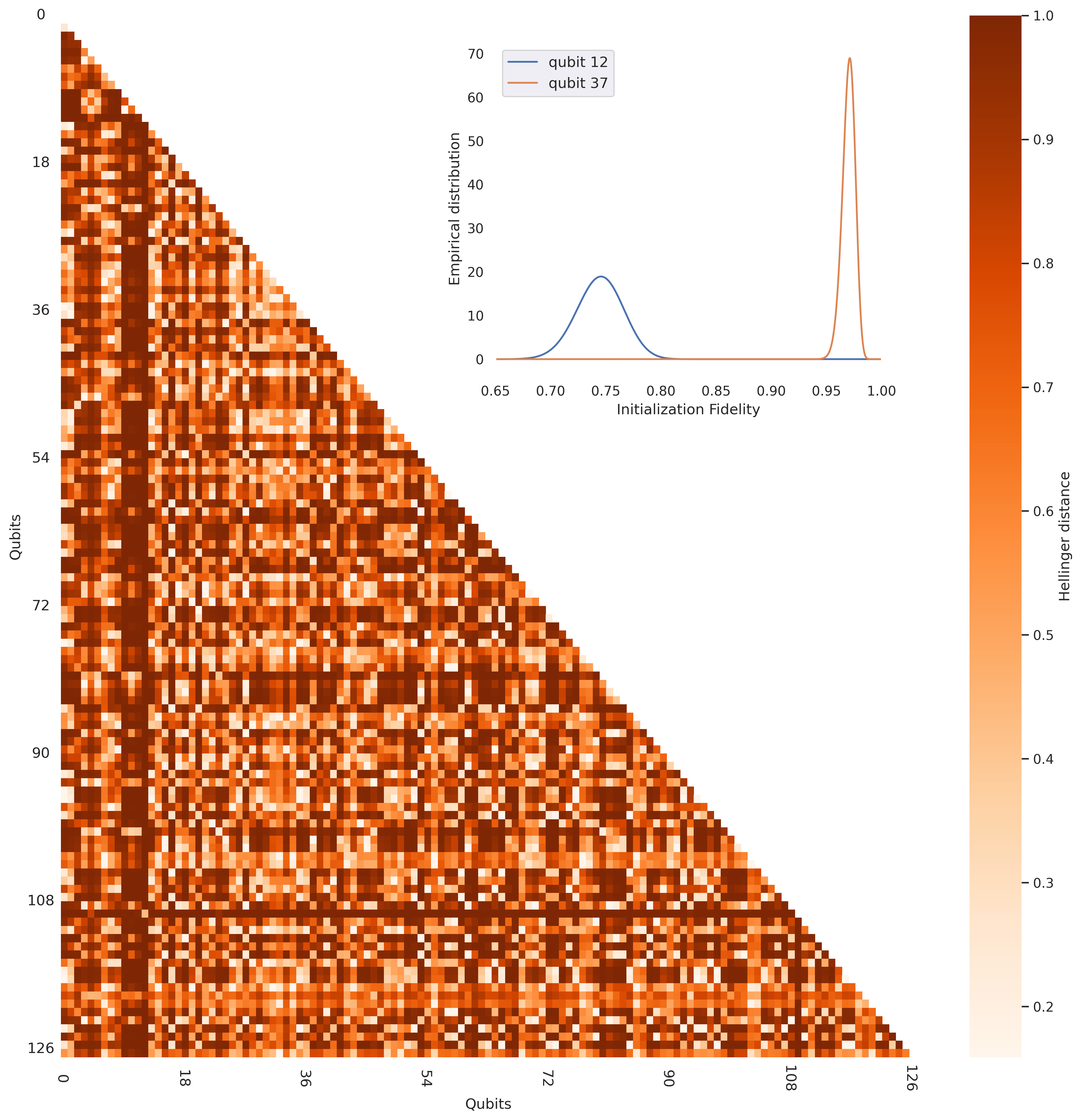}
}
\caption{Spatial reliability of the SPAM fidelity $F_\text{SPAM}$ for the \washington device. The heat map shows the Hellinger distance between the distributions for each register pair. This metric captures the probability that the quantum processor register is spatially dissimilar. The inset shows the distributions of $F_\text{SPAM}$ for registers 12 and 37, which represent distance=0.99 due to minimal overlap. The estimation of the distribution for qubit-$i$ utilizes data for Sep-2022.}
\label{fig:washington_FI_hellinger_spatial}
\end{figure}

On the other hand, spatial reliability refers to the similarity of the densities between different locations. 
We find that the SPAM fidelity can vary significantly depending on location, even on the same device!  
This could be due to variations in the physical properties of the individual qubits.
Fig.~\ref{fig:willow}(f) shows the distance between the spatial densities of \toronto, Fig.~\ref{fig:yorktown_FI_spatial_birch} for \yorktown and Fig.~\ref{fig:washington_FI_hellinger_spatial} for \washington. 
The large distance measures show that spatial reliability is poor within the same device.
\subsection{Reliability of CNOT noise characterization}
We next analyze the reliability of CNOT gates by studying density similarity. 
The underlying random variable is the CNOT gate fidelity $F_G$ \cite{xie202399}. 

\begin{figure*} 
\centering
\subfloat[CNOT (0, 1)]{\includegraphics[width=\figurewidth]{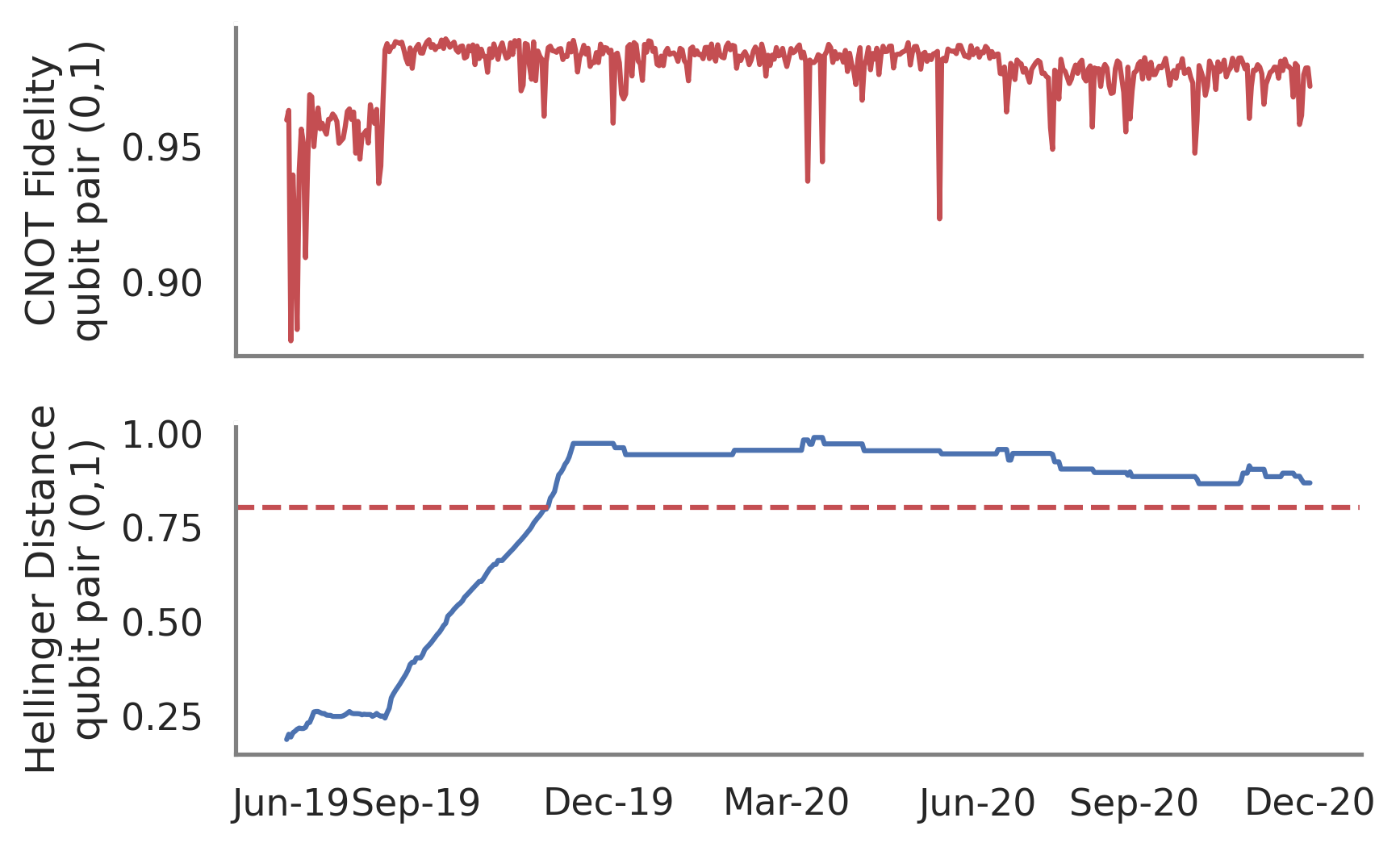}\label{fig:yorktown_FG_temporal_maple_01}}
\hspace{0.5in}\subfloat[CNOT (0, 2)]{\includegraphics[width=0.4\columnwidth]{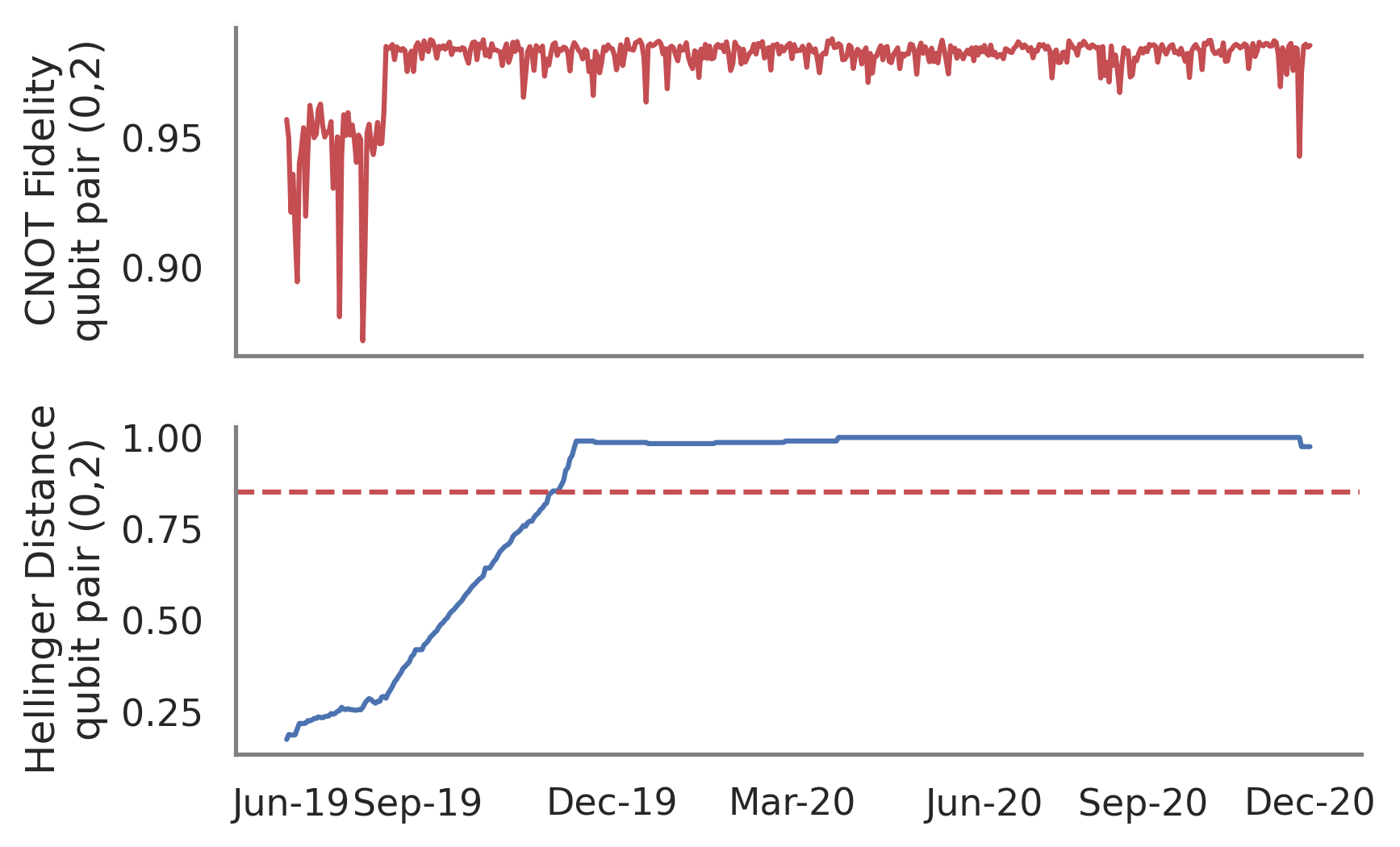}\label{fig:yorktown_FG_temporal_maple_02}}
\hspace{0.5in}\subfloat[CNOT (1, 2)]{\includegraphics[width=0.4\columnwidth]{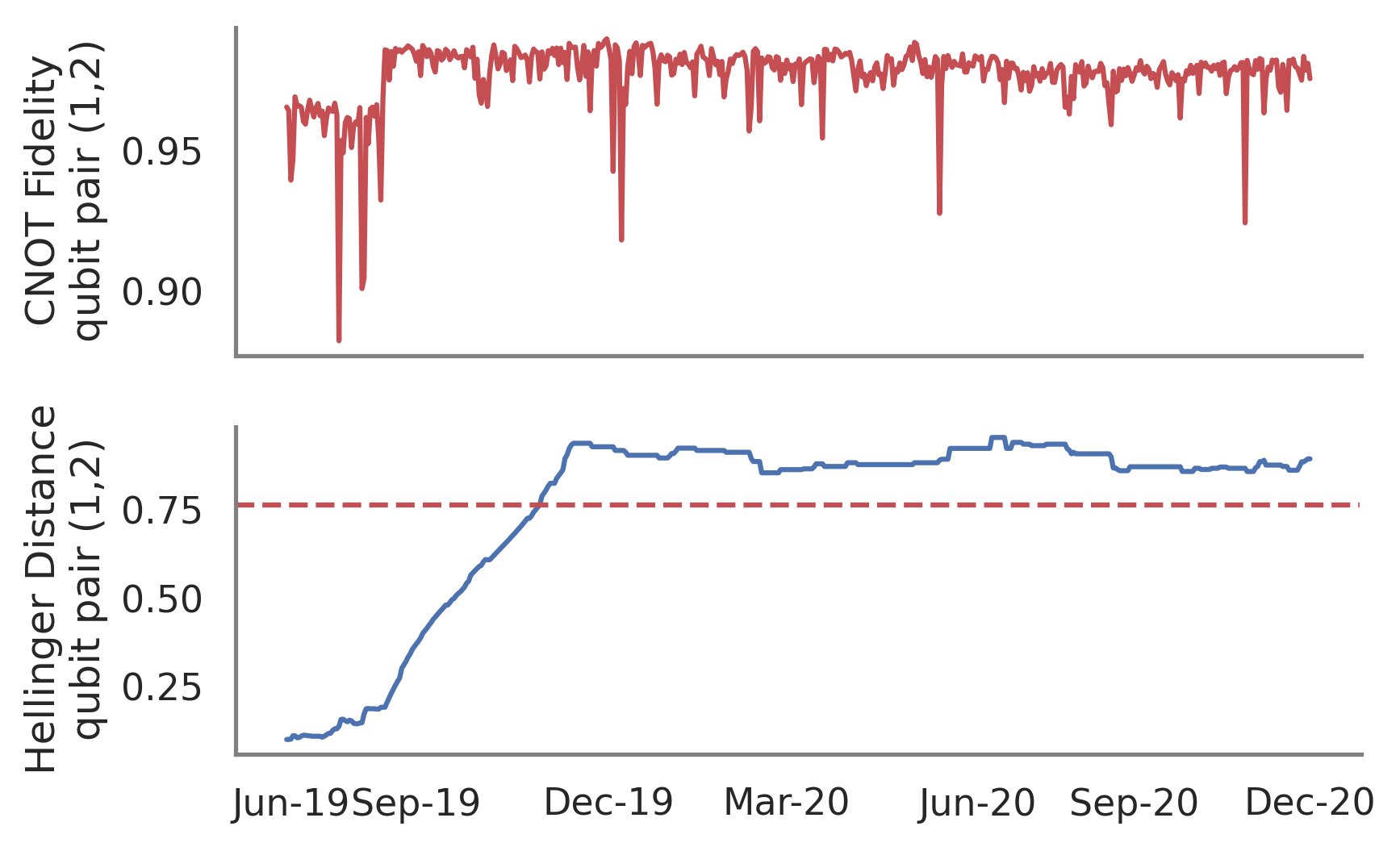}\label{fig:yorktown_FG_temporal_maple_12}}
\hspace{0.5in}\subfloat[CNOT (3, 2)]{\includegraphics[width=0.4\columnwidth]{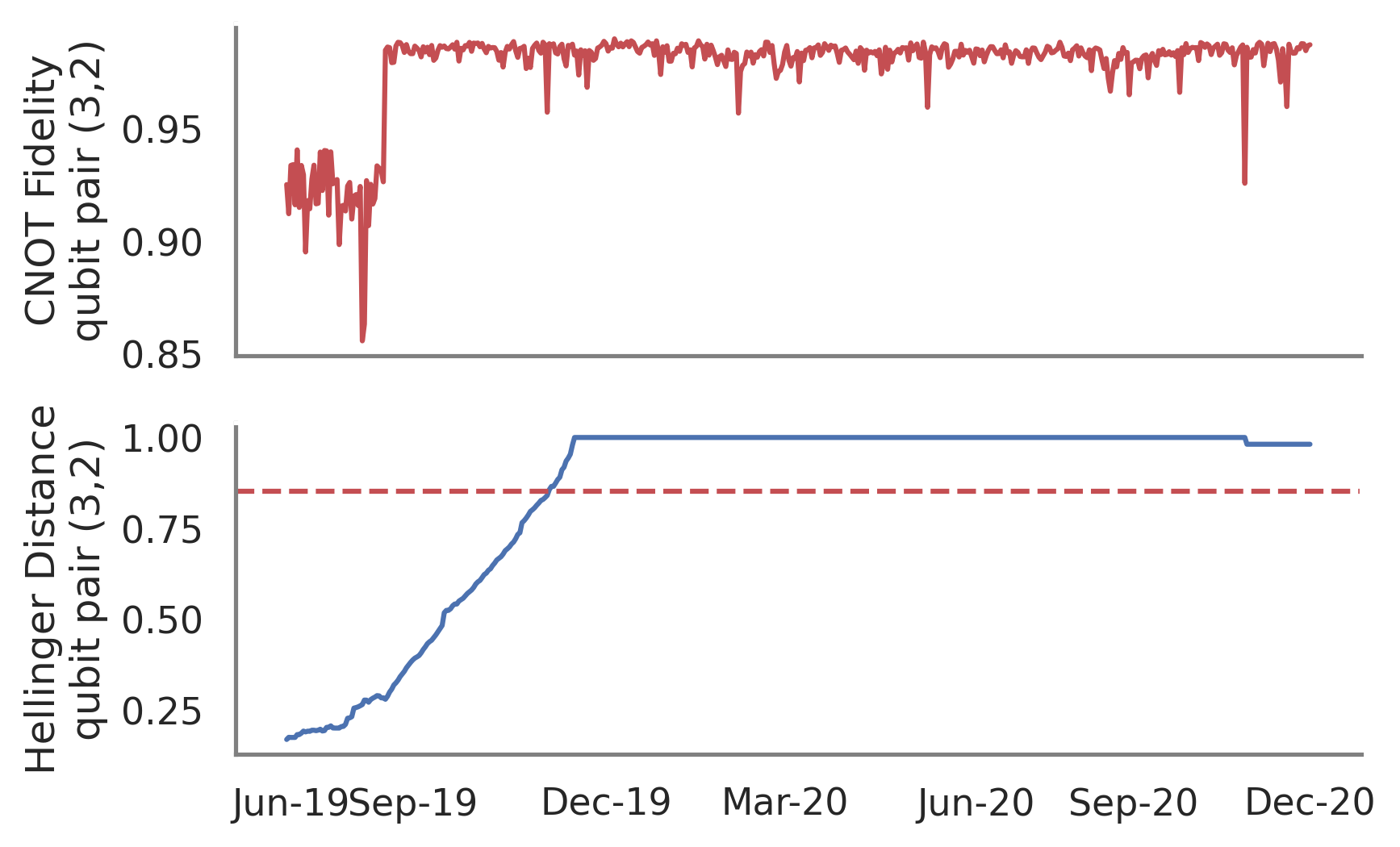}\label{fig:yorktown_FG_temporal_maple_32}}
\hspace{0.5in}\subfloat[CNOT (4, 2)]{\includegraphics[width=0.4\columnwidth]{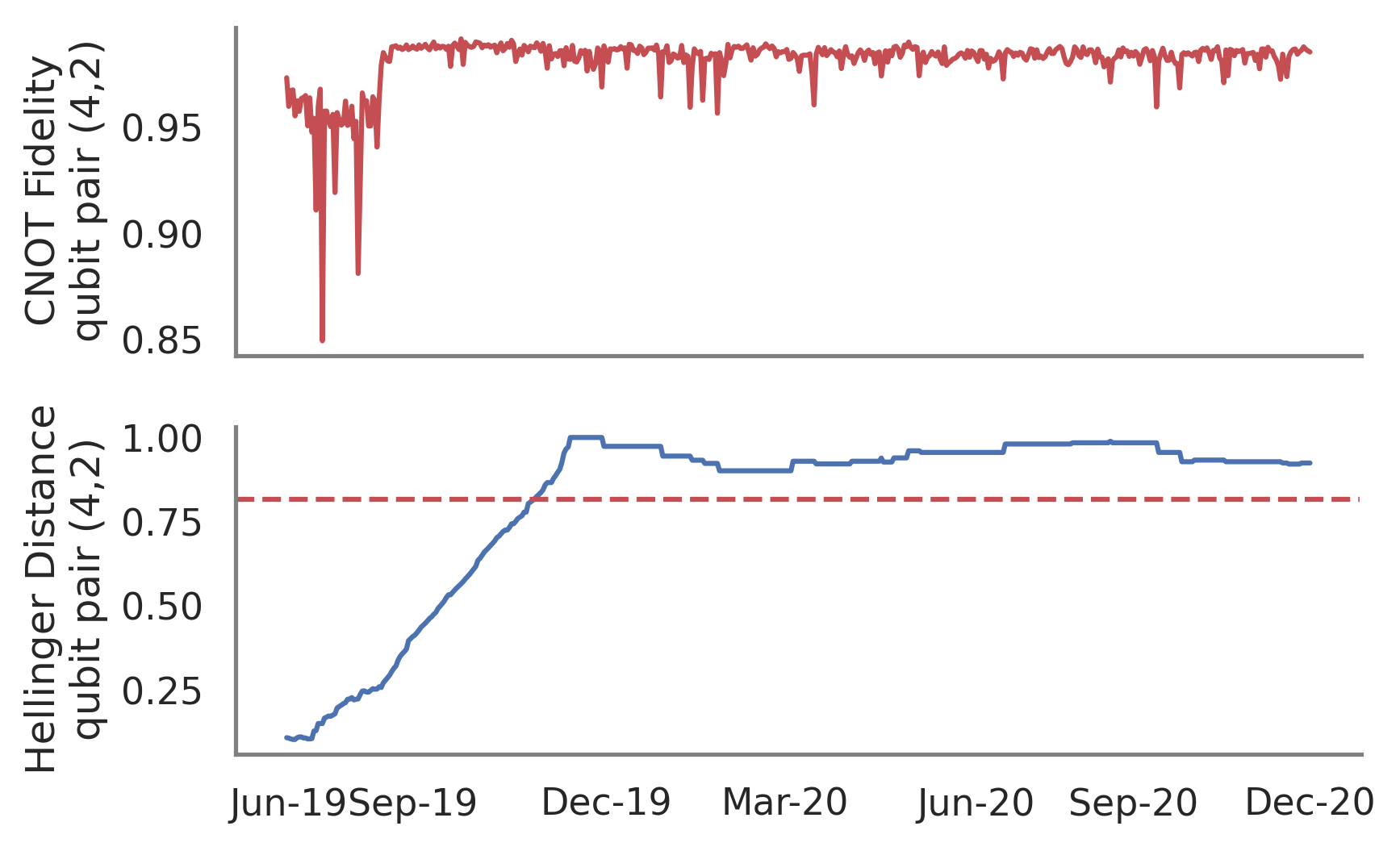}\label{fig:yorktown_FG_temporal_maple_42}}
\caption{
(a)-(e) Temporal stability of the gate fidelity $F_G$ for the CNOT gate for sequential register pairs in the \yorktown device from March 2019 to December 2020. The top panel shows the average $F_G$ of the register pair and the bottom panel shows a running calculation of the Hellinger distance with respect to May 2019. The dashed red line is the median value.
}
\label{fig:fir}
\end{figure*}

\begin{figure}[htbp]
\centering
\includegraphics[width=\textwidth]{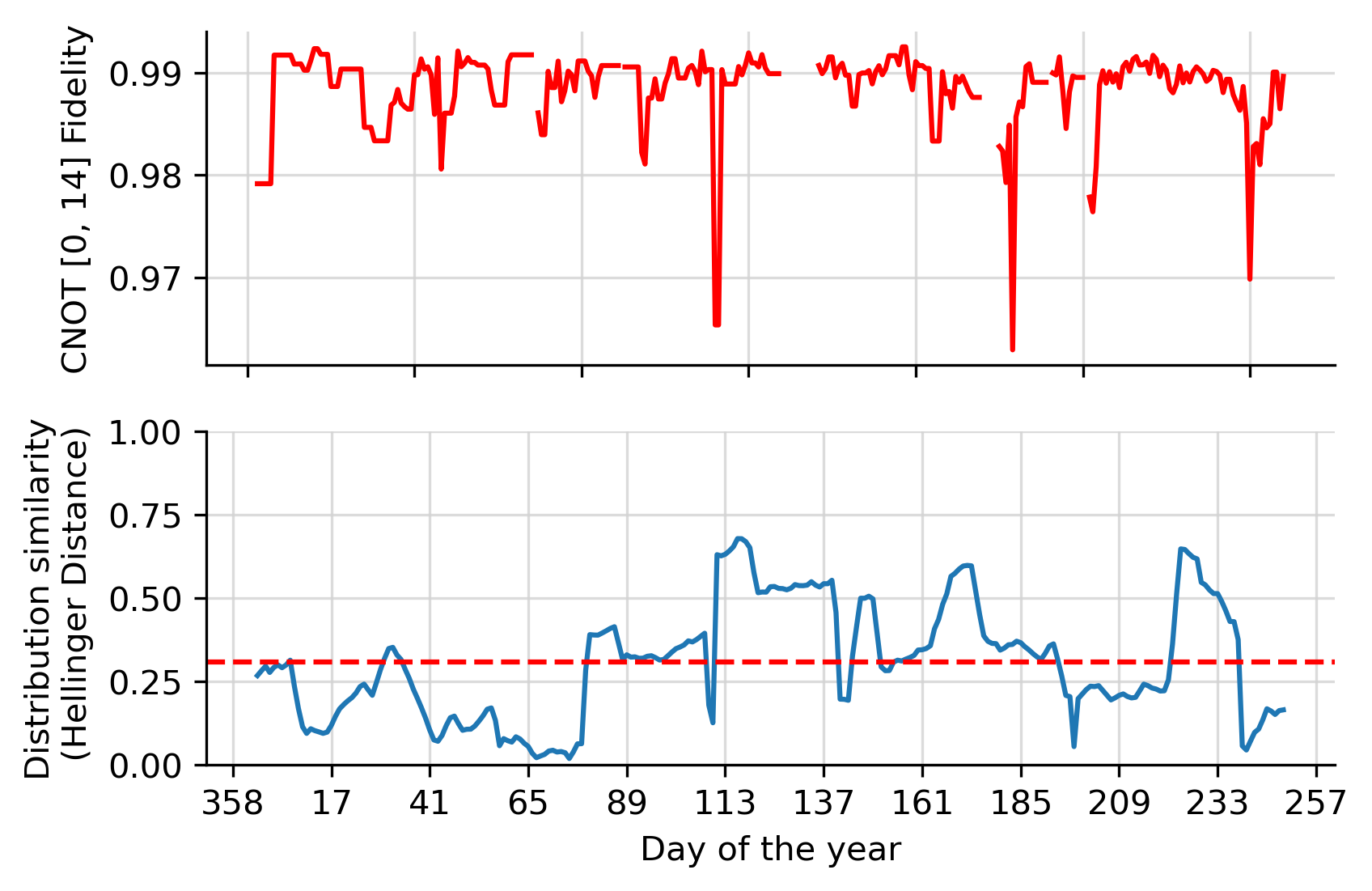}
\caption{Temporal reliability of the gate fidelity $F_G$ for the CNOT gate for register pairs $0$ and $14$.}
\label{fig:washington_FG_hellinger_temporal}
\end{figure}

Fig.~\ref{fig:fir}(a)-(e) show the results for \yorktown between Mar-19 and Dec-20. 
The reference density (for distance computation) is Mar-19. 
We see that the metric diverged sharply between Jun-19 and Dec-19, but fluctuated much less in the next 12 months. 
It stands to reason that the CNOT operations performed in Mar-19 are quite different from those performed in Dec-19! 
Fig.~\ref{fig:washington_FG_hellinger_temporal} shows similar results for \washington for a CNOT between qubits 0 and 14.  The distance reaches as high as 0.7 (max allowed is 1.0) in the second half of 2022.

\begin{figure}[htbp]
\centering
\includegraphics[width=\figurewidth]{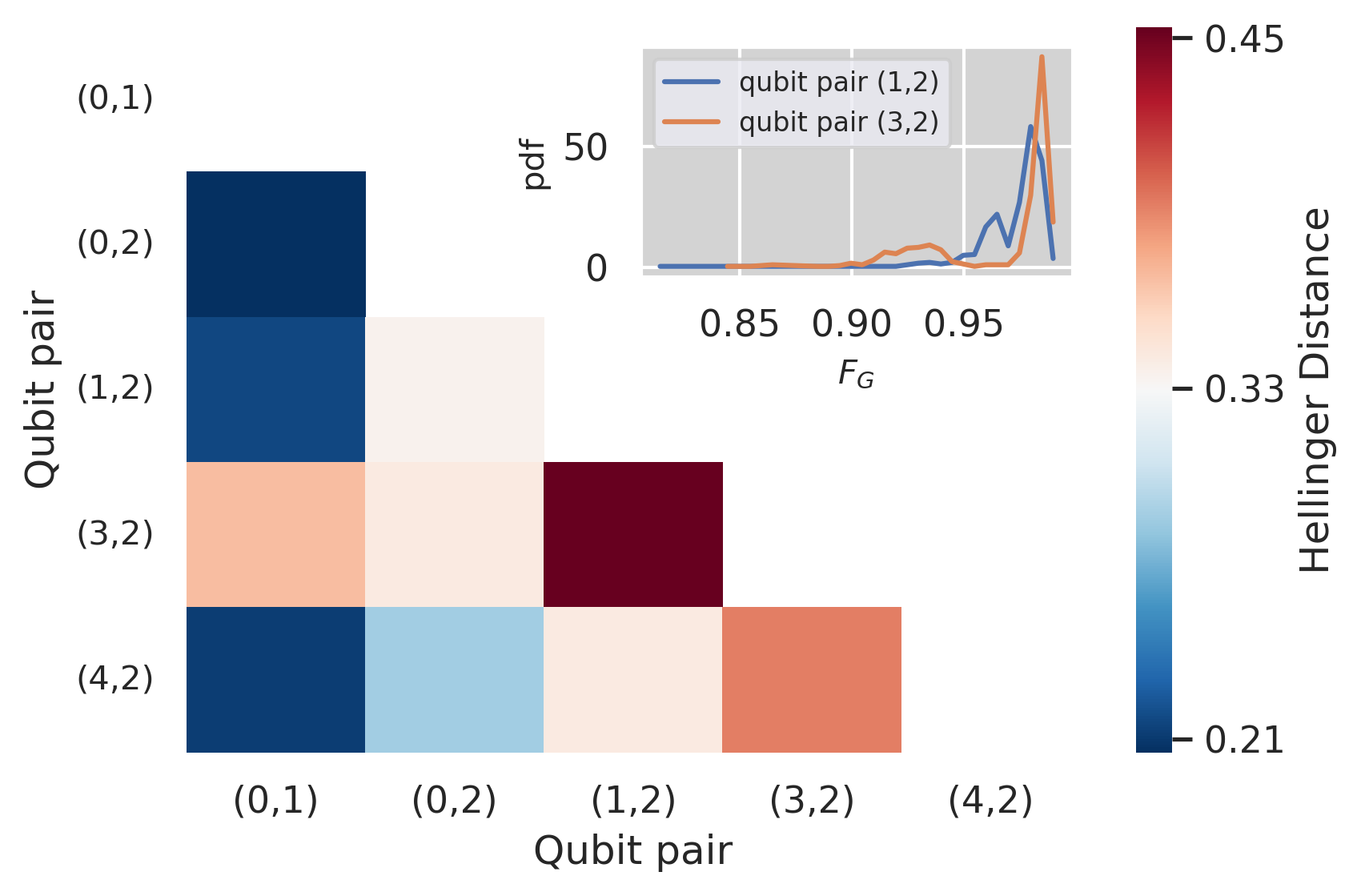}
\caption{Spatial stability of the gate fidelity $F_G$ for the CNOT gates of the \yorktown device from March 2019 to December 2020. The inset shows the distribution of gate fidelities for pairs (1,2) and (3,2), which yield a Hellinger distance 0.467.}
\label{fig:yorktown_FG_spatial_fir}
\end{figure}

The previous discussion was around temporal reliability of CNOT. The spatial reliability of \yorktown is shown in Fig.~\ref{fig:yorktown_FG_spatial_fir}. 
(Note: our spatial calculation used the entire temporal dataset.) 
The worst gates were found to be between qubits (1,2) and (3,2) which yielded a distance of 0.467. 
The worst densities are shown in the inset.  
Fig.~\ref{fig:washington_FG_hellinger_spatial} is the same plot but for \washington for Sep-22. 
The most dissimilar pairs in this case were (11, 12) and (19, 20) for which the distance exceeded 0.99. The inset shows the worst case densities. 
The lookup table for the 144 connections (which specifies which qubit-pairs are being referred to in the CNOT gate) is provided in Table~\ref{tab:cnot_gate_lookup_table}. 
One of the takeaways from these analyses is that CNOT fidelities may show misleadingly similar means despite having starkly dissimilar densities. 
Another takeaway is that there exist certain qubits (such as qubit 3 for yorktown) which can be associated with generally worse reliability outcomes. 
\subsection{Reliability of duty cycle characterization}
\begin{figure}[htbp]
\centering
\includegraphics[width=\figurewidth]{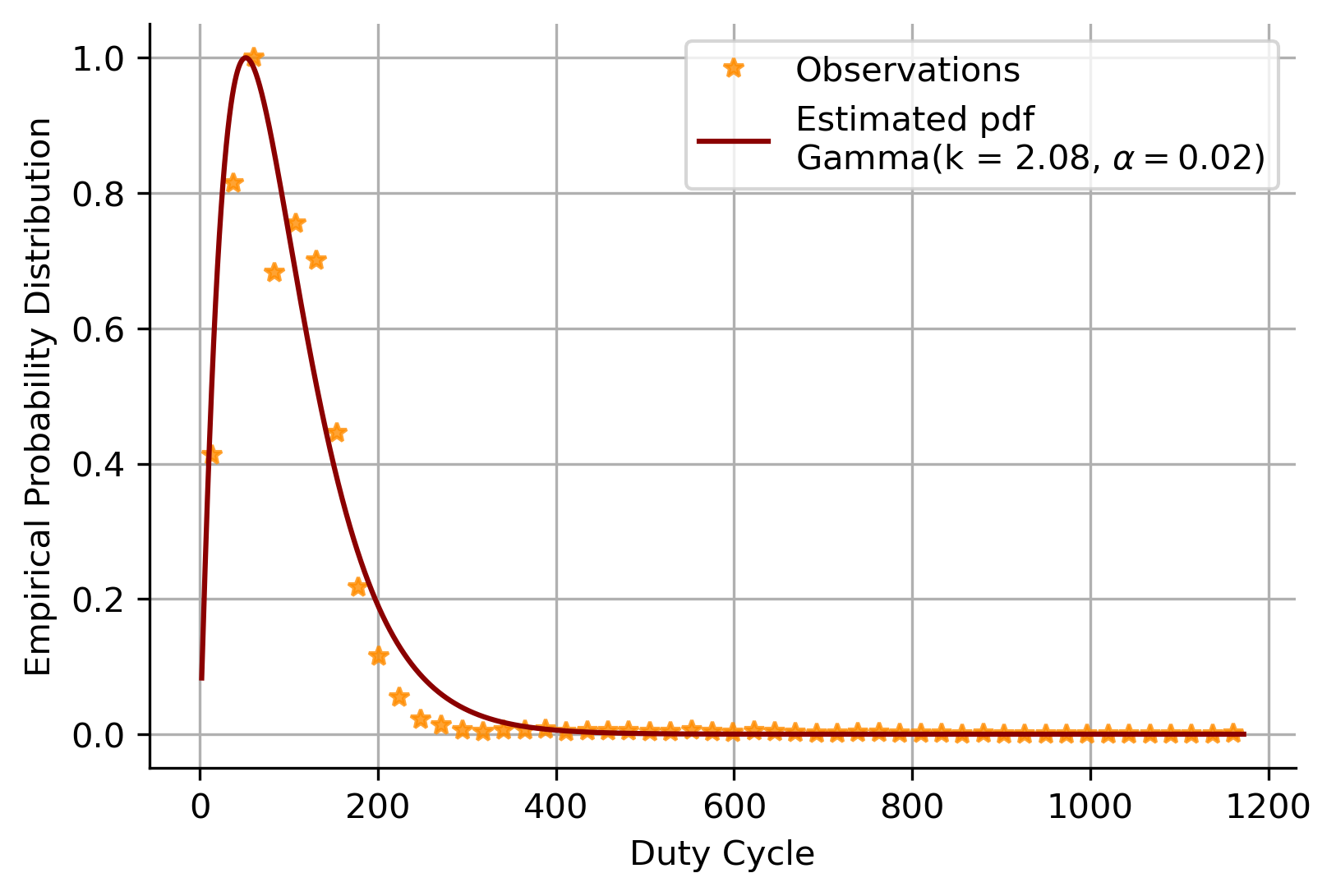}
\caption{Experimentally observed duty cycle ($\tau$) for the CNOT gate for washington.}
\label{fig:washington_tau_gamma_fit_example}
\end{figure}

The CNOT duty cycle \cite{dasgupta2023reliability} was defined in Sec.~\ref{sec:reliability_metrics}. 
It is a random variable with an observable density. 
An example density, shown in Fig.~\ref{fig:washington_tau_gamma_fit_example}, uses data from \washington between Dec-21 and May-22. 

\begin{figure*} 
\centering
\subfloat[Register pair (0,1)]{\includegraphics[width=0.4\columnwidth]{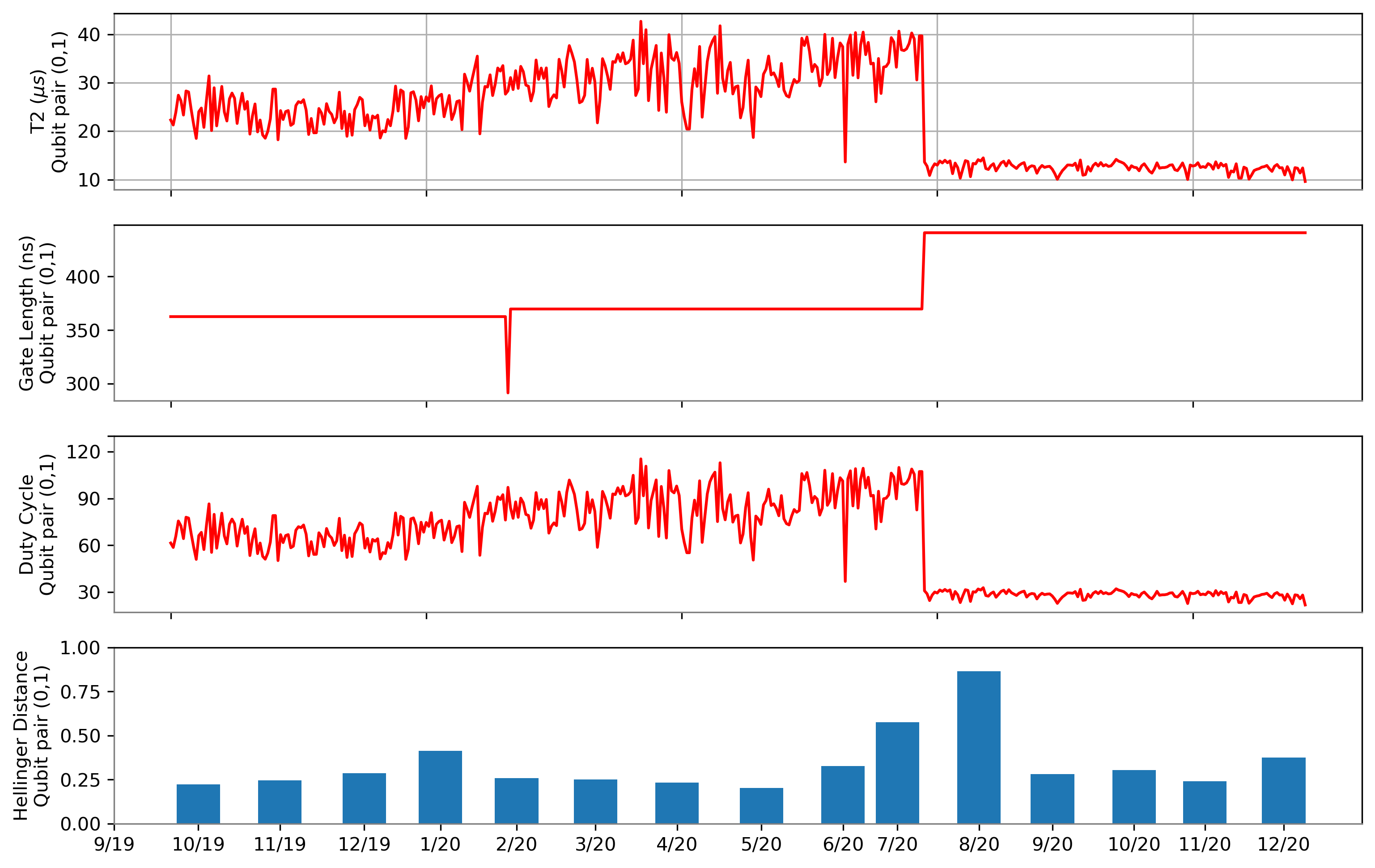}\label{fig:yorktown_tau_temporal_banyan_01}}
\hspace{0.5in}\subfloat[Register pair (1,2)]{\includegraphics[width=0.4\columnwidth]{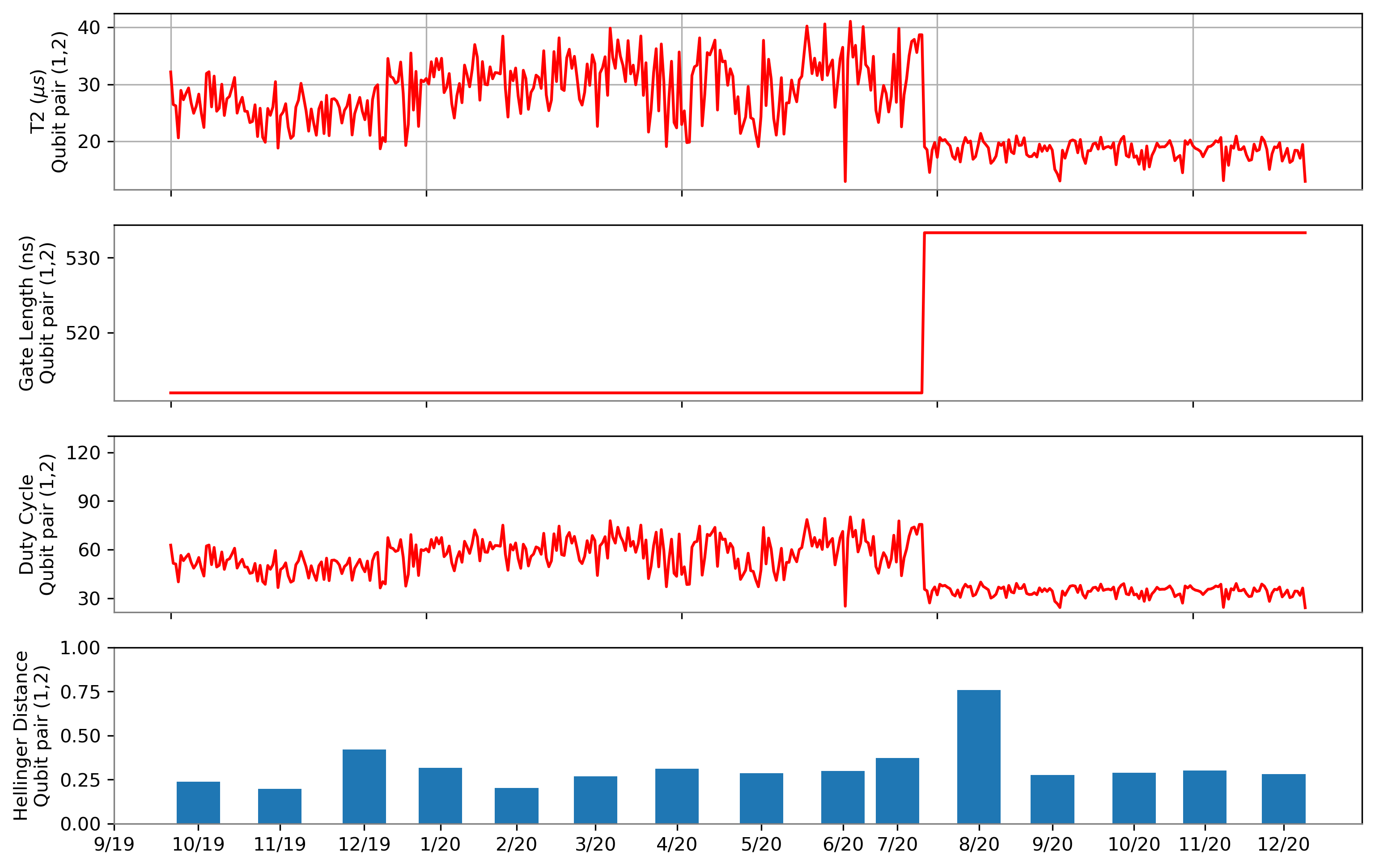}\label{fig:yorktown_tau_temporal_banyan_12}}
\hspace{0.5in}\subfloat[Register pair (2,3)]{\includegraphics[width=0.4\columnwidth]{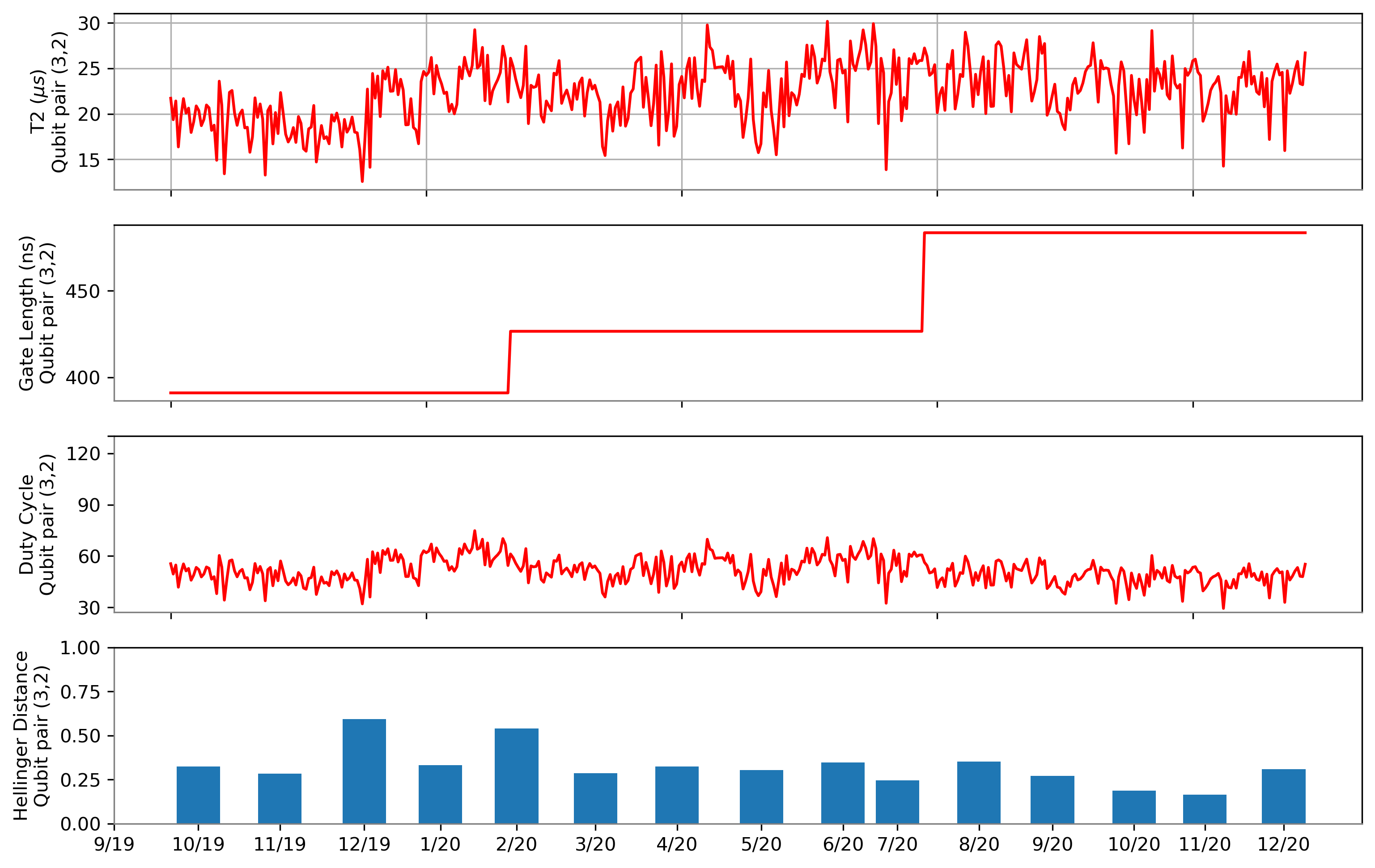}\label{fig:yorktown_tau_temporal_banyan_32}}
\hspace{0.5in}\subfloat[Register pair (2,4)]{\includegraphics[width=0.4\columnwidth]{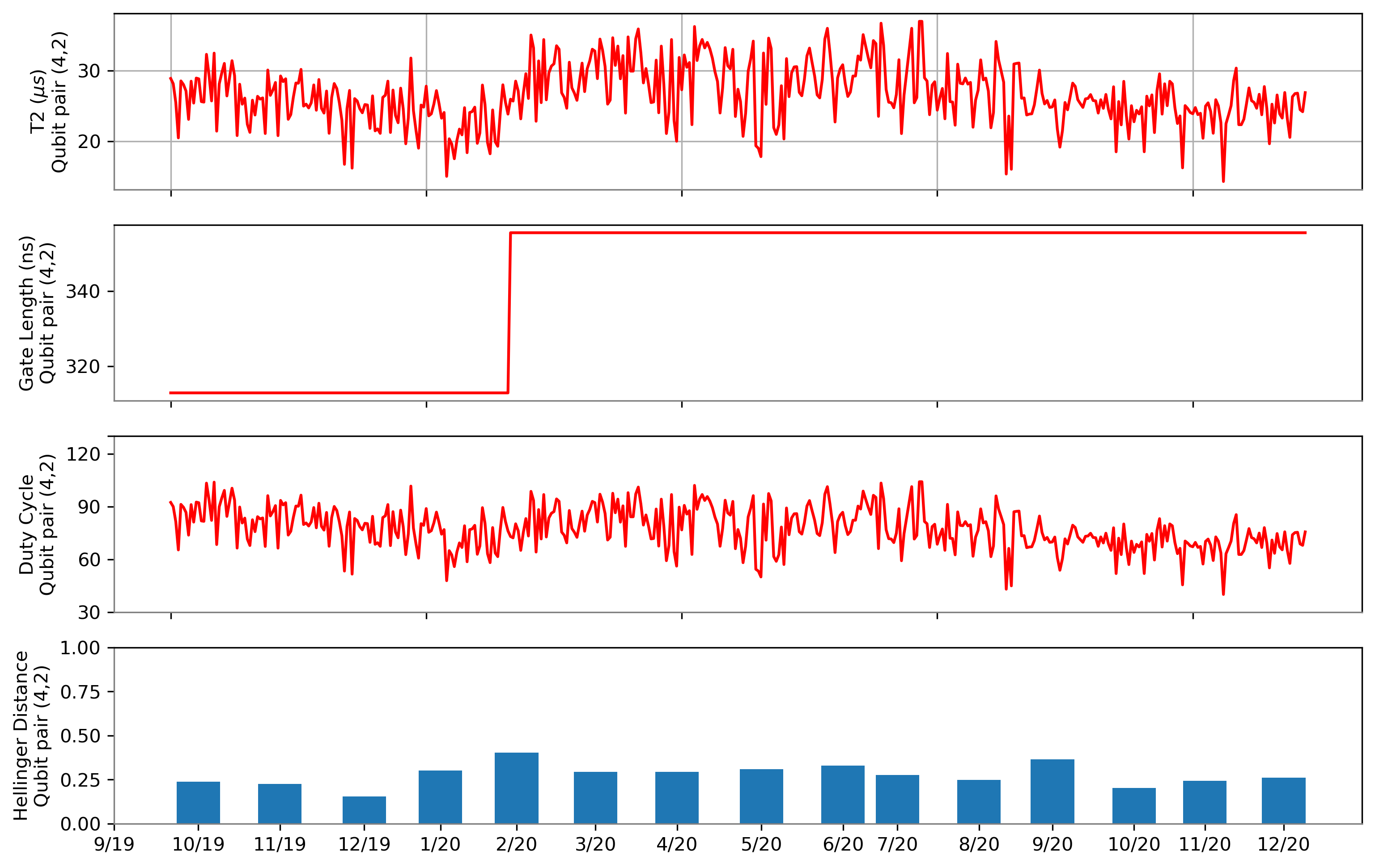}\label{fig:yorktown_tau_temporal_banyan_42}}
\caption{
(a)-(d) Temporal stability of the CNOT duty cycle for sequential register pairs in the yorktown device. The top panel shows the harmonic mean of the register decoherence time $T_2$ for the elements, the upper-middle panel shows the gate duration $T_G$, the lower-middle panel plots the corresponding duty cycle $\tau$, and the bottom panel presents the Hellinger distance for the duty cycle averaged over a one-month window. The dashed red line is the median value.
}
\label{fig:yorktown_tau_temporal_banyan}
\end{figure*}

The temporal reliability analysis for \yorktown is shown in Fig.~\ref{fig:yorktown_tau_temporal_banyan}. 
It has the time-series for the mean decoherence time $T_2$ (to be precise, this is the harmonic mean across the two qubits of the CNOT gate). 
It also has the time-series for the tunable gate duration and distance metric (that quantifies reliability). The latter uses densities based on running $3$-month data. 
We can see, for example, in Fig.~\ref{fig:yorktown_tau_temporal_banyan}~(a) on July $24, 2020$, the $T_2$ time decreased abruptly from $77~\mu$s to $31~\mu$s for register $0$ and from $82~\mu$s to $24~\mu$s for register $1$. 
A corresponding random sharp increase is seen in the gate duration from $370$~ns to $441$~ns. 
These random changes led to a sudden sharp decrease in the duty cycle from $107.2$ to $30.9$.

\begin{figure}[htbp]
\centering
\includegraphics[width=\figurewidth]{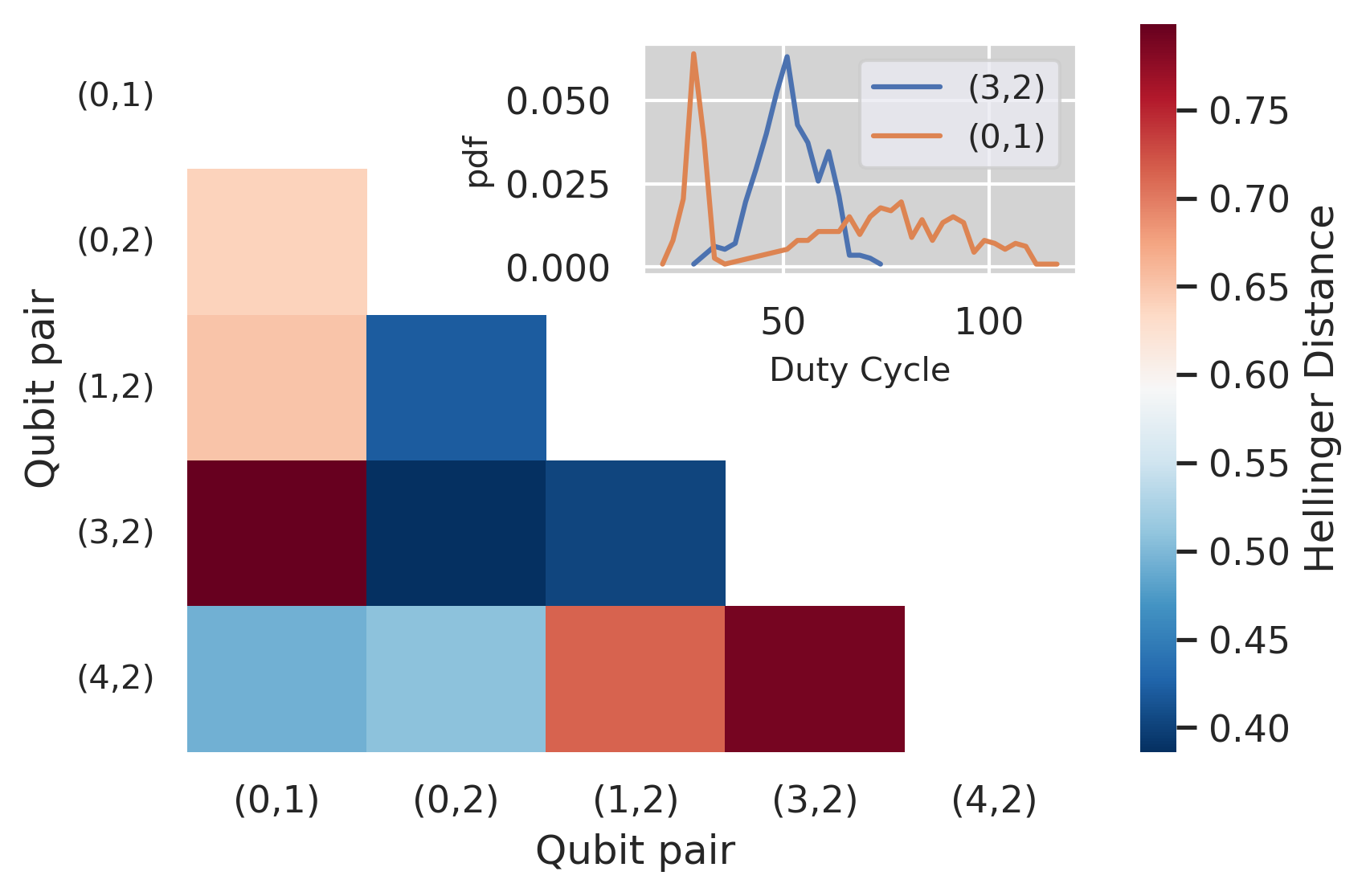}
\caption{The spatial stability of the duty cycle $\tau$ for \yorktown. The inset shows the experimental histograms for register pairs (0,1) and (2,3) which are separated by the largest Helligner distance of 0.789.}
\label{fig:yorktown_tau_spatial_spruce}
\end{figure}

\begin{figure}[htbp]
\centering
\includegraphics[width=\figurewidth]{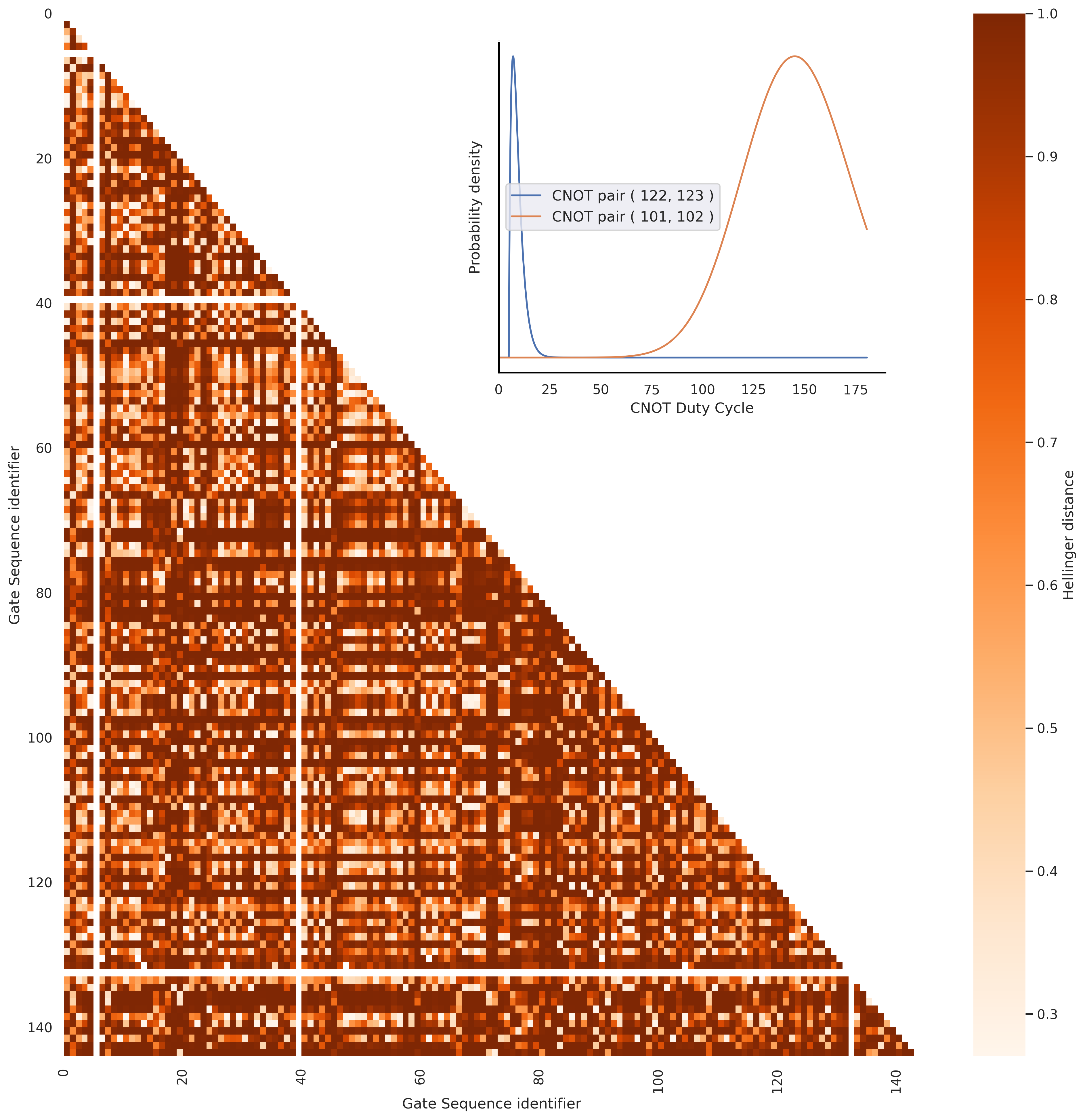}
\caption{Spatial reliability of the duty cycle $\tau$ for the CNOT gates of the \washington device. 
The heat map shows the Hellinger distance between the distributions for the duty cycles of nearest-neighbor connections. The inset shows the distribution of duty cycle for pairs (46, 47) and (96, 109), which yield a Hellinger distance 0.99. This metric captures the probability that the quantum processor register is spatially dissimilar. The estimation of the distribution for CNOT-$i$ utilizes duty-cycle data for Sep-2022.}
\label{fig:washington_tau_hellinger_spatial}
\end{figure}

The spatial reliability of \yorktown is depicted in Fig.~\ref{fig:yorktown_tau_spatial_spruce} 
and for \washington in Fig.~\ref{fig:washington_tau_hellinger_spatial}. 
For \washington, the most dissimilar duty cycles were CNOT$(46,47)$ and CNOT$(96,109)$. 
For \yorktown, the most dissimilar duty cycles were CNOT$(0,1)$ and CNOT$(3,2)$. 

In general, the temporal reliability of the duty cycle improved after July $2020$, but the spatial reliability remained poor.
\subsection{Reliability of addressability characterization}
Addressability \cite{dasgupta2020characterizing} (defined in Sec.~\ref{sec:reliability_metrics}) is a measure of how well register qubits can be addressed individually. 
The ideal addressability is 1, but in practice it can be lower due to hardware noise. 

\begin{figure}[htbp]
\centering
\includegraphics[width=\figurewidth]{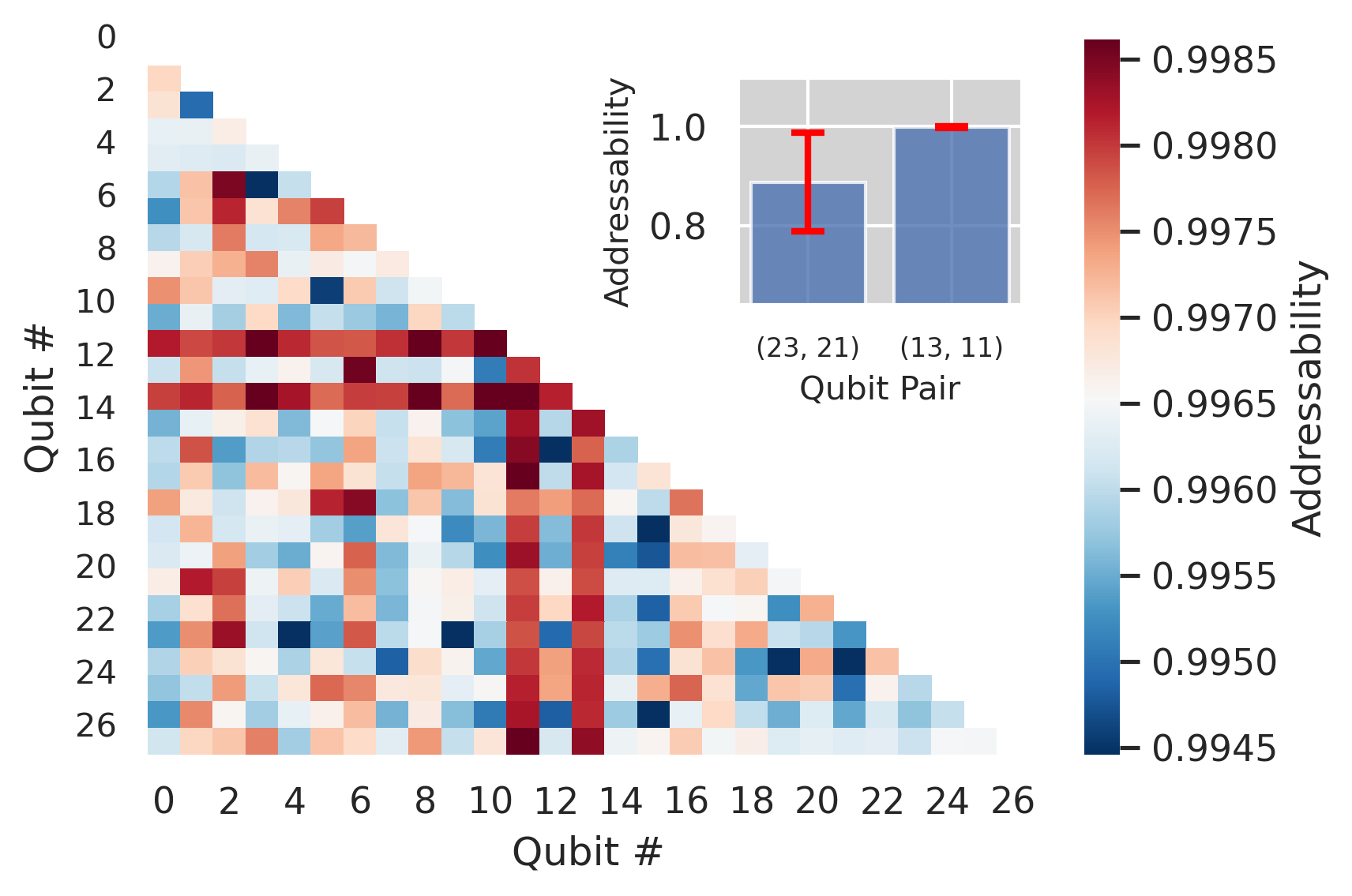}
\caption{Addressability of register pairs in the \toronto device sampled 08:00-08:30 AM (UTC-5) on 11 December 2020. This data corresponds to the register prepared in the separable fiducial state. The inset shows the range i.e. the lowest and highest values for addressability. The average of (23,21) is the lowest value at 0.887 while all other values lie in the range [0.992, 1). The outlier is the only value that does not appear in the plot.}
\label{fig:addr}
\end{figure}

Fig.~\ref{fig:addr} plots the addressability of the \toronto device when tested by encoding a fiducial separable state $\ket{00}$ in each register pair. 
The heatmap highlights how the addressability varies across the register. 
The inset compares the limits of this behavior by showing the worst case $0.89$ for qubits $(23, 21)$ and the best $\sim 1.00$ for qubits $(11, 13)$. 

\begin{figure}[H]
\center
\includegraphics[width=\columnwidth]{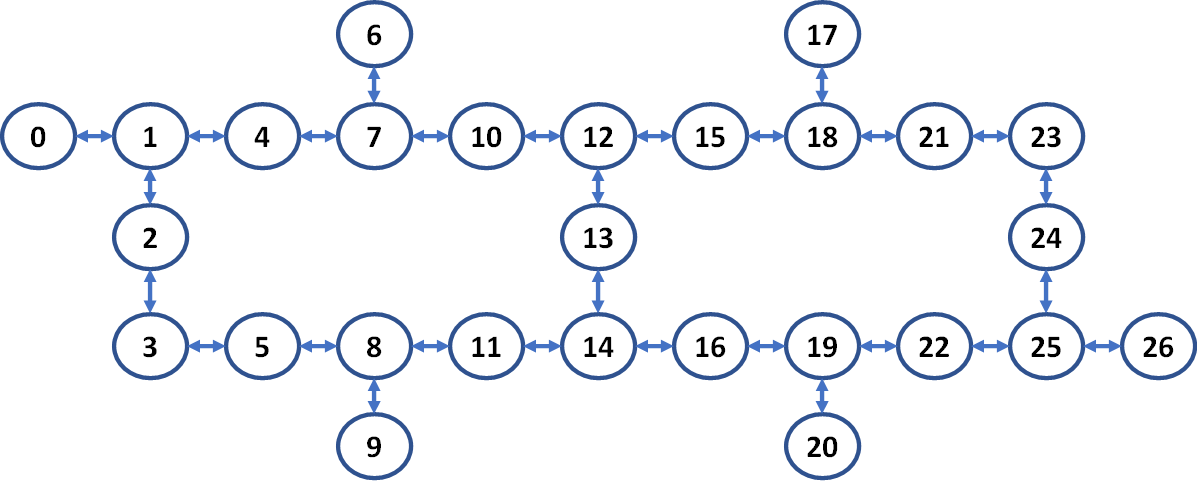}
\caption{Schematic of the \toronto device, produced by IBM. Circles represent register elements, while edges denote the connectivity for performing two qubit operations.}
\label{fig:toronto}
\end{figure}

\begin{figure}[htbp]
\centering
\includegraphics[width=\figurewidth]{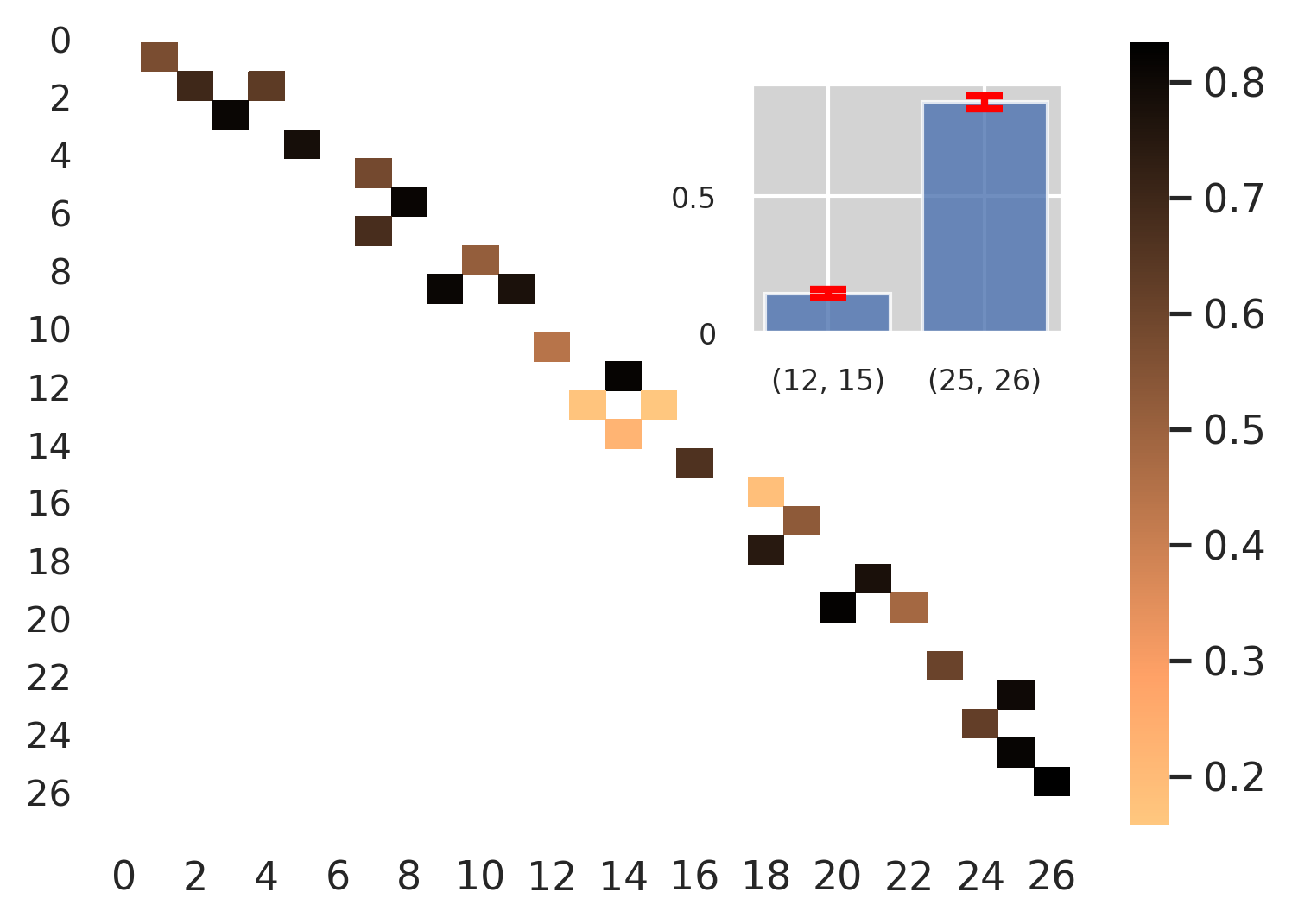}
\caption{Normalized mutual information of register pairs in the toronto device sampled 11:00-11:30 PM (UTC-5) on 11 December 2020. Data corresponds to register pairs prepared in the Bell state and the inset shows the range of the lowest and highest values for the normalized mutual information.}
\label{fig:bell_spatial_nmi_cotton}
\end{figure}

An interesting extension can be performed by encoding a Bell-state within the register pair. 
As a maximally entangled state, the addressability should be $0$. 
Fig.~\ref{fig:bell_spatial_nmi_cotton} shows the results for the 28 register pairs that support direct preparation of a Bell state based on the nearest-neighbor connections shown in Fig.~\ref{fig:toronto}. 
We found that the worst registers for Bell state information preservation were $12$ and $15$ with: 
\begin{equation}
\eta = 1-F_A = 0.14 \pm 0.014, 
\end{equation}
while the best registers were $25$ and $26$ with:
\begin{equation}
\eta = 1-F_A = 0.84\pm 0.023.
\end{equation}
This unreliable device has now been retired.
\section{Circuit reliability}
So far, we focused on analyzing the device components at individual qubit and gate level. 
But how do we measure reliability more holistically at the circuit level? 
Examining thousands (if not millions) of qubits and gates may not help to drawn any conclusions at the application level.  

\begin{figure}[htbp]
\centering
\includegraphics[width=\figurewidth]{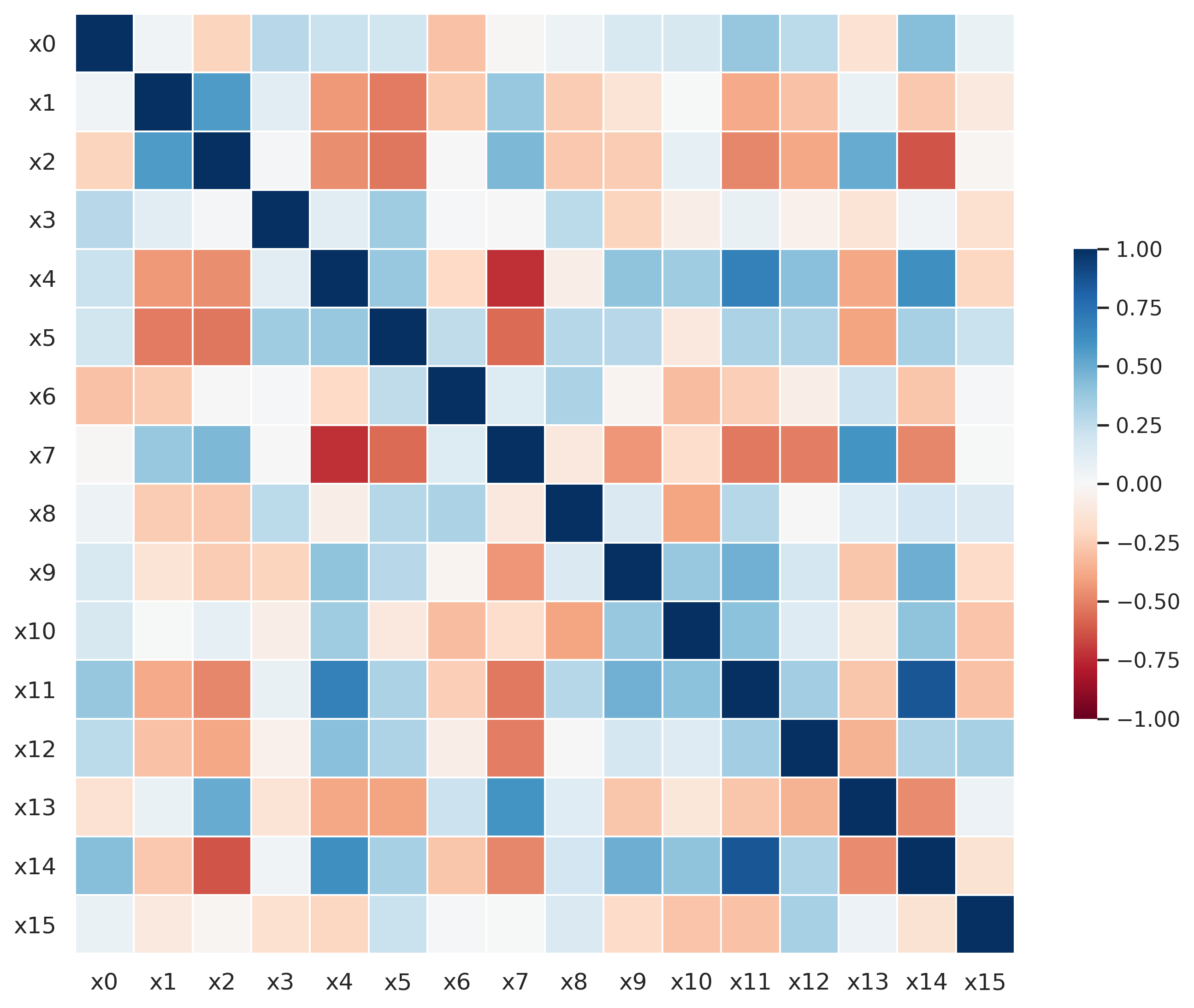}
\caption{
The Pearson correlation coefficients between the 16 characterization parameters as observed in \added{April-2023}. 
Dark blue and dark red colors represent the Pearson coefficients $1$ and $-1$, respectively. 
The axes labels $(x0, x1, \cdots x15)$ correspond to various noise sources as listed in Table~\ref{tab:noiseParameters}. 
}
\label{fig:correlation_matrix}
\end{figure}

To analyze circuit reliability, we study the stationarity of the multi-variate noise vector $X$ associated with a quantum circuit. Conducting a holistic analysis requires studying the time-variation \cite{etxezarreta2021time} of the joint distribution of the quantum noise. However, significant correlations exist amongst the noise parameters characterizing a quantum circuit, and these correlations can have a substantial impact on the performance of quantum error correction and validation methods. The correlation structure can also change over time (see Fig.~\ref{fig:correlation_matrix} for example). For example, in \cite{10.1103/physrevlett.97.040501}, the authors examine the decoherence of a quantum computer in a temporally and spatially correlated environment, finding that minor adjustments to error correction codes can systematically reduce the impact of long-range correlations on the quantum system. In \cite{10.1103/physreva.93.022303}, researchers discuss the limitations of single-metric approaches for quantum characterization and study the influence of noise correlations on randomized benchmarking (RB). They demonstrate that temporal noise correlations affect the probability density function of RB outcomes, described by a gamma distribution with parameters dependent on the correlation structure, while also noting potential finite-sampling issues and deviations in mean RB outcomes from worst-case errors when noise correlations are present. 

The selection of the noise metrics is based on the DiVincenzo criterion \cite{divincenzo2000physical}, like we discussed in Ch.~\ref{ch:statistical_taxonomy} and it spans qubit-specific SPAM fidelity, gate fidelity, duty cycle, and addressability. 
A test circuit is used for reporting, specifically the Bernstein-Vazirani circuit \cite{bernstein1993quantum}, which is a canonical algorithm with proven quantum advantage in noiseless limit. 
These metrics are shown in Table~\ref{tab:noiseParameters}. 

Specifically, we compute the distance between the time-varying densities for the random variable:
$$X = (X_0, \cdots, X_{15}),$$ 
where $X_0, \cdots, X_{15}$ are described in Table~\ref{tab:noiseParameters}. 
The distance is denoted by $H_X(t ,t_0)$ where $t$ denotes the months ranging from  \added{Jan-2022 to Apr-2023}. 
The reference time $t_0$ is set to \added{Jan-$2022$}. 
Table~\ref{tab:marginal_hellinger19} shows the distance from the Jan-2022 density. 
The first 16 columns present the distance data for univariate (marginal) distributions, while the last three columns contain the distances for the composite densities. 
Specifically, the $H_r$ column shows the distance between the joint densities modeled using copulas \cite{10.1177/1748006x13481928}, $H_a$ represents the average over the marginals, and $H_n$ represents the normalized distance (as discussed in Chapter~\ref{ch:statistical_taxonomy}).

\begin{figure}[htbp]
\centering
\includegraphics[width=\figurewidth]{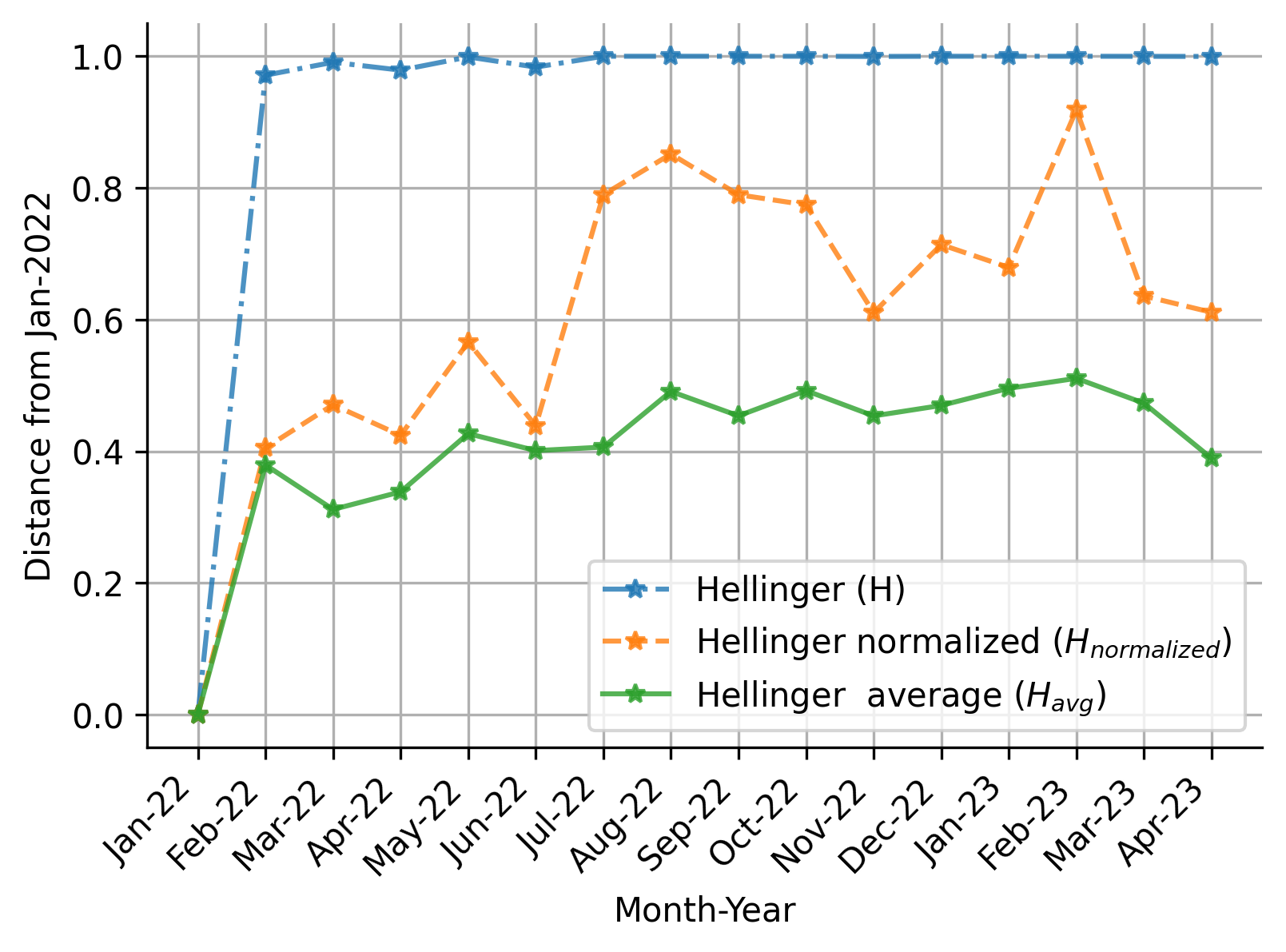}
\caption{The figure presents results of reliability testing on a transmon platform \added{over 16 months}. Plots of the three Hellinger distance measures, with the unmodified measure $H$ (blue line) being insensitive due to dimensionality issues. The blue and orange lines both capture the correlation structure of the joint distribution (with the orange line being normalized to enhance discrimination power), while the green line lacks correlation capture. \added{The normalized measure $H_\textrm{normalized}$ ranges between 0.41 and 0.92, while the average measure $H_\textrm{avg}$ varies between 0.431 and 0.51.} The latter captures monthly variations in the marginal Hellinger distance for each of the 16 error parameters but fails to account for correlations.}
\label{fig:distance_from_ref}
\end{figure}

\begin{figure}[!hb]
\centering
\includegraphics[width=\figurewidth]{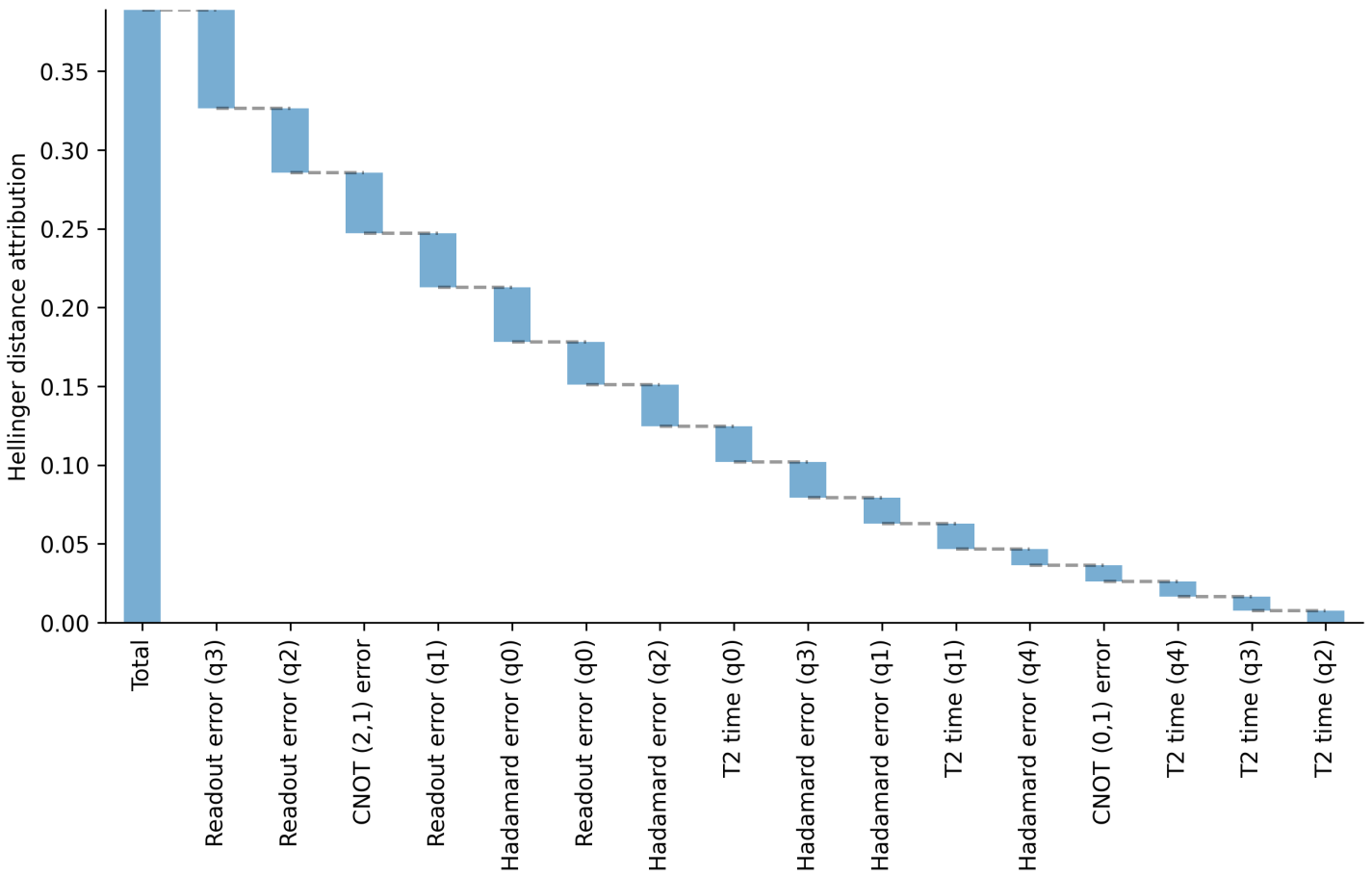}
\caption{
Decomposition of the sources of quantum noise non-stationarity. 
The degree of non-stationarity varies amongst the sources. 
The plot shows the contributions made by each noise type to the composite Hellinger distance (a measure of the degree of non-stationarity). 
The Hellinger distance measures the statistical distance between the joint distribution of the noise observed in \added{Apr-2023} to the joint distribution of the noise observed in  \added{Jan-2022}. 
The various noise types contribute varying percentages but no single term dominates the sum.}
\label{fig:distance_attribution_ind}
\end{figure}

Fig.~\ref{fig:distance_from_ref} plots the the last three columns containing the distances for the composite densities. 
The unmodified distance per Eqn.~\ref{eq:hellinger_unmodified} is less sensitive due to the curse of dimensionality \cite{verleysen2005curse}. 
The orange line represents the normalized distance per Eqn.~\ref{eq:hellinger_normalized}, while the green line represents the distance averaged over all the marginal distributions per Eqn.~\ref{eq:hellinger_avg}. 
The contributions to the average distance \added{for Apr-2023 distribution compared to that of Jan-2022} from various noise sources are compared and contrasted in Fig.~\ref{fig:distance_attribution_ind}. 
It is apparent that no single parameter dominates the average though SPAM noise accounts for the largest contribution to circuit non-stationarity. 

It should be emphasized that the average measure does not consider correlations between parameters, and a specific correlation structure can cause the joint distance to increase by reducing overlap in a subset of dimensions. 
\added{The normalized and average distances, represented by the orange and green lines respectively, demonstrate greater discriminatory power with the observed data ranges of 0.51 and 0.20, respectively. These ranges are considerably higher compared to the unmodified distance with an observed data range of 0.029.}
\clearpage
\vspace{0.5in}
\vspace{0.5in}
\vspace{0.5in}
\vspace{0.5in}
\vspace{0.5in}
\vspace{0.5in}
\vspace{0.5in}
\vspace{0.5in}
\vspace{0.5in}
\vspace{0.5in}
\vspace{0.5in}
\vspace{0.5in}
\vspace{0.5in}
\begin{table}[htp]
\centering
\begin{tabular}{|p{1.4cm}|p{6.0cm}|p{1.4cm}|}
    \hline 
\textit{Parameter} & \textit{Description} &\textit{Model} \\ \hline
$\xrm_{0}$ & SPAM  fidelity, register  0 & ABC  \\ \hline
$\xrm_{1}$ & SPAM  fidelity, register  1 & ABC  \\ \hline
$\xrm_{2}$ & SPAM  fidelity, register  2 & ABC  \\ \hline
$\xrm_{3}$ & SPAM  fidelity, register  3 & ABC  \\ \hline
$\xrm_{4}$ & CNOT fidelity, control 0, target 1 & DP$\otimes$DP  \\ \hline
$\xrm_{5}$ & CNOT fidelity, control 2, target 1  & DP$\otimes$DP \\ \hline
$\xrm_{6}$ & $T_2$ time, register 0 & TR  \\ \hline
$\xrm_{7}$ & $T_2$ time, register 1& TR   \\ \hline
$\xrm_{8}$ & $T_2$ time, register 2& TR   \\ \hline
$\xrm_{9}$ & $T_2$ time, register 3& TR   \\ \hline
$\xrm_{10}$ & $T_2$ time, register 4& TR   \\ \hline
$\xrm_{11}$ & $H$ fidelity, register 0& CP   \\ \hline
$\xrm_{12}$ & $H$ fidelity, register 1& CP   \\ \hline
$\xrm_{13}$ & $H$ fidelity, register 2& CP \\ \hline
$\xrm_{14}$ & $H$ fidelity, register 3  & CP  \\ \hline
$\xrm_{15}$ & $H$ fidelity, register 4  & CP \\ \hline
\end{tabular}
\medskip\medskip\medskip
\captionsetup{font=small}
\caption{The 16-parameter model derived from the \washington data set has four types of quantum noise processes: (i) `ABC': asymmetric binary channel model, (ii) `CP': coherent phase error model, (iii) `DP': depolarizing noise model and (iv) `TR': thermal relaxation noise model. Note that the two-qubit model `DP$\otimes$DP' is a tensor product of depolarizing noise.}
\label{tab:noiseParameters}
\end{table}
\vspace{0.5in}
\begin{table*}[htbp]
\begin{center}
\caption{Hellinger distance values for the device parameters.}
\label{tab:marginal_hellinger19} 
\scalebox{0.6}{
\small
\begin{tabular}{|l|c|c|c|c|c|c|c|c|c|c|c|c|c|c|c|c|c|c|c|}
\hline
Month & $H_{X_0}$ & $H_{X_1}$ & $H_{X_2}$ & $H_{X_3}$ & $H_{X_4}$ & $H_{X_5}$ & $H_{X_6}$ & $H_{X_7}$ & $H_{X_8}$ & $H_{X_9}$ & $H_{X_{10}}$ & $H_{X_{11}}$ & $H_{X_{12}}$ & $H_{X_{13}}$ & $H_{X_{14}}$ & $H_{X_{15}}$ & $H_{\textrm{n}}$ & $H_{\textrm{a}}$ & $H_{\textrm{r}}$\\\hline
Jan-22&0.0&0.0&0.0&0.0&0.0&0.0&0.0&0.0&0.0&0.0&0.0&0.0&0.0&0.0&0.0&0.0&0.0&0.0&0.0\\\hline Feb-22&0.82&0.08&0.43&0.38&0.3&0.35&0.28&0.43&0.45&0.39&0.32&0.24&0.62&0.66&0.04&0.26&0.41&0.38&0.971439\\\hline Mar-22&0.97&0.22&0.31&0.3&0.07&0.05&0.31&0.17&0.6&0.22&0.32&0.31&0.11&0.36&0.48&0.17&0.47&0.31&0.99084\\\hline Apr-22&0.64&0.03&0.23&0.53&0.27&0.45&0.21&0.06&0.95&0.11&0.1&0.45&0.33&0.37&0.07&0.61&0.42&0.34&0.978632\\\hline May-22&0.8&0.27&0.61&0.77&0.11&0.4&0.64&0.21&0.38&0.36&0.26&0.65&0.21&0.34&0.34&0.49&0.57&0.43&0.99897\\\hline Jun-22&0.81&0.4&0.43&0.9&0.26&0.34&0.53&0.34&0.28&0.1&0.16&0.36&0.43&0.69&0.16&0.23&0.44&0.4&0.983197\\\hline Jul-22&0.74&0.42&0.96&1.0&0.3&0.44&0.15&0.25&0.48&0.37&0.2&0.16&0.38&0.17&0.14&0.32&0.79&0.41&1.0\\\hline Aug-22&0.89&0.5&0.9&1.0&0.26&0.41&0.53&0.35&0.97&0.31&0.21&0.55&0.22&0.21&0.27&0.3&0.85&0.49&1.0\\\hline Sep-22&0.82&0.48&0.93&1.0&0.45&0.22&0.44&0.34&0.91&0.1&0.08&0.46&0.27&0.31&0.31&0.13&0.79&0.45&1.0\\\hline Oct-22&0.72&0.55&0.9&1.0&0.05&0.42&0.32&0.18&0.95&0.07&0.27&0.43&0.36&0.66&0.74&0.25&0.77&0.49&1.0\\\hline Nov-22&0.36&0.63&0.65&1.0&0.4&0.13&0.55&0.53&0.98&0.22&0.14&0.4&0.29&0.39&0.29&0.3&0.61&0.45&0.999713\\\hline Dec-22&0.42&0.64&0.58&1.0&0.27&0.53&0.45&0.24&0.99&0.17&0.27&0.7&0.65&0.03&0.19&0.37&0.72&0.47&0.999995\\\hline Jan-23&0.46&0.59&0.46&1.0&0.06&0.5&0.31&0.3&0.91&0.53&0.26&0.69&0.55&0.34&0.53&0.46&0.68&0.5&0.999975\\\hline Feb-23&0.45&0.61&0.65&1.0&0.44&0.46&0.44&0.26&1.0&0.12&0.52&0.53&0.4&0.33&0.61&0.34&0.92&0.51&1.0\\\hline Mar-23&0.47&0.5&0.79&1.0&0.22&0.21&0.46&0.33&0.51&0.21&0.4&0.71&0.62&0.09&0.71&0.31&0.64&0.47&0.999876\\\hline Apr-23&0.43&0.55&0.65&1.0&0.16&0.61&0.37&0.25&0.12&0.14&0.15&0.55&0.26&0.42&0.36&0.17&0.61&0.39&0.999721\\\hline
\end{tabular}
}
\end{center}
\end{table*}
\vspace{0.5in}
\begin{table*}[htbp]
\caption{Gate lookup table}
\begin{center}
\scalebox{0.7}{
\small
\begin{tabular}{|c|c|c|c|c|}
\hline
Label 0 = (0, 14) & Label 1 = (0, 1) & Label 2 = (100, 101) & Label 3 = (100, 110) & Label 4 = (100, 99)\\
\hline
Label 5 = (100, 101) & Label 6 = (101, 102) & Label 7 = (101, 102) & Label 8 = (102, 103) & Label 9 = (102, 92)\\
\hline
Label 10 = (102, 103) & Label 11 = (103, 104) & Label 12 = (103, 104) & Label 13 = (104, 105) & Label 14 = (104, 111)\\
\hline
Label 15 = (104, 105) & Label 16 = (105, 106) & Label 17 = (105, 106) & Label 18 = (106, 107) & Label 19 = (106, 93)\\
\hline
Label 20 = (106, 107) & Label 21 = (107, 108) & Label 22 = (107, 108) & Label 23 = (108, 112) & Label 24 = (109, 114)\\
\hline
Label 25 = (109, 96) & Label 26 = (10, 11) & Label 27 = (10, 9) & Label 28 = (100, 110) & Label 29 = (110, 118)\\
\hline
Label 30 = (104, 111) & Label 31 = (111, 122) & Label 32 = (108, 112) & Label 33 = (112, 126) & Label 34 = (113, 114)\\
\hline
Label 35 = (109, 114) & Label 36 = (113, 114) & Label 37 = (114, 115) & Label 38 = (114, 115) & Label 39 = (115, 116)\\
\hline
Label 40 = (115, 116) & Label 41 = (116, 117) & Label 42 = (116, 117) & Label 43 = (117, 118) & Label 44 = (110, 118)\\
\hline
Label 45 = (117, 118) & Label 46 = (118, 119) & Label 47 = (118, 119) & Label 48 = (119, 120) & Label 49 = (10, 11)\\
\hline
Label 50 = (11, 12) & Label 51 = (119, 120) & Label 52 = (120, 121) & Label 53 = (120, 121) & Label 54 = (121, 122)\\
\hline
Label 55 = (111, 122) & Label 56 = (121, 122) & Label 57 = (122, 123) & Label 58 = (122, 123) & Label 59 = (123, 124)\\
\hline
Label 60 = (123, 124) & Label 61 = (124, 125) & Label 62 = (124, 125) & Label 63 = (125, 126) & Label 64 = (112, 126)\\
\hline
Label 65 = (125, 126) & Label 66 = (11, 12) & Label 67 = (12, 13) & Label 68 = (12, 17) & Label 69 = (12, 13)\\
\hline
Label 70 = (0, 14) & Label 71 = (14, 18) & Label 72 = (15, 22) & Label 73 = (15, 4) & Label 74 = (16, 26)\\
\hline
Label 75 = (16, 8) & Label 76 = (12, 17) & Label 77 = (17, 30) & Label 78 = (14, 18) & Label 79 = (18, 19)\\
\hline
Label 80 = (18, 19) & Label 81 = (19, 20) & Label 82 = (0, 1) & Label 83 = (1, 2) & Label 84 = (19, 20)\\
\hline
Label 85 = (20, 21) & Label 86 = (20, 33) & Label 87 = (20, 21) & Label 88 = (21, 22) & Label 89 = (15, 22)\\
\hline
Label 90 = (21, 22) & Label 91 = (22, 23) & Label 92 = (22, 23) & Label 93 = (23, 24) & Label 94 = (23, 24)\\
\hline
Label 95 = (24, 25) & Label 96 = (24, 34) & Label 97 = (24, 25) & Label 98 = (25, 26) & Label 99 = (16, 26)\\
\hline
Label 100 = (25, 26) & Label 101 = (26, 27) & Label 102 = (26, 27) & Label 103 = (27, 28) & Label 104 = (27, 28)\\
\hline
Label 105 = (28, 29) & Label 106 = (28, 35) & Label 107 = (28, 29) & Label 108 = (29, 30) & Label 109 = (1, 2)\\
\hline
Label 110 = (2, 3) & Label 111 = (17, 30) & Label 112 = (29, 30) & Label 113 = (30, 31) & Label 114 = (30, 31)\\
\hline
Label 115 = (31, 32) & Label 116 = (31, 32) & Label 117 = (32, 36) & Label 118 = (20, 33) & Label 119 = (33, 39)\\
\hline
Label 120 = (24, 34) & Label 121 = (34, 43) & Label 122 = (28, 35) & Label 123 = (35, 47) & Label 124 = (32, 36)\\
\hline
Label 125 = (36, 51) & Label 126 = (37, 38) & Label 127 = (37, 52) & Label 128 = (37, 38) & Label 129 = (38, 39)\\
\hline
Label 130 = (33, 39) & Label 131 = (38, 39) & Label 132 = (39, 40) & Label 133 = (2, 3) & Label 134 = (3, 4)\\
\hline
Label 135 = (39, 40) & Label 136 = (40, 41) & Label 137 = (40, 41) & Label 138 = (41, 42) & Label 139 = (41, 53)\\
\hline
Label 140 = (41, 42) & Label 141 = (42, 43) & Label 142 = (34, 43) & Label 143 = (42, 43) & Label 144 = (43, 44)\\
\hline
Label 145 = (43, 44) & Label 146 = (44, 45) & Label 147 = (44, 45) & Label 148 = (45, 46) & Label 149 = (45, 54)\\
\hline
Label 150 = (45, 46) & Label 151 = (46, 47) & Label 152 = (35, 47) & Label 153 = (46, 47) & Label 154 = (47, 48)\\
\hline
Label 155 = (47, 48) & Label 156 = (48, 49) & Label 157 = (48, 49) & Label 158 = (49, 50) & Label 159 = (49, 55)\\
\hline
Label 160 = (15, 4) & Label 161 = (3, 4) & Label 162 = (4, 5) & Label 163 = (49, 50) & Label 164 = (50, 51)\\
\hline
Label 165 = (36, 51) & Label 166 = (50, 51) & Label 167 = (37, 52) & Label 168 = (52, 56) & Label 169 = (41, 53)\\
\hline
Label 170 = (53, 60) & Label 171 = (45, 54) & Label 172 = (54, 64) & Label 173 = (49, 55) & Label 174 = (55, 68)\\
\hline
Label 175 = (52, 56) & Label 176 = (56, 57) & Label 177 = (56, 57) & Label 178 = (57, 58) & Label 179 = (57, 58)\\
\hline
Label 180 = (58, 59) & Label 181 = (58, 71) & Label 182 = (58, 59) & Label 183 = (59, 60) & Label 184 = (4, 5)\\
\hline
Label 185 = (5, 6) & Label 186 = (53, 60) & Label 187 = (59, 60) & Label 188 = (60, 61) & Label 189 = (60, 61)\\
\hline
Label 190 = (61, 62) & Label 191 = (61, 62) & Label 192 = (62, 63) & Label 193 = (62, 72) & Label 194 = (62, 63)\\
\hline
Label 195 = (63, 64) & Label 196 = (54, 64) & Label 197 = (63, 64) & Label 198 = (64, 65) & Label 199 = (64, 65)\\
\hline
Label 200 = (65, 66) & Label 201 = (65, 66) & Label 202 = (66, 67) & Label 203 = (66, 73) & Label 204 = (66, 67)\\
\hline
Label 205 = (67, 68) & Label 206 = (55, 68) & Label 207 = (67, 68) & Label 208 = (68, 69) & Label 209 = (68, 69)\\
\hline
Label 210 = (69, 70) & Label 211 = (5, 6) & Label 212 = (6, 7) & Label 213 = (69, 70) & Label 214 = (70, 74)\\
\hline
Label 215 = (58, 71) & Label 216 = (71, 77) & Label 217 = (62, 72) & Label 218 = (72, 81) & Label 219 = (66, 73)\\
\hline
Label 220 = (73, 85) & Label 221 = (70, 74) & Label 222 = (74, 89) & Label 223 = (75, 76) & Label 224 = (75, 90)\\
\hline
Label 225 = (75, 76) & Label 226 = (76, 77) & Label 227 = (71, 77) & Label 228 = (76, 77) & Label 229 = (77, 78)\\
\hline
Label 230 = (77, 78) & Label 231 = (78, 79) & Label 232 = (78, 79) & Label 233 = (79, 80) & Label 234 = (79, 91)\\
\hline
Label 235 = (6, 7) & Label 236 = (7, 8) & Label 237 = (79, 80) & Label 238 = (80, 81) & Label 239 = (72, 81)\\
\hline
Label 240 = (80, 81) & Label 241 = (81, 82) & Label 242 = (81, 82) & Label 243 = (82, 83) & Label 244 = (82, 83)\\
\hline
Label 245 = (83, 84) & Label 246 = (83, 92) & Label 247 = (83, 84) & Label 248 = (84, 85) & Label 249 = (73, 85)\\
\hline
Label 250 = (84, 85) & Label 251 = (85, 86) & Label 252 = (85, 86) & Label 253 = (86, 87) & Label 254 = (86, 87)\\
\hline
Label 255 = (87, 88) & Label 256 = (87, 93) & Label 257 = (87, 88) & Label 258 = (88, 89) & Label 259 = (74, 89)\\
\hline
Label 260 = (88, 89) & Label 261 = (16, 8) & Label 262 = (7, 8) & Label 263 = (8, 9) & Label 264 = (75, 90)\\
\hline
Label 265 = (90, 94) & Label 266 = (79, 91) & Label 267 = (91, 98) & Label 268 = (102, 92) & Label 269 = (83, 92)\\
\hline
Label 270 = (106, 93) & Label 271 = (87, 93) & Label 272 = (90, 94) & Label 273 = (94, 95) & Label 274 = (94, 95)\\
\hline
Label 275 = (95, 96) & Label 276 = (109, 96) & Label 277 = (95, 96) & Label 278 = (96, 97) & Label 279 = (96, 97)\\
\hline
Label 280 = (97, 98) & Label 281 = (91, 98) & Label 282 = (97, 98) & Label 283 = (98, 99) & Label 284 = (100, 99)\\
\hline
Label 285 = (98, 99) & Label 286 = (10, 9) & Label 287 = (8, 9) & &\\
\hline
\end{tabular}
}
\end{center}
\label{tab:cnot_gate_lookup_table}
\end{table*}
\vspace{0.5in}
\vspace{0.5in}
\begin{figure}[htbp]
\centering
\includegraphics[width=\figurewidth]{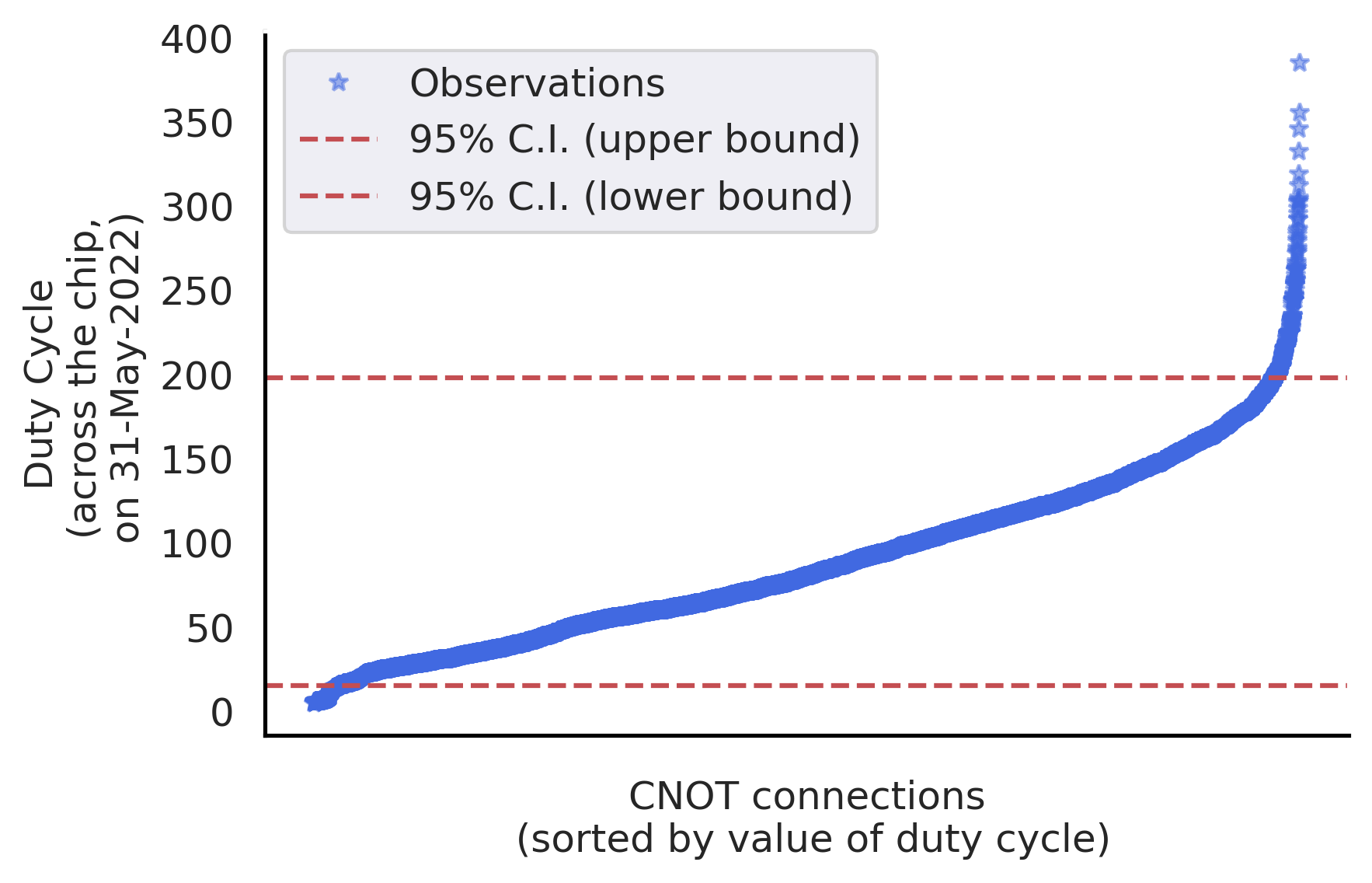}
\caption{The 95\% temporal confidence interval (vertical orange lines) for CNOT duty cycle ($\tau$) for the physical nearest-neighbor connections of \washington. The data-set contains the values for all the 144 physical CNOT gates of \washington between 1-Dec-2021 to 31-May-2022. The dashed black line represents the mean.}
\label{fig:washington_tau_CI_across_time}
\end{figure}
\vspace{0.5in}
\vspace{0.5in}
\vspace{0.5in}
\begin{figure}[htbp]
\centering
\includegraphics[width=\figurewidth]{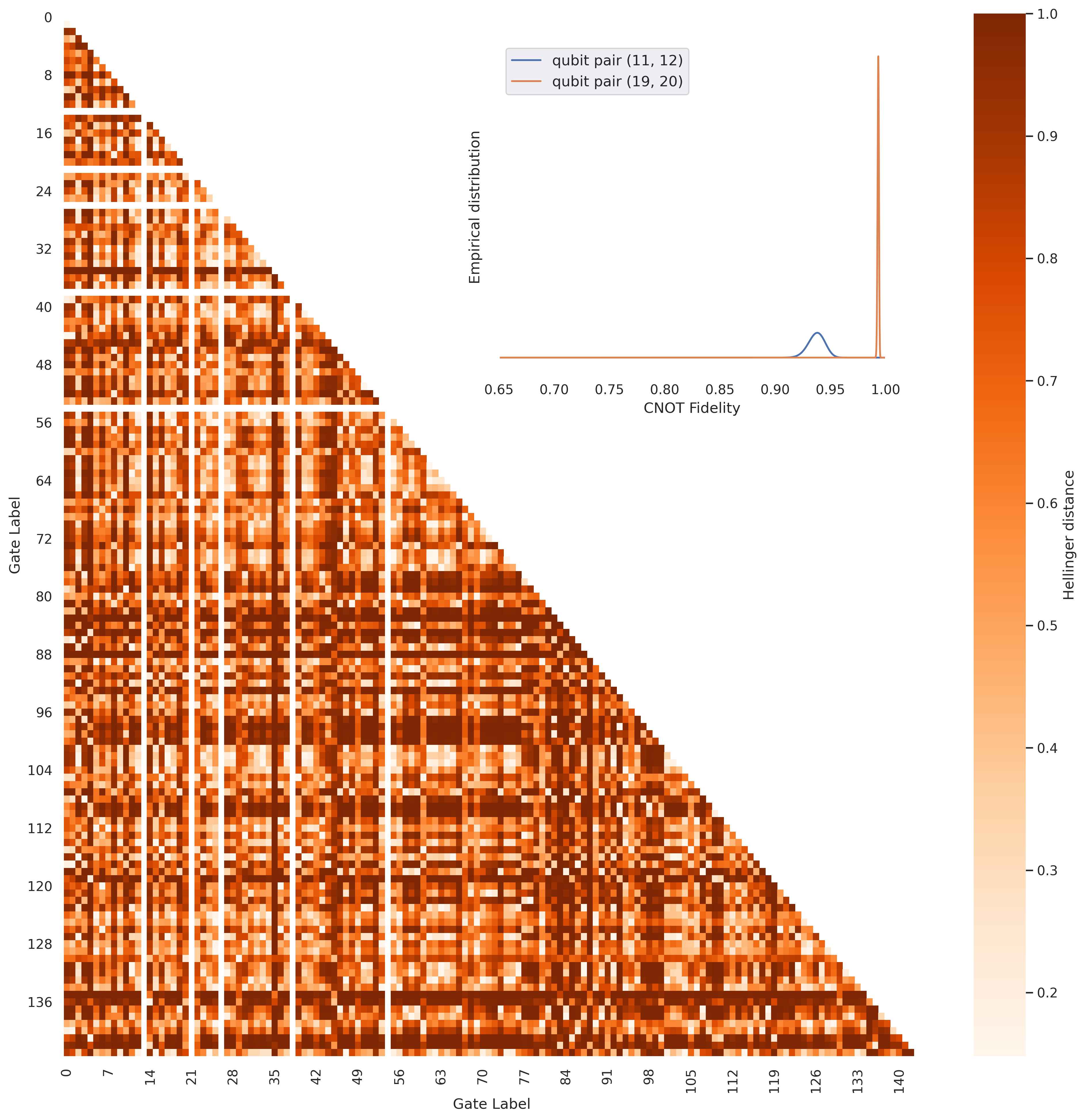}
\caption{Spatial reliability of the gate fidelity $F_G$ for the CNOT gates of the \washington device. The heat map shows the Hellinger distance between the distributions for the gate fidelities of nearest-neighbor connections. Only the lower triangular matrix is shown to avoid redundancy. The upper triangular matrix as well as any data gaps in the lower triangular matrix are colored white. The inset shows the distribution of gate fidelities for pairs $(11, 12)$ and $(19, 20)$, which yield a Hellinger distance 0.99. This metric captures the probability that the quantum processor register is spatially dissimilar. The estimation of the distribution for CNOT-$i$ utilizes data for Sep-2022.}
\label{fig:washington_FG_hellinger_spatial}
\end{figure}
\vspace{0.5in}
\vspace{0.5in}
\vspace{0.5in}
\begin{figure}[htbp]
\centerline{
\includegraphics[width=\figurewidth]{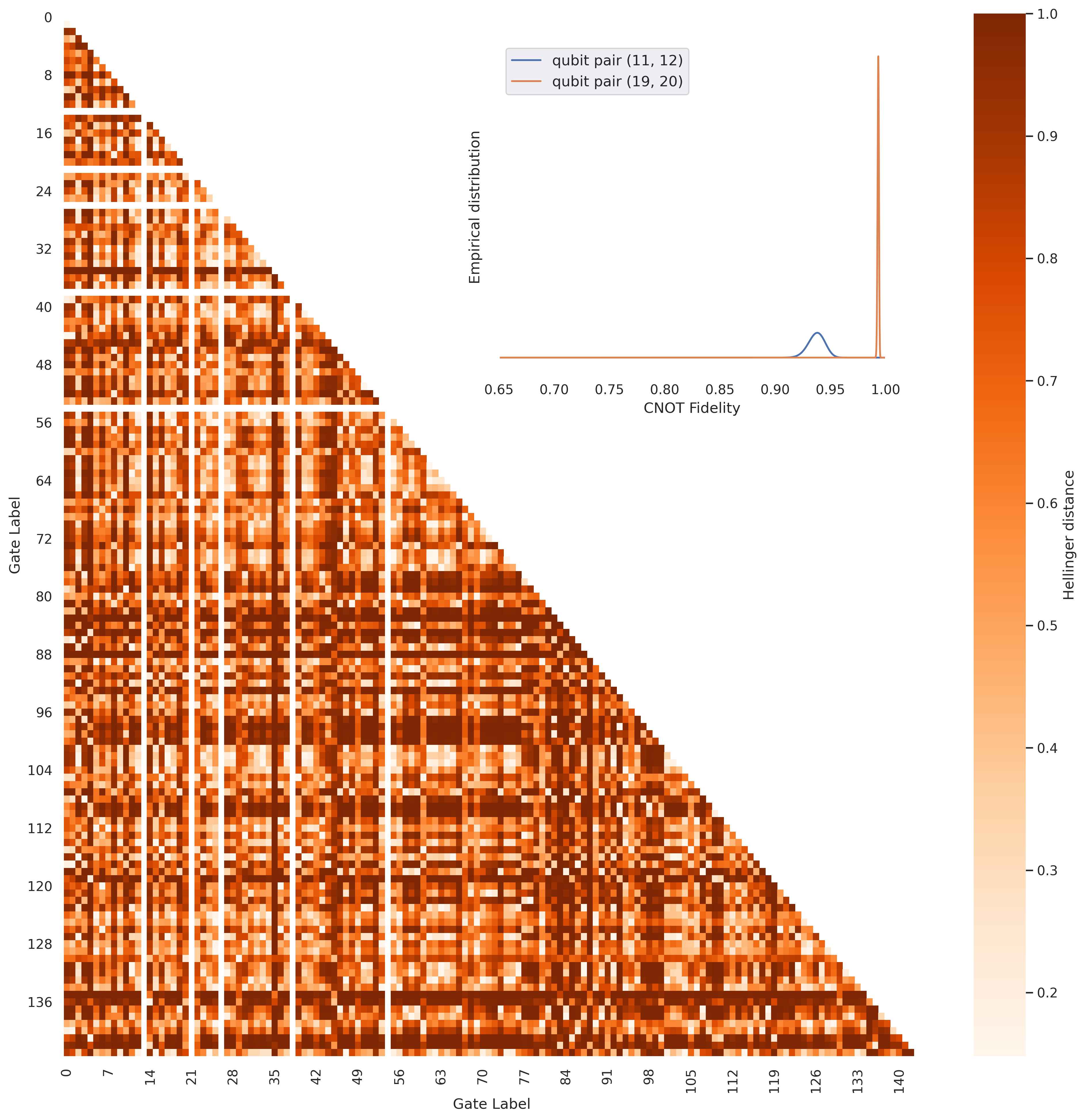}
}
\caption{
Spatial reliability of the gate fidelity $F_G$ for the CNOT gates of the \washington device. The heat map shows the Hellinger distance between the distributions for the gate fidelities of nearest-neighbor connections. Only the lower triangular matrix is shown to avoid redundancy. The upper triangular matrix as well as any data gaps in the lower triangular matrix are colored white. The inset shows the distribution of gate fidelities for pairs $(11, 12)$ and $(19, 20)$, which yield a Hellinger distance 0.99. This metric captures the probability that the quantum processor register is spatially dissimilar. The estimation of the distribution for CNOT-$i$ utilizes data for Sep-2022.}
\label{fig:washington_FG_hellinger_spatial}
\end{figure}
\vspace{0.5in}
\removefigs
\chapter{Bounds on stability of program outcomes}\label{ch:analytical_bounds}
In Chapter 3, we developed a performance assessment framework for noisy quantum outputs, using accuracy, reproducibility, device reliability, and outcome stability. In this chapter, we aim to establish bounds for this framework using available device characterization data. Section 1 asks, given an $\epsilon$-bound on histogram accuracy (measured in computational basis), how to bound a proxy parameter that encapsulates multiple device noise parameters. Section 2 asks, what is the minimum sample size required to ensure histogram reproducibility with $1-\delta$ confidence. Section 3 bounds outcome stability in terms of the distance between time-varying noise densities. Section 4 bounds device reliability metric to attain an $\epsilon$-stable outcome.

\section{Sample bounds on accuracy}
In this section, our aim is to address the following challenge posed by an $\epsilon$-bound on the accuracy of digital histograms obtained from a noisy test circuit (measured in the computational basis). Can we formulate and bound a proxy parameter $\gamma_D$ that encompasses the array of noise parameters characterizing a noisy device $D$? 
\begin{equation}
\gamma_{D} \leq \gamma_{\textrm{max}}
\label{eq:gamma_tau}.
\end{equation}
The importance of such a proxy lies in its ability to streamline high-dimensional noise analysis. 

Consider an $n$ qubit state $\ket{\psi}$ prepared as a uniform superposition across the $2^n$ computational basis states $\{\ket{v}\}$ as: 
\begin{equation}
\ket{\psi} = 2^{-n/2}\sum\limits_{v \in \{0,1\}^n}\ket{v},
\end{equation}
which is the output (in the noiseless limit) from the circuit shown in Fig.~\ref{fig:27_hadamards_with_n_1} i.e. $\ket{\psi} = \mathds{H}^{\otimes n} \ket{0}^{\otimes n}$. For our experiments, we use a $n=27$ qubit register. The noiseless distribution for this circuit is:
$p_v^{\textrm{noiseless}} = 2^{-n} \;\;\;\; \forall v \in \{0, \cdots, 2^n-1\}.$ We assume that gate errors and SPAM noise capture the principal sources of noise in this circuit and ignore inter-qubit cross-talk. Let $\mathds{I}, \mathds{X}, \mathds{Y},$ and $\mathds{Z}$ denote the $2\times 2$ identity matrix, Pauli-$X$ matrix, Pauli-$Y$ matrix, and the Pauli-$Z$ matrix, respectively:
\begin{equation}
\mathds{I}=\begin{pmatrix}
1& 0\\
0& 1\\
\end{pmatrix}, \;\;\;
\mathds{X}=\begin{pmatrix}
0& 1\\
1& 0\\
\end{pmatrix}, \;\;\;
\mathds{Y}=\begin{pmatrix}
0& -i\\
i& 0\\
\end{pmatrix}, \;\;\;
\mathds{Z}=\begin{pmatrix}
1&  0\\
0& -1\\
\end{pmatrix}\\
\end{equation}
Let $R_Y(\theta)$ denote rotation by an angle ($\theta$) about the Y-axis on the Bloch sphere\cite{nielsen2002quantum}:
\begin{equation}
R_Y(\theta) = e^{-i\frac{\theta}{2}Y} 
= \cos \frac{\theta}{2} \mathds{I} - i \sin \frac{\theta}{2} \mathds{Y}
=\begin{pmatrix}
\cos\frac{\theta}{2}  & -\sin\frac{\theta}{2} \\
\sin\frac{\theta}{2} & \cos\frac{\theta}{2} \\
\end{pmatrix}
\end{equation}
A noiseless Hadamard gate is given by:
\begin{equation}
H = R_Y\left( \frac{\pi}{2} \right) \mathds{Z} =
\begin{pmatrix}
\cos\frac{\pi}{4} &  \sin\frac{\pi}{4} \\
\sin\frac{\pi}{4} & -\cos\frac{\pi}{4} \\
\end{pmatrix}
= \frac{1}{\sqrt{2}}
\begin{pmatrix}
1&1\\
1&-1\\
\end{pmatrix}.
\end{equation}
We model an over- or under-rotated Hadamard gate ($\tilde{\mathds{H}}$) by the unitary: 
\begin{equation}
\tilde{\mathds{H}} = 
\begin{pmatrix}
\cos\left(\frac{\pi}{4}+\xrm\right) &  \sin\left(\frac{\pi}{4}+\xrm\right) \\
\sin\left(\frac{\pi}{4}+\xrm\right) & -\cos\left(\frac{\pi}{4}+\xrm\right) \\
\end{pmatrix}
\end{equation}
where $\xrm$ is a small implementation error in radians ($\xrm\ll \pi/4$). The operator representation for a unitary control error \cite{emerson2005scalable} has only one term which can be seen as follows. Write the noisy unitary $\tilde{\mathcal{U}}$ as: $\tilde{\mathcal{U}} = \tilde{\mathcal{U}} \mathcal{U}^\dagger \mathcal{U}$
where $\mathcal{U}$ is the noiseless unitary. Thus, 
$
\rho^\prime = \tilde{\mathcal{U}} \rho  \tilde{\mathcal{U}}^\dagger
= (\tilde{\mathcal{U}} \mathcal{U}^\dagger)  (\mathcal{U} \rho \mathcal{U}^\dagger) (\mathcal{U}  \tilde{\mathcal{U}}^\dagger)
= M (\mathcal{U} \rho \mathcal{U}^\dagger) M^\dagger
$ where $M$ is the operator representing the unitary control error that arises due to imperfections in the control system. For our circuit,
\begin{equation}
M = \tilde{\mathds{H}}\mathds{H}^\dagger = 
\begin{pmatrix}
\cos\xrm &  -\sin\xrm \\
\sin\xrm & \cos\xrm \\
\end{pmatrix},\;\;\;\;
\tilde{\mathds{H}} =\frac{1}{\sqrt{2}}
\begin{pmatrix}
\cos\xrm - \sin\xrm &  \cos\xrm + \sin\xrm \\
\cos\xrm + \sin\xrm & -\cos\xrm + \sin\xrm \\
\end{pmatrix}
\end{equation}
In the absence of SPAM noise, when we initialize a qubit in the ground state, subject it to a noisy Hadamard gate, and measure in the $\mathds{Z}$-basis, we get the probabilities for observing the outputs 0 and 1 as:
\begin{equation}
\textrm{Pr}\left( \ket{v}=\ket{i} \right) = \left(1 - (-1)^i \sin2\xrm \right)/2, \;\;\;\; i \in \{0,1\}
\end{equation}

We next consider what happens when the Hadamard gate is followed by a noisy measurement. The SPAM channel can be characterized as a quantum channel\cite{oszmaniec2019simulating, bravyi2021mitigating, geller2020rigorous} 
using two parameters $f_0$  and $f_1$ for each qubit. The first parameter ($f_0$) defines the probability of observing $0$ post readout when the channel input state is $\ket{0}$, and the second ($f_1$) defines the probability of observing $1$ post readout when the channel input state is $\ket{1}$.

A classical representation for the single qubit SPAM channel is:
\begin{equation}
\mathds{P}^{\textrm{out}} = \Lambda \mathds{P}^{\textrm{in}}    
\end{equation}
where $\Lambda$ is the SPAM error matrix with elements $\Lambda_{ij}$ = probability of observing $\ket{i}$ when the input to channel is $\ket{j} (i, j \in \{0, 1\})$:
\begin{equation}
\Lambda =  
\begin{pmatrix}
f_0 &  1-f_1 \\
f_1 & 1-f_0 \\
\end{pmatrix}
\label{eq:Lambda}    
\end{equation}

Equivalently, the quantum channel representation for a single qubit noisy measurement has two Kraus operators, $M_0$ and $M_1$, and can be specified by a super-operator ($\mathcal{E}$), whose action on the quantum state is as follows \cite{smith2021qubit}:
\begin{equation}
\begin{split}
\mathcal{E}(\rho) =& M_0 \rho M_0^\dagger + M_1 \rho M_1^\dagger\\
M_0 =& \sqrt{f_0}\ket{0}\bra{0}+\sqrt{1-f_1}\ket{1}\bra{1}\\
M_1 =& \sqrt{1-f_0}\ket{0}\bra{0}+\sqrt{f_1}\ket{1}\bra{1}\\
\end{split}
\end{equation}

This is equivalent to Eqn.~\ref{eq:Lambda}, when you consider the action of the measurement operators $\{\ket{i}\bra{i}\}$, given by:
\begin{equation}
Pr( \ket{v} = \ket{i}) = \textrm{Tr} \left[ \ket{i}\bra{i} \mathcal{E}(\rho) \right].
\end{equation}

Let Pr$(0)$ be the probability of observing 0 when we prepare a qubit in the ground state, subject it to a Hadamard gate, and then measure it. Let Pr$(1)$ be the corresponding probability of observing 1 for the same experiment (i.e., we prepare a qubit in the ground state, subject it to a Hadamard gate, and then measure it). Additionally, let $f$ be the average SPAM fidelity and $\varepsilon^\text{SPAM}$ be the SPAM fidelity asymmetry:
\begin{equation}
f = \frac{f_0+f_1}{2}
\label{eq:f_def}
\end{equation}
\begin{equation}
\varepsilon^\text{SPAM} = f_0 - f_1
\label{eq:eps_def}.
\end{equation}

Thus, in the presence of SPAM noise, the probability of observing 0 and 1 for each qubit in Fig.~\ref{fig:27_hadamards_with_n_1} is given by \cite{smith2021qubit}:
\begin{equation}
Pr( \ket{v} = \ket{i}) = \textrm{Tr} \left[ \ket{i}\bra{i} \mathcal{E}(\rho) \right]
= \frac{1+(-1)^i\gamma}{2}
\label{eq:pr0}
\end{equation}
where
\begin{equation}
\gamma = \varepsilon^\text{SPAM} - 2\sin2\xrm\left(f-\frac{1}{2}\right)
\label{eq:gamma_def}.
\end{equation}

Let $(v_{n-1} v_{n-2} \cdots v_0)$ represent the $n$-bit string with $v_i \in \{0,1\}$, and let $v = \sum\limits_{i=0}^{n-1}2^i v_i$ be the decimal integer equivalent. In the absence of cross-talk between gates, 
$\mathds{P}^\text{noisy} = \{p_v^\text{noisy}\}$ where:
\begin{equation}
p_v^\text{noisy}
=\prod\limits_{i=0}^{n-1} 
\left( \frac{1+\gamma_i}{2} \right)^{1-v_i}
\left( \frac{1-\gamma_i}{2} \right)^{v_i}\\
\end{equation}
and $\gamma_i$ refers to the $\gamma$-parameter from Eqn.~\ref{eq:gamma_def} for the $i$-th register element. 

It follows then that the Bhattacharya coefficient\cite{beran1977minimum} is given by:
\begin{equation}
\begin{split}
BC( \mathds{P}^\text{noiseless}, \mathds{P}^\text{noisy}) =&
\sum\limits_{v=0}^{2^n-1}\sqrt{\frac{1}{2^n} \prod\limits_{i=0}^{n-1}
\left( \frac{1+\gamma_i}{2} \right)^{1-v_i}
\left( \frac{1-\gamma_i}{2} \right)^{v_i}
}\\
=& \frac{1}{2^n} \sum\limits_{v=0}^{2^n-1} \prod\limits_{i=0}^{n-1} \sqrt{ 1+\gamma_i }^{1-v_i}
\sqrt{ 1-\gamma_i}^{v_i}\\
\end{split}
\end{equation}

For meeting an $\epsilon$-accuracy constraint, we need:
\begin{equation}
\frac{1}{2^n}
\sum\limits_{v=0}^{2^n-1}
\prod\limits_{i=0}^{n-1}
\sqrt{ 1+\gamma_i }^{1-v_i}
\sqrt{ 1-\gamma_i}^{v_i}
\leq \epsilon^2
\label{eq:hadamard_gamma}.
\end{equation}

When assuming that $\epsilon$ is small, this yields:
\begin{equation}
\left| \varepsilon^\text{SPAM} - 2\sin2\xrm\left(f-\frac{1}{2}\right)\right|
\leq 
\gamma_{\textrm{max}} = 2(1-\epsilon^2)^{1/n} \sqrt{1-(1-\epsilon^2)^{2/n}}
\end{equation}
The left side is the proxy $\gamma_D$ and the right side is the upper bound under $\epsilon$-accuracy constraint for the digital histogram, in the context of the superposition circuit.

Conversely, when armed with noise characterization data,
we can estimate the best possible accuracy as:
\begin{equation}
\epsilon_\text{min} \geq \frac{1}{2}\sqrt{\frac{n}{2}}\gamma_D.
\end{equation}
\subsection{Validation using device characterization data}
\begin{figure}[H]
\center
\includegraphics[width=\figurewidth]{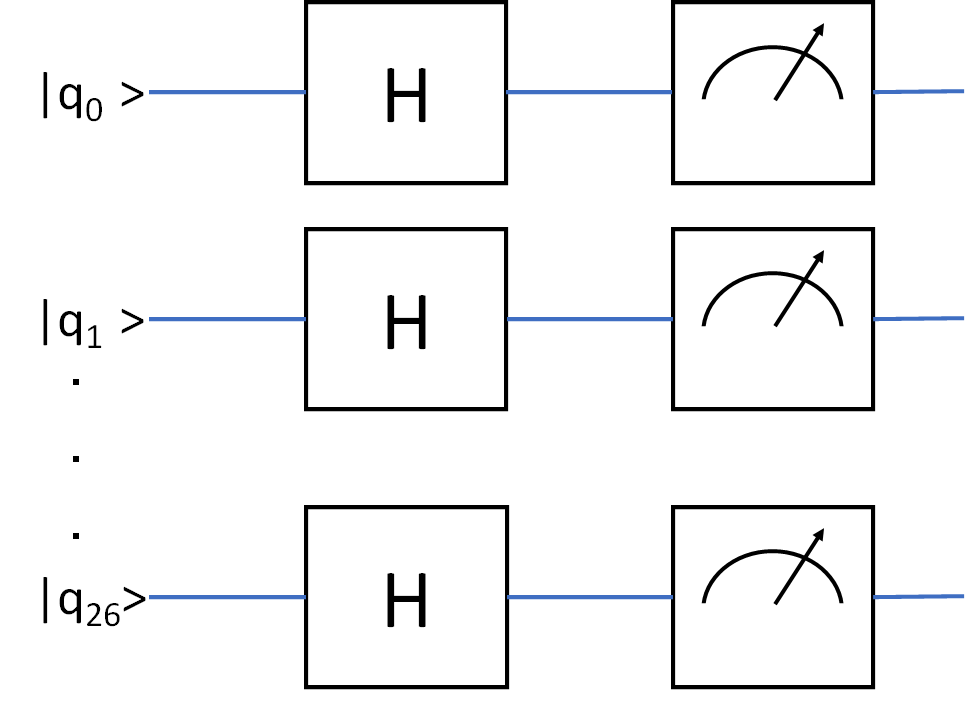} 
\caption{Circuit used for our experiment. In this figure, $H$ represents the Hadamard gate. The meter symbol denotes measurement gate.}
\label{fig:27_hadamards_with_n_1}
\end{figure}

We validated our bound using the device \toronto, whose schematic is shown in Fig.~\ref{fig:toronto}. The test circuit shown in Fig.~\ref{fig:27_hadamards_with_n_1} was programmed using the IBM Qiskit toolkit \cite{ibm_quantum_experience_website} and compiled and executed remotely on 8 April, 2021. 

To estimate the noise parameters, we repeated our experiments $L$ times. Let $l$ denote the index of the $l$-th experiment. For any instance of circuit execution, the device prepared an ensemble of $N_s$ identical circuits, where $N_s$ denotes the number of shots and $k$ denotes the $k$-th shot in a particular experiment. In the tests reported below, $L=203$ was the number of repetitions successfully executed during a 30-min reservation-window, and $N_s=8,192$ was the number of shots, the maximum allowed by the device. We separately analyzed the results for the case $n = 1$ using each of the 27 register elements available.

In this section, we use the convention that a caret sign denotes a particular realization of a random variable. We first characterized SPAM fidelity, in which SPAM(0) denotes an experiment with a register element, prepared as $\ket{0}$ and measured. Similarly, SPAM(1) denotes an experiment in which a register element, prepared as $\ket{1}$ is measured. 

Let $b^{SPAM(0)}_{l,k,q}$ denote the binary outcome of measuring in the computational basis, when collecting the $k$-th shot of the $l$-th experiment of the SPAM(0) circuit on the $q$-th register element. Additionally, let $f_1^q(l)$ denote the initialization fidelity observed in the $l$-th SPAM(1) experiment for the $q$-th register element. Similarly, let $f_0^q(l)$ denote the initialization fidelity observed in the $l$-th SPAM(0) experiment for the $q$-th register element. Thus:
\begin{equation}
f_1^q(l) = \frac{\sum\limits_{k=1}^{N_s} b^{SPAM(1)}_{l,k,q}}{N_s}
,\;\;\;\;
\hat{f}_1^q(l) = \frac{\sum\limits_{k=1}^{N_s} \hat{b}^{SPAM(1)}_{l,k,q}}{N_s}
\end{equation}
\begin{equation}
f_0^q(l) = 1-\frac{\sum\limits_{k=1}^{N_s} b^{SPAM(0)}_{l,k,q}}{N_s}
,\;\;\;\;
\hat{f}_0^q(l) = 1-\frac{\sum\limits_{k=1}^{N_s} \hat{b}^{SPAM(0)}_{l,k,q}}{N_s}.
\end{equation}

Let $\varepsilon^\text{SPAM}_{q,l}$ denote the realized fidelity asymmetry of the $q$-th register element in the $l$-th experiment. Thus: 
\begin{equation}
\varepsilon^\text{SPAM}_{q,l} = f_0^q(l) - f_1^q(l).
\end{equation}

Let $\bar{\varepsilon}^\text{SPAM}_q$ denote the mean of the fidelity asymmetry for the $q$-th register element over the $L$ experiments, and let $\hat{\bar{\varepsilon}}^\text{SPAM}_q$ be the corresponding observed value.
Thus:
\begin{equation}
\bar{\varepsilon}^\text{SPAM}_q = \frac{\sum\limits_{l=1}^{L} \varepsilon^\text{SPAM}_{q,l} }{L}
,\;\;\;\;
\hat{\bar{\varepsilon}}^\text{SPAM}_q = \frac{\sum\limits_{l=1}^{L} \hat{\varepsilon}^\text{SPAM}_{q,l}}{L}.
\end{equation}

To quantify the error on these measurements, we define $\sigma( \bar{\varepsilon}^\text{SPAM}_q )$ as the standard deviation of population mean $\bar{\varepsilon}^\text{SPAM}_q$, such that:
\begin{equation}
\hat{\sigma}^2( \bar{\varepsilon}^\text{SPAM}_q ) =  \frac{\hat{\sigma}^2( \varepsilon^\text{SPAM}_q )}{L}
= \frac{1}{L(L-1)} \sum\limits_{l=1}^{L} \left( \hat{\varepsilon}^\text{SPAM}_{q,l} - \hat{\bar{\varepsilon}}^\text{SPAM}_{q,l} \right)^2.
\end{equation}
The average SPAM fidelity $f^q$ for each qubit $q$ is then calculated using Eqn.~\ref{eq:f_def}.  The initialization fidelities of the computational states are not the same. The asymmetric nature of the single qubit noise channel is brought out starkly by the negligible overlap between the distributions of $f_0$ and $f_1$ for qubit~$5$. Additionally, observe the significant spread in values in qubit 3, relative to the others. These results show that the naive approach of assuming a single value for SPAM error for a qubit is fallacious. Not only do we have to characterize $f_0$ and $f_1$ separately, our work must also take into account the significant dispersion around the mean. 

\begin{figure}[H]
\center
\includegraphics[width=\figurewidth]{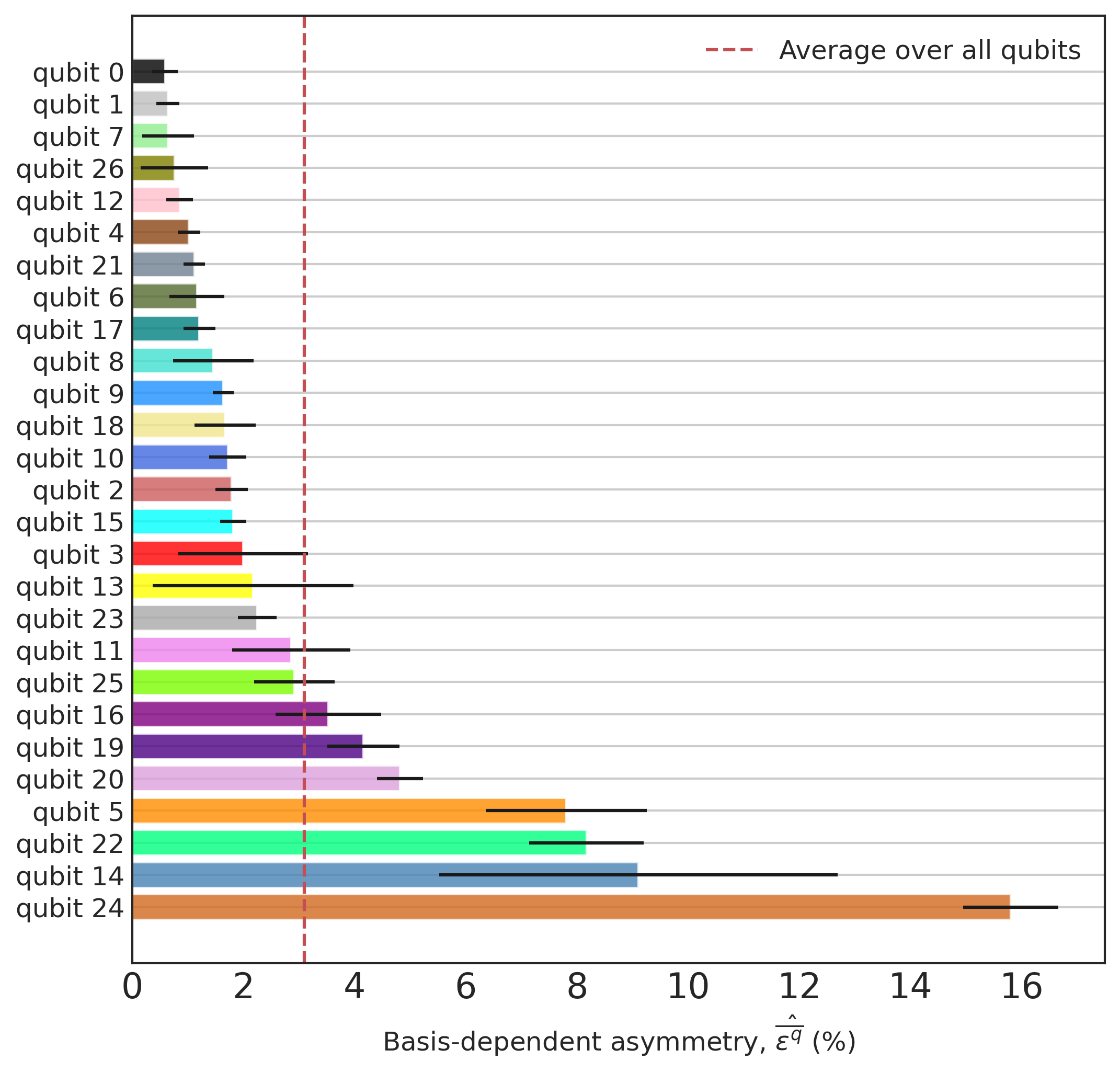}
\caption{Plot illustrating significant spatial non-stationarity in the register-wise variation of the SPAM asymmetry for \toronto.}
\label{fig:asymmetry_toronto_04082021_spruce}
\end{figure}

\begin{figure}[H]
\centering
\includegraphics[width=\figurewidth]{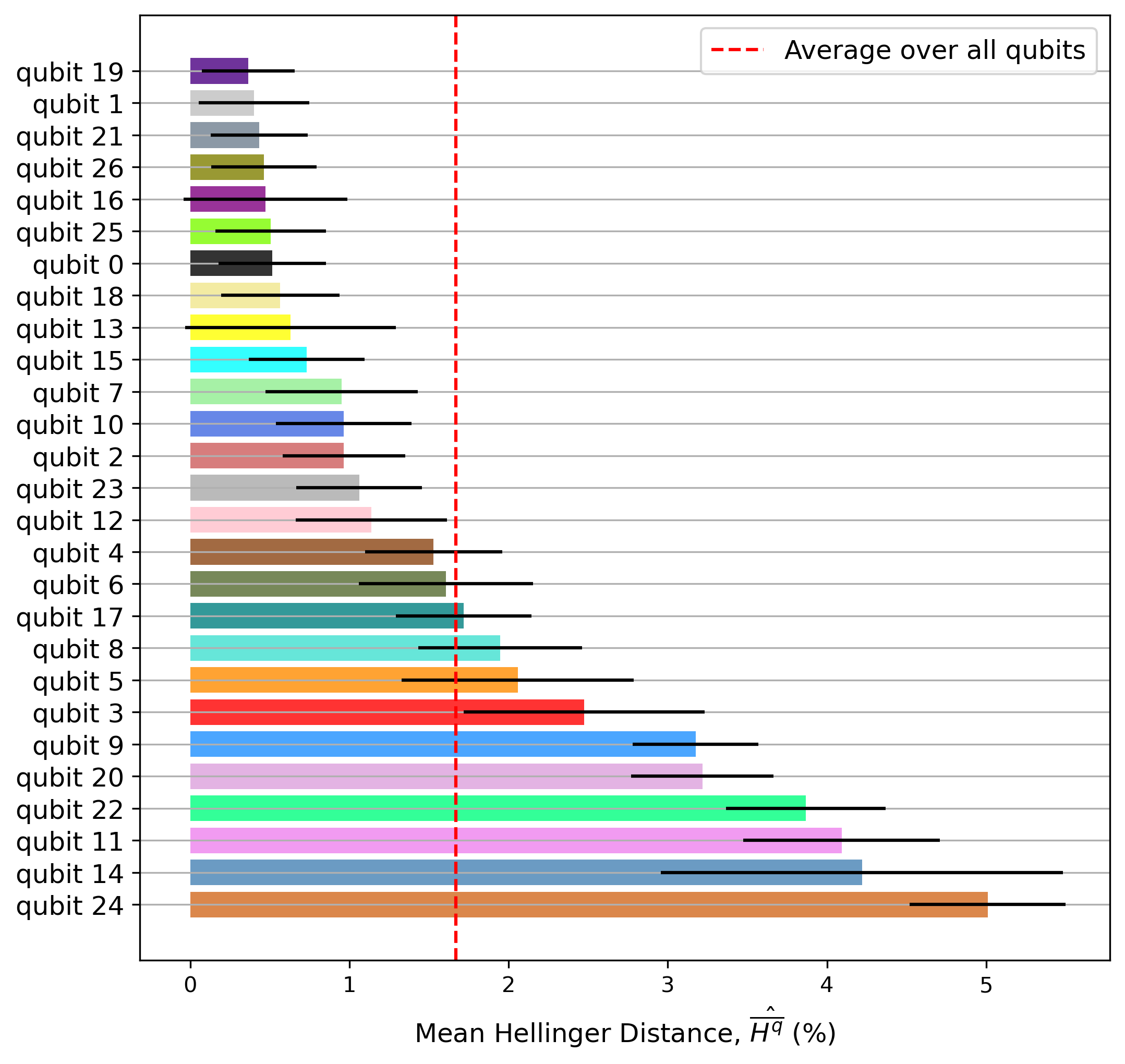}
\captionof{figure}{The plot depicts distance variation across the $27$ register elements of \toronto device, illustrating the impact of spatially non-stationary noise on program outcomes and the dependence of output on register choice.}
\label{fig:hellinger_toronto_04082021_spruce}
\end{figure}

\begin{figure}[H]
\centering
\includegraphics[width=\figurewidth]{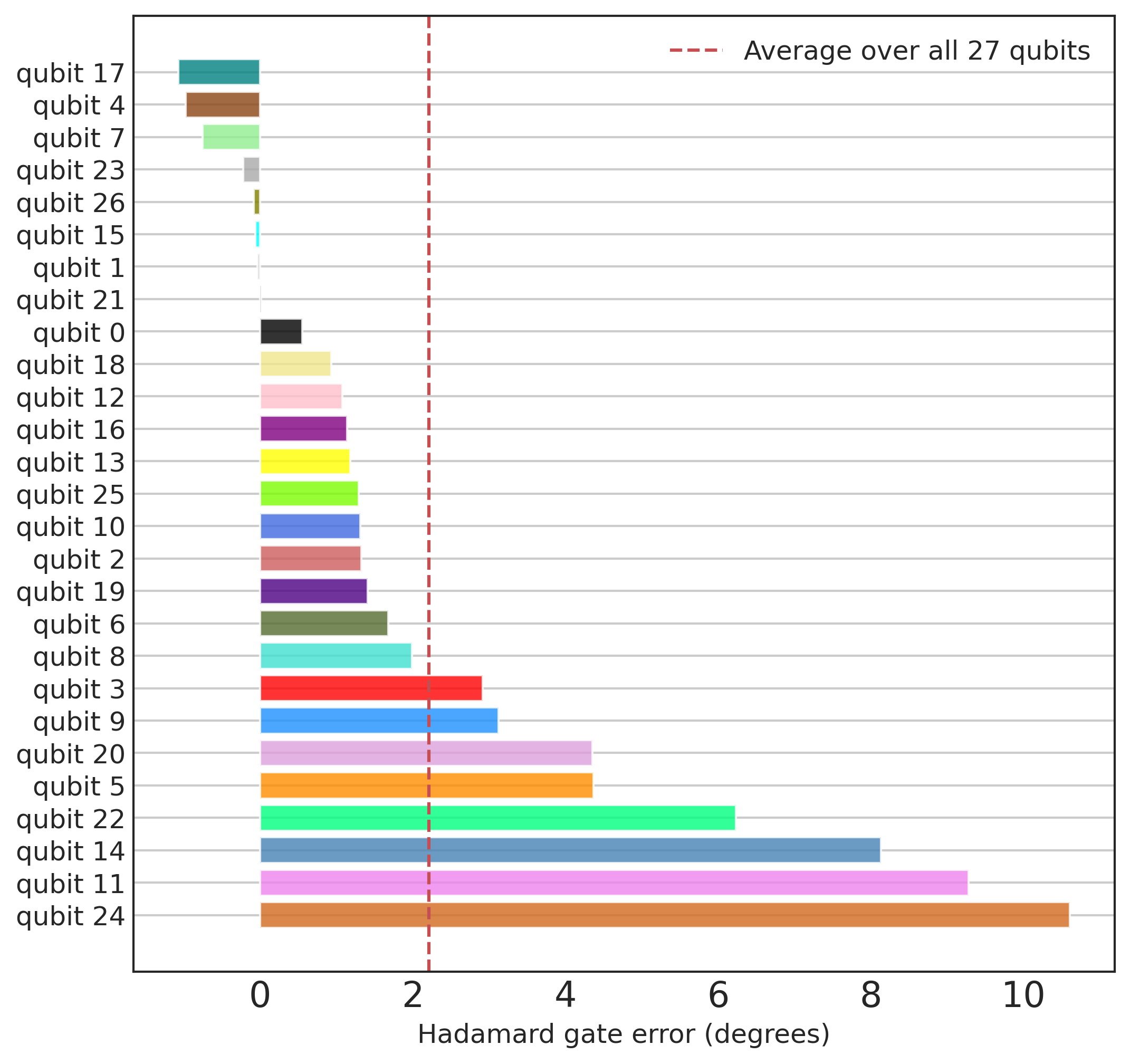}
\captionof{figure}{The plot illustrates the register-wise Hadamard gate angle error (in degrees) within \toronto's quantum register of $27$ elements, emphasizing spatial non-stationarity and cautioning against averaging qubit values for coherent noise mitigation.}
\label{fig:theta_sq_qubitwise}
\end{figure}

\begin{figure}[H]
\centering
\includegraphics[width=\figurewidth]{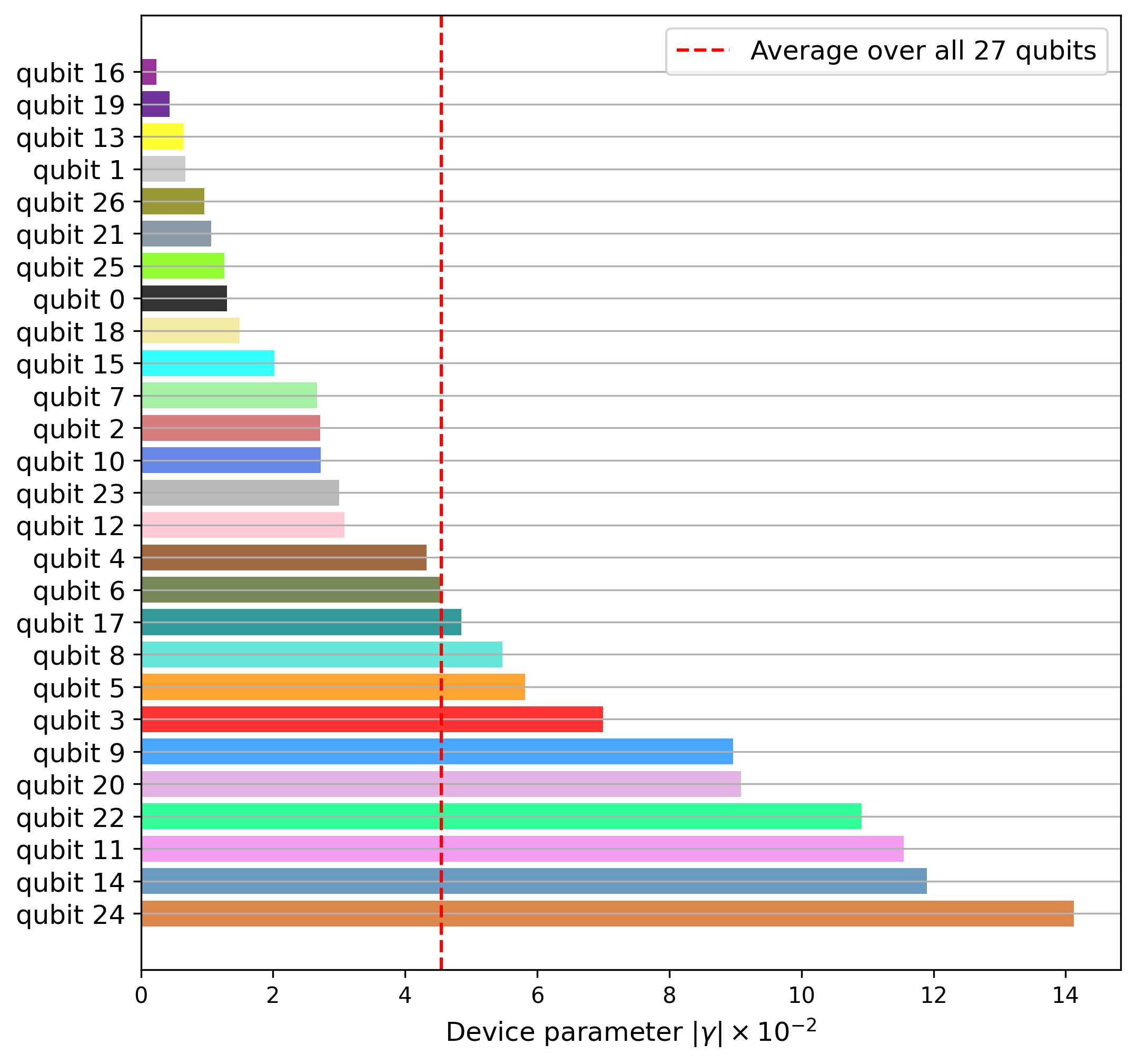}
\captionof{figure}{This plot shows the register-wise variation of the composite accuracy metric $\gamma_D$ for \toronto device with 27 qubits, where a higher value significantly impacts program precision. The graph highlights the necessity of re-estimating the metric due to temporal non-stationary noise in unreliable devices, emphasizing the crucial role of analyzing noise parameter interactions for desired accuracy.}
\label{fig:gamma_qubitwise}
\end{figure}

The register-wise variation of the SPAM asymmetry is depicted in Fig.~\ref{fig:asymmetry_toronto_04082021_spruce}. 
The plot illustrates the SPAM asymmetry $(f_0-f_1)$ for \toronto, revealing substantial spatial non-stationarity across the register. The y-axis arranges individual qubits in ascending order of SPAM asymmetry magnitude, while the x-axis represents the mean SPAM fidelity asymmetry expressed as a percentage. Evidently, qubit 0 demonstrates the best performance in this regard, while qubit 24 exhibits the least favorable outcome. Given that SPAM errors are a predominant source of quantum computer errors, this variation is particularly concerning for mitigation procedures\cite{bharti2022noisy}, as precise knowledge of SPAM noise parameters $f_0$ and $f_1$ is essential. The figure features 27 qubits (register elements), with the dashed red line indicating the mean SPAM asymmetry value (averaged across all qubits). The error bars represent the standard deviation of population means across $L=203$ experiments. Consistency in the register's color scheme is maintained across Figs.~\ref{fig:asymmetry_toronto_04082021_spruce}, \ref{fig:hellinger_toronto_04082021_spruce}, \ref{fig:theta_sq_qubitwise}, and \ref{fig:gamma_qubitwise}.

The probability Pr$_q(0)$, for each qubit, as defined by Eqn.~\ref{eq:pr0}, was estimated by executing the circuit $\mathcal{C}$ in Fig.~\ref{fig:27_hadamards_with_n_1} and counting the fraction of zeros in the 8192-bit long binary string, returned by the remote server. Let $b_{l,k,q}^{\mathcal{C}}$ denote the random measurement outcome (a classical bit) when we conduct an experiment and measure the $q$-th register element in the $l$-th experiment's $k$-th shot (measurement done in the computational $Z$-basis). Let $\hat{b}_{l,k,q}^{\mathcal{C}}$ denote the corresponding observed value. Similarly, let $\textrm{Pr}_{l, q}(1)$ denote the probability of observing 1 as the outcome in the $l$-th experiment for the $q$-th register element. Similarly, let $\textrm{Pr}_{l, q}(0)$ denote the probability of observing 0 as the outcome in the $l$-th experiment for the $q$-th register element. Thus:
\begin{equation}
\textrm{Pr}_{l, q}(1) = \frac{\sum\limits_{k=1}^{N_s} b_{l,k,q}^{\mathcal{C}}}{N_s}
,\;\;\;\;
\hat{\textrm{Pr}}_{l, q}(1) = \frac{\sum\limits_{k=1}^{N_s} \hat{b}_{l,k,q}^{\mathcal{C}}}{N_s}
\end{equation}
and
\begin{equation}
\textrm{Pr}_{l, q}(0) = 1 - \textrm{Pr}_{l, q}(1)
,\;\;\;\;
\hat{\textrm{Pr}}_{l, q}(0) = 1 - \hat{\textrm{Pr}}_{l, q}(1).
\end{equation}

Let $H_{l, q}$ denote the Hellinger distance\cite{beran1977minimum} between the noisy and noiseless outcomes in the $l$-th experiment for the $q$-th register element. Let $\hat{H}_{l, q}$ be the corresponding observed value (or realization). Let $\bar{H}^q$ denote the mean (a random variable) of the distance for the $q$-th register element over $L$ experiments. Let $\hat{\bar{H}}^q$ be the corresponding observed realization. Thus:
\begin{equation}
H_{l, q} = \sqrt{ 1- \sqrt{\frac{1}{2} \textrm{Pr}_{l, q}(0)} - \sqrt{\frac{1}{2} \textrm{Pr}_{l, q}(1)} }
,\;\;\;\;
\hat{H}_{l, q} = \sqrt{ 1- \sqrt{\frac{1}{2} \hat{\textrm{Pr}}_{l, q}(0)} - \sqrt{\frac{1}{2} \hat{\textrm{Pr}}_{l, q}(1)  } }
\end{equation}
\begin{equation}
\bar{H}^q = \frac{\sum\limits_{l=1}^{L} H_{l, q} }{L}
,\;\;\;\;
\hat{\bar{H}}^q = \frac{\sum\limits_{l=1}^{L} \hat{H}_{l, q} }{L}
\end{equation}

To quantify the error on these measurements, define $\sigma( \bar{H}^q)$ as the standard deviation of population mean $\bar{H}^q$. Thus:
\begin{equation}
\hat{\sigma}^2( \bar{H}^q ) =  \frac{\hat{\sigma}^2( H^q )}{L}
= \frac{1}{L(L-1)} \sum\limits_{l=1}^{L} \left( \hat{H}_{l, q} - \hat{\bar{H}}_q \right)^2.
\end{equation}

Fig.~\ref{fig:hellinger_toronto_04082021_spruce} pertains to the quantum register of \toronto, comprising $27$ register elements. It shows the experimentally derived distance variation across the register. The dashed red line represents the mean distance across the register (averaged over all qubits). Among the qubits, qubit $19$ demonstrates the closest proximity to the noiseless state, while qubit $24$ exhibits the greatest deviation. The error bars depict the standard deviation of the population mean from $203$ experiments. This graph highlights the impact of spatially non-stationary noise on program outcomes. The y-axis arranges qubits in ascending order of the distance between the obtained and noiseless outputs while the x-axis denotes the mean distance, computed as the population mean across $L=203$ experiments.

The Hadamard gate angle error was subsequently estimated using Eqn.~\ref{eq:gamma_def}. The register variation of the the Hadamard gate angle error (in degrees) is shown in Fig.~\ref{fig:theta_sq_qubitwise}. Among the qubits, qubit $21$ demonstrates the closest proximity to the noiseless, while qubit $24$ exhibits the greatest deviation. This graph serves to highlight spatial non-stationarity in Hadamard gate noise, revealing the inappropriateness of averaging qubit values for coherent noise error mitigation. The y-axis arranges qubits in ascending order of Hadamard gate error, while the x-axis quantifies the Hadamard gate error in degrees.

\begin{figure}[H]
\centering
\includegraphics[width=\figurewidth]{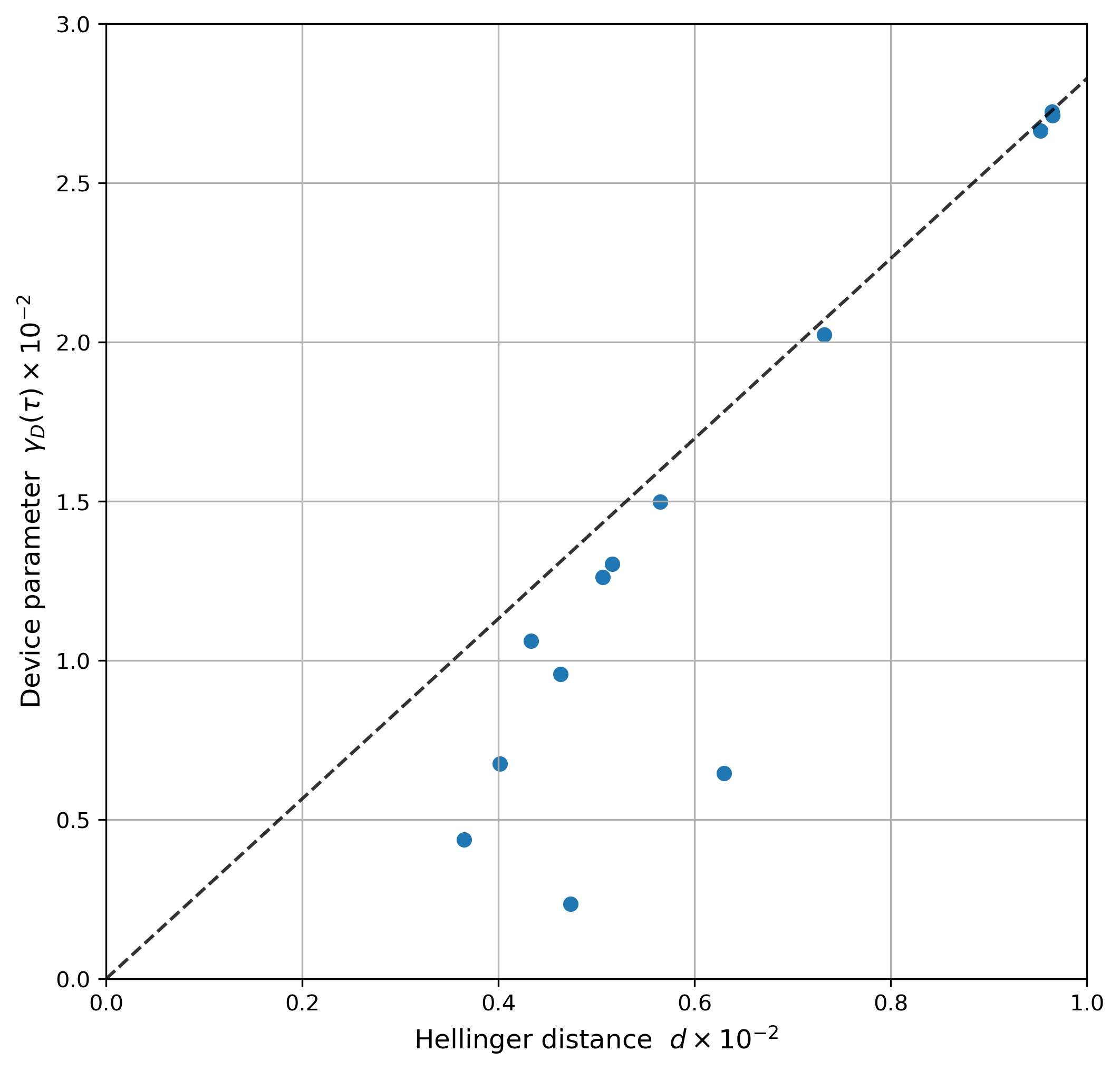}
\captionof{figure}{Characterizing circuit accuracy on toronto. Plot of $\gamma_{\textrm{max}}$ (dashed line) and $\gamma_D$ (blue dots) for \toronto on 8 April 2021.
The blue dots are experimentally-observed data plotted using the characterization data versus the actual observed distance $(d)$ for each register element. Only a subset of qubits are shown.}
\label{fig:model_verification_corrected_lower}
\end{figure}

Fig.~\ref{fig:gamma_qubitwise} displays the register-wise variation of the proxy $\gamma_D$. Qubit $16$ outperforms others, while qubit $24$ fares the worst. Fig.~\ref{fig:model_verification_corrected_lower} shows the values for $\gamma_{\textrm{max}}$ and $\gamma_D$ for \toronto on 8 April, 2021, when $\epsilon$ is set to be the observed distance. The blue dots are experimentally-observed data for each register element (see Table~\ref{tab:gamma_vals} for the full list), using the characterization data versus the actual observed distance. It validates our noise model as Eqn.~\ref{eq:gamma_tau} holds. {The dashed line in Fig.~\ref{fig:model_verification_corrected_lower} provides the decision boundary to test circuit accuracy, using characterization data as a proxy. Given an $\epsilon$-accuracy bound on the statistical distance between the observed distribution and reference to be accurately generated, the plot provides an upper bound for the proxy $\gamma_D$ (the register variation of $\gamma_D$ is shown in Fig.~\ref{fig:gamma_qubitwise}. The latter must lie below this boundary for accuracy by the device.} We conjecture that the magnitude of $|\gamma_{\textrm{max}}-\gamma_D|$ serves as a reliability indicator, i.e., higher value provides greater cushion against temporal fluctuations. Table~\ref{tab:gamma_vals} can serve as a basis for register selection.
\section{Sample bounds on reproducibility}
This section aims to establish the minimum sample size ($L$) required to achieve reproducibility in generated outputs with a specified statistical confidence level ($1-\delta$), using the reproducibility condition:
\begin{equation}
\textrm{Pr}( \hat{H} \leq \epsilon) \geq 1 - \delta,
\end{equation}
Consider an $n$-qubit Bernstein-Vazirani problem (Sec.~\ref{sec:BV}) where the secret string is denoted as $r$. 
In a noiseless, the probability of obtaining the string $r$ is certain:
\begin{equation}
p_r^\text{noiseless} = 1.
\end{equation}
When a circuit is executed once, it yields a single classical bit string $v$. 
We introduce the indicator variable $Y_r$, assigned the value 1 if $v$ matches $r$, and 0 otherwise. 

Let $Y_r(l)$ indicate the outcome of $Y_r$ from the $l$-th execution. Upon conducting the $L$ runs of the circuit, the experimental sample estimate for the success probability:
\begin{equation}
\hat{p}_r = \frac{\sum\limits_{l=0}^{L-1} Y_r(l)}{L},
\end{equation}
The mean of $\hat{p}_r$ equals $p_r$, while its variance is $\frac{p_r(1-p_r)}{L}$. The experimental distance obtained from the noiseless histogram is:
\begin{equation}
\hat{H} = 1 - \sqrt{\hat{p_r}}
\end{equation}
Then, the $\delta$-reproducibility condition translates to:
\begin{equation}
\begin{split}
\textrm{Pr}(\hat{H} \leq \epsilon) &\geq 1 - \delta \\
\Rightarrow \textrm{Pr}\left[ \hat{p_r} \geq (1-\epsilon)^2 \right] &\geq 1 - \delta
\end{split}
\end{equation}
This can be reformulated as:
\begin{equation}
\textrm{Pr}\left[ \frac{\hat{p_r} - p_r}{\sqrt{ \frac{p_r(1-p_r)}{L} }} \geq \frac{(1-\epsilon)^2 - p_r}{ \sqrt{ \frac{p_r(1-p_r)}{L} } } \right] \geq 1 - \delta
\label{eq:BV_reproducibility}
\end{equation}
Defining the standard normal variable as $z$ with $z \sim \mathcal{N}(0,1)$, the central limit theorem establishes that $\frac{\hat{p_r} - p_r}{\sqrt{ \frac{p_r(1-p_r)}{L} }}$ follows the standard normal distribution. Given that $\text{Pr}(z \geq z_{\delta}) = 1-\delta$, where $z_{\delta}$ corresponds to a constant dependent on $\delta$ for the one-sided confidence interval, we can set:
\begin{equation}
\frac{\hat{p_r} - p_r}{\sqrt{ \frac{p_r(1-p_r)}{L} }} = z_{\delta}
\end{equation}
This satisfies Eqn.\ref{eq:BV_reproducibility}. Solving for $L$ yields the minimum bound as:
\begin{equation}
L_\text{min} = z_\delta^2 \frac{p_r^{-2}-1}{p_r^{-2}(1-\epsilon)^2-1}
\end{equation}
Hence, the minimum sample size exhibits an inverse relation with the accuracy threshold $\epsilon$ and a direct, non-linear correlation with the confidence level $1-\delta$.
\section{General bounds on stability}
The goal of this section is understanding the extent to which non-stationary noise\cite{muller2015interacting, klimov2018fluctuations, etxezarreta2021time} affects the stability of outcomes generated from a noisy quantum device.

In this section, $\braket{O_{ \xrm}}$ denotes the mean of a quantum observable $O$ in presence of a sample realization of the circuit noise $\xrm$. It is well know that $f_X(\xrm; t)$ varies with time. We define the average of $\braket{O_{ \xrm}}$ with respect to the $f_X(\xrm; t)$ as:
\begin{equation}
\braket{O}_t = \int\limits \braket{O_{ \xrm}}  f_X(x; t) d\xrm.
\label{eq:O_td_def}
\end{equation}
In the absence of knowledge of the exact realization of x at time t, $\braket{O}_t$ is an estimate for the mean of the observable in presence of time-varying noise channels. Let $s(t_1, t_2)$ be the absolute difference in the mean of the observable obtained from the noisy quantum device at times $t_1$ and $t_2$:
\begin{equation}
s(t_1, t_2) = | \braket{O}_{t_1} - \braket{O}_{t_2} |.
\end{equation}
We will refer to $s(t_1, t_2)$ as the stability  of the observable \cite{dasgupta2021stability}. Now,
\begin{equation}
\begin{aligned}
s^2(t_1, t_2) &= \left( \braket{O}_{t_1} - \braket{O}_{t_2} \right)^2\\
&= \left( \int \braket{O_{ \xrm}} f_X(\xrm; t_1)\textrm{dx} - \int \braket{O_{ \xrm}} f_X(\xrm; t_2)\textrm{dx} \right)^2\\
&= \left( \int \braket{O_{ \xrm}}  \{ f_X(\xrm; t_1)\textrm{dx} - f_X(\xrm; t_2)\} \textrm{dx} \right)^2\\
&\leq \left( \int \left| \braket{O_{ \xrm}}  \{ f_X(\xrm; t_1)\textrm{dx} - f_X(\xrm; t_2)\} \right| \textrm{dx} \right)^2.
\end{aligned}
\label{eq:abs_integrand}
\end{equation}
In the last step, the inequality stems from the absolute value on the integrand. Now, per H\"older's inequality, if $m, n \in [1, \infty)$ and $1/m+1/n=1$, then:
\begin{equation}
\int \left| f(x) g(x) \right|dx  \leq  \left( \int |f(x)|^m dx\right)^{1/m} \left( \int |g(x)|^n dx\right)^{1/n}.
\end{equation}
Thus, our inequality becomes:
\begin{equation}
\begin{aligned}
&\left( \int \left| \braket{O_{ \xrm}}  \{ f_X(\xrm; t_1)\textrm{dx} - f_X(\xrm; t_2)\} \right| \textrm{dx} \right)^2 \\
&\leq \left[ \left( \int  | \braket{O_{ \xrm}} |^m \textrm{dx}\right)^{1/m} \left( \int | f_X(\xrm; t_1) - f_X(\xrm; t_2) |^n \textrm{dx}\right)^{1/n} \right]^2.
\end{aligned}
\end{equation}
Now, let $m \rightarrow \infty, n=1$ and define:
\begin{equation}
c= \underset{\xrm}{\textrm{sup}} |\braket{O_{ \xrm}}|.
\end{equation}
Clearly, 
\begin{equation}
\lim\limits_{m \rightarrow \infty} \left( \int |\braket{O_{ \xrm}}|^m\textrm{dx}\right)^{1/m}\leq \lim\limits_{m \rightarrow \infty} \left( \int c^m\textrm{dx}\right)^{1/m} = c \left( \lim\limits_{m \rightarrow \infty}  \left( \int \textrm{dx}\right)^{1/m} \right)= c.
\label{eq:c_factor}
\end{equation}
Thus, we have
\begin{equation}
\begin{aligned}
s(t_1, t_2)^2 
&\leq  \lim\limits_{m\rightarrow \infty, n = 1} \left( \left( \int  | \braket{O_{ \xrm}} |^m \textrm{dx}\right)^{1/m} \left( \int | \{ f_X(\xrm; t_1)\textrm{dx} - f_X(\xrm; t_2)\} |^n \textrm{dx}\right)^{1/n} \right)^2\\
&=c^2 \left( \int \left| \sqrt{f_X(\xrm; t_1)} - \sqrt{f_X(\xrm; t_2)} \right| \left( \sqrt{f_X(\xrm; t_1)} + \sqrt{f_X(\xrm; t_2)} \right) \textrm{dx} \right)^2\\
&=c^2 \int \left( \sqrt{f_X(\xrm; t_1)} - \sqrt{f_X(\xrm; t_2)} \right)^2 \textrm{dx}\int \left( \sqrt{f_X(\xrm; t_1)} + \sqrt{f_X(\xrm; t_2)} \right)^2 \textrm{dx}\\
&\textrm{ (applying H\"older's inequality with m=n=2)} \\
&=c^2\int \left( f_X(\xrm; t_1)+ f_X(\xrm; t_2) - 2\sqrt{f_X(\xrm; t_1)}\sqrt{f_X(\xrm; t_1)} \right) \textrm{dx} \\
&\int \left( f_X(\xrm; t_1)+ f_X(\xrm; t_2) + 2\sqrt{f_X(\xrm; t_1)}\sqrt{f_X(\xrm; t_1)} \right) \textrm{dx}\\
&=c^2 \left( 1+ 1 - 2\int \sqrt{f_X(\xrm; t_1)}\sqrt{f_X(\xrm; t_1)}\textrm{dx} \right) \left( 1+ 1 + 2\int \sqrt{f_X(\xrm; t_1)}\sqrt{f_X(\xrm; t_1)}\textrm{dx} \right)\\
&=4c^2H_{X}^2(2-H_{X}^2),
\end{aligned}
\end{equation}
where, for clarity, we use:
\begin{equation}
H_X^2 = H^2_X(t_1, t_2) = 1-\int \sqrt{f_X(\xrm; t_1)}\sqrt{f_X(\xrm; t_2)}dx
\end{equation}
Thus the observable stability $s$ is always upper bounded by 
\begin{equation}
s^2_{\textrm{max}} = 4c^2H_{X}^2(2-H_{X}^2),
\end{equation}
an upper bound determined by the degree of time-variation of the device parameters. Thus,
\begin{equation}
\begin{split}
\frac{s}{s_\textrm{max}} &\leq 1.
\end{split}
\label{eq:smax}
\end{equation}
The upper bound on the observable stability can also be expressed as:
\begin{equation}
s_{\textrm{max}} = 
2c\sqrt{1-(1-H_\textrm{normalized}^2)^{2d}}.
\label{eq:smax_normalized}
\end{equation}
using Eq.~\ref{eq:hellinger_normalized}.
\subsection{Validation using synthetic data}\label{sec:bound_verification}
\begin{figure}[htbp]
\centering
\includegraphics[width=\figurewidth]{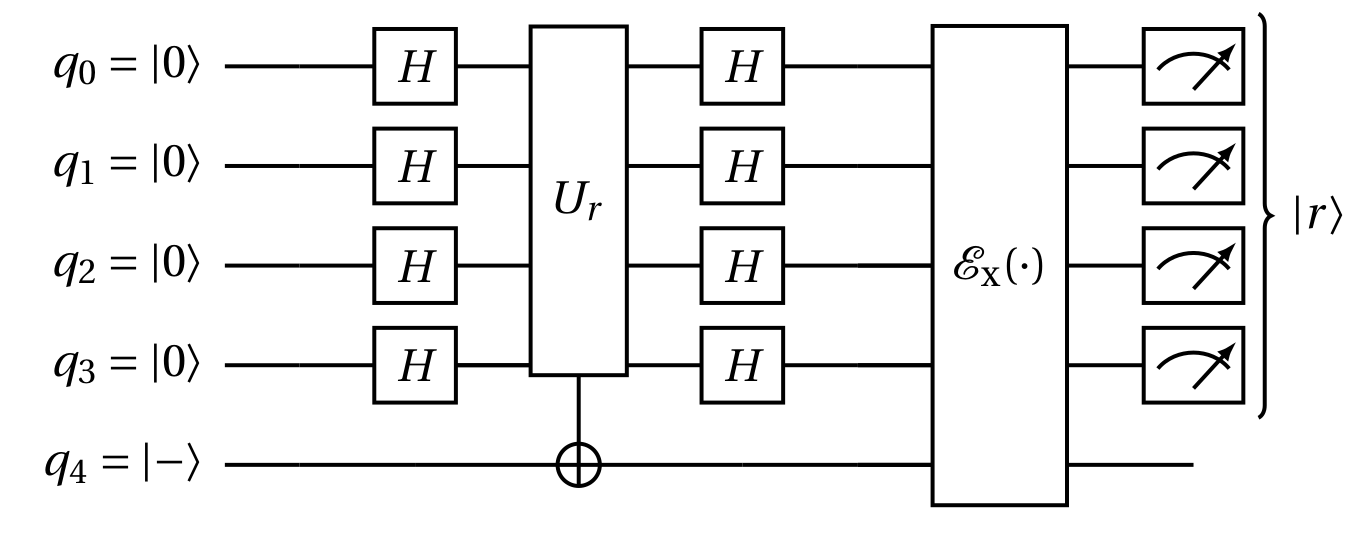}
\caption{A quantum circuit implementation of the Bernstein-Vazirani algorithm that employs 5 qubits, denoted $q_0$ to $q_4$. The first four qubits are used to compute the 4-bit secret string, while the fifth qubit serves as an ancilla and initially resides in the $\ket{-}$ superposition state. The symbol $H$ denotes the Hadamard gate while the oracle unitary ($U_r$) implements the secret string ($r$). The depolarizing noise channel is denoted by $\mathcal{E}_\xrm(\cdot)$. A quantum measurement operation is represented by the meter box symbol at the circuit's end. 
}
\label{fig:bv_depol}
\end{figure}

Our synthetic example evaluates the bound on a noisy \BV \cite{bernstein1993quantum} (Sec.~\ref{sec:BV}). 
We are interested in the probability of success to compute the secret bit string $r$ where 
\begin{equation}
\ket{r} = \bigotimes\limits_{q=0}^{n-1} \ket{r_q}
\end{equation}
with $r_q \in \{0,1\}$. The observable for the problem is:
\begin{equation}
\begin{split}
O &= \Pi_r= \ket{r}\bra{r}.
\end{split}
\end{equation}
The state for the noiseless, noiseless circuit is $\rho_{\textrm{out}}^{\textrm{noiseless}} = \ket{r}\bra{r}$ and, hence, the corresponding probability of success for the noisy circuit describe in Fig.~\ref{fig:bv_depol} is
\begin{equation}
\textrm{ Pr(r) = Tr} \left[\Pi_r \mathcal{E}_{\xrm}(\ket{r}\bra{r}) \right].
\end{equation}
This synthetic simulation, our noise model assumes each register element is acted upon by depolarizing noise, such that the super-operator $\mathcal{E}_{\xrm}(\cdot)$ represents the tensor product of independent single-qubit depolarizing channels. The $i$-th qubit is acted upon by the de-polarizing noise channel:
\begin{equation}
\begin{split}
\mathcal{E}_{\xrm_i}(\rho) &= \left( 1-\frac{3\xrm_i}{4} \right)\rho + \frac{\xrm_i}{4} (\mathds{X}_i\rho \mathds{X}_i + \mathds{Y}_i\rho \mathds{Y}_i + \mathds{Z}_i\rho \mathds{Z}_i)
\end{split}
\end{equation}
where $\xrm_i$ denotes the depolarizing parameter for the $\textrm{i}$-th qubit's noise channel and 
$\mathds{X}_i$, 
$\mathds{Y}_i$, and 
$\mathds{Z}_i$ are the Pauli matrices. 
Further, let $\xrm_i$ be a particular realization of the random variable  $X_i$, sampled from the multi-variate joint distribution $f_{X}(\xrm;t)$ which has $d$ random variables characterizing the noise in circuit $C$. We will further assume that the $\{X_i\}$ can have correlations in their values\cite{ben2011approximate}. The univariate marginal distribution for the random variable $X_i$ is denoted by $f_{X_i}(\xrm_i; t)$ where $i \in (1, \cdots, d)$. In this specific example, $d=n$.

Assuming the noise channel is separable but correlated:
\begin{equation}
\begin{split}
\braket{O_{ \xrm}}
&= \textrm{Tr}\left[ O \mathcal{E}_{\xrm} \left( \rho_{\textrm{noiseless}}^{\textrm{out}} \right)\right]\\
&= \textrm{Tr}\left[ O \mathcal{E}_{\xrm} \left( \ket{r}\bra{r} \right)\right] \textrm{ (for Bernstein-Vazirani)}\\
&= \textrm{Tr}\left[ \ket{r}\bra{r} \mathcal{E}_{\xrm} \left( \ket{r}\bra{r} \right)\right]\\
&= \textrm{Tr}\left[ \ket{r}\bra{r} \mathcal{E}_{\xrm} \left( 
\ket{r_1}\bra{r_1} \otimes \cdots \ket{r_n}\bra{r_n}
\right)\right]\\
&=\prod_{i=1}^{n} \textrm{Tr}\left[ \ket{r_i}\bra{r_i} \mathcal{E}_{\xrm_{i}} (\ket{r_i}\bra{r_i} )\right]\\
&=\prod_{i=1}^{n} \textrm{Tr}\left[ \ket{r_i}\bra{r_i} \left[ (1-x_i)\ket{r_i}\bra{r_i} + x_i\frac{\mathds{I}}{2} \right]\right]\\
\braket{O_{ \xrm}}&=\prod_{i=1}^{d} \left( 1-\frac{\xrm_i}{2}\right).
\end{split}
\end{equation}
As a specific instance of a time-varying depolarizing channel, suppose the noise marginals stay constant in the mean while the variance increases linearly with time:
\begin{equation}
\begin{split}
\mathds{E}(\xrm_i)   =& \mu_0 \;\;\;\;\forall i, t\\
\textrm{Var}(\xrm_i) = \sigma_t^2 =& \sigma_0^2 \left( 1+(\omega-1)\frac{t}{T}\right) \;\;\;\;\forall i,
\end{split}
\label{eq:depol_requirements}
\end{equation}
where $\omega = \sigma^2_T / \sigma^2_0$ is a constant capturing how volatile the distribution becomes at time $T$ compared to initial time t=0 and $i$ denotes the register number. 
Classical correlation in the noise is modeled by the correlation matrix $\Sigma$ where $\Sigma_{ij}$ represents the correlation coefficient between the depolarizing parameter $X_i$ acting on register element $i$ and $X_j$ acting on register element $j$.

We use a beta distribution to represent the marginal distribution of the depolarizing parameter $\xrm_i$ as
\begin{equation}
f_{X_i}(\xrm_i; t) = \frac{\xrm_i^{\alpha_t-1}(1-\xrm_i)^{\beta_t-1}}{\textrm{Beta}(\alpha_t, \beta_t)}, 0 \leq \xrm_i \leq 1,
\label{eq:beta_dist}
\end{equation}
with time-varying parameters $\alpha_t$ and $\beta_t$:
\begin{equation}
\alpha_t = \frac{\alpha_0}{k_0+t},\;\;\;\;
\beta_t  = \frac{\beta_0}{k_0+t},
\end{equation}
and the Beta function, by definition:
\begin{equation}
\textrm{Beta}(\alpha_t, \beta_t) = \int\limits_0^1 y^{\alpha_t-1}(1-y)^{\beta_t-1}dy, \;\;\;\; y\in[0,1].
\end{equation}
We will show later how to estimate the constants $\alpha_0, \beta_0, k_0$ from observed data. This choice of model is appropriate if the parameter value ranges between 0 and 1, and the observed data follows a skewed bell-shaped distribution. 
For simplicity, we will assume the distribution parameters do not vary with register location and the constants $\alpha_0, \beta_0$ and $k_0$ can be estimated from the model requirements in Eqn.~\ref{eq:depol_requirements} as:
\begin{equation}
\begin{aligned}
k_0 =& T\left( \omega \left( 1 + (\omega-1)\left( 1-\frac{\mu_0(1-\mu_0)}{\sigma_0^2}\right)^{-1}\right)^{-1} -1\right)^{-1}\\
\alpha_0 =& \frac{\mu_0(\mu_0-\mu_0^2-\sigma_0^2)}{\sigma_0^2}k_0\\
\beta_0 =& \frac{(1-\mu_0)(\mu_0-\mu_0^2-\sigma_0^2)}{\sigma_0^2}k_0.
\end{aligned}
\end{equation}
It is verified by substitution that this model satisfies the requirements of Eqn.~\ref{eq:depol_requirements}. The higher moments of the depolarizing parameter under the beta distribution given by:
\begin{equation}
\mathds{E}(\xrm_i^k) = \frac{\textrm{Beta}(\alpha_t+k, \beta_t)}{\textrm{Beta}(\alpha_t, \beta_t)} = \prod\limits_{n=0}^{k-1} \frac{\alpha_t+n}{\alpha_t+\beta_t+n}
\end{equation}
We next construct a joint distribution for the $d$-dimensional distribution using a copula structure, a direct application of Sklar's theorem \cite{sklar1959fonctions}, to model the correlation $\Sigma$ between the register elements. 
The use of copulas to study empirical correlation is well-established\cite{10.1111/j.1467-9868.2009.00707.x, 10.1177/1748006x13481928, 10.1243/1748006xjrr226, 10.21608/cjmss.2023.205330.1007}. 
Various choices for copulas exist including the Gaussian copula, elliptical copulas, Archimedean copulas, Ali-Mikhail-Haq copula, Clayton copula, Gumbel copula, Independence copula, and Joe copula \cite{zhu2022generative, de2012multivariate, nelsen2007introduction, mcneil2015quantitative, wilkens2023quantum, genest2011inference}. 
They offer different types of modeling capabilities for tail-risk correlations. 
We chose the Gaussian copula for its simplicity and ease of interpretation. 
\begin{equation}
f_{X}(\xrm;t) = \Theta \left[ F_{X_1}(\xrm_1; t), \cdots F_{X_{d}}(\xrm_{d}; t) \right]
\prod\limits_{j=1}^{d} f_{X_j}(\xrm_j; t),
\label{eq:copulas}
\end{equation}
where $\Theta(\cdot)$ is the copula function. We use $F_X(\xrm;t)$ as the joint cumulative distribution function for the multi-variate random variable $X$ at time $t$. Thus, 
\begin{equation}
F_{X}(\xrm; t) = \int\limits_{-\infty}^{\xrm} f_{X}(y; t) dy.
\end{equation}
Also, $F_{X_i}(\xrm_i; t)$ is the cumulative distribution function for the univariate random variable $X_i$ at time $t$. The Gaussian copula is simply the standard multi-variate normal distribution with correlation matrix $\Sigma$:
\begin{equation}
\Theta(y) = \Theta(y_{1},\cdots, y_{d}) = \frac{\exp\left( -\frac{1}{2}(y-\mu_y)^T \Sigma^{-1} (y-\mu_y)\right)}{(2\pi)^{n/2}|\Sigma|^{1/2}},
\end{equation}
where the vector $\mu_y$ is the mean of $y$. 

Having specified the statistics of the time-evolution of the depolarizing noise, we now turn to the task of estimating the distance of the distribution at time $t$ relative to a distribution at time 0. The univariate case has an analytical solution: 
\begin{equation}
H_{\xrm_i}(0, t) = \sqrt{1 - \frac{\textrm{Beta}\left( 
\frac{\alpha_0}{2}\left( \frac{1}{k_0}+\frac{1}{k_0+t} \right), 
\frac{\beta_0}{2} \left( \frac{1}{k_0}+\frac{1}{k_0+t} \right)
\right)}{\sqrt{\textrm{Beta}(\frac{\alpha_0}{k_0}, \frac{\beta_0}{k_0} )}\sqrt{\textrm{Beta}( \frac{\alpha_0}{k_0+t}, \frac{\beta_0}{k_0+t})}}},
\end{equation}
while the general multi-variate correlated case is analytically intractable. However, the distance can also be computed using Monte Carlo methods. Let $H_{X}(t_1, t_2)$ be the distance between the $d$-dimensional multi-variate correlated distributions. Drawing $N$ samples from the distribution $f_{X}(\xrm; t_1)$ yields $\{\xrm^j\}_{j=1}^N$ and, assuming $N$ is large enough to ensure convergence, we numerically approximate the integral as:
\begin{equation}
\frac{1}{N}\sum\limits_{j=1}^{N} \sqrt{ \frac{f_{X} (\xrm^j_{1,\cdots,d}; t_2)}{f_{X} (\xrm^j_{1,\cdots,d}; t_1) }}
\approx \mathds{E} \left( \sqrt{
\frac{f_{X} (\xrm; t_2)}{ f_{X} (\xrm; t_1)}}\right) \\
= \int \sqrt{ f_{X} (\xrm; t_1) f_{X} (\xrm; t_2) } d\xrm
= 1-H^2_{X}.
\label{eq:mcmc_hellinger}
\end{equation}
We now demonstrate the validity of Eqn.~\ref{eq:bound} using simulations of the noisy quantum circuit under the correlated depolarizing channel, for which the constant
\begin{equation}
c = \underset{\xrm}{\textrm{sup}} | \braket{O_{ \xrm}}| = \underset{(\xrm_1, \cdots, \xrm_{n})}{\textrm{sup}} \;\; \prod\limits_{i=1}^d \left( 1-\frac{\xrm_i}{2} \right) = 1,
\end{equation}
is maximal in the absence of noise and the noisy, time-dependent observable is modeled as
\begin{equation}
\braket{O}_t = \int \langle O_{\xrm} \rangle  \Theta \left[ 
F_1(\xrm_1; t ), 
\cdots 
F_{d}(\xrm_{d}; t ) \right] f_1(\xrm_1; t)\cdots f_{d}(\xrm_{d}; t)d\xrm_1 \cdots d\xrm_{d}.
\end{equation}
We estimate this observable through Monte Carlo sample of numerical simulations of the noisy quantum circuit. Our correlated depolarizing noise model assumes the variance of the univariate noise distribution increases linearly each month while the correlation between the isotropic single-qubit depolarizing coefficients is fixed as \added{$\Sigma_{i,j} = 0.80 \textrm{ for } i \neq j$}. 

Fig.~\ref{fig:depol_sbysmax} plots the ratio of the simulated stability $s(t)$ to the upper bound $s_\textrm{max}$ with respect to the simulated month for the cases of 4, 8, and 12-bit secret-strings. 
The results confirm the analytical upper bound. 
\subsection{Validation using device data}\label{sec:bound_verification}
\begin{figure}
\centering
\includegraphics[width=\figurewidth]{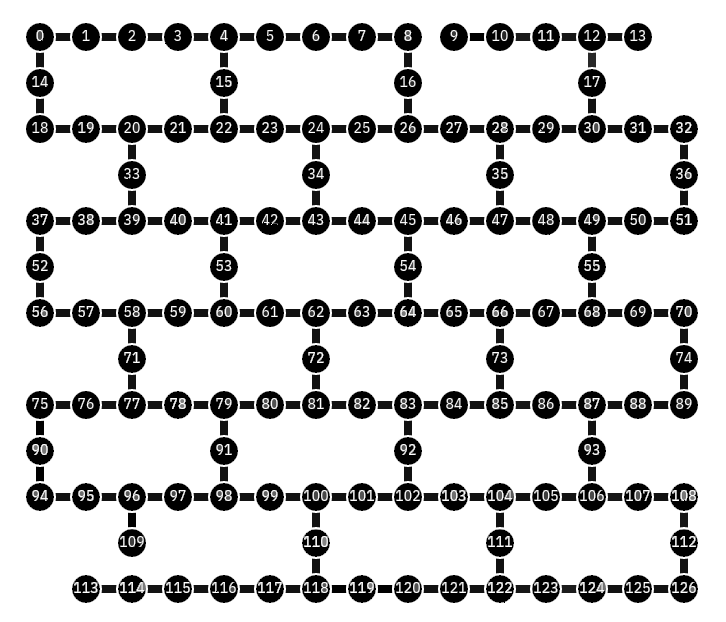}
\caption{Schematic layout of the $127$-qubit washington device produced by IBM. Circles denote register elements and edges denote connectivity of 2-qubit operations. The register elements $0$, $1$, $2$, $3$, and $4$ are mapped to the physical qubits $4$, $3$, $2$, $1$, and $0$, respectively, in the diagram above. The CNOT gates used in the circuit connect the physical qubits $(0,1)$ and $(2,1)$, in the diagram above, where the first number represents the control qubit and the second one represents the target qubit.}
\label{fig:wash_bw}
\end{figure}

We now verify the analytical upper bound using data from the \washington device. The register elements $0$, $1$, $2$, $3$, and $4$ in the algorithm are mapped to the physical qubits $4$, $3$, $2$, $1$, and $0$, respectively, as shown in Fig.~\ref{fig:wash_bw}. The CNOT gates used in the circuit connect the physical qubits $(0,1)$ and $(2,1)$. The data spanned \added{from 1-Jan-2022 to 30-Apr-2023}.

Fig.~\ref{fig:correlation_matrix} shows the correlation between the 16 device parameters taken from  Table~\ref{tab:noiseParameters}. Axes index the corresponding parameters. The data correspond to \added{daily observations made in Apr-2023}. The figure presents the Pearson coefficients with blue shades indicative of positive correlation and red shades indicative of negative correlation. Our estimate for the error bars on these coefficients is approximately $1/\sqrt{30-1} \approx 0.18$. 

\begin{figure}[htbp]
\centering
\includegraphics[width=0.5\columnwidth]{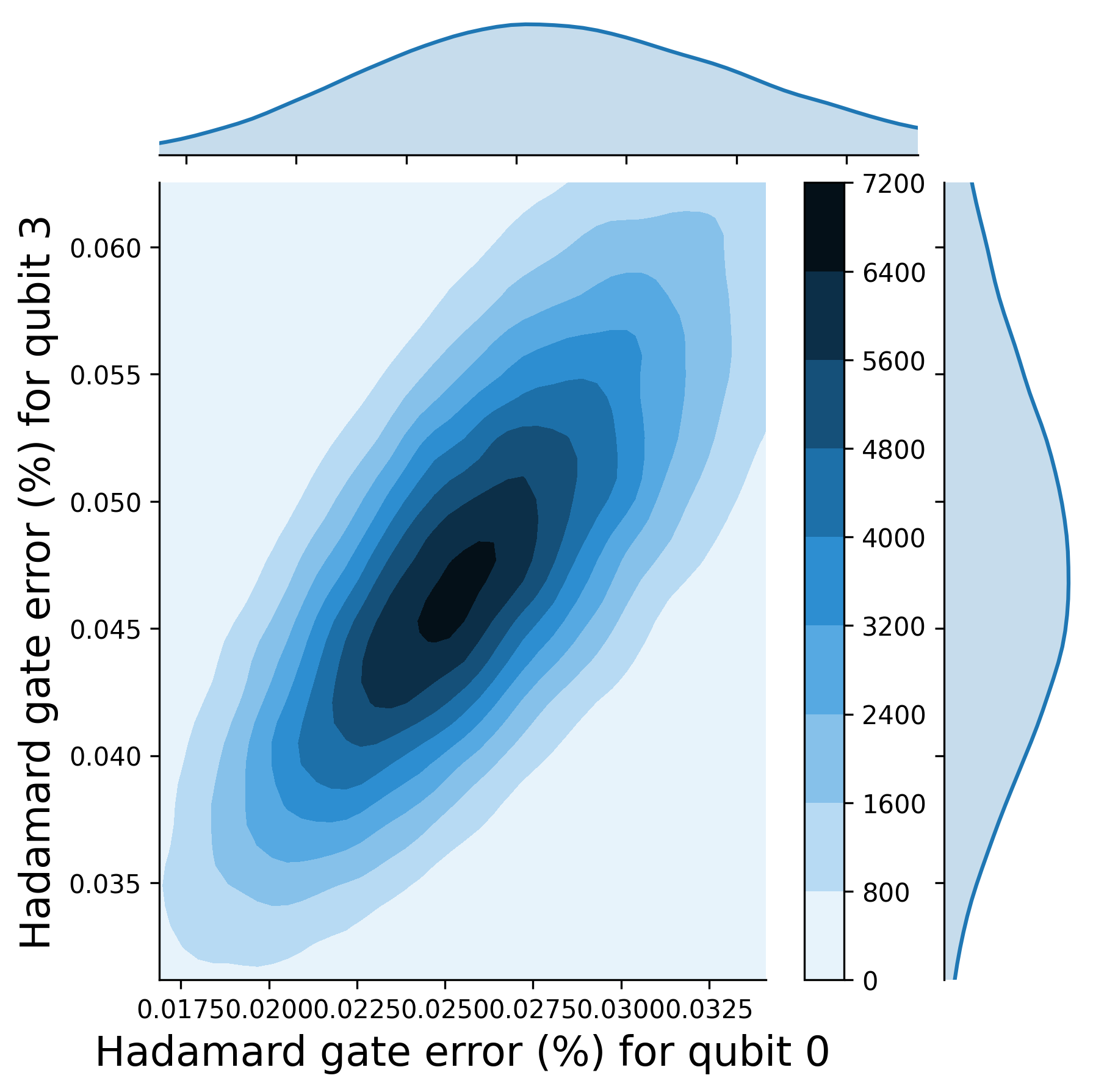}
\caption*{(a)}
\vspace{0.5in}
\includegraphics[width=0.5\columnwidth]{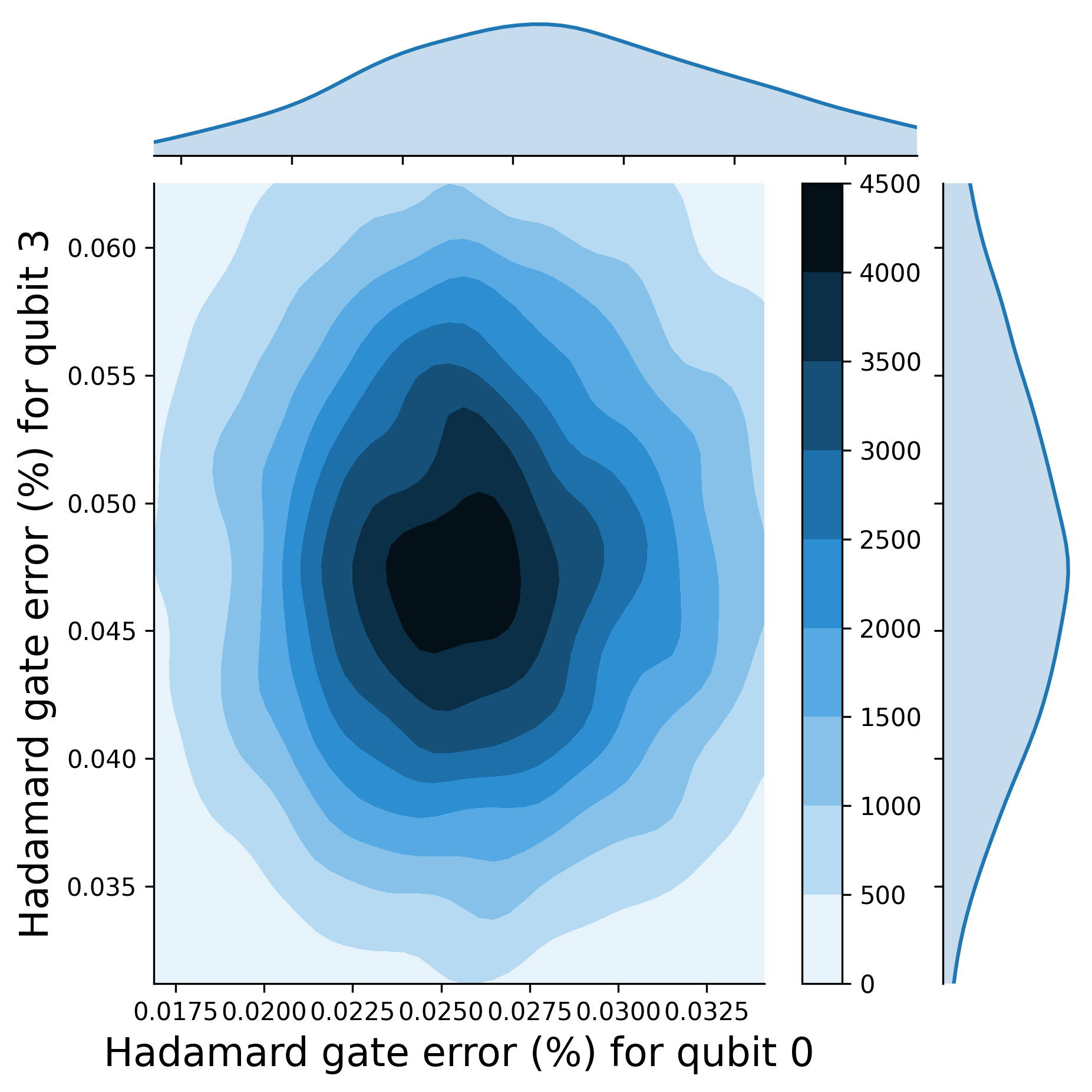}
\caption*{(b)}
\caption{Contour plots to compare the probability densities of a two-dimensional subset of Hadamard gate errors for qubit 0 and 3 in Apr-2023, (a) with and (b) without correlation modeling using a copula function.}
\label{fig:dist_with_copula}
\end{figure}

We constructed the joint density using the copulas method \cite{sklar1959fonctions} discussed in Eqn.~\ref{eq:copulas}. The full 16-dimensional distribution cannot be visualized but the significance of these correlations is apparent from the example of a bi-variate marginal distribution shown in Fig.~\ref{fig:dist_with_copula}, which compares the constructed probability distribution with and without correlation. Importantly, the correlation structure itself changes  monthly with the characterization data. 
 
The full 16-dimensional problem requires a high Monte Carlo sampling overhead for convergence as per Eqn.~\ref{eq:mcmc_hellinger}. To address this issue, we determined that our machine's configuration allows for a Monte Carlo sampling size of 100,000, which corresponds to a program runtime of approximately six hours including IBM Qiskit \cite{alexander2020qiskit} simulations and Monte Carlo sampling overhead\cite{calderhead2014general}. Introducing a threshold for correlation enables the clustering of variables and reduces the effective problem dimensionality\cite{vapnik1999nature}.
 
As the correlations between device parameters varies each month, the number of clusters and their composition also changes. For example, \added{in May 2022, our method identified 13 clusters with the biggest cluster comprising 3 device parameters, while in April 2023, we found 16 independent clusters.} 

Generally, given $d$ device parameters that form $K$ independent clusters at time $t$, denote the $i$-th cluster as $\mathcal{B}_i(t)$. 
The cardinality of $\mathcal{B}_i(t)$ is denoted by $m_i(t)$, such that $\sum_i m_i(t) = d$ for all $t$. Let the elements of $\mathcal{B}_i(t)$ be given by $\{ \xrm_{(1,i)}, \cdots, \xrm_{(m_i(t), i)} \}$ and let $\Theta_i(t)$ from Eqn.~\ref{eq:copulas} denote the copula function for cluster $\mathcal{B}_i(t)$, i.e., 
\begin{equation}
\Theta_i(t) = \Theta\left[
F_{X_{(1     ,i)}} \left( \xrm_{(1     ,i)};t \right), 
\cdots, 
F_{X_{(m_i(t),i)}} \left( \xrm_{(m_i(t),i)};t \right) 
\right].
\label{eq:shorthand}
\end{equation}
Then, Eqn.~\ref{eq:mcmc_hellinger} becomes:
\begin{equation}
\begin{aligned}
1-H^2_{X} =&
\mathds{E}\left( \sqrt{
\frac{ \underset{i \in t_1\textrm{ clusters}}{\prod} \Theta_i(t_2)}
{\underset{j \in t_2\textrm{ clusters}}{\prod} \Theta_j(t_1)}
\prod\limits_{k=1}^{d}
\frac{f_{X_k}(\xrm_k;t_2)}
{f_{X_k}(\xrm_k;t_1)}
}\right),
\end{aligned}
\label{eq:time_varying_copula}
\end{equation}
which we will approximate through Monte Carlo sampling.

We \added{use 100,000} Markov-Chain Monte Carlo simulations to estimate $\braket{O}_t$ for a given month. This sample size was chosen based on numerical convergence by using the Qiskit Aer numerical simulator to calculate noisy simulations of the circuit. From these estimates, we then calculated the monthly average observable value, $\braket{O}_t$ and the observable stability, $s = |\braket{O}_t-\braket{O}_0|$. Moreover, we performed these simulation 100 times for each month to estimate the underlying distribution for the stability itself.  
 
\begin{figure}[htbp]
\centering
\includegraphics[width=\figurewidth]{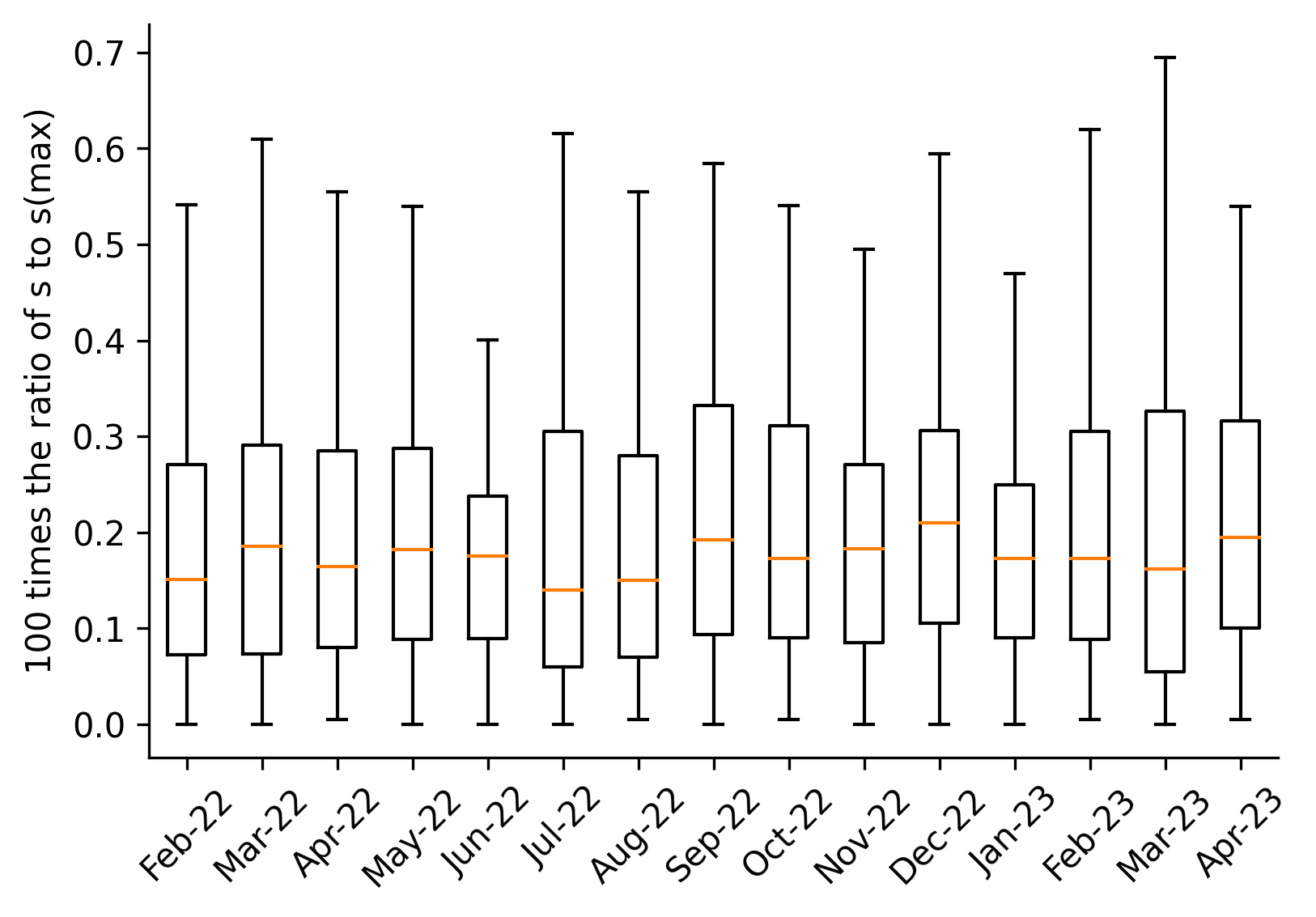}
\caption{ Simulations of the stability ratio $s/s_{max}$ times 100 for a 4-qubit Bernstein-Vazirani circuit using the noise characterization from the \washington platform. The box-and-whisker plot of the monthly statistics are based on noisy circuit simulations using the joint probability distribution derived from data \added{from 1-Jan-2022 to 30-Apr-2023}. Ratio values below unity confirm that the upper bound is never exceeded. 
}
\label{fig:s_by_smax}
\end{figure}

Fig.~\ref{fig:s_by_smax} presents the the observable stability $s$ to s$_{\textrm{max}}$ ratio from these simulations for each month. In this box-and-whisker plot, the central box at each point signifies the interquartile range (IQR), with its lower and upper edges representing the first (Q1) and third quartiles (Q3), respectively. The median is indicated by a line within the box. Notably, all ratios remain well below unity and verify that the upper bound calculated from characterization data is never surpassed. 

From Fig.~\ref{fig:s_by_smax}, we also see that our upper bound for the temporal variations of the quantum observable is 100 times higher than the experimentally observed values. Although looser bounds are symptomatic of an overestimation of the device noise, that is acceptable because underestimating the noise is not an option for performance improvement roadmap\cite{roadmap, acin2018quantum, hughes2004quantum} and exact bounds is impossible.

\begin{figure}[htbp]
\centering
\includegraphics[width=\figurewidth]{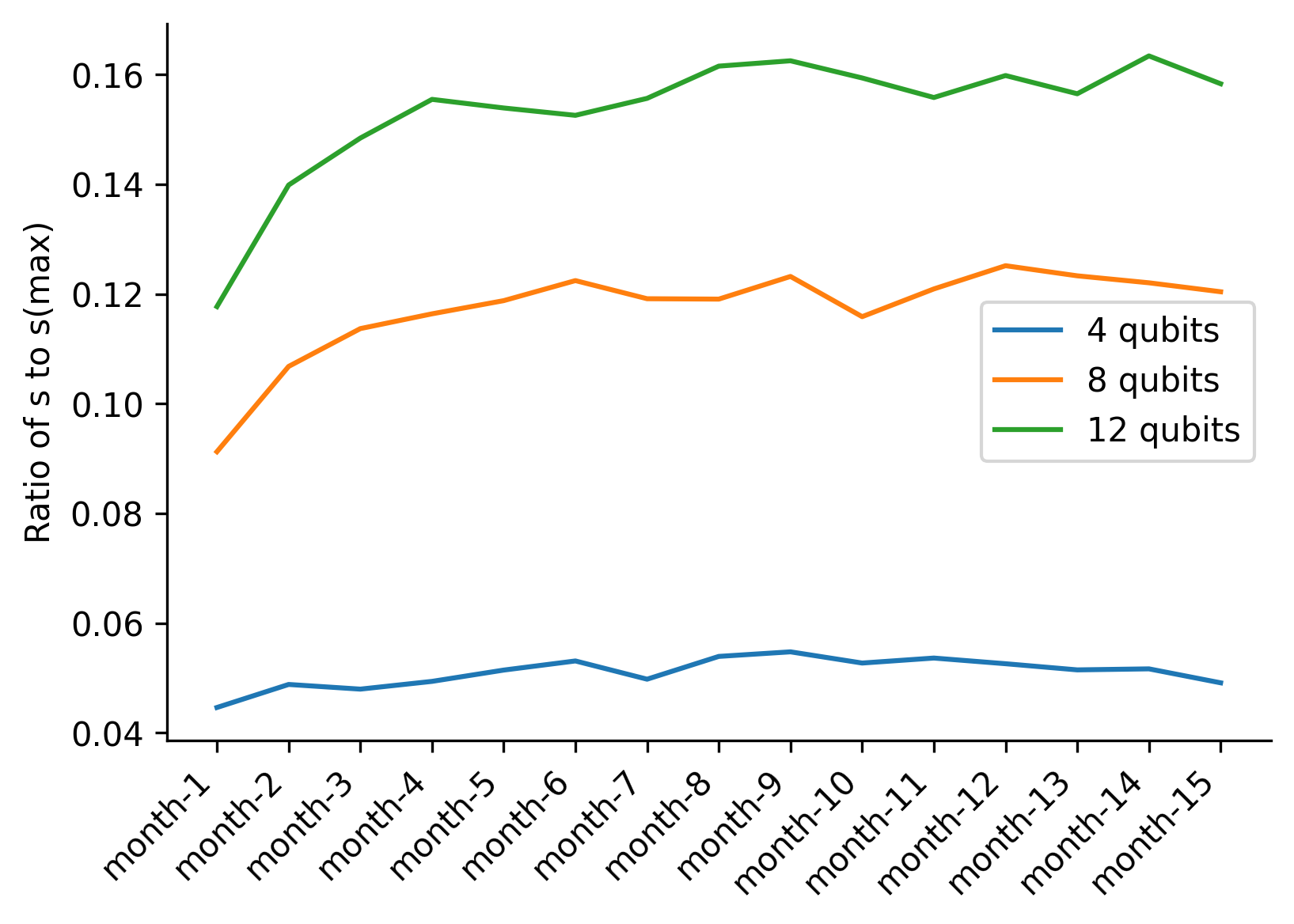}
\caption{The ratio $s/s_\textrm{max}$ for a simulated time-varying noisy circuit is plotted with respect to the increasing noise variance across  15 months. The results from noisy simulations of the Bernstein-Vazirani circuit with a secret string of 4, 8, or 12 bits validate the bound expected.}
\label{fig:depol_sbysmax}
\end{figure}

Note that Eqn.~\ref{eq:bound} does not provide a tight bound due to three sources. Firstly, we can make Eqn.~\ref{eq:abs_integrand} tighter by restricting ourselves to scenarios where the noise distribution function at a later time is consistently lower than at an earlier time, which often occurs in between calibrations. In fact, the reason that Fig.~\ref{fig:depol_sbysmax} was able to achieve a more accurate estimate of the temporal variations of the quantum observable is because we had modeled an in-between calibrations scenario. Secondly, the use of H\"older's inequality introduces additional loss of tightness, since the equality holds only when the two functions are linearly dependent. Thirdly, Eqn.~\ref{eq:c_factor} leads to a looser bound for observables that heavily fluctuate with platform characterization metrics. This approximation, found in the appendix, employs the maximum value of $\braket{O_\xrm}$ to set the integral's bound. The accuracy of this approximation diminishes as $\braket{O_\xrm}$ fluctuates more with x, while it improves with less variation in x. In our Bernstein-Vazirani application, where $\braket{O_\xrm}$ ranges from 0 to 1, this introduces significant approximation, contributing to the observed loose bound.

Despite not being very tight, our bound in Eqn.~\ref{eq:bound} is still useful for several reasons. Firstly, it helps estimate the maximum temporal variations and ensures result reproducibility. Secondly, if the platform noise stays within the bounds determined by the analysis, experimental reproducibility can be guaranteed with a high degree of certainty. Finally, numerical simulations using real data allow us to scale down the requirements to be less restrictive. 
\section{General bounds on reliability}
The purpose of this section is to determine the bound on the reliability metric for a noisy quantum device in achieving an $\epsilon$-stable outcome.

Eqn.~\ref{eq:smax_normalized} can be re-arranged to yield the upper bound on distance:
\begin{eqnarray}
H_{\textrm{max}}(t_1, t_2) = \sqrt{1-\sqrt{1-\phi}}
\label{eq:bound}
\end{eqnarray}
with $\phi = s_\textrm{tol}^2 / (4c^2)$.

We validate the bound using a numerical simulation of of the \BV circuit like before. 
To validate this bound, first, we mapped the 16 noise parameters essential for our simulation of the 5-qubit Bernstein-Vazirani circuit shown Fig.~\ref{fig:bv_qiskit_ckt} to specific independent noise processes. The parameters mapped to gate and register specific noise model in Table~\ref{tab:noiseParameters}. For example, the asymmetric binary channel for register $0$ flips the measured output bit $b_0$ to $b_0\oplus 1$ with probability $x_0$, while the coherent phase error channel\cite{khatri2020information} for the Hadamard gate $H$ applied to register 0  transforms the underlying quantum state as $CP(H \rho H) = R_z(\theta) H \rho H R_z^\dagger(\theta)$. Thermal relaxation\cite{10.1142/s0219477522500602} is modeled by an exponential dephasing process that depends on the $T_2$ time and the duration of the underlying gate not shown here. While the 16 noise processes above act independently, the underlying noise parameters are assumed to be correlated. We construct a joint distribution of to describe these parameters using the method of Gaussian copula\cite{nelsen2007introduction}. 

We generate an ensemble of noisy simulations by drawing samples from the multi-parameter noise distribution. We initially establish a joint distribution from the daily data gathered in January 2022 for the \washington device, utilizing copulas. Over the next 15 months, we introduce minor perturbations to this distribution, ensuring that the distance never exceeds $H_\text{max}$ between the perturbed and original January 2022 distributions. In this perturbation scheme, the marginal distribution of the CNOT gate error between qubits 1 and 2 is modeled using a beta distribution, which is based on the aforementioned January 2022 daily data. Small, random perturbations to the beta distribution parameters are incorporated over 15 months for the CNOT error, with distance constraint maintained. For each perturbed distribution, we generate 100,000 noise metric samples, and execute 100 Qiskit simulations (each with 8192 shots). The stability metric is then computed from the obtained output.

\begin{figure}[!t]
\centering
\includegraphics[width=\figurewidth]{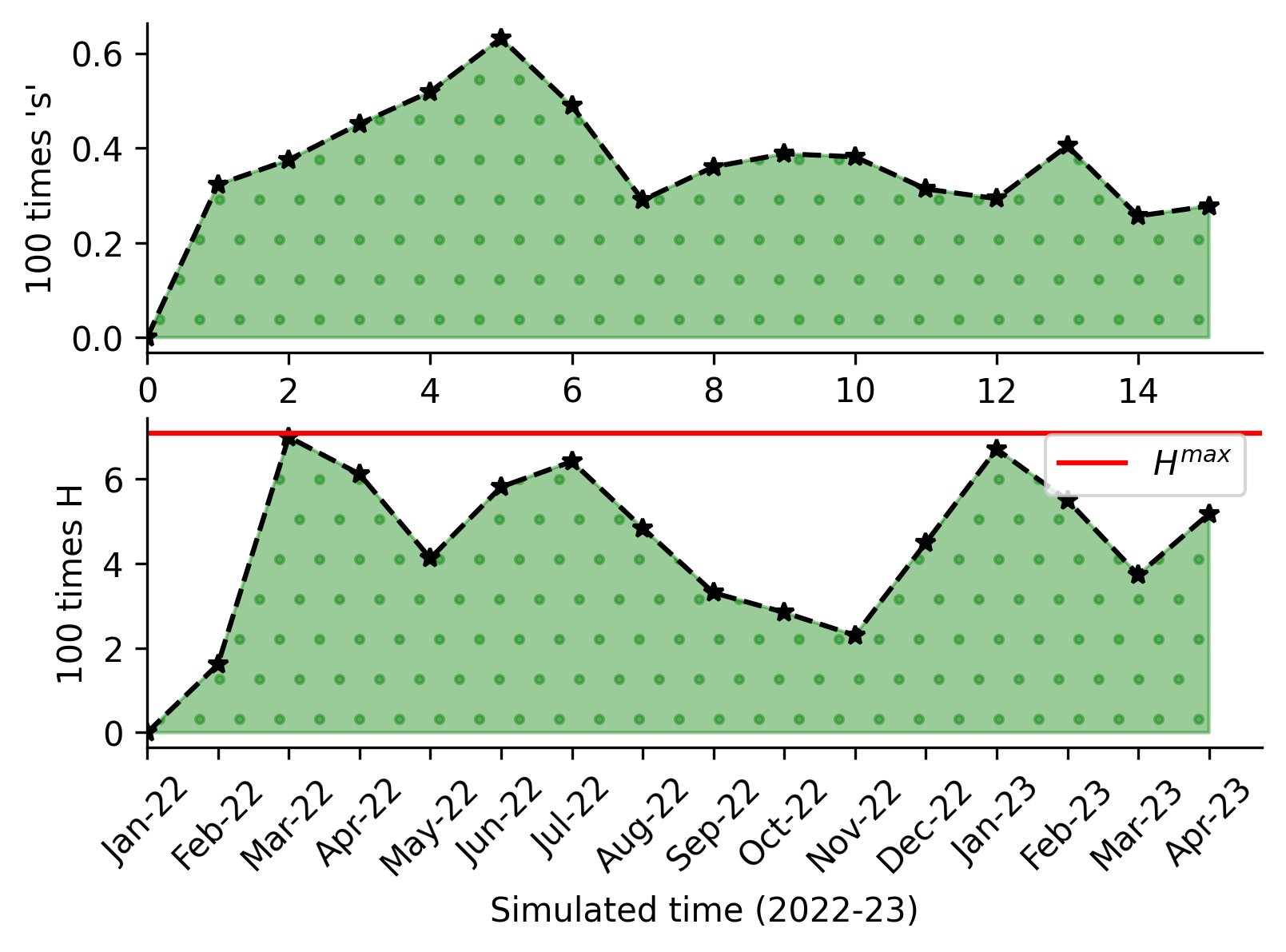}
\caption{Simulation demonstrating that when $H\leq H_\textrm{max}$ (i.e. a reliable, slowly varying noise platform), then $s\leq s_\textrm{max}$ (i.e. the temporal difference of the observable stays within the predicted upper bound).}
\label{fig:slowly-varying}
\end{figure}

Figure~\ref{fig:slowly-varying} presents the simulation results illustrating the relationship between the stability metric ($s$) and the reliability of a quantum device characterized by the distance ($H$). The results demonstrate that when $H\leq H_\textrm{max}$ the device is reliable such that the temporal difference of the observable ($s$) remains within the specified upper bound ($s\leq s_\textrm{max}$). In our simulations, we set the tolerance threshold $s_\text{tol} = 20\%$, which limits the maximum acceptable deviation in the expectation value over time. According to Eqn.~\ref{eq:bound}, this results in an upper limit of 7.1\% for the device reliability metric $H_\text{max}$ for the Bernstein-Vazirani circuit. The lower panel presents the distance between the noise processes. These calculations show how noise can fluctuate on a monthly basis while still respecting the $H_\text{max}$ constraint. While time varying, these process emulate the behavior of a reliable device. The upper panel of Fig.~\ref{fig:slowly-varying} presents the corresponding stability metric, which never exceeds the 20\% tolerance. Moreover, we find the stability is nearly two orders of magnitude smaller than the tolerance, with an average of 0.6\%. By selecting a reliable device, we can ensure the stability of quantum output.
\clearpage
\begin{table}[htp]
\centering
\caption{Register values for $\gamma_D(\tau)$ and $\gamma_{\textrm{max}}$}
\label{tab:gamma_vals}
\begin{tabular}{|p{2cm}|p{2cm}|p{2cm}|}
\hline 
Register No.	& \textbf{$\gamma_{\textrm{max}}$} & \textbf{$\gamma_D(\tau)$} \\ \hline
0&      1.4590&     1.3040\\ \hline
1&      1.1365&     0.6755\\ \hline
2&      2.7284&     2.7118\\ \hline
3&      6.9946&     6.9931\\ \hline
4&      4.3229&     4.3226\\ \hline
5&      5.8171&     5.8157\\ \hline
6&      4.5425&     4.5325\\ \hline
7&      2.6946&     2.6649\\ \hline
8&      8066&     5.4724\\ \hline
9&      8.9672&     8.9666\\ \hline
10&     2.7272&     2.7231\\ \hline
11&    11.5502&    11.5486\\ \hline
12&     3.2212&     3.0797\\ \hline
13&     1.7818&     0.6460\\ \hline
14&    11.9104&    11.9038\\ \hline
15&     2.0713&     2.0228\\ \hline
16&     1.3392&     0.2359\\ \hline
17&     4.8557&     4.8553\\ \hline
18&     1.5986&     1.4980\\ \hline
19&     1.0322&     0.4378\\ \hline
20&     9.0893&     9.0886\\ \hline
21&     1.2259&     1.0620\\ \hline
22&    10.9146&    10.9136\\ \hline
23&     3.0018&     3.0017\\ \hline
24&    14.1254&    14.1241\\ \hline
25&     1.4325&     1.2624\\ \hline
26&     1.3103&     0.9567\\ \hline
\end{tabular}
\end{table}
\vspace{0.5in}
\begin{table}[htp]
\centering
\caption{Device parameters}
\label{tab:noiseParameters}
\begin{tabular}{|p{1.4cm}|p{8cm}|}
\hline 
\textit{Parameter} & \textit{Description} \\ \hline
$\xrm_{0}$ & SPAM  fidelity for register element 0  \\ \hline
$\xrm_{1}$ & SPAM  fidelity for register element 1  \\ \hline
$\xrm_{2}$ & SPAM  fidelity for register element 2  \\ \hline
$\xrm_{3}$ & SPAM  fidelity for register element 3  \\ \hline
$\xrm_{4}$ & CNOT gate fidelity for control 0, target 1  \\ \hline
$\xrm_{5}$ & CNOT gate fidelity for control 2, target 1  \\ \hline
$\xrm_{6}$ & $T_2$ de-coherence time for register element 0  \\ \hline
$\xrm_{7}$ & $T_2$ de-coherence time for register element 1  \\ \hline
$\xrm_{8}$ & $T_2$ de-coherence time for register element 2  \\ \hline
$\xrm_{9}$ & $T_2$ de-coherence time for register element 3  \\ \hline
$\xrm_{10}$ & $T_2$ de-coherence time for register element 4  \\ \hline
$\xrm_{11}$ & Hadamard gate fidelity for register element 0  \\ \hline
$\xrm_{12}$ & Hadamard gate fidelity for register element 1  \\ \hline
$\xrm_{13}$ & Hadamard gate fidelity for register element 2  \\ \hline
$\xrm_{14}$ & Hadamard gate fidelity for register element 3  \\ \hline
$\xrm_{15}$ & Hadamard gate fidelity for register element 4  \\ \hline
\end{tabular}
\end{table}
\vspace{0.5in}
\vspace{0.5in}
\vspace{0.5in}
\vspace{0.5in}
\vspace{0.5in}
\vspace{0.5in}
\vspace{0.5in}
\vspace{0.5in}
\vspace{0.5in}
\vspace{0.5in}
\vspace{0.5in}
\vspace{0.5in}
%

\removefigs
\chapter{Enhancing histogram accuracy}\label{ch:adaptive_algorithms}
The previous chapter focused on modeling the stability of outcomes \cite{dasgupta2023reliability} from noisy quantum computers in presence of non-stationary quantum noise \cite{etxezarreta2021time}. 
It developed analytical bounds leveraging device characterization data to enable estimation of outcome stability. 
The bounds were validated using noise characterization data collected from IBM transmon processors \cite{roth2021introduction}.

In this chapter, we study Bayesian techniques \cite{lukens2020practical, zheng2020bayesian, gordon1993novel, kotecha2003gaussian} to improve the accuracy of histograms obtained from a noisy quantum computer using a uniform superposition circuit \cite{wong2017exceptional} is utilized as a test case. 
The performance metric in this chapter is the Hellinger distance ($H_\xrm$) \cite{spehner2017geometric} between the noisy histogram in the computational $\mathds{Z}$-basis observed at time $t$ and the noiseless histogram. 
The chapter is divided into two sections: the first section focuses on improving histogram accuracy in presence of uncorrelated noise, while the second deals with correlated noise. 

\section{Uncorrelated noise}
Suppose we want to execute a noisy quantum circuit $L$ times, indexed by $l$. 
In each execution, the number of repetitions allowed by a remote cloud computer \cite{devitt2016performing} (also called n-shots) is $N_s$ and is indexed by k. 
Let us call the $n$-bit digital output of the noisy quantum computer as $v$.
This $n$-bit digital output is measured in the computational basis. 
To be precise, $v(k,l)$ denotes the output of the $k$-th shot for the $l$-th circuit execution. 
Thus,
\begin{equation}
v(k,l) = [v_{n-1}(k,l) \cdots v_0(k,l)]
\end{equation}
When represented as a decimal integer, $v(k,l)$ takes values in $\{0, \cdots, 2^n-1\}$. 
When represented in binary, each $v_i(k,l)$ denotes a classical bit $\in \{0,1\}$. In Dirac notation, the classical bit $v_i$ can be written as as $\ket{v_i} \in \{ \ket{0}, \ket{1} $ and the output can be written as:
\begin{equation}
v(k,l) = \ket{v_{n-1}}(k,l) \otimes \cdots \otimes \ket{v_0}(k,l)
\end{equation}

Note that $l$ can also be thought of as a proxy for a short time-window during which the noise can be assumed to stay constant. 
During this short time-window, we are able to collect $N_s$ outcomes (each of length $n$ bits). These $N_s$ observations are denoted by $\{ v(k,l)\}_{k=0}^{N_s-1}$ and they correspond to the $l$-th3 circuit execution instantiation. 

The noise parameter $\xrm$ characterizes a quantum noise channel $\mathcal{E}_\xrm(\cdot)$. For instance, for a single-qubit  depolarizing channel, where $ \xrm \in [0, 1]$, the effect of the noise on an input density matrix $\rho$ is represented by:
\begin{equation}
\mathcal{E}_\xrm(\rho) = (1-\xrm)\rho + \frac{\xrm}{3} \left( \mathds{X}\rho \mathds{X} + \mathds{Y}\rho\mathds{Y} + \mathds{Z}\rho\mathds{Z} \right)
\end{equation}
where $\mathds{X}, \mathds{Y}, \mathds{Z}$ are the Pauli matrices. Also, let the probability density for $\xrm$ at the l-th instant be denoted by $f_X(\xrm; l)$.

Using Bayes' theorem \cite{stigler1982thomas}:
\begin{align}
\text{Pr}[ \xrm; l \mid \{ v(k,l) \}_{k=0}^{N_s-1}] &\propto \text{Pr}[ \{ v(k,l) \}_{k=0}^{N_s-1} \mid \xrm] f_X(\xrm; l)
\label{eq:bayes_theorem}
\end{align}
where $\text{Pr}[ \{ v(k,l) \}_{k=0}^{N_s-1} \mid \xrm]$ is the likelihood, 
$f_X(\xrm; l)$ is the prior, and 
$\text{Pr}[\xrm; l\mid \{ v(k,l) \}_{k=0}^{N_s-1}]$ is the posterior.
The prior can be assumed from available old device characterization data. 
In absence of available old data, it is also okay to assume that the prior is a uniform distribution (to indicate a lack of information about the quantum channel) \cite{mezghani2008achieving}. 
Note that the missing proportionality constant $\tilde{c}$ in Eqn.~\ref{eq:bayes_theorem} is given by:
\begin{align}
\frac{1}{c} =& \int\limits_{\xrm} f_X( \{ v(k,l) \}_{k=0}^{N_s-1} \mid \xrm) f_X(\xrm; l)d\xrm
\end{align}
Only in rare cases, this proportionality constant can be calculated analytically. It is usually computationally intractable. However, in Markov Chain Monte Carlo based methods \cite{blei2017variational}, $\tilde{c}$ is often not required.

\begin{figure}
\centering
\includegraphics[width=\figurewidth]{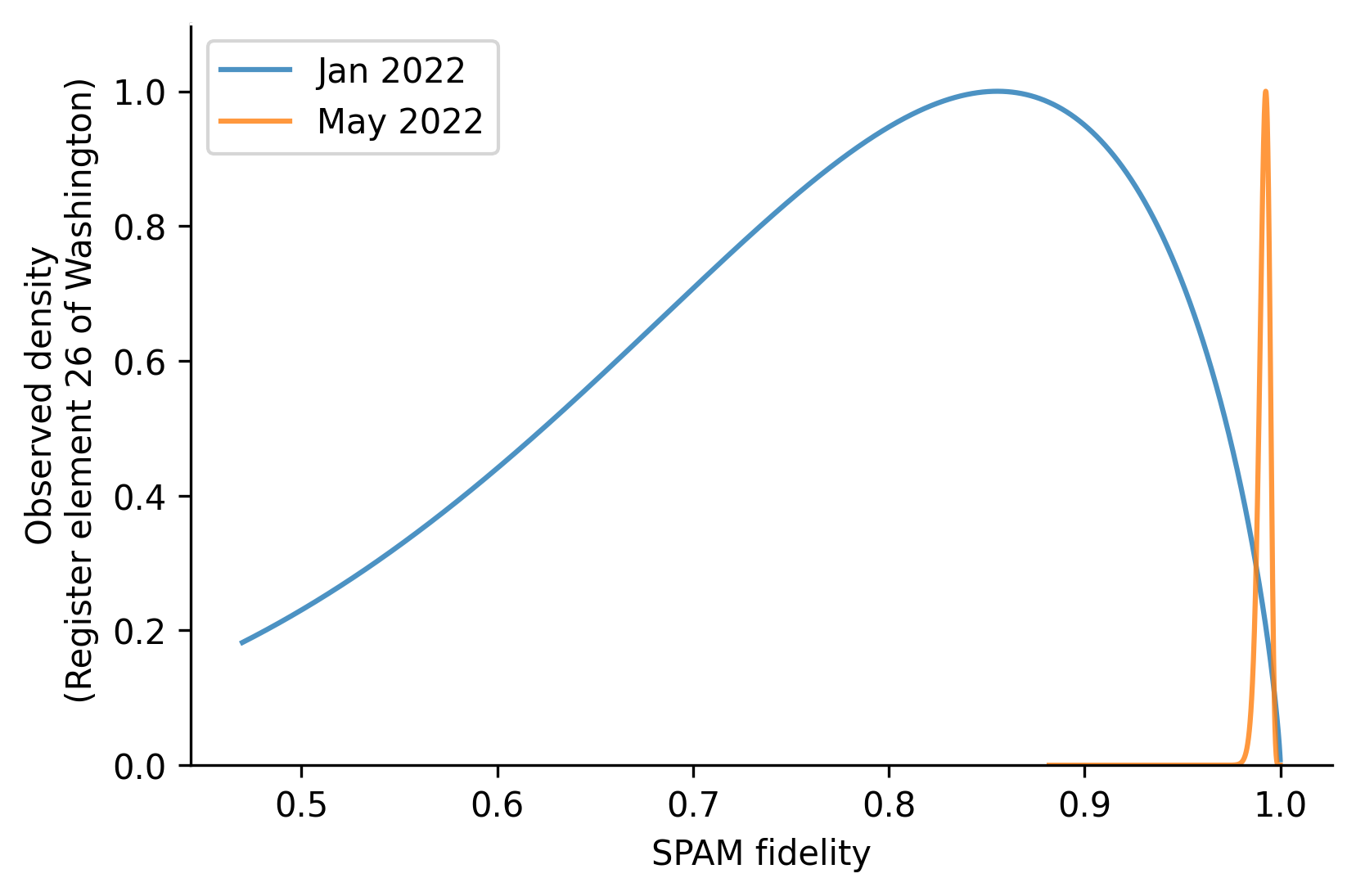}
\caption{
Time-varying density of SPAM fidelity. Data shown for register element \#26. 
}
\label{fig:q26_washington_jan_may}
\end{figure}

\begin{figure}
\centering
\includegraphics[width=\figurewidth]{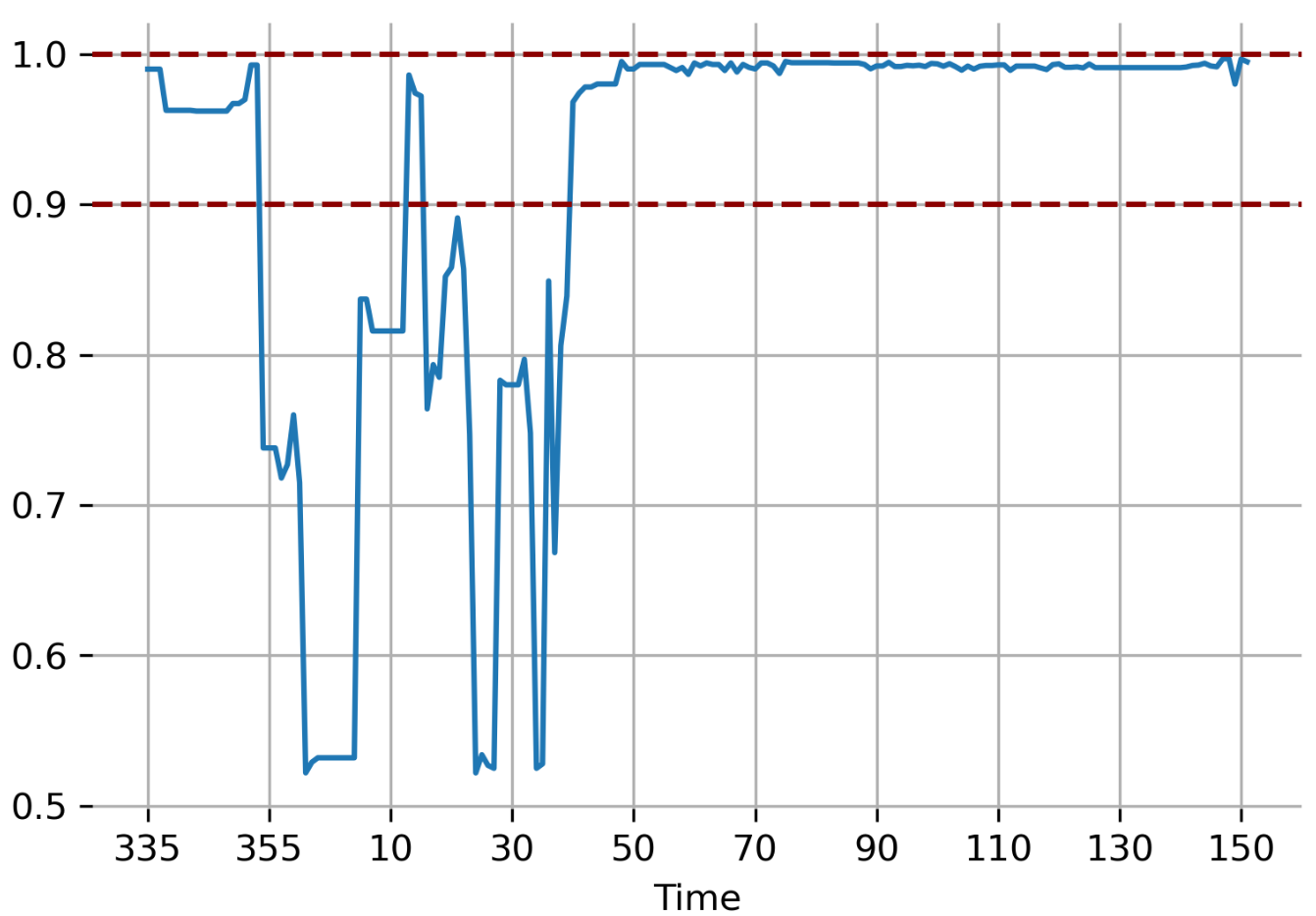}
\caption{
SPAM fidelity time-series for qubit \#26 for Dec-May 2022.
}
\label{fig:q26_washington_dec_may2022}
\end{figure}

The next step in the algorithm is to find the maximum-a-posteriori (MAP) \cite{bassett2019maximum} estimate for $\xrm$:
\begin{align}
\hat{\xrm} =& \underset{\xrm}{\text{ argmax }} f_X(\xrm \mid \{ v(k,l) \}_{k=0}^{N_s-1})
\end{align}
For purposes of quantum error mitgation, $\hat{\xrm}$ is our best guess for $\xrm$ in a non-stationary noise environment (such as depicted in Fig.~\ref{fig:q26_washington_jan_may} 
and 
Fig.~\ref{fig:q26_washington_dec_may2022}
). 

An advantage of this method is that it helps to mitigate noise in gate operations from the software interface without having to do pulse-level programming \cite{alexander2020qiskit}. 
However, Bayesian methods are notorious for not being rapidly scaleable. 
The scalability depends on the noise model granularity. 
Using numerous noise parameters that exponentially increase with register elements might not aid in efficient statistical estimation and can lead to poorer outcomes. 
Usually, embracing simpler models reduces bias and prevents over-fitting \cite{ziegel2003elements}.

To illustrate, consider a 4-qubit register initialized to $\ket{0000}$. Each  register element is subjected to a Hadamard gate to produce four-qubits in uniform superposition. The noise parameter $\xrm$ for this circuit has 12 elements:
\begin{itemize}
\item the SPAM error, $ \xrm_{q,0}$, characterizing the SPAM noise for register element $q$, when the input state is $\ket{0}$ $(q \in \{0,1,2,3\})$.
\item the SPAM error, $ \xrm_{q,1}$, characterizing the SPAM noise for register element $q$, when the input state is $\ket{1}$ $(q \in \{0,1,2,3\})$.
\item the Hadamard gate error, $ \xrm_{q,2}$, characterizing the single-qubit rotation error for register element $q$ $(q \in \{0,1,2,3\})$.
\end{itemize}
As a side note, in other chapters, we have denoted SPAM fidelity with the letter $f$. However, in this chapter, we use a slightly different notation for the sake of clarity.

Next, we will specify the Bayesian model \cite{blei2017variational}. Since this section addresses the case of independent noise sources, it implies no entanglement between register elements. This enables us to use a separable noise channel \cite{bennett1998quantum}:
\begin{equation}
\mathcal{E}_\xrm ( \ket{v} ) = \bigotimes\limits_{i=n-1}^{0} \mathcal{E}_{ \xrm_i } ( \ket{v_i} )
\end{equation}
Note that $i$ is in descending order to reflect register endianness \cite{kapl2020endicheck}. The prior density is:
\begin{align}
f_X(\xrm) &= \prod\limits_{j=0}^{2} \prod\limits_{i=0}^{n-1} f_X( \xrm_{i,j} )
\end{align}
where $j$ iterates through the circuit noise types (SPAM errors, gate error) and $i$ through the qubits.

Each of the independent univariate parameters $\xrm_{i,j}$ is modeled using a time-varying beta distribution \cite{gupta2004handbook}:
\begin{align}
f( \xrm_{i, j}; t) &= \frac{ \xrm_{i,j}^{\alpha_{{}_{i,j}}(t)-1} (1-\xrm_{i,j})^{\beta_{{}_{i,j}}(t)-1}}{\text{Beta}(\alpha_{{}_{i,j}}(t), \beta_{{}_{i,j}}(t))}
\end{align}
where $\text{Beta}(\cdot, \cdot)$ is the Beta function:
\begin{equation}
\text{Beta}\left[ \alpha_{{}_{i,j}}(t), \beta_{{}_{i,j}}(t) \right] =\int\limits_0^1 t^{\alpha_{{}_{i,j}}(t)-1}(1-t)^{\beta_{{}_{i,j}}(t)-1}dt, \;\;\;\; \forall \alpha_{{}_{i,j}}(t), \beta_{{}_{i,j}}(t) > 0
\end{equation}
The choice of the beta distribution is motivated by the bounded nature of the experimental data (upper bounded at 1 for SPAM noise and $\pi/4$ for hadamard gate noise) and its skewed characteristics, often exhibiting a peak. We remind the reader that the discrete letter $l$ (describing the quantum circuit execution instant) will be used interchangeably with the continuous time $t$.

The time-varying likelihood function \cite{eliason1993maximum} is given by:
\begin{align}
\mathcal{L} &= \text{Pr}[ \{ v(k,l) \}_{k=0}^{N_s-1} \mid \xrm ]\\
&= \prod\limits_{k=0}^{N_s-1} \text{Pr} (v(k,l) \mid \xrm)\\
&= \prod\limits_{k=0}^{N_s-1} \prod\limits_{i=0}^{n-1} [ \pi_{i}(l) ]^{1-v_i(k,l)}[ 1 - \pi_{i}(l) ]^{v_i(k,l)}
\end{align}
where,
\begin{align}
\pi_{i}(l) &= \frac{ 1 - \xrm_{i,0}(l) + \xrm_{i,1}(l)}{2} + \sin\left[2\xrm_{i,2}(l)\right] \frac{1-\xrm_{i,0}(l)-\xrm_{i,1}(l)}{2}
\end{align}
is the probability of observing $\ket{0}$ for qubit $i$ when measured in the computational basis after the $l$-th execution.

The posterior distribution is estimated (using Metropolis-Hastings algorithm \cite{calderhead2014general}) per:
\begin{align}
\text{ Posterior } &\propto \text{ Likelihood } \times \text{ Prior }\nonumber \\
\Rightarrow \text{Pr}[\xrm; l \mid \{ v(k,l) \}_{k=0}^{N_s-1}] &\propto 
\prod\limits_{i=0}^{n-1} \left[ 
\pi_{i}(l)^{N_s - \sum\limits_k v_i(k,l) }
(1-\pi_{i}(l))^{\sum\limits_k v_i(k,l)}
\prod\limits_{j=0}^{2}
\frac{\xrm_{i, j}^{\alpha_{{}_{i,j}}(l-1)-1} (1-\xrm_{i, j})^{\beta_{{}_{i,j}}(l-1)-1}}{\text{Beta}\left[\alpha_{{}_{i,j}}(l-1), \beta_{{}_{i,j}}(l-1)\right]}
\right]
\end{align}

Lastly, the maximum-a-posteriori (MAP) estimate is obtained using a log maximization:
\begin{align}
\hat{\xrm}(l) &= \underset{\xrm}{\text{ argmax }} \log \text{Pr}[ \xrm; l \mid \{ v(k,l) \}_{k=0}^{N_s-1}] 
\end{align}
$\hat{\xrm}(l)$ is our best estimate for the time-varying quantum noise when the noise terms are independent. This time-varying noise estimate is then used for quantum error mitigation at time $t$. Specifically, $\hat{\xrm}_{i,0}(l)$ and $\hat{\xrm}_{i,1}(l)$ define the time-varying SPAM noise matrix for qubit $i$ for readout mitigation using matrix inversion \cite{yang2022efficient}. $\hat{\xrm}_{i,2}(l)$ is our best estimate for Hadamard calibration noise \cite{nielsen2002quantum}. We mitigate this noise by using $\pi/4 - \hat{\xrm}_{i,2}$ as the input in the software for the single qubit rotation. This helps us to avoid pulse level programming \cite{alexander2020qiskit}.

\begin{figure}
\centering
\includegraphics[width=\figurewidth]{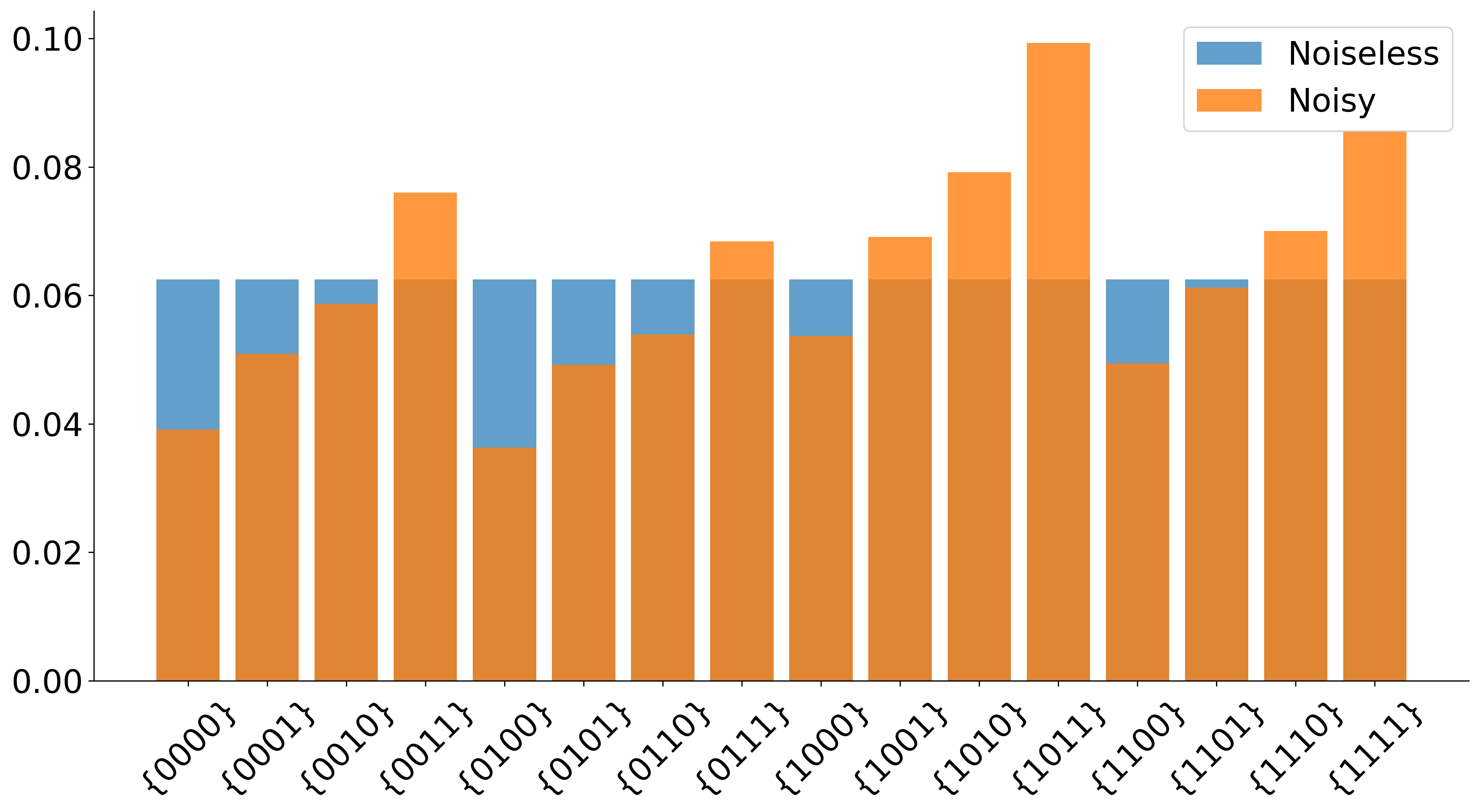}
\caption{We simulated SPAM and gate error channels for a quantum circuit with 4 qubits that creates a uniform superposition across all the computational basis states using Hadamard gates. The blue bars represent the probability distribution across the measurement outcomes for the ideal, noiseless circuit while the orange bars represent the same for a realization of an execution on an unstable device.}
\label{fig:q_stabilization_a}
\end{figure}

\begin{figure}
\centering
\includegraphics[width=\figurewidth]{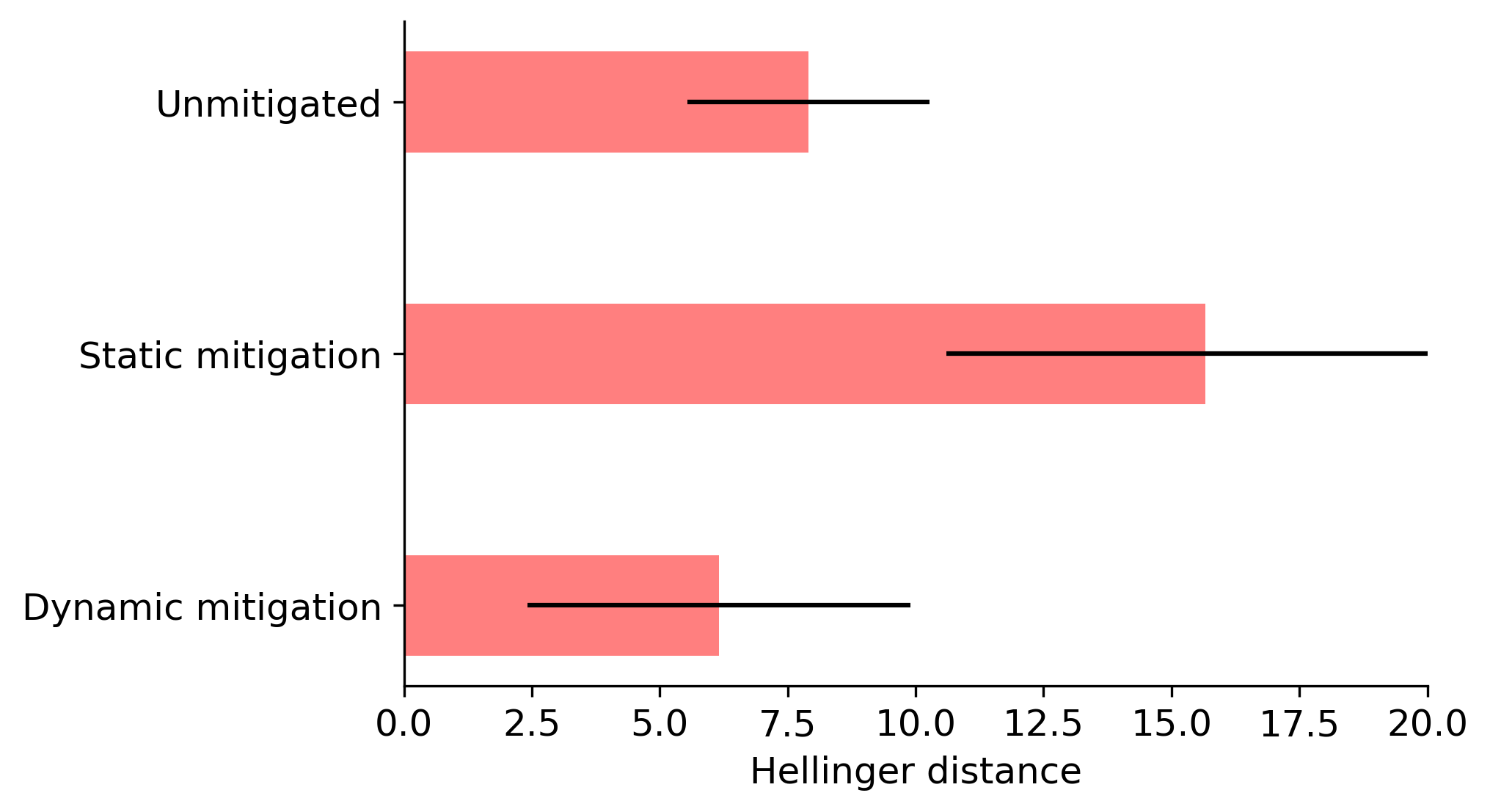}
\caption{The plot shows improved accuracy achieved using Bayesian optimization in presence of  time-varying noise. Mitigation using average parameters deteriorates accuracy. The number of samples used in the Bayesian estimation process is held constant at $10^{4}$.}
\label{fig:q_stabilization_b}
\end{figure}

\begin{figure}
\centering
\includegraphics[width=\figurewidth]{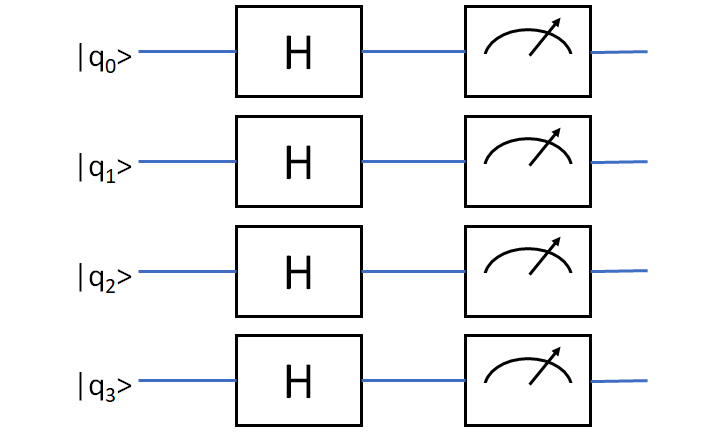}
\caption{The 4-qubit quantum circuit used for the simulation experiment. Each qubit is assumed to have a different, independent SPAM error process. Each Hadamard gate is similarly assumed to have a different, independent 
  gate error process.}
\label{fig:setup_a}
\end{figure}

We used Qiskit \cite{ibm_quantum_experience_website} for our simulations. 
The circuit layout is depicted in Fig.~\ref{fig:setup_a}. 
For simulating the SPAM noise channel, we employed a binary asymmetric channel 
for each register element. 
The SPAM parameters were drawn from a beta distribution. 
For the initial state $\ket{0}$, the mean of the SPAM fidelity distributions were $(0.9, 0.8, 0.85, 0.75)$ for the respective qubits, with the standard deviation being one-tenth of the mean. 
Similarly, for the initial state $\ket{1}$, the mean of the SPAM fidelity distributions were $(0.85,0.75,0.80,0.70)$ for each register element, with the standard deviation remaining one-tenth of the mean. 
For simulating the Hadamard noise, we used Qiskit's $U3$ gate \cite{ibm_quantum_experience_website} which is parameterized by three angles: $\theta, \phi$ and $\lambda$. For our noise simulation, we used 
$\theta = \frac{\pi}{2} + \xrm_{i,2}$, 
$\phi = 0$, and 
$\lambda = \pi$. 
Here, $\xrm_{i,2}$ is stochastic Hadamard noise drawn from a beta distribution with mean (in degrees) given by: $3.1, 4.1, 4.9, 2.9$ for the four qubits respectively. The standard deviation remained one-tenth of the mean, similar to previous cases.
The circuit was repeated $L=10$ times to obtain acceptable error bars. 
A sample execution is shown in Fig.~\ref{fig:q_stabilization_a}. 
Each execution comprised $N_s=8192$ shots. 
Three scenarios were investigated: 
\begin{itemize}
\item Unmitigated: Raw results with no mitigation.
\item Static Mitigation: Traditional method disregarding time-varying noise (uses average numbers).
\item Adaptive Mitigation: Using Bayesian optimization for inferring time-varying estimates (as detailed in this section).
\end{itemize}
The result of our experiment is shown in Fig.~\ref{fig:q_stabilization_b}. It clearly demonstrates a decrease in error (quantified by the Hellinger distance) with an adaptive approach. Importantly, it also illustrates that mitigation using average noise parameters can sometimes increase the error in presence of time-varying quantum noise, compared to raw results with no mitigation.
\section{Correlated noise}
In this section, we will explore a correlated multi-qubit noise model. 
We will use a $n$-qubit Pauli noise channel. 
The Pauli oise model, although not a completely general noise model, still manages to model many practical situations. 
It is widely used because of two reasons: (a) it is efficiently simulatable on a classical computer (per the Gottesman-Knill theorem) and (b) when used as a proxy for physically accurate noise models (such as the amplitude and phase damping noise) which are not efficiently simulatable on a classical computer, it still manages to preserve important properties like entanglement fidelity \cite{horodecki1999general}. 
The channel coefficients constitute a probability simplex \cite{krstovski2013efficient} (i.e. they add up to 1 and remain positive at all times). 
Thus, these coefficients are not independent and introduce correlations between the terms.

We will use a Dirichlet distribution \cite{wong1998generalized} to model a stochastic Pauli noise channel. 
We will deploy Bayesian techniques to improve the accuracy of probabilistic error cancellation (PEC) \cite{temme2017error, endo2018practical} under time-varying noise. 
Our results will show that Bayesian PEC can outperform non-adaptive approaches by a factor of 4.5x when measured using Hellinger distance from the ideal distribution. 

Consider a single-qubit amplitude and phase damping channel (APD) \cite{khatri2020information}. 
Upon Pauli twirling \cite{cai2020mitigating}:
\begin{align}
\mathcal{E}_\textrm{twirl}(\rho) =& \frac{1}{4} \sum\limits_{A \in \{I, X, Y, Z\}} A^\dagger \mathcal{E}_\text{APD} \left( A\rho A^\dagger \right)A \\
=& \sum\limits_{k=0}^3 c_k \sigma_k \rho \sigma_k
\end{align}
an APD channel becomes a Pauli noise channel. 
Here, $\{ \sigma_k \}_{k=0}^3 = \{I, X, Y, Z\}$ are the Pauli matrices. 
Thus, the coefficients of the Pauli noise channel are functions of the coefficients of the original APD channel, which in turn are functions of the decoherence times $T_1$ and $T_2$ \cite{etxezarreta2021time}: 
\begin{align}
c_1 = c_2 = & \frac{1}{4}\left[1-\exp\left(-t/T_1\right)\right]\\
c_3 =& \frac{1}{4}\left[1-\exp\left(-t/T_2\right)\right]\\
c_0 =& 1- (c_1 + c_2 + c_3)
\label{eq:depol_coeffs}
\end{align}

We model the decoherence times $T_1$ and $T_2$ as random variables dependent on register location $i$ and time $\tau$:
\begin{equation}
T_1 = T_1(i,\tau) \quad \text{and} \quad T_2 = T_2(i,\tau) \; .
\label{eq:T_i_t}
\end{equation}
It follows from Eq.~(\ref{eq:depol_coeffs}) that the coefficients of the single-qubit Pauli noise channel are temporally and spatially varying stochastic processes \cite{etxezarreta2021time} which also depend on register location $i$ and time $\tau$:
\begin{equation}
c_0 = c_0(i,\tau), \;\; c_1 = c_1(i,\tau), \;\; c_2 = c_2(i,\tau), \;\; c_3 = c_3(i,\tau) \; .
\label{eq:x_i_t}
\end{equation}
Having modeled the stochasticity of the single-qubit Pauli noise channel, let us generalize to the $n$-qubit case. The quantum noise channel model for an n-qubit register is given by:
\begin{equation}
\mathcal{E}_\xrm(\rho) = \sum\limits_{i=0}^{N_p-1} \xrm_i P_i(n) \rho P_i(n)^\dagger \; ,
\end{equation}
where $n$ is the register size, $N_p=4^n$ is the total number of Pauli coefficients and $P_i(n)$ are the n-qubit Pauli operators. The  channel coefficients are subject to the conditions:
\begin{equation}
\sum\limits_{i=0}^{N_p-1}\xrm_i=1, \;\;\;\; \xrm_i \geq 0 \; .
\label{eq:simplex}
\end{equation}
The $N_p$ coefficients of the Pauli operators are then:
\begin{equation}
\xrm_k = \xrm_k(i, \tau)
\end{equation}
where $i = (i_1, \cdots, i_n)$ identifies the register location(s) and $\tau$ is time. 

A prior hypothesis for the channel can be obtained by assuming channel separability. The $N_p$ coefficients can be obtained using a direct product:
\begin{equation}
\begin{split}
\xrm &= 
\begin{pmatrix}
c_0(i=0,\tau)\\
c_1(i=0,\tau)\\
c_2(i=0,\tau)\\
c_3(i=0,\tau)\\
\end{pmatrix}
\times \cdots \times 
\begin{pmatrix}
c_0(i=n-1,\tau)\\
c_1(i=n-1,\tau)\\
c_2(i=n-1,\tau)\\
c_3(i=n-1,\tau)\\
\end{pmatrix} \; ,
\label{eq:direct_product}
\end{split}
\end{equation}
where $\times$ refers to the direct product. 

Because the Pauli channel coefficients form a probability simplex, the natural way to model the probability distribution function  $f_X(\xrm)$ of the multi-dimensional Pauli channel distribution is the Dirichlet distribution:
\begin{equation}
f_X(\xrm) \equiv \textrm{Dirichlet}(\xrm; \eta)  \coloneqq  
\frac{ \Gamma \left( \sum\limits_{i=0}^{N_p-1} \eta_i \right)}
{\prod\limits_{i=0}^{N_p-1}\Gamma(\eta_i)}\left(\prod\limits_{i=0}^{N_p-1}\xrm_i^{\eta_i-1}\right) \; ,
\end{equation}
where $\eta_i \ge 0$ are the Dirichlet hyper-parameters, $\Gamma$ is the Gamma function:
\begin{equation}
\Gamma(y) = \int\limits_0^\infty t^{y-1} e^{-t}dt,\;\;\;\; y > 0
\end{equation}
and,
\begin{equation}
\int\limits_{\xrm} \textrm{Dirichlet(x; $\eta$) dx} = 1.
\end{equation}

Recall that we experimentally observe that the distribution of the decoherence times $T_1$ and $T_2$ fluctuates with time. Consequently, the distribution of the Pauli coefficients i.e. the Dirichlet distribution, varies with time. To be precise, the coefficients $\eta_i(\tau)$ vary with time $\tau$. The distribution may be represented as $f_X(\xrm; \tau)$. 

At any circuit execution instance $\tau$, the noise channel is obtained as a specific realization of the random variable X (the Pauli coefficients) which is just a sample from $f_X(\xrm; \tau)$. The degree of channel non-stationarity can be quantified using the Hellinger distance $H_\xrm$ between the distributions at time $\tau$ and $\tau^\prime$:
\begin{equation}
\begin{aligned}
& H_d = \sqrt{1-B C} \; , \\
& BC = \int\limits_\xrm \sqrt{f(\xrm; \tau) g(\xrm; \tau^\prime)} d\xrm\\
& =\int\limits_\xrm 
\sqrt{
\frac{\Gamma\left(\eta_{0}(\tau)+\cdots+\eta_{N-1}(\tau)\right)}
{\prod\limits_{i=0}^{N-1} \Gamma\left(\eta_{i}(\tau)\right)}}
\prod\limits_{i=0}^{N-1} \xrm_{i}^{\left(\eta_{i}(\tau)-1\right) / 2}
\times\sqrt{
\frac{\Gamma\left(\eta_{0}(\tau^\prime)+\cdots+\eta_{N_p-1}(\tau^\prime)\right)}
{\prod\limits_{i=0}^{N-1} \Gamma\left(\eta_{i}(\tau^\prime)\right)}}
\prod\limits_{i=0}^{N-1} \xrm_{i}^{\left(\eta_{i}(\tau^\prime)-1\right) / 2}  d\xrm\\
&=\frac{
\sqrt{
\Gamma\left( \sum\limits_{i=0}^{N_p-1} \eta_{i}(\tau) \right)
\Gamma\left( \sum\limits_{i=0}^{N_p-1} \eta_{i}(\tau^\prime)\right)
}}
{\prod\limits_{i=0}^{N_p-1} 
\sqrt{
\Gamma(\eta_{i}(\tau)) \Gamma(\eta_{i}(\tau^\prime))
}}\times \frac{
\prod\limits_{i=0}^{N_p-1}
\Gamma\left(\frac{\eta_{i}(\tau)+\eta_{i}(\tau^\prime)}{2}\right)}
{\Gamma\left(\frac{ \sum\limits_{i=0}^{N_p-1}  \eta_{i}(\tau)+\eta_{i}(\tau^\prime)}{2}\right)}
\end{aligned}
\end{equation}
where $N_p=4^{n}$ and BC is the Bhattacharya coefficient.

The time-varying Pauli noise channel can be estimated using Bayes' rule:
\begin{equation}
\begin{aligned}
f_X(\xrm | \textrm{observed data}) &\propto f_X(\textrm{observed data} | \xrm )  f^\textrm{prior}_X(\xrm) \; .
\end{aligned}
\end{equation}
The prior is given by:
\begin{equation}
\begin{aligned}
\textrm{Dirichlet}(\xrm; \eta) &= \frac{ \Gamma \left( \sum\limits_{j=0}^{N_p-1} \eta_j \right)}
{\prod\limits_{j=0}^{N_p-1}\Gamma(\eta_j)}\left(\prod\limits_{j=0}^{N_p-1}\xrm_i^{\eta_j-1}\right)
\end{aligned}
\end{equation}
where the $\eta_i$ are estimated from the experimental data generated using Eqn.~\ref{eq:direct_product}. This data can be old stale data and serves only to inform the prior.

To update the prior with new knowledge, we need to obtain a current dataset of circuit outcomes. This new new dataset will be used to obtain the Bayesian posterior using the likelihood.

This new dataset can be generated by executing the quantum circuit L times and recording the outcome $v$. The dataset is represented by $\{v_l\}$ where $l \in 0, \cdots, L-1$. The probability of observing $v$ is given by:
\begin{equation}
p_v = \textrm{Tr}[\Pi_v \tilde{\mathcal{G}}_{\hat{\xrm}}(\rho_\textrm{test})] \;,\;\;\;\; v \in \{0, \cdots, N-1\}
\end{equation}
Here, $\Pi_v$ is the orthogonal projection operator, $\tilde{\mathcal{G}}_\xrm$ is a noisy implementation of the ideal quantum operation $\mathcal{G}$, $\xrm$ is the noise parameterization and $\rho_\textrm{test}$ is a known density matrix used for channel characterization.

The likelihood function is given by:
\begin{equation}
\begin{aligned}
\mathcal{L} = \Pr( \{ v_\ell \} | \xrm) &= \Prob( V_0 = v_0, \cdots, V_{L-1} = v_{L-1}|\xrm) \\
&=\prod_{l=0}^{L-1}\Prob( V_l = v_l | \xrm)\\
&=\prod_{l=0}^{L-1}\textrm{Categorical}_{0, \cdots, N-1}(v_l; p_0, \cdots, p_{N-1})\\
&=\prod_{l=0}^{L-1} p_0^{\delta_0(v_l)}\cdots p_{N-1}^{\delta_{N-1}(v_l)}\\
&=p_0^{C_0(\textrm{data})} \cdots p_{N-1}^{C_{N-1}(\textrm{data})}\\
&=\prod_{i=0}^{N-1} \left[ \textrm{Tr}[\Pi_i \tilde{\mathcal{G}}_\xrm(\rho_\textrm{test})] \right]^{C_i(\{ v_\ell \})} \; , \\
\end{aligned}
\end{equation}
where, $C_i(\{ v_\ell \}) = \sum\limits_{\ell=0}^{L-1} \delta_i(v_\ell)$ (with $i \in \{0, 1, \cdots N-1\}$) is simply a counter function that counts how many times $i$ appeared in the experimentally observed data post-measurement ($\delta_i(v)$ is Kronecker delta function which is 1 if $i=v$ and zero otherwise). 

The log of the posterior is given by:
\begin{equation}
\begin{aligned}
f^\textrm{posterior}_X(\xrm | \textrm{data}) 
&= \sum\limits_{i=0}^{N-1} C_i(\textrm{data}) \log \left[ \textrm{Tr}[\Pi_i \tilde{\mathcal{G}}_\xrm(\rho_\textrm{test})] \right]
+ \log \Gamma\left(\sum\limits_{j=0}^{N_p-1} \eta_j \right) - \sum\limits_{j=0}^{N_p-1}\log \Gamma(\eta_j)\\
&+\sum\limits_{j=0}^{N_p-1}(\eta_j-1)\log \xrm_j + \textrm{terms independent of x} \; .
\end{aligned}
\end{equation}
In the final step, the maximum-a-posterior (MAP) estimate is obtained as:
\begin{equation}
\begin{aligned}
\hat{\xrm}(\tau) =& \underset{\xrm}{\textrm{argmax}} \left[ \sum\limits_{i=0}^{N-1} C_i(\textrm{data}) \log \left[ \textrm{Tr}[\Pi_i \tilde{\mathcal{G}}_\xrm(\rho_\textrm{test})] \right] +\sum\limits_{j=0}^{N_p-1}(\eta_j-1)\log \xrm_j\right] \; .
\end{aligned}
\end{equation}
$\hat{\xrm}(\tau)$ is our best guess for the time-varying noise at time $\tau$. We will use this updated estimate for error mitigation using probabilistic error cancellation (PEC).

Probabilistic error cancellation (PEC) is a well-known error mitigation method \cite{temme2017error, endo2018practical, zhang2020error}. The four broad steps of the PEC workflow are as follows. We use the convention that calligraphic symbols denote super-operators acting on density matrices: 
\begin{equation}
\mathcal{G}(\rho)=G\rho G^\dagger.
\end{equation}

First, expand an ideal unitary gate $\mathcal{G}$ as a (noise-model dependent) linear combination of implementable noisy gate set $\{\tilde{\mathcal{G}}_j\}$ (with its ideal counterpart $\{\mathcal{G}_j\}$), as follows:
\begin{equation}
\mathcal{G} = \sum\limits_{j=0}^{N_p-1} \theta_j \tilde{\mathcal{G}}_j,
\end{equation}
where $\theta_j$ are real coefficients, and $\mathcal{E}_\xrm$ is an error channel (such as Pauli noise channel), and $\tilde{\mathcal{G}_j}\equiv \mathcal{E}_\xrm \circ \mathcal{G}_\textrm{j}$. 
The implementable noisy gate set $\{\tilde{\mathcal{G}}_j\}$ is also called noisy basis circuit set. 
For example, if $G=H\otimes H$ is a two-qubit Hadamard gate, then the noisy basis circuits are given by 
\begin{equation}
\tilde{\mathcal{G}}_{\sigma \sigma^{\prime}} = \mathcal{E}_\xrm \circ \mathcal{P}_{\sigma \sigma^{\prime}} \circ \mathcal{G}
\end{equation}
where 
\begin{equation}
\mathcal{P}_{\sigma \sigma^{\prime}}(\cdot)\equiv (\sigma \otimes \sigma^{\prime})(\cdot)(\sigma \otimes \sigma^{\prime})\;\;\;\;,
\end{equation}
and 
$\sigma, \sigma^{\prime}$ are picked from the set of Pauli matrices $\{I,X,Y,Z\}$.

The second step of PEC involves estimating the expectation value of the noise-mitigated observable as:
\begin{equation}
\sum\limits_{j=0}^{N_p-1} \theta_j \braket{\tilde{\mathcal{G}}_j}.
\end{equation}
The ideal gate can be approximated as:
\begin{equation}
\mathcal{G} = \sum\limits_{w=0}^{N_p-1} p(w) [\gamma \textrm{sgn}(\theta_w) \tilde{\mathcal{G}}_w],
\end{equation}
where 
\begin{equation}
\gamma=\sum\limits_{w=0}^{N_p-1} |\theta_w|
\end{equation}
and $p(w) = |\theta_w|/\gamma$ and $w$ is a random variable such that $w \in {0,1,\cdots,N_p-1}$. Said differently, the super-operator $\gamma \textrm{sgn}(\theta_w) \tilde{\mathcal{G}}_w$ is an unbiased estimator for the ideal super-operator $\mathcal{G}$ since: 
\begin{align}
\mathcal{G} =&\sum\limits_{w=0}^{N_p-1} \textrm{sgn}(\theta_w) |\theta_w| \tilde{\mathcal{G}}_w \label{eq:pec_sum}\\
 =&\sum\limits_{w=0}^{N_p-1} \textrm{sgn}(\theta_w) \frac{|\theta_w|}{\sum |\theta_w|} (\sum |\theta_w|)\tilde{\mathcal{G}}_w\\
 =&\sum\limits_{w=0}^{N_p-1} p(w) \left[ \gamma \textrm{sgn}(\theta_w) \tilde{\mathcal{G}}_w\right]\\
=&\mathds{E}_w \left[ \gamma \textrm{sgn}(\theta_w) \tilde{\mathcal{G}}_w\right].
\end{align}

The third step of PEC involves sampling from each of the noisy implementable circuits and computing the mean of the observable for these noisy basis circuits.

The final step of PEC involves inferring the (hopefully) noiseless observable  using a weighted sum of the mean of the observable obtained from the noisy basis circuits in previous step (as per Eqn.~\ref{eq:pec_sum}).

\begin{figure}
\centering
\includegraphics[width=4in]{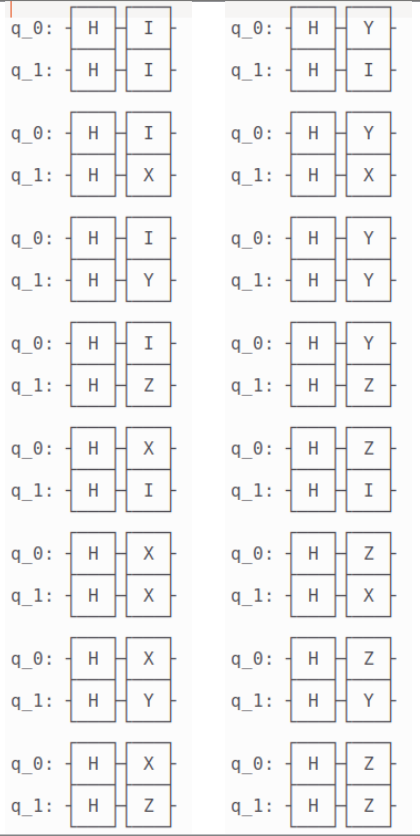} 
\caption{
The figure displays the noisy basis circuits for the linear combination step (the first step) in PEC, where each figure represents the noisy operations $\mathcal{E}_\xrm \circ \mathcal{P}_{\sigma, \sigma^\prime} \circ \mathcal{G}$ and $G = H \otimes H$ is the desired, ideal operation. Note that different basis set choices are possible, depending on the hardware.}
\label{fig:basis_ckt_diagrams}
\end{figure}

In our 2-qubit application ($\mathds{H}\otimes \mathds{H}$), the observables are the projection operators $\Pi_0=\ket{00}\bra{00}$, $\Pi_1=\ket{01}\bra{01}$, $\Pi_2=\ket{10}\bra{10}$, and $\Pi_3=\ket{11}\bra{11}$. The complete noisy basis circuit set is shown in Fig.~\ref{fig:basis_ckt_diagrams}.

For our noise simulation, we assume that the mean of the stochastic $T_1$ coherence time for the first qubit decreases uniformly in a simple step-function-like manner over five time periods, deteriorating from 150 to 60 $\mu$s. Similarly, for the second qubit, we assume that the mean of the $T_1$ time deteriorates from 200 to 10 $\mu$s in five time periods. Additionally, we assume a simple step-function-like decrease in the mean of the $T_2$ coherence time for the first qubit, deteriorating from 70 to 50 $\mu$s in five time periods. Finally, for the second qubit, we assume that the mean of the $T_2$ coherence time deteriorates from 130 to 62.5 $\mu$s in five time periods. The coefficients of the Pauli channel are then computed using Eq.~(\ref{eq:depol_coeffs}). This setup mimics intra-calibration deterioration of the noise in a quantum circuit. We assume a typical execution time of 100 $\mu$s for the Hadamard gate on the IBM transmon platform.

\begin{figure}[htbp]
\centering
\includegraphics[width=\figurewidth]{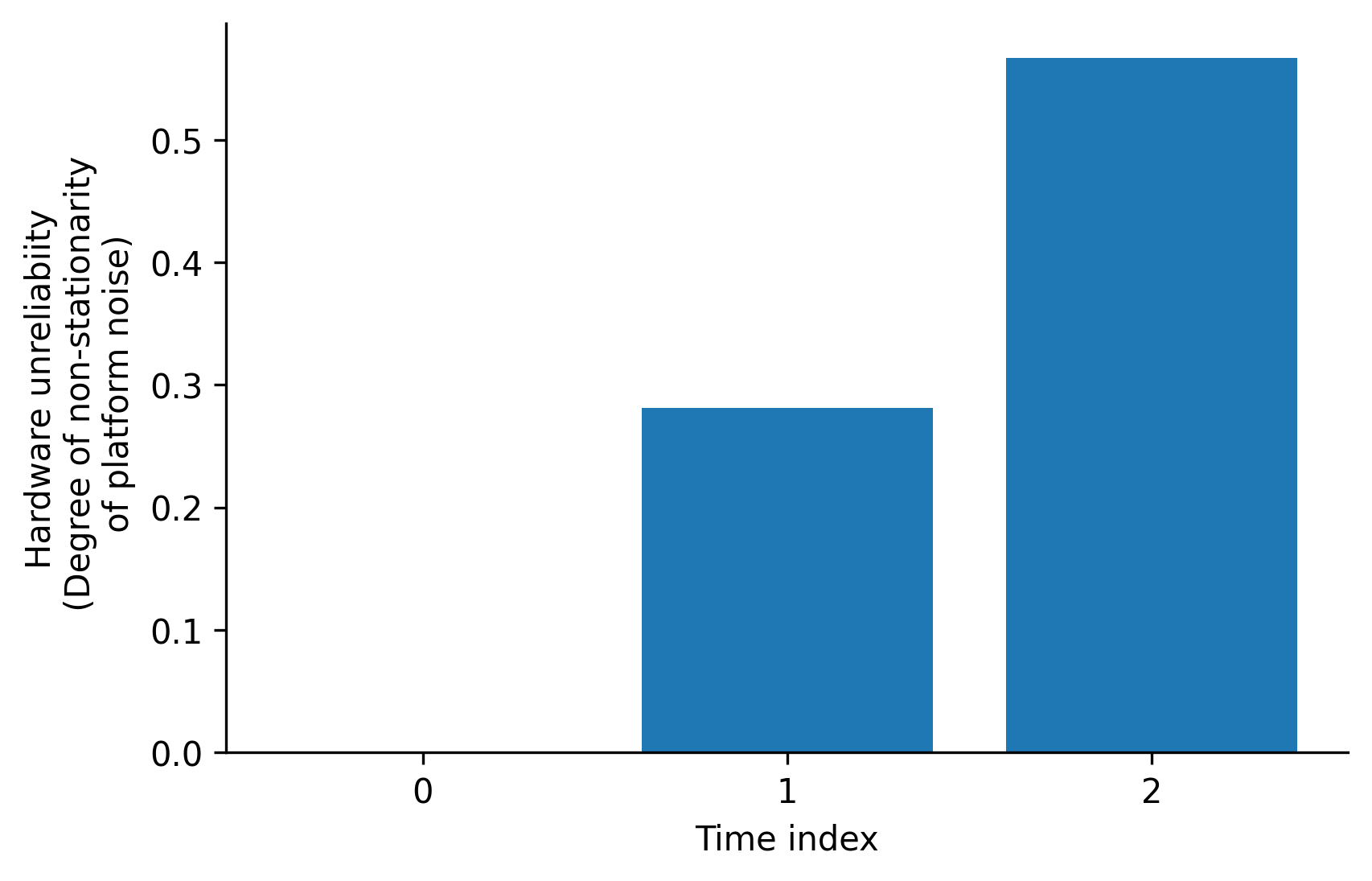}
\caption{We model the degradation of a non-stationary Pauli noise channel by assuming the coherence time steadily decreases over time. Using a non-stationary Dirichlet distribution, we model the joint distribution of coefficients for a two-qubit circuit, which fluctuate as coherence times deteriorate. The y-axis represents the degree of non-stationarity, and the x-axis shows four time-periods. The Hellinger distance between the Dirichlet distributions at time $\tau$=0 and a later time is a measure of non-stationarity, increasing from 0 to 57\%. This model is based on transmon platforms and is used as an experimental setup for our simulation experiments.}
\label{fig:channel_quality}
\end{figure}

As described before, we model the distribution of the Pauli coefficients using a time-varying Dirichlet distribution. The Hellinger distance between the density at time $\tau$=0 and a later time is a measure of the degree of non-stationarity, increasing from 0 to 57\% as shown in Fig.~\ref{fig:channel_quality}. The true means of the time-varying noise Pauli channel coefficients are shown in Table~\ref{tab:coefficients_true}. In period 0, the coefficient of the identity term in the Pauli noise channel is 38\%, but by period 2, it degrades to 26\%. This change is driven by the deterioration in the coherence times for qubit 0 and qubit 1, respectively, as described before. 

\begin{figure}[htbp]
\centering
\includegraphics[width=\figurewidth]{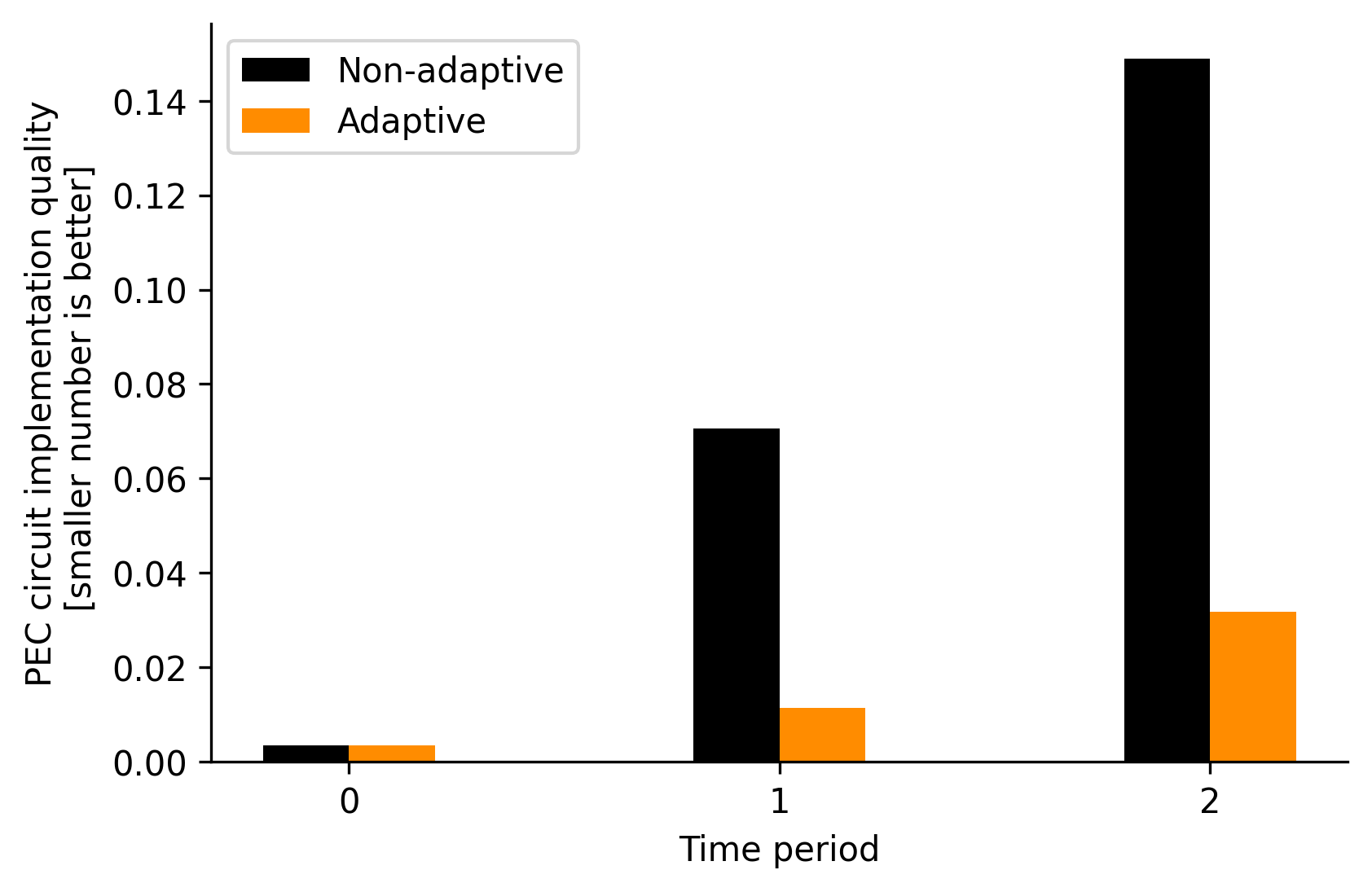}
\caption{This graph compares adaptive and non-adaptive PEC implementations over four time-periods with time-varying noise, using Hellinger distance to measure the difference from the ideal distribution. The black and orange bars represent the observed distributions with non-adaptive and adaptive PEC, respectively. Adaptive PEC significantly outperforms non-adaptive PEC in reducing Hellinger distance to 1.1\%, and 3.2\% compared to 7\%, and 15\% for non-adaptive PEC across time-periods. Adaptive PEC uses adaptive estimation of noise super-operators to improve accuracy compared to non-adaptive PEC. Time-varying noise underscores the need for adaptive PEC.}
\label{fig:output_quality_b}
\end{figure}

At the outset, we assume we are in period 0, equipped with accurate knowledge of the Pauli noise channel. With this knowledge, we obtain the super-operator expression for the noisy basis circuits, which we linearly combine to estimate the ideal operation ($H \otimes H$). The reconstructed operation is a weighted average using a quasi-probability distribution that uses the true noisy basis. Our results, shown in Fig.~\ref{fig:output_quality_b}, indicate an accurate ideal gate implementation in the presence of noise in period 0, with a Hellinger distance between the expected and observed output of 0.34\%. This small error stems from the shot noise due to finite sample size.

In subsequent time periods, the noise characteristics of the Pauli channel change as shown in Fig.~\ref{fig:channel_quality} and Table~\ref{tab:coefficients_true}. This change renders the previously implemented PEC approach invalid as the super-operators characterizing the noisy basis circuits are no longer accurate. Consequently, the PEC coefficients are also invalid. 

\begin{figure}[htbp]
\centering
\includegraphics[width=\figurewidth]{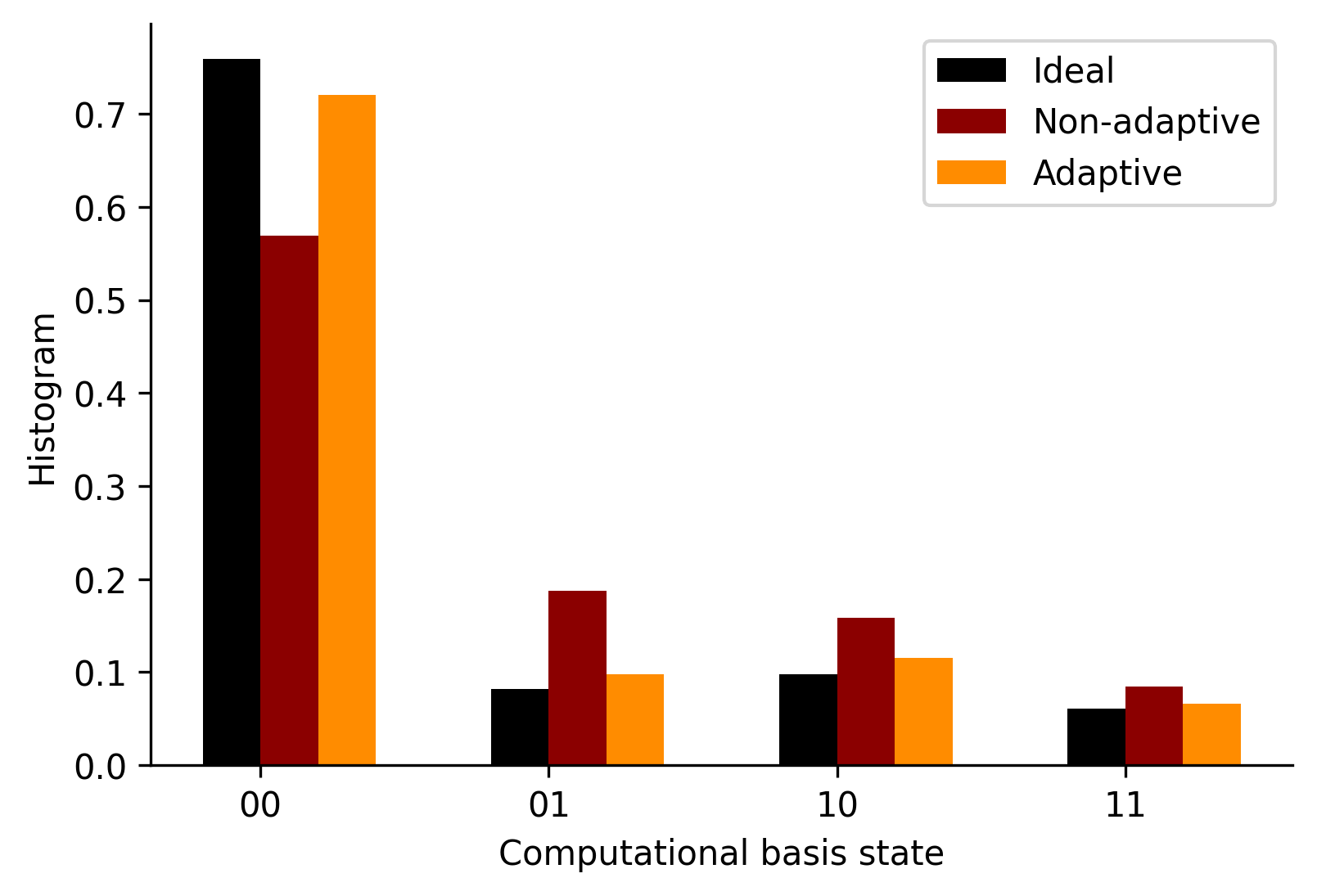}
\caption{This graph compares adaptive and non-adaptive PEC for the $H\otimes H$ gate under time-varying noise. The y-axis shows the probability of observing a basis state. The black bars represent the ideal histogram for the test input, while red/orange bars are non-adaptive PEC results. Adaptive PEC improves accuracy. For example, the $\ket{00}$ probability increases from 57\% to 72\%.}
\label{fig:output_quality_a}
\end{figure}

The black bars in Fig.~\ref{fig:output_quality_b} indicate that the Hellinger distance between the output and ideal increases from 0.34\% to 7\%, and 15\% in periods 1, and 2, respectively. To examine the raw data of the obtained histograms for the non-adaptive case, refer to the crimson colored bars in Fig.~\ref{fig:output_quality_a}.

For the Bayesian update, we use the histogram of projective measurements obtained from applying the PEC circuits to the input density matrix shown in Eq.~(\ref{eq:random_input}). The resulting observation stream of 2-bit strings belongs to one of four possibilities: ${00, 01, 10, 11}$ with probabilities $0.76, 0.08, 0.10,$ and $0.06$, respectively. The input density matrix must have sufficient off-diagonal components to produce an observation rich histogram (as opposed to a bland uniform histogram which will make it impossible to differentiate between the Pauli coefficients).
\begin{equation}
\rho_\textrm{test} =
\begin{pmatrix}
0.2&0.22-0.02j&0.15-0.09j&0.16-0.1j\\
0.22+0.02j&0.24&0.16-0.08j&0.19-0.1j\\
0.15+0.09j&0.16+0.08j&0.36&0.14+0.06j\\
0.16+0.1j&0.19+0.1j&0.14-0.06j&	0.21\\
\end{pmatrix} \; .
\label{eq:random_input}
\end{equation}

\begin{figure}[htbp]
\centering
\includegraphics[width=\figurewidth]{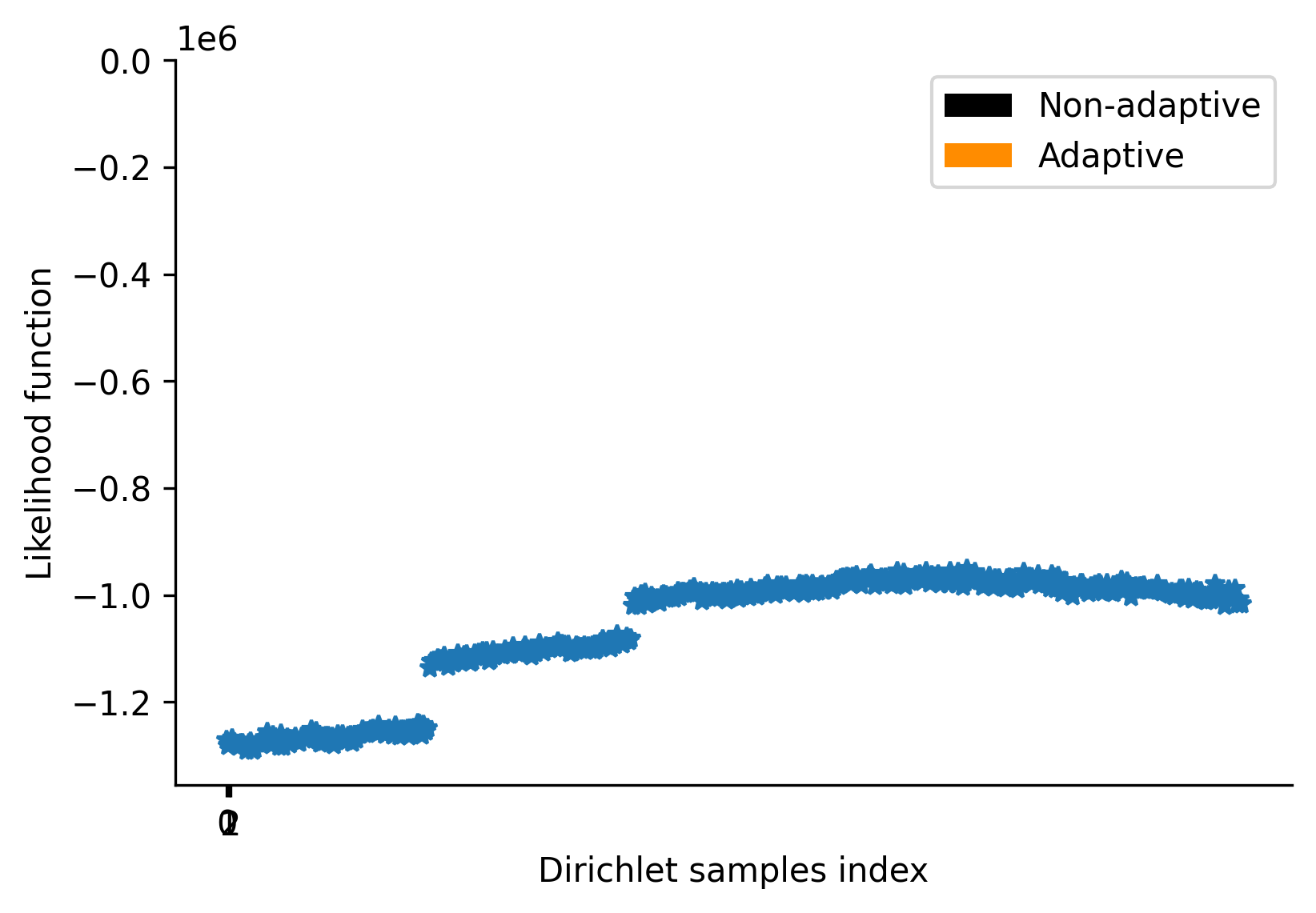}
\caption{The likelihood function used in the Bayesian inference procedure for the second period. }
\label{fig:likelihood}
\end{figure}

\begin{figure}[htbp]
\centering
\includegraphics[width=\figurewidth]{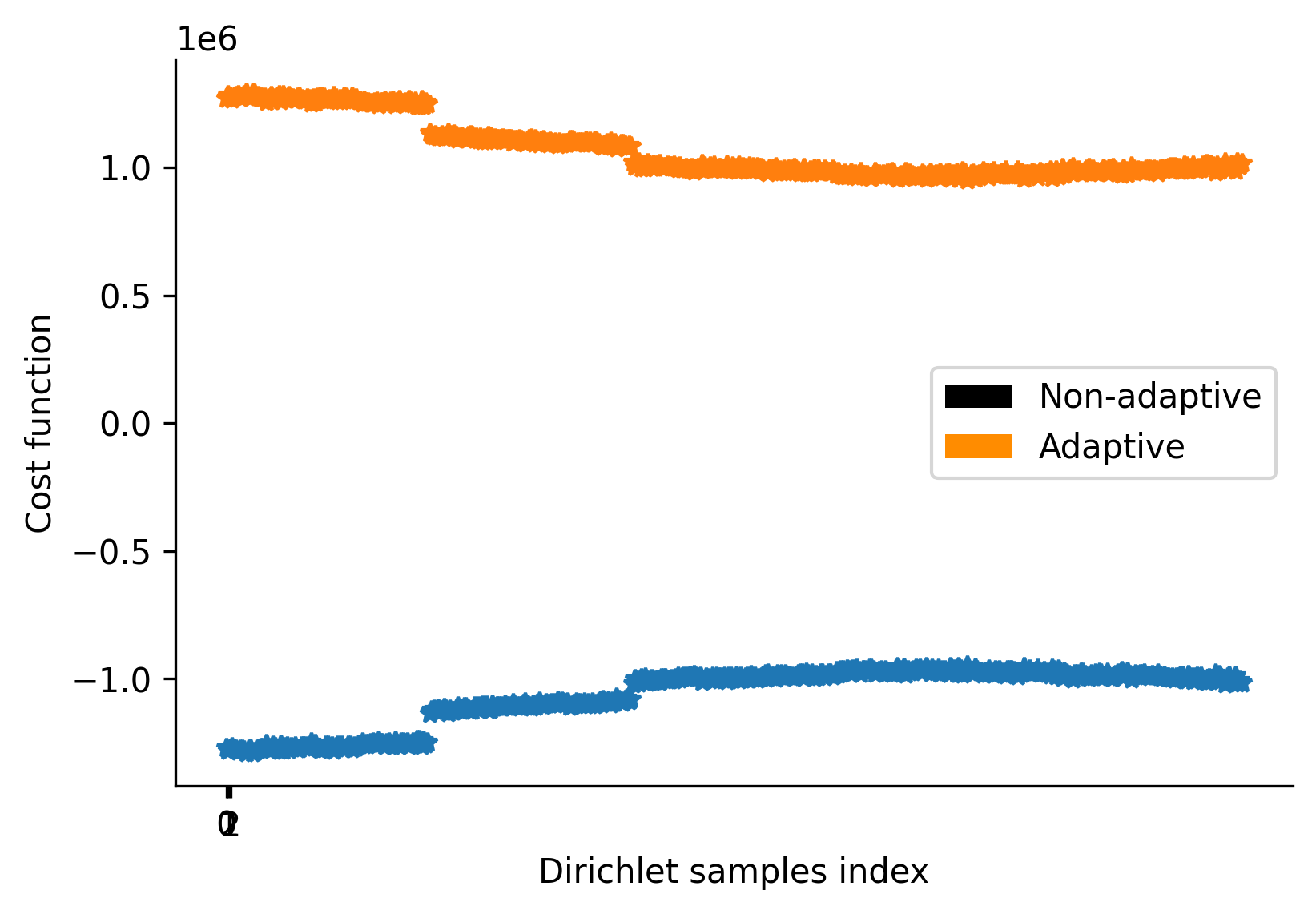}
\caption{The cost function used in the Bayesian inference procedure for the second period.}
\label{fig:cost_function}
\end{figure}

Table~\ref{tab:coefficients2} and Table~\ref{tab:coefficients3} demonstrate the MAP estimation of Pauli coefficients and illustrates the quality of Bayesian estimation. 
With such updated estimates, we re-compute the super-operators for the noisy basis circuits and the linear combination coefficients.  PEC is then implemented using the updated super-operators, resulting in improved output quality, as shown in Fig.~\ref{fig:output_quality_a}. 
In this figure, the y-axis represents the probability of observing a particular computational basis state. In this two-qubit case, there are four possible states, and their probabilities  sum to 1. The black bars indicate the ideal probabilities, while the red and orange bars represent the probabilities obtained with non-adaptive and adaptive PEC, respectively, for the fourth time-period. The graph demonstrates that adaptive PEC, which used adaptive estimation of the noise super-operators, improves accuracy compared to non-adaptive PEC. Specifically, the probability of observing 00 in period 2 increases from 57\% for non-adaptive PEC to 72\% for adaptive PEC. The likelihood and cost functions used in the Bayesian inference procedure are shown in Figures~\ref{fig:likelihood} and \ref{fig:cost_function}.

Fig.~\ref{fig:output_quality_b} compares the performance of adaptive and non-adaptive PEC implementations across four back-to-back time-periods in the presence of time-varying noise. The y-axis represents the Hellinger distance, which is the distance between two discrete probability distributions over the computational basis states. The black bars show the experimentally observed distribution when non-adaptive PEC is used, while the orange bars show the distribution when adaptive PEC is used. The x-axis represents the four time-periods. When using non-adaptive PEC, the Hellinger distance from the ideal distribution is 7\%, and 15\% for time-periods 1, and 2, respectively. When using adaptive PEC, the Hellinger distance significantly improves to 1.1\%, and 3.1\% for the same time-periods.

Note that we used a Pauli noise channel with all $4^n$ terms. However, in practical applications, it becomes necessary to reduce the number of terms. To achieve this, one can explore the use of a sparse Lindbladian noise model \cite{van2023probabilistic}, which considers noise only in nearest-neighbor connections for Pauli terms with weight greater than 1. This reduction in terms leads to a linear scaling instead of exponential with the number of qubits, making the model more computationally efficient. Also note that while it has been observed that single-qubit gate noise can be more than 10 times smaller than two-qubit gate noise \cite{van2023probabilistic}, it cannot be disregarded in Probabilistic Error Cancellation (PEC) due to error propagation effects. Moreover, in the presence of time-varying quantum noise, it becomes even more critical to account for single-qubit noise to ensure accurate error cancellation.

Note that re-calibrating the noise model does not solve the challenge of dynamic estimation, as experimental evidence from various studies indicates significant fluctuations in decoherence times over time. These fluctuations, observed in studies like \cite{carroll2021dynamics, mcrae2021reproducible}, show that decoherence times can vary by approximately 50\% within an hour due to the presence of oxides on superconductors' surfaces, represented as fluctuating two-level systems (TLS) \cite{muller2015interacting, klimov2018fluctuations}. These fluctuations at the scale of minutes and hours are considered non-systematic noise, which cannot be addressed solely through re-calibrations. This emphasizes the necessity for Bayesian algorithms that can provide reliable error bars (as opposed to erroneous point estimates from MLE).

To summarize, this section delved into the behavior of non-stationary noise channels on real-world quantum computing platforms, with a focus on Pauli noise channels in superconducting qubits. The investigation specifically addressed the spatio-temporal non-stationarity of these noise channels, particularly in the context of \belem transmon device. Spatial correlations within multi-qubit noise were explored by treating it as a collection of single-qubit channels while retaining the spatial correlations between individual qubits' $T_{1}$ and $T_{2}$ times. The Dirichlet distribution modeled the joint distribution of Pauli noise channel coefficients, while the Hellinger distance gauged the reliability of error channel characterization. The impact of time-varying Pauli noise on quantum information encoded in a $n$-qubit register was characterized, and coefficients of a separable 2-qubit Pauli noise channel were obtained. These coefficients directly relate to the decoherence times of individual elements, showing strong correlations among them. PEC is effective when noise is well-characterized, but non-stationary noise necessitates an adaptive approach. An adaptive Bayesian inference strategy is proposed to enhance PEC performance in the presence of time-varying noise. This approach dynamically estimated the time-varying Dirichlet distribution of Pauli coefficients using a Bayesian inference-based rolling update. An application of adaptive PEC for executing a Hadamard operation on two-qubits amidst time-varying noise was presented, revealing the need for adaptability due to changing noise characteristics. Without adaptive mitigation, outdated coefficients lead to inaccuracies and newly introduced noise affects circuit execution, severely impacting output accuracy.

\clearpage
\vspace{0.5in}
\vspace{0.5in}
\vspace{0.5in}
\vspace{0.5in}
\vspace{0.5in}
\vspace{0.5in}
\begin{table}[htbp]
\caption{True means of the time-varying noise Pauli channel coefficients}
\begin{center}
\begin{tabular}{c c c c c}
\textrm{Pauli term [$\textrm{qubit} 0 \otimes \textrm{qubit} 1$]} & \textrm{Period 0} & \textrm{Period 1} & \textrm{Period 2}&\\
\toprule
II&0.379&0.326&0.26&\\
IX&0.056&0.064&0.075&\\
IY&0.056&0.064&0.075&\\
IZ&0.076&0.078&0.079&\\
XI&0.081&0.083&0.082&\\
XX&0.012&0.016&0.024&\\
XY&0.012&0.016&0.024&\\
XZ&0.016&0.02&0.025&\\
YI&0.081&0.083&0.082&\\
YX&0.012&0.016&0.024&\\
YY&0.012&0.016&0.024&\\
YZ&0.016&0.02&0.025&\\
ZI&0.127&0.12&0.108&\\
ZX&0.019&0.024&0.031&\\
ZY&0.019&0.024&0.031&\\
ZZ&0.026&0.029&0.033&\\
\bottomrule
\end{tabular}
\end{center}
\label{tab:coefficients_true}
\end{table}
\vspace{0.5in}
\begin{table}[htbp]
\caption{Estimated Pauli coefficients for peiod 1}
\begin{center}
\begin{tabular}{c c c c}
\textrm{Pauli term [$\textrm{qubit} 0 \otimes \textrm{qubit} 1$]} & \textrm{Estimated value} & \textrm{True value}&\\
\toprule
II&0.311&0.326&\\
IX&0.064&0.064&\\
IY&0.064&0.064&\\
IZ&0.097&0.078&\\
XI&0.078&0.083&\\
XX&0.016&0.016&\\
XY&0.016&0.016&\\
XZ&0.024&0.02&\\
YI&0.078&0.083&\\
YX&0.016&0.016&\\
YY&0.016&0.016&\\
YZ&0.024&0.02&\\
ZI&0.114&0.12&\\
ZX&0.023&0.024&\\
ZY&0.023&0.024&\\
ZZ&0.036&0.029&\\
\bottomrule
\end{tabular}
\end{center}
\label{tab:coefficients2}
\end{table}
\vspace{0.5in}
\begin{table}[htbp]
\caption{Estimated coefficients for period 2}
\begin{center}
\begin{tabular}{c c c c}
\textrm{Pauli term [$\textrm{qubit} 0 \otimes \textrm{qubit} 1$]} & \textrm{Estimated value} & \textrm{True value}&\\
\toprule
II&0.243&0.26&\\
IX&0.073&0.075&\\
IY&0.073&0.075&\\
IZ&0.103&0.079&\\
XI&0.074&0.082&\\
XX&0.022&0.024&\\
XY&0.022&0.024&\\
XZ&0.031&0.025&\\
YI&0.074&0.082&\\
YX&0.022&0.024&\\
YY&0.022&0.024&\\
YZ&0.031&0.025&\\
ZI&0.103&0.108&\\
ZX&0.031&0.031&\\
ZY&0.031&0.031&\\
ZZ&0.044&0.033&\\
\bottomrule
\end{tabular}
\end{center}
\label{tab:coefficients3}
\end{table}
\vspace{0.5in}
\vspace{0.5in}
\vspace{0.5in}
\vspace{0.5in}
\vspace{0.5in}
\vspace{0.5in}
\vspace{0.5in}

\removefigs
\chapter{Adaptive probabilistic error cancellation}\label{ch:adaptPEC}
In this final chapter, we investigate the accuracy and stability of probabilistic error cancellation (PEC) outcomes in the presence of non-stationary noise, which is an obstacle to achieving accurate observable estimates. Leveraging Bayesian methods, we design a strategy to enhance PEC stability and accuracy. 

The practical realization of quantum computing has witnessed rapid advancements~\cite{roadmap}, with quantum devices now operating as systems with hundreds of interacting qubits. However, these real quantum devices~\cite{roth2021introduction} are noisy~\cite{preskill2019quantum}, and practical efforts to realize a quantum computer introduce various noise processes like decay, de-coherence~\cite{kandala2019error}, environmental coupling, intra-register cross-talk~\cite{parrado2021crosstalk, fang2022crosstalk}, and leakage from computational space~\cite{vittal2023eraser}. Physical operations like quantum gates and measurements rely on electromagnetic fields susceptible to pulse distortion, attenuation, jitter, and drift, which further increase noise~\cite{kliesch2021theory,PRXQuantum.3.020344}. Imperfections in thermodynamic controls (such as cryogenic cooling, magnetic shielding, vibration suppression, and imperfect vacuum chambers) can disturb the operating conditions of the quantum computer~\cite{blume2020modeling, blume2010optimal}. 

These lead to computational errors that make it essential to address noise and implement error mitigation strategies~\cite{bharti2022noisy} to improve the accuracy of quantum outcome~\cite{ferracin2021experimental}.

Contemporary quantum computers are not only noisy but they also exhibit non-stationarity. Non-stationary noise processes~\cite{thorbeck2023two, mcewen2022resolving, etxezarreta2021time, muller2015interacting, klimov2018fluctuations} in \superc~qubits~\cite{martinis2015qubit} are well-studied. For example, \SPAM (SPAM) fidelities have been observed to fluctuate significantly showing more than 25\% deviation from their long-term average~\cite{dasgupta2023reliability}. 
Similarly, the fidelity of CNOT gates have been noted to change by over 40\% within similar time frames~\cite{dasgupta2021stability}. 
Moreover, the qubit relaxation times, known as $T_1$, have experienced fluctuations of up to 400\% in just 30 minutes~\cite{dasgupta2020characterizing}. 
Likewise, de-phasing times, denoted as $T_2$, have been recorded to vary by over 50\% within an hour~\cite{carroll2021dynamics, mcrae2021reproducible}. 

\added{The non-stationarity observed in contemporary \superc~quantum computers stems from two primary sources linked to material defects: impurities within the material and ionization induced by cosmic rays.} 
It is theorized that fluctuating two-level systems, possibly stemming from certain oxides on the superconductor's surface, contribute to non-stationarity~\cite{muller2015interacting, klimov2018fluctuations}.
Additionally, cosmic rays~\cite{xu2022distributed, mcewen2022resolving} contribute by ionizing the substrate upon impact, leading to the emission of high-energy phonons, which in turn triggers a burst of quasi-particles. These quasi-particles disrupt qubit coherence across the device. It has been shown that quantum computers can experience catastrophic errors in multi-qubit registers approximately every 10 seconds due to cosmic rays originating from outer space~\cite{mcewen2022resolving}. Studies that address non-stationary noise in \superc~quantum computers include 
investigations on output reproducibility~\cite{proctor2020detecting},  
noise modeling~\cite{etxezarreta2021time}, 
tracking the \ns~profile of quantum noise~\cite{danageozian2022noisy}, and 
quantum error mitigation using continuous control~\cite{majumder2020real}.

Quantum error mitigation is a set of techniques that employs statistical tools from estimation theory to reduce the impact of noise in quantum computations without directly correcting the quantum state~\cite{bharti2022noisy}. Such techniques can become vulnerable to errors stemming from over or under estimation of noise due to the presence of non-stationarity~\cite{henao2023adaptive}. 

\begin{figure}[htbp]
\centering
\includegraphics[width=\linewidth]{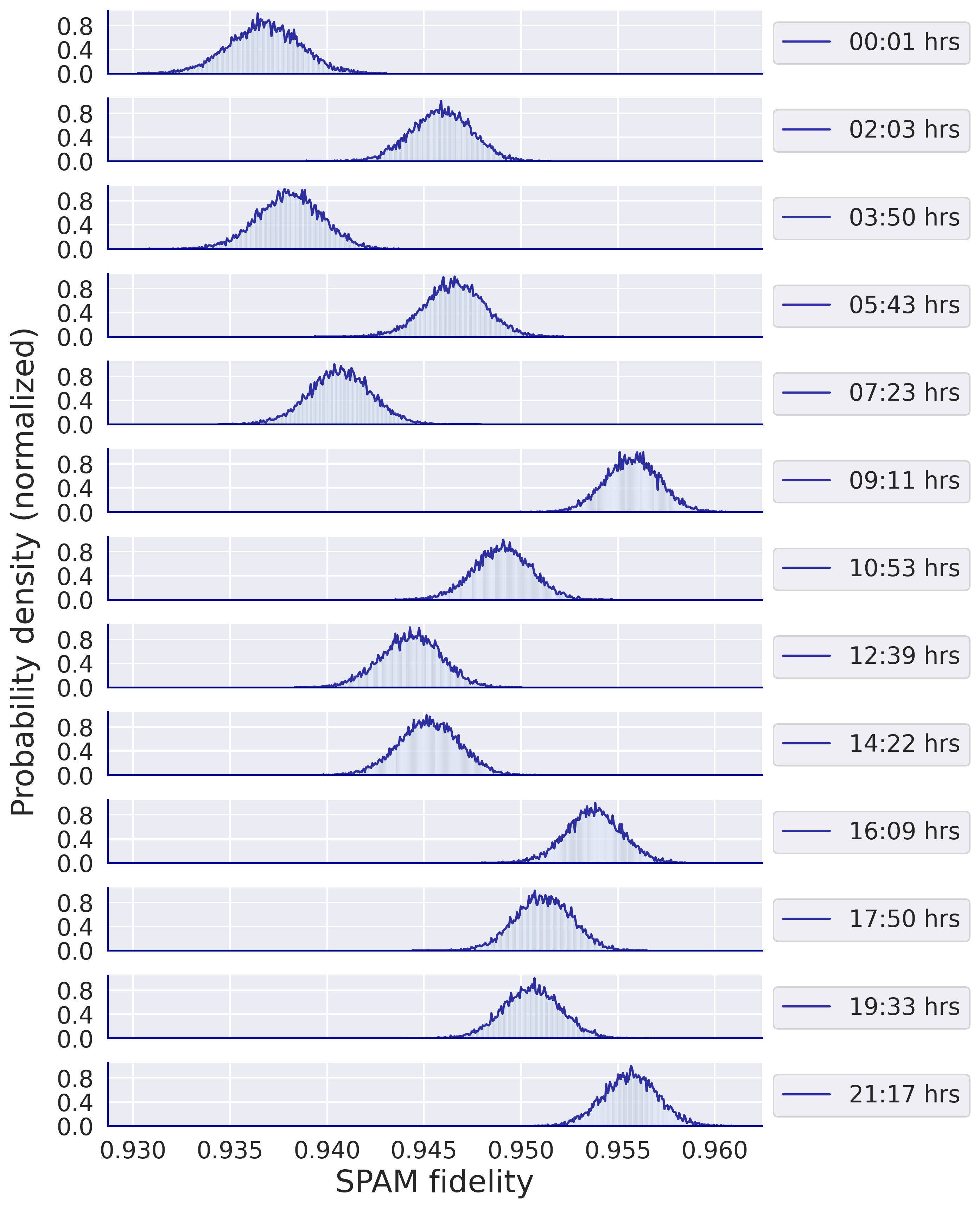}
\caption{Non-stationary distribution functions of the \SPAM (SPAM) fidelity for qubit 2 on \kolkata~\superc~device collected on Jan 15, 2024.}
\label{fig:tv_noise_f2}
\end{figure}

A quantum noise channel~\cite{nielsen2002quantum} can be described as a stochastic process, allowing for the continuous update of its estimated characteristics in response to varying noise conditions. Such an approach treats the channel as a time-varying random variable. Consider Fig.~\ref{fig:tv_noise_f2} which shows the \SPAM (SPAM) fidelity for the second register element on \kolkata~device on Jan 15, 2024. The probability density is clearly changing with time, even though the variance stays consistent. In light of such non-stationary data, a single-qubit SPAM noise channel can be described by the model: $\mathcal{E}(\rho) = f(t) \rho + [1-f(t)] \mathds{X} \rho \mathds{X}$, where $\rho$ represents the single-qubit density matrix, $\mathds{X} = 
\begin{pmatrix}
0 & 1 \\
1 & 0
\end{pmatrix}$ denotes the Pauli-X matrix, and $f(t)$ denotes the SPAM fidelity drawn from a time-dependent distribution. From Fig.~\ref{fig:tv_noise_f2}, non-stationarity of noise in this \superc~device is apparent within 24 hours. The noise parameters estimated during device re-calibration quickly becomes outdated, compromising the accuracy of noise channel information essential for mitigation. Thus re-calibration alone is insufficient and there is a need for adaptive error mitigation techniques that can function in between calibration intervals in the face of changing noise conditions.

This study focuses on probabilistic error cancellation (PEC)~\cite{temme2017error} in the presence of non-stationary noise. PEC is a quantum error mitigation approach that aims to construct unbiased estimates of the means of quantum observables from noisy observations. 
Effective implementation requires an accurate noise characterization~\cite{torlai2020quantum}. 
For example, 
learning correlated noise channels in large quantum circuits on a \superc~quantum processor has proven to be difficult~\cite{van2023probabilistic}. 
Yet, leveraging sparse noise models, PEC has been successful in estimating the mean of observables in circuits comprising 2,880 CNOT gates, executed on a 127-qubit noisy \superc~processor - a task that conventional brute-force computing could not match~\cite{kim2023evidence}.

We demonstrate that adaptive probabilistic error cancellation, which views quantum noise channels as evolving random variables, outperforms its non-adaptive counterpart in devices subject to non-stationary noise. To achieve this aim, we will make use of a 5-qubit implementation of the \BV algorithm~\cite{bernstein1993quantum}. 

The manuscript is organized as follows. 
In Sec.~\ref{sec:background}, we provide background for \pec (PEC). 
Sec.~\ref{sec:adaptPEC} develops a Bayesian~\cite{lukens2020practical, zheng2020bayesian} approach for adapting the method of \pec to \nsn. It also sets up a performance evaluation framework for the accuracy and stability of PEC results. 
Sec.~\ref{sec:PEC_under_Pauli} presents a numerical validation of adaptive probabilistic error cancellation using a 5-qubit implementation of the \BV algorithm, where we treat the noise parameters (qubit-specific \SPAM (SPAM) fidelities and depolarizing parameters characterizing noise in CNOT gate) as \ns~random variables.
In Sec.~\ref{sec:real_data}, we present the results of experiments conducted on a real, noisy quantum device to test the adaptive PEC algorithm. Concluding remarks are provided in Sec.~\ref{sec:conclusion}.
\section{Background}\label{sec:background}
In this section, we provide background for the quantum error mitigation method called probabilistic error cancellation (PEC)~\cite{temme2017error, van2023probabilistic, kim2023evidence} \added{which aims to mitigate errors by approximating the noiseless mean of the observable as a weighted sum of noisy observables. }
We use calligraphic symbols to denote {\sop}s acting on density matrices ($\rho$): 
\begin{equation}
\mathcal{G} \rho = G\rho G^\dagger.
\end{equation}
where $\mathcal{G}$ is the {\sop} and $G$ is a unitary quantum operator. 
For example, if $G$ denotes the CNOT operator, then $\mathcal{G}$ is the {\sop} for the CNOT operation.

Typically we do not have access to a noiseless implementation of $\mathcal{G}$. Let $\tilde{\mathcal{G}}$ denote the \sop~corresponding to the available, noisy implementation of $\mathcal{G}$. We have access to other synthesized implementations of $\tilde{\mathcal{G}}$ by subjecting $\tilde{\mathcal{G}}$ to basis operations available on a noisy device. 
\added{Originally, the noisy basis set was specified as the set of native gates that a quantum computer could implement~\cite{temme2017error}. However, with advancements in \superc~quantum computers, the noise associated with single-qubit Pauli operators has become negligible~\cite{van2023probabilistic, bravyi2021mitigating}. Consequently, Pauli operators, which might be a composition of multiple native gates, can be employed as the basis set~\cite{mitiq-qpr}.} 
We denote by $\{\tilde{\mathcal{G}_k}\}$ the set of all noisy {\sop}s, with $k=0$ denoting the noisy implementation of $\mathcal{G}$ without additional operations from the basis set.

As an example, consider a 2-qubit quantum gate $G$. The set of noisy {\sop}s composed under a Pauli channel assumption are given by: $\{ \mathcal{P} \circ \tilde{\mathcal{G}} \}$ where 
$\mathcal{P} \circ \tilde{\mathcal{G}} (\rho) = \mathcal{P} ( \tilde{\mathcal{G}} (\rho) )$ and 
$\mathcal{P}(\cdot) \equiv [\mathds{P}_0 \otimes \mathds{P}_1](\cdot)[\mathds{P}_0 \otimes \mathds{P}_1]$. 
Here $0$ and $1$ refer to qubit $0$ and qubit $1$ respectively and $\mathds{P}_0, \mathds{P}_1$ are picked from the set of Pauli operators $\{ \mathds{I}, \mathds{X},\mathds{Y},\mathds{Z}\}$. 
Thus, by varying $P_0$ and $P_1$, we obtain the set $\{ \mathcal{P} \circ \tilde{\mathcal{G}} \}$ which forms a basis $\{\tilde{\mathcal{G}_k}\}_{k=0}^{15}$ that spans the \sop~space. In this 2-qubit example, the map for $k$ is derived from the cartesian product of $\{ \mathds{I}, \mathds{X},\mathds{Y},\mathds{Z}\} \times \{ \mathds{I}, \mathds{X},\mathds{Y},\mathds{Z}\}$, with the sequence of this ordered set determining the value of $k$.

In general, the {\sop} $\mathcal{G}$ can be expressed as a linear combination of the basis {\sop}s:
\begin{equation}
\mathcal{G} = \sum\limits_{k=0}^{N_p-1} \eta_\low{k} \tilde{\mathcal{G}}_k,
\label{eq:pec}
\end{equation}
where $N_p$ is the dimension of the \sop~space. The PEC coefficients ${ \eta_\low{k} }$ in the linear combination are determined either analytically, under a noise model assumption for single and two-qubit gates, which can be extended to larger circuits, or numerically, by minimizing the one-norm between high-dimensional matrices~\cite{mitiq-qpr, temme2017error}. We employ the analytical approach.



The circuits corresponding to the {\sop}s $\tilde{\mathcal{G}}_k$ in Eqn.~\ref{eq:pec} are constructed by subjecting each of the gates in the quantum circuit for $\tilde{\mathcal{G}}$ to operations from the noisy basis set. 
If we execute these noisy circuits and collect the mean of the observable, then, from Eqn.~\ref{eq:pec}, we recover the ideal noiseless mean of an observable $\braket{\mathcal{O}}$ as:
\begin{equation}
\braket{\mathcal{O}} = \Tr\left[ \mathcal{O} \mathcal{G} \rho \right] = \Tr\left[ \mathcal{O} \sum\limits_k \eta_\low{k} \tilde{\mathcal{G}}_k \rho \right]
\label{eq:pec_basic}
\end{equation}
where $\rho$ is the input density matrix to the circuit. Thus, an observable $\braket{\mathcal{O}}$ may be estimated as a weighted sum of the mean of the observables from the noisy circuits. 

Eqn.~\ref{eq:pec_basic} can be re-written as:
\begin{equation}
\braket{\mathcal{O}} = \gamma \sum\limits_k \text{sgn}(\eta_\low{k}) \mathds{Q}_\low{k}  \Tr \left[
\mathcal{O} \tilde{\mathcal{G}_\low{k}}\rho
\right]
\label{eq:pec_estimator}
\end{equation}
with $\gamma = \sum |\eta_\low{k}|$ and $\mathds{Q}_\low{k} = |\eta_\low{k}|/\gamma$. 
The sign function, \( \text{sgn}(\cdot) \), returns +1 for positive inputs, -1 for negative inputs, and 0 for an input of 0.

Note that the set $\{\mathds{Q}_k\}$ forms a valid probability distribution because all its elements are positive and sum to 1. 
However, $\{\mathds{Q}_k\}$ is termed a quasi-probability distribution (QPD)~\cite{temme2017error, mitiq-qpr}. To see this, consider a random integer $K \in \{0, \cdots N_p-1\}$ which follows the probability distribution function denoted by $\mathds{Q}_k$. This random integer $K$ can be mapped one-to-one to the random variable 
$\text{sgn}(\eta_\low{K}) \Tr \left[  \mathcal{O}  \tilde{\mathcal{G}_\low{K}}\rho \right]$, 
which inherits the probability distribution function $\mathds{Q}_k$. By repeatedly sampling the random variable $K$, we realize a set of random integers represented as $\{\mathfrak{m}\}$. 
For each specific $\mathfrak{m}$ that is realized, \added{we execute the corresponding noisy quantum circuit $\tilde{\mathcal{G}_\mathfrak{m}}$ multiple times} and obtain a set of noisy means of observables denoted by the set $\left\{ \Tr \left[ \mathcal{O} \tilde{\mathcal{G}_\mathfrak{m}}\rho \right] \right\}$. 

When computing the average over the set of noisy means of observables $\left\{ \Tr \left[ \mathcal{O} \tilde{\mathcal{G}_\mathfrak{m}}\rho \right] \right\}$, we need to 
adjust the sign of each element of the set by the sign of $\eta_\low{m}$ and scale it by $\gamma$.
This average then converges to the mean of the observable from a noiseless gate $\mathcal{G}$, in the asymptotic limit of a large number of repeated samplings of the random variable $K$. The need for adjustment by the sign function leads us to denote $\mathds{Q}_k$ as quasi-probabilities.

\added{Achieving an accuracy of \(O(\epsilon)\) using the empirical mean of the random variable 
\(\gamma \text{sgn}(\eta_\low{K}) \Tr \left[ \mathcal{O} \tilde{\mathcal{G}_\low{K}}\rho \right]\) 
(which is an unbiased estimator of \(\braket{\mathcal{O}}\))
requires \(O(\gamma/ \epsilon)^2\) PEC circuit samples and the result has variance of order \(O(\gamma^2)\) \cite{temme2017error}.}
\section{Adaptive PEC}\label{sec:adaptPEC}
Estimating channel noise parameters is crucial for determining the PEC coefficients $\{\eta_\low{k}\}$. 
However, the non-stationary nature of noise, along with drift and latency in characterization, complicates this task. This in turn makes it difficult to accurately assess the PEC coefficients. In this section, we demonstrate how adaptive parameter estimation can be applied to PEC.

The parameters characterizing the noise during idle time (such as qubit decoherence time~\cite{burnett2019decoherence}) and quantum operations (such as CNOT fidelity~\cite{dasgupta2023reliability}) exhibit random non-stationary behavior in some hardware. Estimating non-stationary stochastic processes is challenging and their predictive value is limited because the patterns identified from historical data may not reliably indicate future behavior, making it difficult to discern underlying trends. 
A model that is effective at one time point can become inaccurate at another. 

However, we can utilize intermittent incremental measurements from quantum circuits to devise a Bayesian~\cite{lukens2020practical, zheng2020bayesian, gordon1993novel, kotecha2003gaussian} update for the current state of the device noise:
\begin{equation}
\Pr(\textbf{x} | \text{data} ) \propto \Pr(\text{data}|\textbf{x}) \Pr(\textbf{x})
\label{eq:basic_bayes}
\end{equation}
where 
data refers to measurements obtained from circuit execution, 
$\textbf{x}$ denotes the multi-dimensional vector of parameters characterizing the device noise (such as connection-specific CNOT fidelity and qubit-specific SPAM fidelities), 
$\Pr(\textbf{x} | \text{data} )$ is the posterior noise distribution, 
$\Pr(\text{data}|\textbf{x} )$ is the likelihood, 
and $\Pr(\textbf{x})$ is the prior noise distribution. 
We use this to update $\{\eta_\low{k}\}$ via updates to model-specific parameters. 

Obtaining the posterior distribution for $\xrm$ is a two-step process: first, we estimate the posterior for uncorrelated parameters, and second, for correlated parameters in the underlying noise model.\\
\subsection{Uncorrelated parameters}
Uncorrelated parameters refer to those model parameters for which there exist datasets (such as the partial trace of the observable on a single qubit) where the observed data is modeled by error due to only one noise parameter. 
Here, the analysis is simpler as univariate priors can be used. 
For example, if we disregard noise from single-qubit rotations, then the measurements obtained from any qubit that was not subjected to entangling operations, can be described by a one-parameter model under symmetric SPAM noise model. 

Let $\mathcal{D}_q = \{b_q(0), \cdots, b_q(L-1)\} $ represent a dataset of $L$ samples obtained for qubit $q$ after measurement in the computational basis. 
Each $b_q(l)$ is a single-bit measured after the $l$-th execution of $\tilde{\mathcal{G}}_0$. We can adopt a beta distribution as the prior for the SPAM fidelity $f_q$ because the beta distribution is well-suited for values restricted to the [0,1] interval and can effectively accommodate the experimentally observed unimodal and skewed density as seen in Fig.~\ref{fig:tv_noise_f2} for qubit 2. The beta distribution's flexibility allows for an accurate fit to these characteristics. The Beta distribution with parameters $\alpha_q, \beta_q$ is given by $\text{Beta}(f_q; \alpha_q, \beta_q) = f_q^{\alpha_q-1}(1-f_q)^{\beta_q-1} / B(\alpha_q,\beta_q)$ where 
the normalizing denominator $B(\alpha_q,\beta_q)$ is the Beta function defined as $B(m,n) = \Gamma(m)\Gamma(n)/\Gamma(m+n)$, and $\Gamma(\cdot)$ is the gamma function given by: $\Gamma(y) = \int\limits_0^\infty t^{y-1} e^{-t}dt$, defined for any positive y.

The prior for the mean $\mu_q^\text{prior}$ and variance $v_q^\text{prior}$ of the SPAM fidelity $f_q$ can be derived from historical characterization data (e.g. using data post calibration). If such data is not accessible, we can obtain the starting point from a small perturbation to the ideal value. 
The prior parameters are then obtained as: $\alpha_q^\text{prior} = \mu_q^\text{prior} \left[ \mu_q^\text{prior}(1-\mu_q^\text{prior}) / v_q^\text{prior} - 1\right]$ and $\beta_q^\text{prior} = (1-\mu_q^\text{prior}) \left[ \mu_q^\text{prior}(1-\mu_q^\text{prior}) / v_q^\text{prior} -1\right]$.

The likelihood is obtained as:
\begin{equation}
\mathcal{L}(f_q)
= f_q^{C_0[\mathcal{D}_q]} (1-f_q)^{L-C_0[\mathcal{D}_q]} / B(\alpha_q, \beta_q)^L
\end{equation}
where $C_0[\mathcal{D}_q]$ counts the number of 0's in the dataset $\mathcal{D}_q$. 
The updated posterior density (indicated by the prime on the updated parameters) for the SPAM fidelity ($f_q$)~\cite{gelman1995bayesian} :
\begin{equation}
f_q | \mathcal{D}_q \sim \text{Beta}(\alpha_q^\prime, \beta_q^\prime)
\end{equation}
where $\alpha_q^\prime = \alpha_q + L - C_0[\mathcal{D}_q]$ and $\beta_q^\prime = \beta_q + C_0[\mathcal{D}_q]$. This shows the influence of incremental measurements on posterior noise density. 
The qubit-wise updated mean~\cite{sivia2006data} of the SPAM fidelity, obtained as $\mu_q^\prime= \alpha_q^\prime / (\alpha_q^\prime+\beta_q^\prime)$, are then used in estimating the PEC coefficients.\\
%
\subsection{Correlated parameters}
The second task involves the adaptive estimation of correlated noise parameters, which is more complex due to the measurements being influenced by multiple noise processes simultaneously. These are the noise parameters for which there exist datasets (typically the partial trace of an observable across a subset of qubits) which reflect errors from various noise parameters jointly. The analysis uses the probability of observing classical bit-strings on this qubit subset as the random variables.

For a subset of $m$ qubits from a total of $n$ qubits, there are $M = 2^m$ possible observed bit-strings, each with a probability denoted by $p_0, p_1, \ldots, p_{M-1}$, where $p_0$ represents the probability of observing all zeros and $p_{M-1}$ that of all ones. These probabilities are treated as random variables which form a probability simplex, as they are all positive and add up to 1. Hence the natural way to model the joint density is using a Dirichlet prior given by:
\begin{equation}
\text{Pr}(p_0, \cdots, p_{M-1}) = \frac{\Gamma(\sum a_i)}{\prod\limits_0^{M-1} \Gamma(a_i)} \prod\limits_0^{M-1} p_i^{a_i-1}
\end{equation}
where the gamma function $\Gamma(\cdot)$ was already defined previously and $\{a_i: a_i > 0, i \in \{0, \cdots M-1\}\}$ are the parameters characterizing the Dirichlet distribution that need to be estimated and updated -- a multi-variate task analogous to the uni-variate task for the beta distribution in the previous section.

We denote the dataset derived from $L$ measurements of the $m$ qubits in the computational basis as $\mathcal{D} = \{w(0), \cdots, w(L-1)\}$, where each $w(i)$ is a binary string of length $m$. 
The likelihood function is given by:
\begin{equation}
\mathcal{L}(p_0, \cdots, p_{M-1}) = 
\prod\limits_{i=0}^{M-1} p_i^{C_{v_i}[\mathcal{D}]}
\end{equation}
where $v_i \in \{0, 1\}^m$, 
$C_{v_i}[\mathcal{D}]$ is the count of occurrences of the string $v_i$ in the dataset $\mathcal{D}$. 
Note that $w(i)$ is a specific realization post-measurement that takes one of the values in the set $\{v_i\}$.

The posterior joint density is also a Dirichlet distribution~\cite{robert2010introducing}:
\begin{equation}
\Pr( p_0, \cdots, p_{M-1} \mid \mathcal{D}) = \frac{\Gamma(\sum a_i^\prime)}{\prod\limits_{i=0}^{M-1} \Gamma(a_i^\prime)} \prod\limits_{i=0}^{M-1} p_i^{a_i^\prime-1}
\end{equation}
with parameters $a_i^\prime = a_i + C_{v_i}(\mathcal{D})$. Upon marginalizing this joint density, the marginals follow a Beta distribution. For example, the random variable $p_i$ is Beta-distributed with parameters $\alpha_i = a_i^\prime$ and $\beta_i = -a_i^\prime+\sum\limits_{j=0}^{M-1} a_j^\prime$~\cite{lee1989bayesian}.

The updated marginals from the Dirichlet distribution give us the evolving densities for each $p_i$, allowing us to calculate their time-varying means and variances. Specifically, the mean for $p_i$ updates to $a_i^\prime /\sum a_i^\prime$, and its variance updates to $a_i^\prime\left(\sum_j a_j^\prime - a_i^\prime\right)/\left(\sum_j a_j^\prime\right)^2/\left(1+\sum_j a_j^\prime\right)$.

At the final stage, the method employs the relationship between the probabilities $p_0, \cdots, p_{M-1}$ and the noise model parameters to derive the time-dependent parameter means and variances of the noise parameters. 
The process concludes by updating the quasi-probability distribution using the updated noise parameters per Eqn.~\ref{eq:basic_bayes}.
\subsection{Accuracy and stability}\label{sec:accuracy_and_stability}
We next evaluate the performance of PEC in presence of \nsn using the lens of accuracy and stability. A mean of a quantum observable $\mathcal{O}$ is represented as $\braket{\mathcal{O}}_\xrm$ (where $\xrm$ labels the noise instance), while the observed mean after mitigation is denoted as $\braket{\mathcal{O}}_\xrm^\text{mit}$, which equals $\braket{\mathcal{O}}$ only in the asymptotic limit of infinite samples and zero noise. When we want to say the mitigated noisy mean at a specific time $t$, we will use $\braket{\mathcal{O}}_\xrm^\text{mit}(t)$.

We say that a PEC mitigated observable is $\epsilon-$accurate if the absolute difference between the mitigated and noiseless observable is upper bounded by $\epsilon$:
\begin{align}
&|\braket{O}_\xrm^\text{mit} - \braket{O}| \leq \epsilon
\end{align}

Now, suppose the underlying noise, being non-stationary, is characterized by a time-dependent density $f(\xrm;t)$. 
We say that a PEC mitigated observable is $\epsilon-$stable between times $t_1$ and $t_2$ if:
\begin{equation}
| \braket{O}_\xrm^\text{mit}(t_1) - \braket{O}_\xrm^\text{mit}(t_2) | \leq\epsilon
\label{eq:stability}
\end{equation}
where 
\begin{equation}
\begin{split}
&\braket{O}_\xrm^\text{mit}(t) = \int\limits_\xrm \braket{O}_\xrm^\text{mit} f(\xrm; t) d\xrm\\
&= \int\limits_\xrm 
\sum\limits_k 
\eta_\low{k}(\xrm, t)
\text{Tr} \left[ \mathcal{O} \tilde{\mathcal{G}}_k \rho \right] 
f(\xrm; t) d\xrm
\end{split}
\end{equation}
Note the dependence of $\eta_\low{k}$ on the random variable $\xrm$ and time $t$.

When measuring $\mathcal{O}_q$ for qubit $q$, the qubit-wise accuracy metric is:
\begin{equation}
\epsilon_q = \left| \braket{\mathcal{O}_q}_\xrm - \braket{ \mathcal{O}_q } \right|
\end{equation}
without mitigation, and
\begin{equation}
\epsilon_q = \left| \braket{\mathcal{O}_q}_\xrm^\text{mit} - \braket{ \mathcal{O}_q } \right|
\end{equation}
with error mitigation. The register average, for a $n$-qubit register, is:
\begin{equation}
\epsilon_R = \sum\limits_{q=0}^{n-1} \epsilon_q / n
\label{eq:eR}
\end{equation}

Similarly, the qubit-wise stability metric is:
\begin{equation}
s_q(t) = \left| \braket{\mathcal{O}_q}_\xrm(t) - \braket{\mathcal{O}_q}_\xrm(0) \right|
\end{equation}
without mitigation, and
\begin{equation}
s_q(t) = \left| \braket{\mathcal{O}_q}_\xrm^\text{mit}(t) - \braket{\mathcal{O}_q}_\xrm^\text{mit}(0) \right|
\end{equation}
with error mitigation. The register average, for a $n$-qubit register, is:
\begin{equation}
s_R(t) = \sum\limits_{q=0}^{n-1} s_q(t)/n.
\label{eq:sR}
\end{equation}
We expect the accuracy ($\epsilon_q, \epsilon_R$) and stability $(s_q, s_R)$ metrics to be smaller when using adaptive PEC because of more accurate estimates for the time-varying PEC coefficients $\{ \eta_\low{k}\}$.

\section{Numerical Validation}\label{sec:PEC_under_Pauli}
For studying the stability and accuracy of PEC in presence of \nsn, we use an implementation of the Bernstein-Vazirani algorithm~\cite{bernstein1993quantum}, a standard benchmarking circuit that requires only a modest number of gates.
The purpose of the algorithm is to recover a $n$-bit secret string $r$, encoded in a black-box oracle function. 
The algorithm identifies the secret with a single query, while the classical method needs $2^n$ queries (worst-case). 

A quantum circuit for the \BV algorithm is shown in Fig.~\ref{fig:bv_circuit}. 
We conceptualize the circuit as having four principal layers ($\mathds{L}_1, \mathds{L}_2, \mathds{L}_3$, and $\mathds{L}_4$) separated by dashed vertical lines. 
Note that the total number of qubits needed by an $n-1$-bit secret string is $n$, comprising $n-1$ data qubits and one ancilla qubit. So, a 5-qubit implementation of the \BV algorithm has $n=5$ but the secret string length is $n-1=4$.

The circuit is initialized with all the qubits in the register in the $\ket{0}$ state. Thus, a pure state description yields an initial state $\ket{\psi_0}=\ket{0}^{\otimes n}$. 
In the first layer, all the qubits are subjected to Hadamard gates. The ancilla qubit (which is the last qubit with index n) is additionally subjected to a $\mathds{Z}$ gate. The output after the first layer $\mathds{L}_1 = \mathds{H}^{\otimes {n-1}} \otimes (\mathds{Z}_n\mathds{H}_n)$ is $\ket{\psi_1} = \ket{+}^{\otimes {n-1}}\ket{-}$. 

The second layer implements the oracle function for which we can use CNOT gates. 
Each CNOT's control qubit corresponds to one of the bits in the secret string $r$, while the target qubit remains fixed at qubit $n$. 
Specifically, if the bit $r_i$ of the secret string $r$ is 1, we add a CNOT between qubit $i$ (control) and qubit $n$ (target). 
The input to the second layer is $\mathds{H}^{\otimes {n-1}}\ket{0} \ket{-}$ while the output is $\ket{\psi_2} = \left( \mathds{H}^{\otimes {n-1}}\ket{r} \right)\ket{-}$ where $r$ is the secret string.

To retrieve the secret string $r$, the third layer requires another layer of Hadamard gates. 
A $\mathds{Z}$ gate is applied to the ancilla qubit (qubit $n$) before the application of the Hadamard layer to make the computing reversible. The output quantum state after the third layer $\mathds{L}_3 = \mathds{H}^{\otimes (n-1)} \otimes (\mathds{H}_n \mathds{Z}_n)$ is $\ket{\psi_3} = \ket{r}\ket{0}$.

\subsection{Observable}
The fourth layer is for measurement in the $\mathds{Z}$-basis and is not a unitary layer. We measure the state of each of the $n$-qubits after projection onto the computational basis states. 
Post-measurement, the observation obtained is a classical bit (either $0$ or $1$) for each of the n-qubits measured. 
Thus, the final observed output is a bit-string of length $n$, including the ancilla.

We make a distinction between the qubit-wise observables $\mathcal{O}_q$, where $q \in \{0, \cdots n\}$ and the measurement operator for the register  $M_Z = \mathds{Z}_0 \otimes \cdots \mathds{Z}_n$ in computational basis. 
The qubit-wise observable $\mathcal{O}_q$ has identity across the $n$-qubit tensor product except in the $q$-th position. 
For example, $\mathcal{O}_0 = \mathds{Z}_0 \otimes \mathds{I}^{\otimes {n-1}}$ and $\mathcal{O}_n = \mathds{I}^{\otimes {n-1}} \otimes \mathds{Z}_n$. 
The eigenvalues of $\mathcal{O}_q$ are +1 (corresponding to classical bit 0) and -1 (corresponding to classical bit 1). 
The theoretical mean of the qubit-wise observable $\mathcal{O}_q$ is denoted by $\braket{\mathcal{O}_q} = \Tr[\mathcal{O}_q \rho ]$ where $q \in \{0, \cdots, n \}$, which reflects the measurement process being fundamentally probabilistic. 

The experimentally observed mean of the observable, denoted by $\braket{\mathcal{O}_q}_\xrm$ is calculated as the sum of eigenvalues weighted by their empirically observed frequencies. In a noiseless circuit, the observed mean asymptotically converges to the theoretical mean as the sample size tends to infinity. However, in presence of shot noise and non-zero variance of $\xrm$, $\braket{\mathcal{O}_q}_\xrm$ may not equal $\braket{\mathcal{O}_q}$. 

The experimentally observed measurements of the $M_Z$ operator in computational basis belong to one of the $2^{n}$ eigenstates, each of which can be represented by a $n$-bit binary string. These observations contain the information necessary for computing $\mathcal{O}_q$ for all $q$. 
In the context of the \BV algorithm, we discard the ancilla bit and declare the search a success when the first $n$-bits of the observed value of the $M_Z$ operator matches the secret string $r$. 
%
\subsection{Modeling circuit noise}
\begin{figure}[htbp]
\centering
\includegraphics[width=\linewidth]{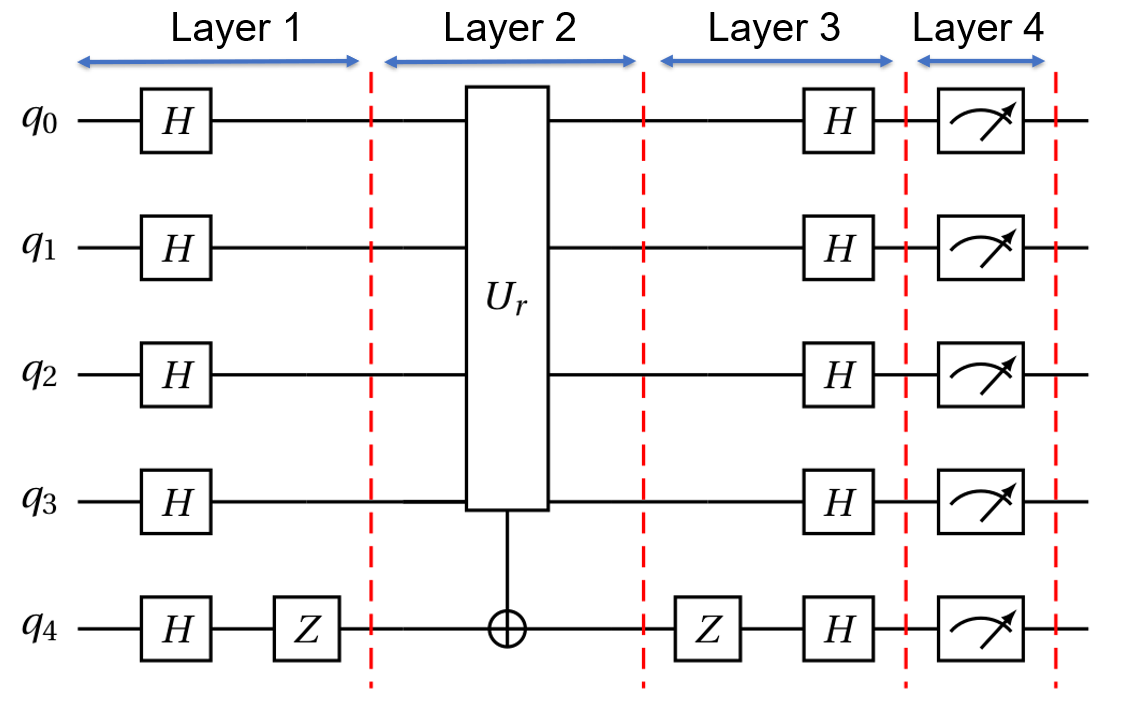}
\caption{A 5-qubit implementation of the Bernstein-Vazirani algorithm with secret bit-string $r$.}
\label{fig:bv_circuit}
\end{figure}

Our experimental focus is on \superc~hardware in next section. The potential sources of noise in a \superc~implementation of the circuit depicted in Fig.~\ref{fig:bv_circuit} are: (i) state preparation noise, (ii) noise in the implementation of the Hadamard gate, (iii) noise in the implementation of the $\mathds{Z}$ gate, (iv) noise in the implementation of the CNOT gate, and (v) measurement noise (also known as readout noise). The error resulting from the first and last noise sources, namely state preparation noise and measurement noise, are often measured collectively due to experimental limitations. This combined noise is commonly referred to as SPAM (state preparation and measurement) noise. An effective model assumes that the state preparation is noiseless, and the noise impacts the measurement (or readout)  process. The relative magnitudes of the different types of noise are different. State preparation and measurement (SPAM) noise typically has the largest contribution to errors~\cite{bravyi2021mitigating}. The next most significant contribution arises from imperfect implementations of entangling gates, such as the CNOT. After that the strength of the noise, for single-qubit rotations, decreases by two orders of magnitude or more~\cite{van2023probabilistic, bravyi2021mitigating}. The $\mathds{Z}$ gate is a software-based operation~\cite{mckay2017efficient} and error-free. After disregarding single-qubit rotation errors, only two predominant types of noise emerge: SPAM noise and CNOT noise. 


Next, we create a quantum channel based description of the circuit noise. 
The circuit noise can be modeled layer-wise. 
The output density matrix prior to measurement can be represented as: 
$
\tilde{\rho} = 
\mathcal{E}_4( 
\mathcal{E}_3( 
\mathds{L}_3 \mathcal{E}_2(
\mathds{L}_2 \mathcal{E}_1(
\mathds{L}_1 \rho \mathds{L}_1^\dagger) 
\mathds{L}_2^\dagger) 
\mathds{L}_3^\dagger )
$
where $\mathcal{E}_k(\cdot)$ denotes the noise channel for the $k$-th layer of the circuit in Fig.~\ref{fig:bv_circuit}. We approximate $\mathcal{E}_1$ as identity channel because we ignore single-qubit errors. 

We model noise in the CNOT gate using a 2-qubit depolarizing model. 
Let $\xrm_C(t)$ and $\xrm_T(t)$ represent the stochastic depolarizing parameters for the control and target qubits at time t. 
The depolarizing noise model for the CNOT gate is represented by: 
\begin{equation}
\begin{split}
&\mathcal{E}_\text{CNOT}(\cdot) = [ 1-\xrm_T(t) ][ 1 - \xrm_C(t) ] (\cdot) +\\
&\frac{1-\xrm_C(t)}{3}\xrm_T \sum \limits_{\mathds{P}_T^\prime}  (\mathds{I}_C \otimes \mathds{P}_T^\prime) (\cdot) (\mathds{I}_C \otimes \mathds{P}_T^\prime)\\
&+\frac{1-\xrm_T(t)}{3}\xrm_C \sum \limits_{\mathds{P}_C^\prime}  (\mathds{P}_C^\prime \otimes I_T ) (\cdot) (\mathds{P}_C^\prime \otimes I_T )\\
&+\frac{\xrm_C(t) \xrm_T }{9} \sum \limits_{\mathds{P}_C^\prime, \mathds{P}_T^\prime}  (\mathds{P}_C^\prime \otimes \mathds{P}_T^\prime) (\cdot) (\mathds{P}_C^\prime \otimes \mathds{P}_T^\prime)
\end{split}
\end{equation}
In this expression, we use $\mathds{P}_C^\prime, \mathds{P}_T^\prime \in \{\mathds{X}, \mathds{Y}, \mathds{Z}\}$ as the single-qubit Pauli operators excluding identity $\mathds{I}$, acting on the control (C) and target (T) qubits respectively. The sum is over all the single-qubit Pauli operators, excluding identity. The effect of the quantum noise channel $\mathcal{E}_2(\cdot)$ for layer 2, on a 5-qubit state, combines the identity channel $\mathcal{I}$ (which acts on qubits without CNOT connection) and $\mathcal{E}_\text{CNOT}$ (which acts on qubits linked by CNOT connections). 

Similar to layer 1, layer 3 also comprises single-qubit rotations only. Hence, we treat $\mathcal{E}_3(\cdot)$ as identity channel. Lastly, the fourth layer is subjected to SPAM noise which we adapt to a noise channel description~\cite{smith2021qubit}. \added{
It effectively handles measurement noise by corrupting the density matrix post execution but pre measurement, and then conducting noise-free projective measurements on the corrupted output. }
The SPAM noise channel for qubit $q$ has two Kraus operators $M_0$ and $M_1$:
\begin{equation}
\begin{split}
M_0 =& \sqrt{ f_q}\ket{0}\bra{0}+\sqrt{1-f_q}\ket{1}\bra{1}\\
M_1 =& \sqrt{1-f_q}\ket{0}\bra{0}+\sqrt{f_q}\ket{1}\bra{1}\\
\end{split}
\end{equation}
where $f_q$ represents the SPAM fidelity of qubit $q$. 
The probability of observing $0$ is given by $\text{Tr}[M_0^\dagger M_0 \rho]$ and the probability of observing $1$ is given by $\text{Tr}[M_1^\dagger M_1 \rho]$. Note that we have assumed a symmetric model for SPAM noise with $f_q$ denoting the average SPAM fidelities for the initial states prepared as $\ket{0}$ and $\ket{1}$.

Neglecting inter-qubit cross-talk, we then have the noise channel representation for the last layer $\mathcal{E}_4$ as a separable SPAM noise channel:
\begin{equation}
\mathcal{E}_4(\cdot) = \left[\bigotimes\limits_{q=0}^{n} \mathcal{E}_q^\text{SPAM}\right](\cdot)
\end{equation}
%
\subsection{PEC coefficients under Pauli noise}\label{sec:analytical_BV_model}
We first discuss CNOT noise mitigation using PEC. 
Following that, we discuss single-qubit SPAM noise mitigation using PEC. 
Then, we integrate the two discussions for noise mitigation in the quantum circuit implementation of the \BV algorithm using PEC. 
%
\subsubsection{CNOT noise}\label{sec:cnotPEC}
Let $\tilde{\mathcal{G}_0}$ denote the \sop~for the noisy CNOT operation. 
Under the Pauli channel assumption, there are 16 basis {\sop}s, indexed by $k$:
\begin{equation}
\tilde{\mathcal{G}_k} \rho = (\mathds{P}_C \otimes \mathds{P}_T) \left( \tilde{\mathcal{G}_0} \rho \right) (\mathds{P}_C \otimes \mathds{P}_T)
\end{equation}
where $\rho$ is a 2-qubit density matrix, 
$\mathds{P}_C, \mathds{P}_T \in \{ \mathds{I},\mathds{X},\mathds{Y},\mathds{Z}\}$ are the single-qubit Paulis acting on the control (C) and target (T) qubits respectively, 
and $k \in \{0, \cdots 15\}$. 
The index $k$ is determined by the specific combination of $\mathds{P}_C$ and $\mathds{P}_T$, starting from $\mathds{I}\otimes\mathds{I}$ for $k=0$ and ending with $\mathds{Z}\otimes\mathds{Z}$ for $k=15$, incrementing $k$ for each subsequent combination in the sequence $\mathds{I}, \mathds{X}, \mathds{Y}, \mathds{Z}$ applied to $\mathds{P}_C$ and $\mathds{P}_T$. 

\begin{figure}[htbp]
\centering
\includegraphics[width=.2\linewidth]{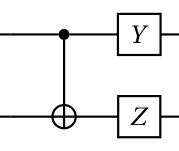}
\caption{Circuit diagram for implementing one of the 16 noisy {\sop}s for CNOT noise mitigation using PEC. The initial noisy CNOT gate is followed up by $\mathds{Y}$ and $\mathds{Z}$ gates on the control and target qubits respectively.}
\label{fig:CNOT_YZ}
\end{figure}

When $\mathds{P}_C = \mathds{P}_T = \mathds{I}$, the circuit corresponds to the noisy \sop~$\tilde{\mathcal{G}_0}$ denoting the noisy CNOT gate available. 
An example of one of the remaining 15 PEC circuits for CNOT noise mitigation is shown in Fig.~\ref{fig:CNOT_YZ}.

Using the requirement that the linear combination $\sum \eta_{ {}_k} \tilde{\mathcal{G}}_k $ should equal the noiseless CNOT operation, we derive the PEC coefficients as follows. 

Let $c_0 = 3/[\xrm_T (1-\xrm_C)]^2+[\xrm_C (1-\xrm_T)]^2-3c_1^2]$ and, $c_1 = \xrm_C + \xrm_T - \xrm_C \xrm_T - 1$. 

When $\mathds{P}_C = \mathds{P}_T = \mathds{I}$, 
$$\eta_\low{0}  = c_0c_1.$$ 
When $\mathds{P}_C=\mathds{I}$ and $\mathds{P}_T \in \{\mathds{X}, \mathds{Y}, \mathds{Z}\}$, 
$$\eta_\low{k} = c_0\frac{\xrm_T (1-\xrm_C)}{3}, \;\;\;\; k \in \{1,2,3\}.$$ 
When $\mathds{P}_T=\mathds{I}$ and $\mathds{P}_C \in \{\mathds{X}, \mathds{Y}, \mathds{Z}\}$, 
$$\eta_\low{k} = c_0\frac{\xrm_C (1-\xrm_T)}{3}, \;\;\;\; k \in \{4,8,12\}.$$
For all the remaining terms, 
$$\eta_\low{k} = c_0\frac{\xrm_C \xrm_T}{9}, \;\;\;\; k \in \{9, 10, 11, 13, 14, 15\}.$$

This yields the quasi-probability distribution as: 
$
\{\mathds{Q}_k\} = \left\{ |\eta_{{}_0}|/\gamma_{{}_\text{CNOT}}, \cdots, |\eta_{{}_{15}}|/\gamma_{{}_\text{CNOT}} \right\}
$ 
where $\gamma_{{}_\text{CNOT}} = |\eta_{{}_0}| + \cdots + |\eta_{{}_{15}}|.$\\
\subsubsection{SPAM noise for qubit q}\label{sec:SPAMpec}
Consider a single-qubit SPAM noise channel for qubit $q$. The four noisy {\sop}s that can be implemented in this case are: 
(i) the SPAM noise channel: $\tilde{\mathcal{G}}_\mathds{I}(\rho) = f_q\rho + (1-f_q)\mathds{X}\rho \mathds{X}$, 
(ii) the SPAM noise channel followed by an X error: $\tilde{\mathcal{G}}_\mathds{X}(\rho) = f_q\mathds{X}\rho \mathds{X} + (1-f_q)\rho$, 
(iii) the SPAM noise channel followed by a Y error: $\tilde{\mathcal{G}}_\mathds{Y}(\rho) = f_q\mathds{Y}\rho \mathds{Y} + (1-f_q)\mathds{Z}\rho \mathds{Z}$ and, 
(iv) the SPAM noise channel followed by a Z error: $\tilde{\mathcal{G}}_\mathds{Z}(\rho) = f_q\mathds{Z}\rho \mathds{Z} + (1-f_q)\mathds{Y}\rho \mathds{Y}$.

Solving the linear equation: 
\begin{equation}
\mathcal{I} = \eta_0 \tilde{\mathcal{G}}_I + \eta_1 \tilde{\mathcal{G}}_X + \eta_2 \tilde{\mathcal{G}}_Y + \eta_3 \tilde{\mathcal{G}}_Z
\label{eq:pec_eq_spam}
\end{equation}
we get the quasi-probability distribution as:
$
\left\{ |\eta_\low{0}|/\gamma_{{}_\text{SPAM}}, |\eta_\low{1}|/\gamma_{{}_\text{SPAM}}, 0, 0 \right\}
$
with
\begin{equation}
\begin{split}
&\eta_\low{0} = \frac{f_q}{2f_q-1}, \;\;\;\; \eta_\low{1} = -\frac{1-f_q}{2f_q-1}, \;\;\;\; \eta_\low{2} = \eta_\low{3} = 0\\
&\text{sgn}(\eta_\low{0}) = +1, \;\;\;\; \text{sgn}(\eta_\low{1}) =-1\\
&\gamma_{{}_\text{SPAM}}(q) = |\eta_\low{0}|+|\eta_\low{1}|\\
\end{split}
\label{eq:SPAM_PEC}
\end{equation}
Estimating $f_q$ therefore provides a complete specification of the PEC coefficients.\\
\subsubsection{Composite noise in 5-qubit implementation of \BV algorithm}
The two noisy basis {\sop}s for each of the 5 distinct SPAM noise channels and the 16 noisy basis {\sop}s for the CNOT noise channel leads to 512 ($16\times 2^5$) noisy basis circuits $\{ \mathcal{G}_k \}$ for the 5-qubit \BV circuit, where $k$ runs from 0 to 511. 


Under the SPAM noise separability assumption, $\gamma$ is obtained as:
\begin{equation}
\gamma = \gamma_{{}_{CNOT}} \prod\limits_{q=0}^{4} \gamma_{{}_{SPAM}} (q)
\end{equation}
where $\gamma_{{}_{SPAM}} (q)$ refers to the $\gamma_{{}_{SPAM}}$ for the $q$-th qubit. The PEC coefficients for the \BV circuit are the elements of:
\begin{equation}
\left\{ \eta_\low{0}^\text{CNOT}, \cdots, \eta_\low{15}^\text{CNOT} \right\} 
\times \prod\limits_{q=0}^{4}
\left\{ \eta_\low{0}^\text{SPAM}(q),  \eta_\low{1}^\text{SPAM}(q) \right\}
\label{eq:cartesian_product}
\end{equation}
The quasi-probability distribution then follows as: $\left\{\mathds{Q}_k\right\} = \left\{|\eta_\low{k}^\text{BV}| /\gamma \right\}$.\\

%

\subsection{Adaptive noise model}\label{sec:bayesian-stabilization}
In the context of our 5-qubit Bernstein-Vazirani setup, the noise parameters $f_0$, $f_1$, and $f_2$ are estimated using the method for uncorrelated parameters as they do not have CNOT correlations. The adaptive estimation of parameters $f_3, f_4, \xrm_C,$ and $\xrm_T$ employs the method for handling correlated parameters, detailed in Sec.~\ref{sec:adaptPEC}. 
The process is initiated within the Bayesian inference framework, which considers the probabilities of observing outcomes 00, 01, 10, and 11 on qubits 3 and 4, as depicted in Fig.~\ref{fig:bv_circuit}, to be random variables. 
The estimation of the time-varying means and variances of these correlated noise parameters ($f_3, f_4, \xrm_C, \xrm_T$) is achieved by associating the mean values of the estimated densities directly with the parameters of the correlated noise model using:
\begin{equation}
\Pr(i) = \Tr\left[
\Pi_i
\mathcal{E}_4( 
\mathds{L}_3 \mathcal{E}_2
(\mathds{L}_2 \mathds{L}_1 \rho \mathds{L}_1^\dagger \mathds{L}_2^\dagger) 
\mathds{L}_3^\dagger
)
\right]
\end{equation}
where $\Pi_i = \ket{i}\bra{i}$ are the projection operators and $i \in \{00, 01, 10, 11\}$.
The probabilities are given by:
\begin{equation}
\begin{split}
\text{Pr}(00) =& f_3 f_4 (-1 - \xrm_C \xrm_T + \xrm_C+\xrm_T)\\
&+ f_3 (\xrm_C \xrm_T /2-\xrm_T/2) \\
&+ f_4 (1 -\xrm_T - \xrm_C /2  + \xrm_C \xrm_T /2)\\
&- \xrm_C \xrm_T/4 +\xrm_T/2\\
\end{split}
\end{equation}

\begin{equation}
\begin{split}
\text{Pr}(01) = & f_3 f_4 (1 -\xrm_C -\xrm_T + \xrm_C \xrm_T)\\
&+ f_3 (-1 -\xrm_C \xrm_T /2 + \xrm_C   - \xrm_T /2)\\
&+ f_4 (-1 -\xrm_C \xrm_T /2 + \xrm_C   +\xrm_T)\\
&+1 +\xrm_C \xrm_T /4  -\xrm_C/2 -\xrm_T /2\\
\end{split}
\end{equation}

\begin{equation}
\begin{split}
\text{Pr}(10) = & f_3 f_4 (1-\xrm_T -\xrm_C +\xrm_C \xrm_T )\\
&+f_3 (\xrm_T /2-\xrm_C \xrm_T /2)\\
&+f_4 (\xrm_C /2 -\xrm_C \xrm_T /2)\\
&+ \xrm_C \xrm_T /4\\
\end{split}
\end{equation}

\begin{equation}
\begin{split}
\text{Pr}(11) = & f_3 f_4 (-1 +\xrm_C +\xrm_T - \xrm_C\xrm_T)\\
&+ f_3 (1 -\xrm_T/2 -\xrm_C +\xrm_C\xrm_T /2)\\
&+ f_4 (\xrm_C\xrm_T/2 -\xrm_C/2)\\
&+\xrm_C/2 - \xrm_C\xrm_T /4\\
\end{split}
\end{equation}

In the last step, the updated PEC coefficients are obtained using Eqn.~\ref{eq:cartesian_product}. 
\subsection{Numerical simulation}
For our 5-qubit circuit implementing the \BV algorithm, we used secret string $r=``1000"$. 
Thus, the qubit-wise mean of the $\mathds{Z}$ observable for the noiseless case is given by: $\braket{ \mathcal{O}_0 } = +1, \braket{ \mathcal{O}_1 } = +1, \braket{ \mathcal{O}_2 } = +1, \braket{ \mathcal{O}_3 } = -1, \braket{ \mathcal{O}_4 } = +1$, qubit 4 being the ancilla. 

To validate our method, we used a numerical experiment that conducts a density matrix simulation of a 5-qubit noisy quantum circuit implementing the Bernstein-Vazirani algorithm using the Qiskit~\cite{alexander2020qiskit} software. 
The simulation begins with the mean of the beta distributions characterizing the SPAM fidelities for qubits 0-4 set at $0.96, 0.95, 0.94, 0.93, 0.92$, respectively, and the mean of the depolarizing channel parameters for the control and target qubits in the CNOT gate both fixed at $0.017$. Over the course of ten simulated time periods, the average SPAM fidelity for each qubit decreased by 0.01 per period, resulting in final mean SPAM fidelities of $0.86, 0.85, 0.84, 0.83, 0.82$ for qubits 0-4, respectively. Similarly, the average depolarizing parameter for the CNOT gate also declined by 0.01 per time period, leading to a final mean value of $0.117$ for both control and target qubits by the simulation's end. The noise parameters were adaptively estimated, per the methodology described in previous sections, using data generated by executing each of the 512 PEC circuits using $10,000$ shots.

\begin{figure*}
\centering
\begin{minipage}{0.49\textwidth}
  \centering
  \includegraphics[width=\linewidth]{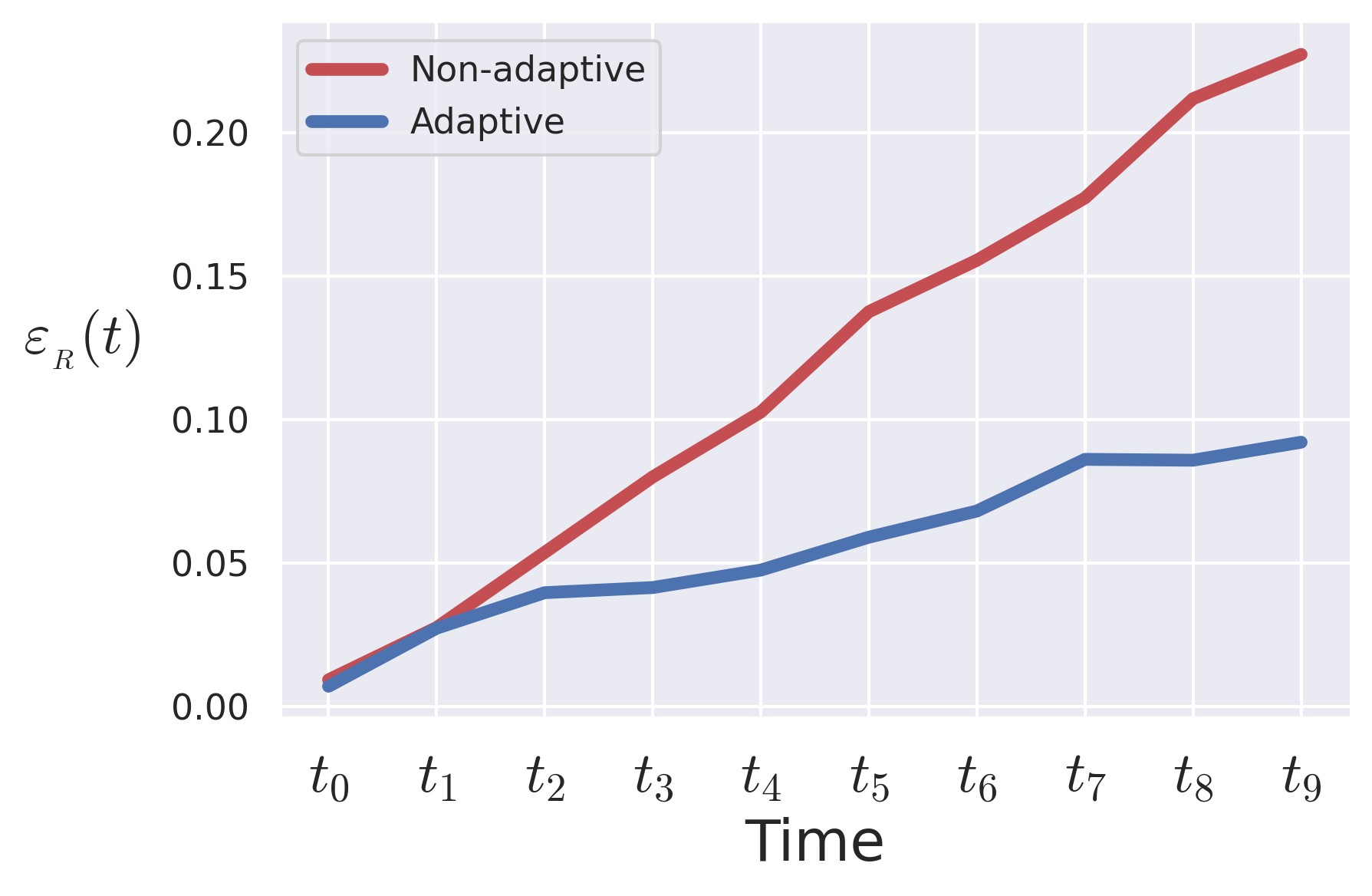}
  \caption*{(a)}
\end{minipage}
\hfill
\begin{minipage}{0.49\textwidth}
  \centering
  \includegraphics[width=\linewidth]{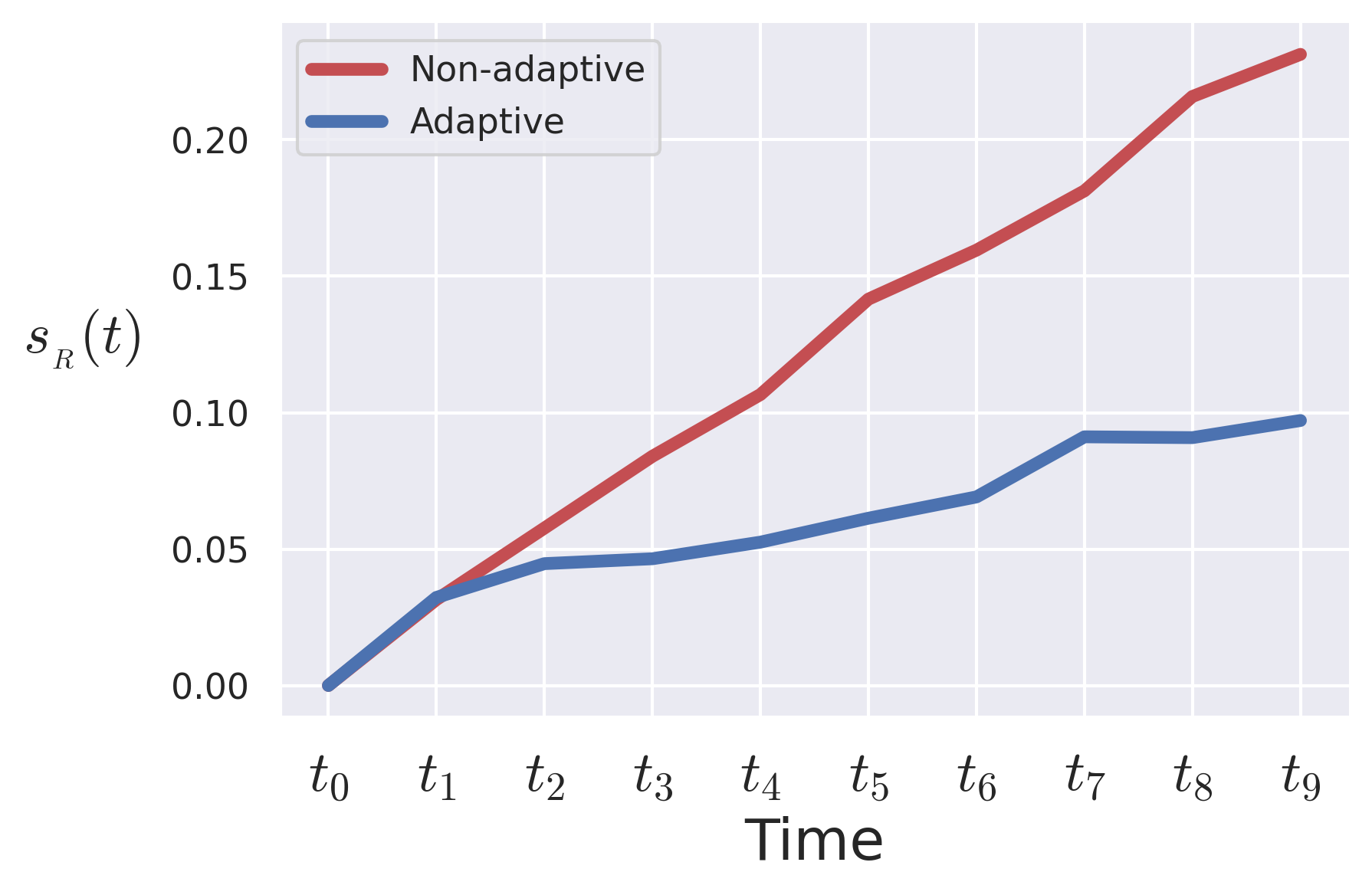}
  \caption*{(b)}
\end{minipage}
\caption{
Comparison of the adaptive PEC algorithm with a non-adaptive approach for a density matrix simulation that implements a noisy 5-qubit circuit for solving the Bernstein Vazirani problem. (a) Average accuracy across the 5-qubits and (b) Average stability across the 5-qubits.
}
\label{fig:num_sim_stability_accuracy}
\end{figure*}

Fig.~\ref{fig:num_sim_stability_accuracy} demonstrates the simulated efficacy of the adaptive PEC algorithm. The plot compares the register accuracy and stability (averaged over the 5 qubits) achieved with the adaptive approach against a non-adaptive approach. It shows an improvement in accuracy of 59.5\% in the final time period, when device noise is at its peak, with an average accuracy improvement of 53.4\% across all ten periods. Similarly, stability improved by 58.0\% in the last period, and by an average of 51.5\% over the entire span of ten periods. The improvement in accuracy and stability of the outcomes from adaptive PEC occur due to more accurate noise characterizations using Bayesian inference.
\section{Experimental Testing}\label{sec:real_data}
\begin{figure}[htbp]
\centering
\includegraphics[width=\linewidth]{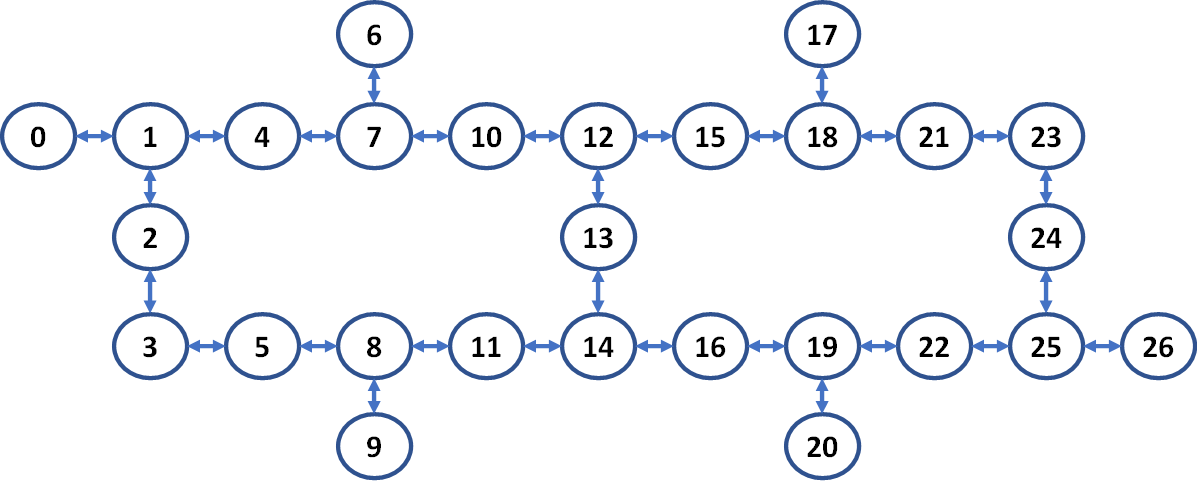}
\caption{Qubit layout of the 27-qubit \superc~device \kolkata. Circles represent superconducting transmons and lines indicate possible gate operations between sites.}
\label{fig:kolkata4}
\end{figure}

We tested the adaptive PEC method on the 27-qubit \superc~device called \kolkata.  
Qubits 0,1,2,3,4 in Fig.~\ref{fig:bv_circuit} map to physical qubits 0,1,2,3,5 on the device shown in Fig.~\ref{fig:kolkata4}. 
The CNOT gate is between the physical qubits 3 (control) and 5 (target) on \kolkata. 
Our dataset spans 24 hours and comprises 13 complete PEC datasets. 
It was collected on January 15, 2024, and have the following time-stamps: 00:01 hrs, 02:03 hrs, 03:50 hrs, 05:43 hrs, 07:23 hrs, 09:11 hrs, 10:53 hrs, 12:39 hrs, 14:22 hrs, 16:09 hrs, 17:50 hrs, 19:33 hrs, and 21:17 hrs. 
Each dataset is derived from measurements made in the computational basis, with each observation being a 5-bit string. 
The observations were obtained from the 512 noisy basis circuits, 
as described in Sec.~\ref{sec:analytical_BV_model}, 
with each circuit repeated using $L=10,000$ shots, 
resulting in approximately 67 million observations in total.
\subsection{Non-stationary noise estimates}
Fig.~\ref{fig:tv_bv_depol_spam} illustrates how noise in the quantum computer changed over time. The blue line in plot (a) shows the depolarizing parameter for the target qubit of the CNOT gate, while the black line represents the depolarizing parameter for the control qubit. Plot (b) shows five lines, each representing the SPAM fidelity for the register elements. The x-axis denotes intra-calibration timestamps. 

The graph shows periods where the noise levels in the depolarizing parameter for qubit 3 (the control qubit) are steady, notably between 2:00 am and 7:30 am, contrasting with times of significant fluctuation, as observed between 9:00 am and 2:30 pm. The depolarizing parameter fluctuates between 1\% and 4\% for qubit 4 (the target qubit) and between 1\% and 3\% for qubit 3, both peaking sharply at 10:53 am.

Qubit 4 experiences the most significant impact from SPAM noise, with its values fluctuating between 0.99 and 0.94, a notable range given the sensitivity of PEC  to accurate noise estimations. 
In contrast, Qubit 3 maintains a consistent SPAM fidelity throughout the same period, indicated by small error margins and a steady average value of 0.99. Meanwhile, Qubit 2 demonstrates a gradual drift in its values, starting from below 0.94 and rising to 0.96. This progressive change suggests a systematic, non-random trend that might be rectifiable with bias shift corrections. However, such patterns are not uniform across the entire register, implying the necessity to consider non-stationary statistics for modeling the system. Conducting experiments in times of significant non-stationary activity, like from 7:30 am to 2:30 pm, lead to more unstable outcomes when using non-adaptive PEC.
\subsection{Non-stationary quasi-probability distribution}
Fig.~\ref{fig:tv_qpr} underscores the importance of considering the \ns~nature of the quasi-probability distribution when implementing PEC, especially given the lengthy data collection process required for a single PEC mitigation (approximately 2 hours in our example). The abrupt change observed at 12:39 p.m. in the quasi-probability distribution directly correlates with the sharp change in the noise parameters characterizing the quantum circuit at the same time, as illustrated in Fig.~\ref{fig:tv_bv_depol_spam}.

To maintain clarity in Fig.~\ref{fig:tv_qpr}, we have not plotted all 512 bins of the histogram in one plot. 
Fig.~\ref{fig:tv_qpr}~(a) displays the time-varying weight $\mathds{Q}_0$ for the Bernstein-Vazirani circuit as-is without any additional Pauli-gate added (we also refer to this as the raw \BV circuit). 
If the circuit were noiseless, then the weight for the raw circuit will be a constant 1. 
We observe a decrease in weight for the raw circuit, dipping below 78\% around 10:53 am from a peak of almost 83\%, coinciding with a peak in circuit noise as seen in Fig.~\ref{fig:tv_bv_depol_spam}. This decrease in weight is expected as the circuit noise peaks.

The values of the quasi-probability bins for the next 10 basis circuits are shown in plot (b), with values approximately 10 times lower than the first circuit. 
Ignoring seemingly small coefficients in the quasi-probability distribution without considering the precision of final reported results can be risky. Subsequent basis circuits, not shown here, have significantly smaller quasi-probability weights (around $10^{-4}$). Yet their collective impact in a sum of 500 can be substantial, contributing up to 0.05. Given our reported accuracy and stability are around $10^{-2}$, these coefficients, though small, can significantly influence the results. Therefore, in our analysis, we included all basis circuits without approximation, focusing on the effects of non-stationary noise on PEC, rather than on resource optimization.
\subsection{Non-stationary PEC outcomes}
\begin{figure*}
\centering
\begin{minipage}{.7\textwidth}
  \centering
  \includegraphics[width=\linewidth]{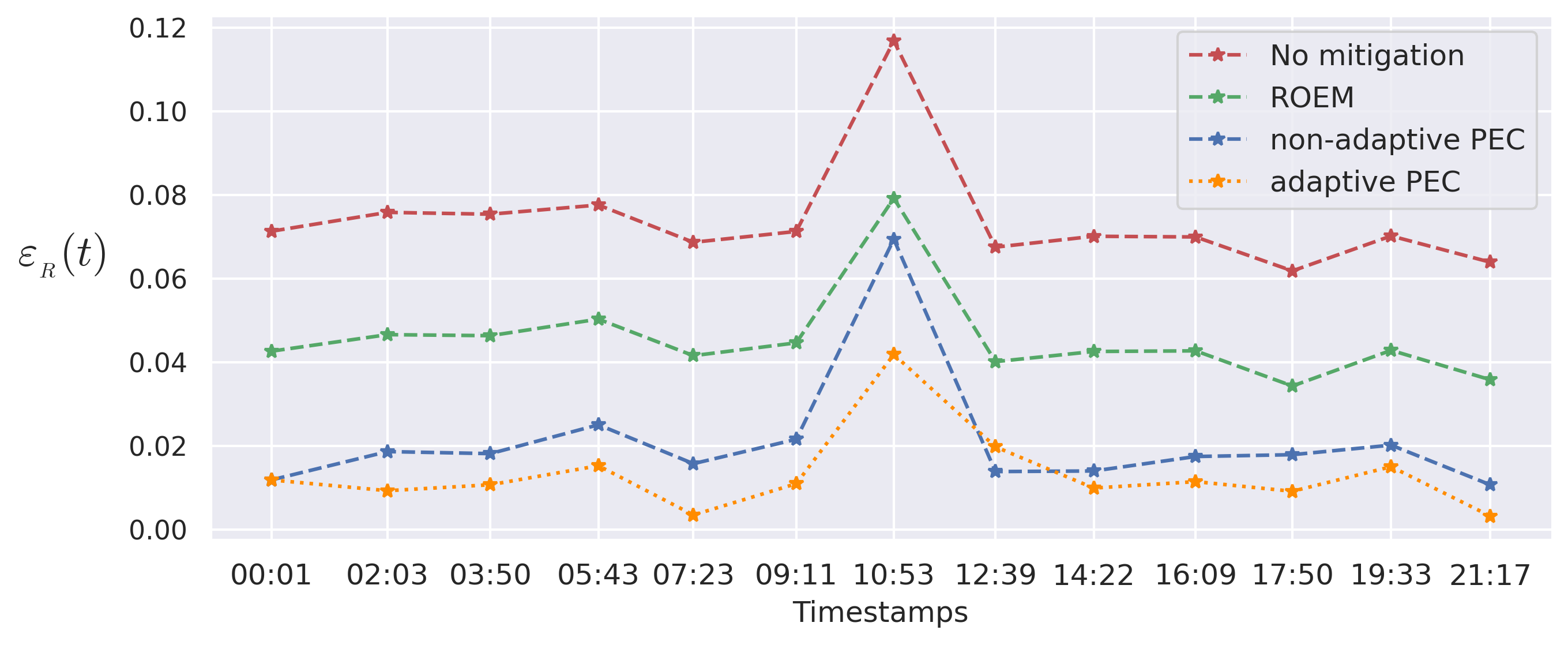}
  \caption*{(a)}
\end{minipage}
\hfill
\begin{minipage}{.7\textwidth}
  \centering
  \includegraphics[width=\linewidth]{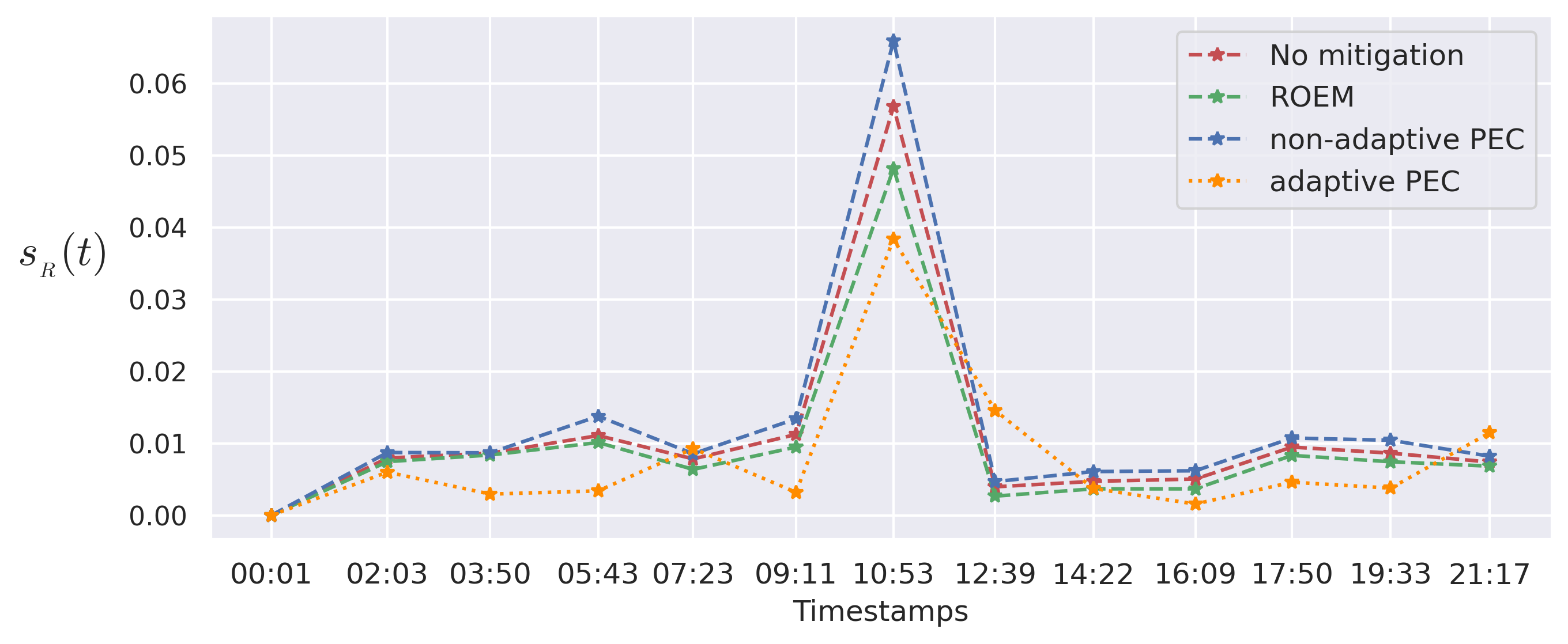}
  \caption*{(b)}
\end{minipage}
\caption{
The figure comprises two plots, each depicting four graphs: (1) "No mitigation" presents raw Bernstein-Vazirani metrics without error mitigation, (2) "ROEM" shows metrics after readout error mitigation, with constant SPAM noise parameters, (3) "Non-adaptive PEC" displays metrics for PEC, and (4) "Adaptive PEC" exhibits metrics for PEC with adaptive noise mitigation. The x-axis denotes intra-calibration time-stamps (UTC) for Jan 15. \added{Plot (a) illustrates the time-varying accuracy metric from Eqn.~\ref{eq:eR}.} It demonstrates the adaptive PEC's 42\% accuracy improvement over non-adaptive PEC. \added{Plot (b) shows the time-varying stability metric from Eqn.~\ref{eq:sR}.} It illustrates the adaptive PEC's 60\% stability enhancement compared to non-adaptive PEC. These plots underscore the significant impact of non-stationary noise on PEC resilience. Due to PEC's lengthy completion time (a couple of hours), adaptive methods are able to handle non-stationary noise conditions better.
}
\label{fig:stability_accuracy}
\end{figure*}

The impact of the non-stationary noise can be seen in Fig.~\ref{fig:stability_accuracy}. The first plot, labeled No mitigation, presents the accuracy and stability metrics defined in Eqns.~\ref{eq:eR} and \ref{eq:sR} respectively, for the raw Bernstein Vazirani circuit without any form of quantum error mitigation. The second plot, labeled ROEM, which stands for readout error mitigation, displays the metrics after performing SPAM noise mitigation. In this case, the SPAM noise parameters are held constant after initial device characterization. It deploys the standard matrix inversion~\cite{bravyi2021mitigating} technique for mitigation. The third plot, labeled non-adaptive PEC, exhibits the accuracy and stability metrics for the Bernstein Vazirani circuit with non-adaptive PEC. This method incorporates SPAM noise mitigation within the PEC framework, as detailed in Sec.~\ref{sec:analytical_BV_model}. The fourth plot, labeled adaptive PEC, presents the metrics for the adaptive PEC method, as discussed in Sec.~\ref{sec:adaptPEC}.

Fig.~\ref{fig:stability_accuracy}~(a) shows the effectiveness of the adaptive PEC in enhancing result accuracy. It reveals approximately a 42\% improvement in accuracy on average compared to the non-adaptive method. The observed accuracy benefit ranges from a minimum of 25\% to a maximum of 78\%.  Fig.~\ref{fig:stability_accuracy}~(b) shows the impact on result stabilization. It shows an approximately 60\% enhancement in stability on average compared to the non-adaptive method. The observed stability benefit ranges from a minimum of 8\% to a maximum of 200\%.

\begin{figure*}
\centering
\begin{minipage}{0.8\textwidth}
  \centering
  \includegraphics[width=\linewidth]{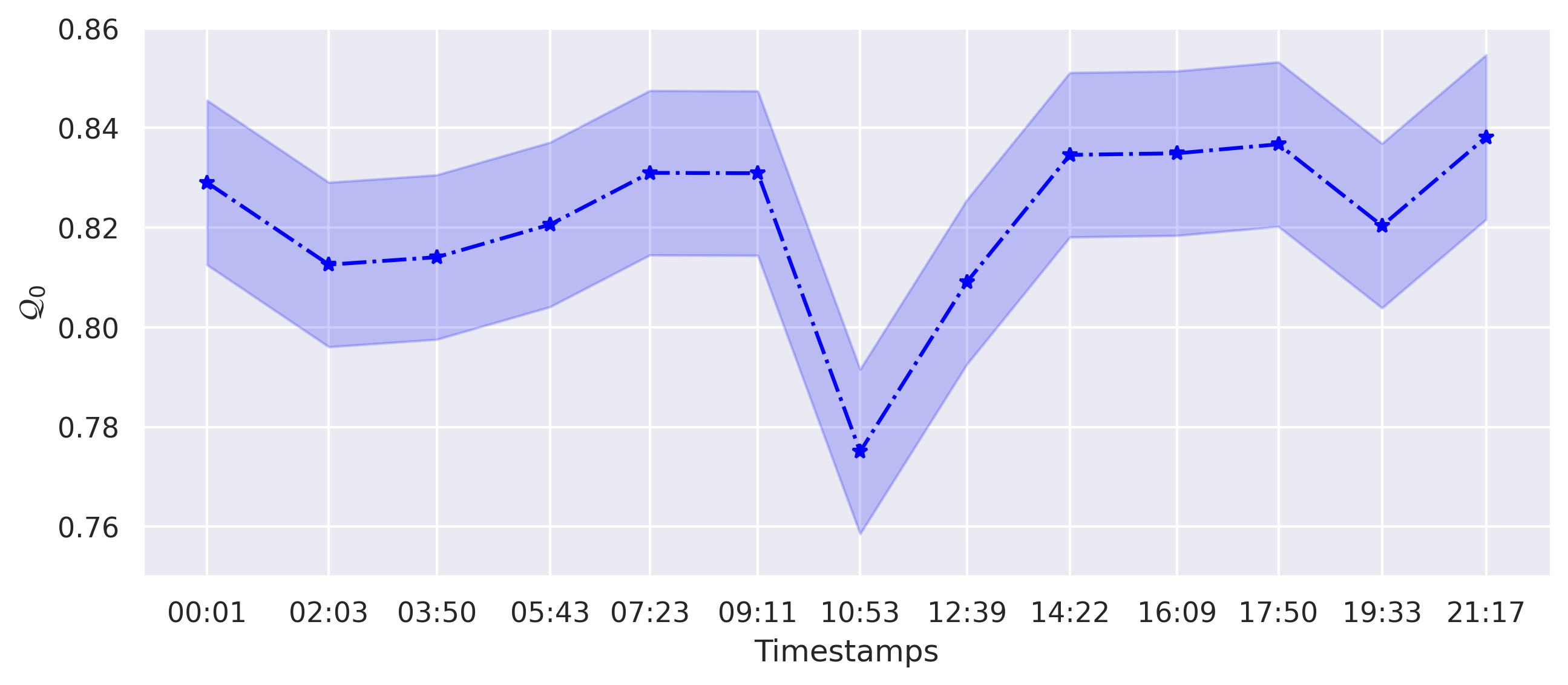}
  \caption*{(a)}
\end{minipage}
\hfill
\begin{minipage}{0.8\textwidth}
  \centering
  \includegraphics[width=\linewidth]{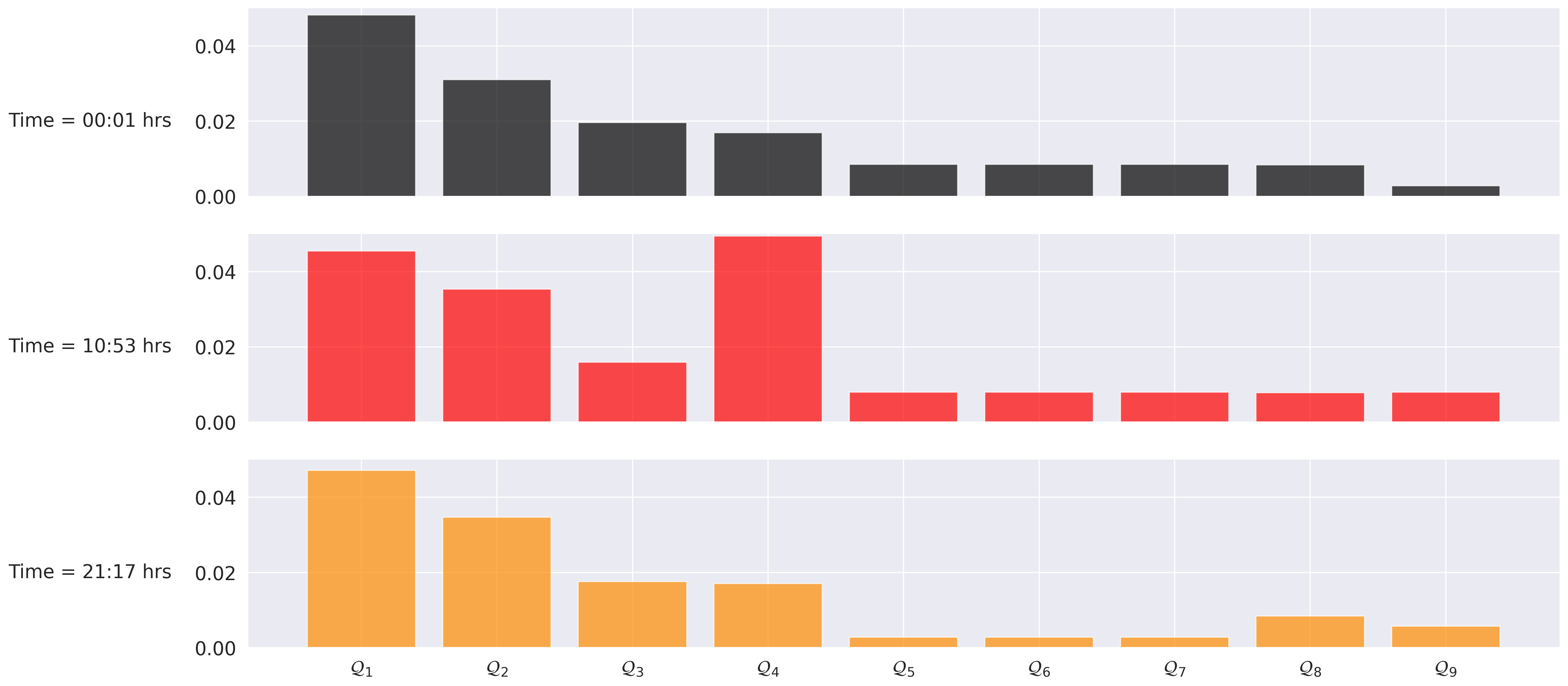}
  \caption*{(b)}
\end{minipage}
\caption{
These plots depict the \ns~nature of the quasi-probability distribution. To maintain clarity, we opted not to plot all 512 bins of the histogram in a single plot. In (a), we observe the time-varying weight $\mathds{Q}_0$ for the raw Bernstein-Vazirani circuit which carries the most substantial weight, accounting for almost 80\% of the distribution. In (b), we show the values of the quasi-probability bins for the subsequent 10 basis circuits, which are more than 10 times lower in magnitude compared to the first circuit. The \ns~nature of the quasi-probability distribution becomes crucial given the lengthy data collection process required for PEC mitigation because the noise estimation becomes inaccurate in these time-frames. Our experiment took approximately 2 hours for each dataset comprising 512 circuits.
}
\label{fig:tv_qpr}
\end{figure*}

Observing the plots, it is evident that adaptive PEC significantly outperforms standard PEC. Additionally, all four methods (no mitigation, ROEM, PEC, and adaptive PEC) exhibit a time-series trend that deteriorates notably at 10:53 am. The observation correlates with the abrupt change in the underlying quasi-probability distribution at 10:53 am as seen in Fig.~\ref{fig:tv_qpr}. 

\begin{figure*}
\centering
\begin{minipage}{0.8\textwidth}
  \centering
  \includegraphics[width=\linewidth]{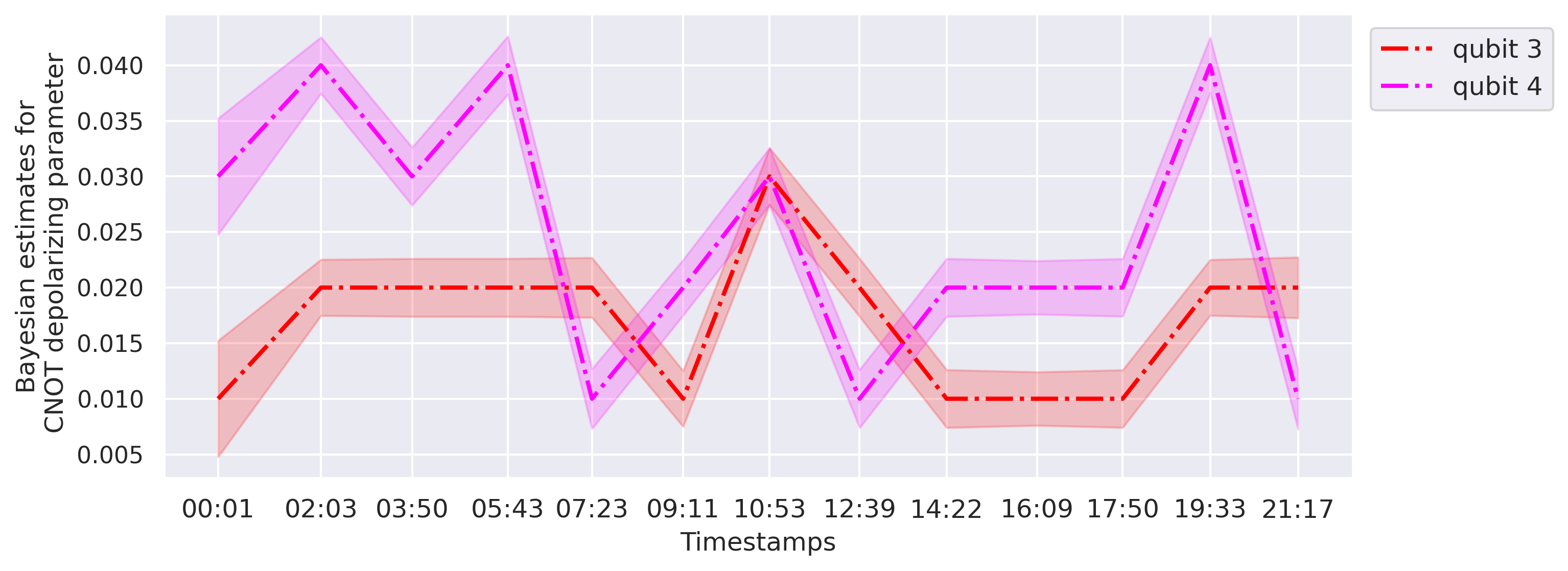}
  \caption*{(a)}
\end{minipage}
\hfill
\begin{minipage}{0.8\textwidth}
  \centering
  \includegraphics[width=\linewidth]{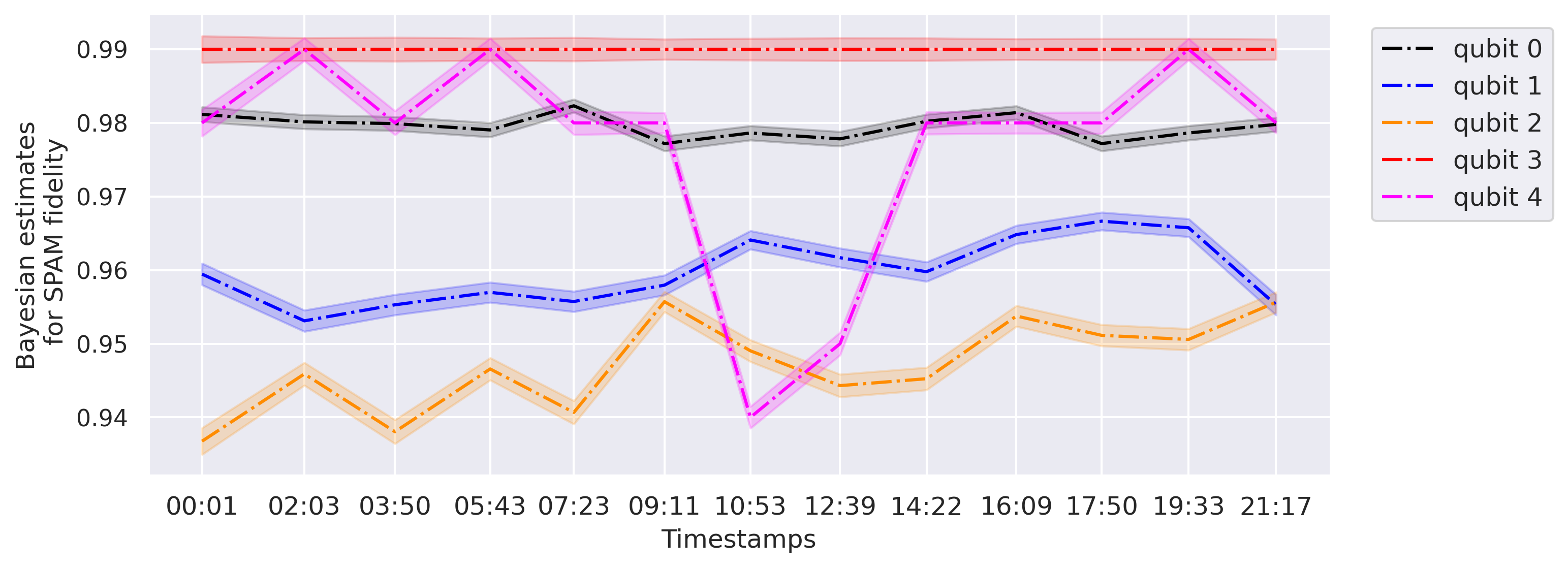}
  \caption*{(b)}
\end{minipage}
\caption{
This figure depicts the \nsn on the experimental device \kolkata. In plot (a), the blue line represents the depolarizing parameter for the target qubit, while the black line denotes the depolarizing parameter for the control qubit in the CNOT gate. Plot (b) illustrates five lines, each indicating the SPAM fidelity for the register elements. The x-axis corresponds to intra-calibration timestamps for January 15. The shaded regions denote the time-varying standard deviations.
}
\label{fig:tv_bv_depol_spam}
\end{figure*}

Both the accuracy and stability metrics at 12:39 p.m. are slightly worse for adaptive PEC compared to non-adaptive PEC. This discrepancy stands out as the only instance where adaptive PEC performs sub-optimally. It seems, surprisingly, that the stale data serves as a better noise estimate for this specific time-point. However, the noise at 12:39 p.m. is not necessarily closer to the noise at the starting time-stamp of 00:01 a.m., as demonstrated in Fig.~\ref{fig:tv_bv_depol_spam}. While adaptive PEC succeeds in most cases, the learning process is not instantaneous. There exists a slight lag in learning due to the influence of prior information on the final estimate. This phenomenon reflects a fundamental aspect of learning methods: the presence of memory, which can aid learning but also slows down adaptation to the fast, spiky changes. The sharp change in noise at 10:53 a.m. leads to an overestimation of the noise estimate compared to when using the initial value at 00:01 a.m. as a reference point. Residual errors in the noise estimates likely arises from inaccurate models and non-stationary processes changing at a faster rate than sampled here.

\clearpage
\vspace{0.5in}
\vspace{0.5in}
\vspace{0.5in}
\vspace{0.5in}
\begin{figure}[htbp]
\centering
\includegraphics[width=\linewidth]{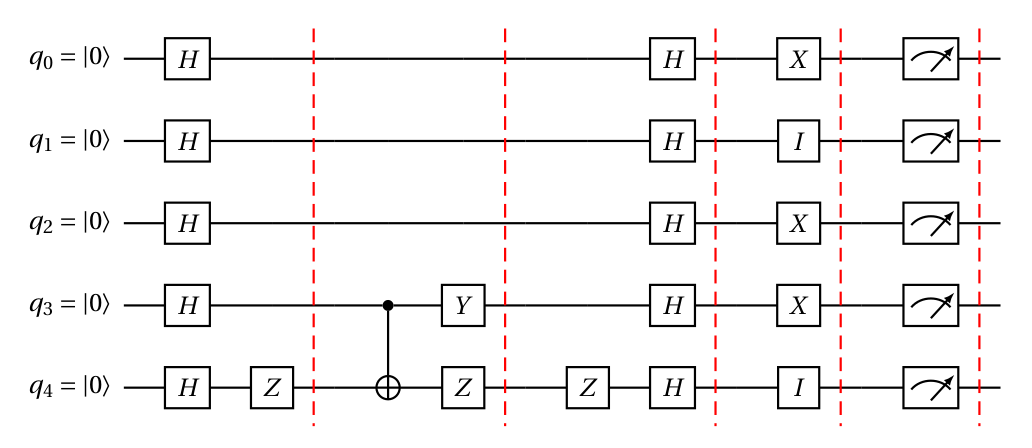}
\caption{One of the 512 noisy basis circuits for mitigation using PEC in the 5-qubit implementation of the \BV circuit. The initial noisy CNOT gate is followed up with $\mathds{Y}$ and $\mathds{Z}$ gates on the control and target qubits respectively. The readout lines for qubit $0, 1, 2, 3, 4$, and $5$ are subjected to the Pauli gates $\mathds{X}, \mathds{I}, \mathds{X}, \mathds{X}, \mathds{I}$ respetively, prior to measurement, in this specific noisy basis circuit for PEC.}
\label{fig:bv_circuit_XIXXIYZ}
\end{figure}
\vspace{0.5in}
\vspace{0.5in}
\vspace{0.5in}
\vspace{0.5in}
\vspace{0.5in}
\vspace{0.5in}
\begin{figure}[htbp]
\centering
\includegraphics[width=\linewidth]{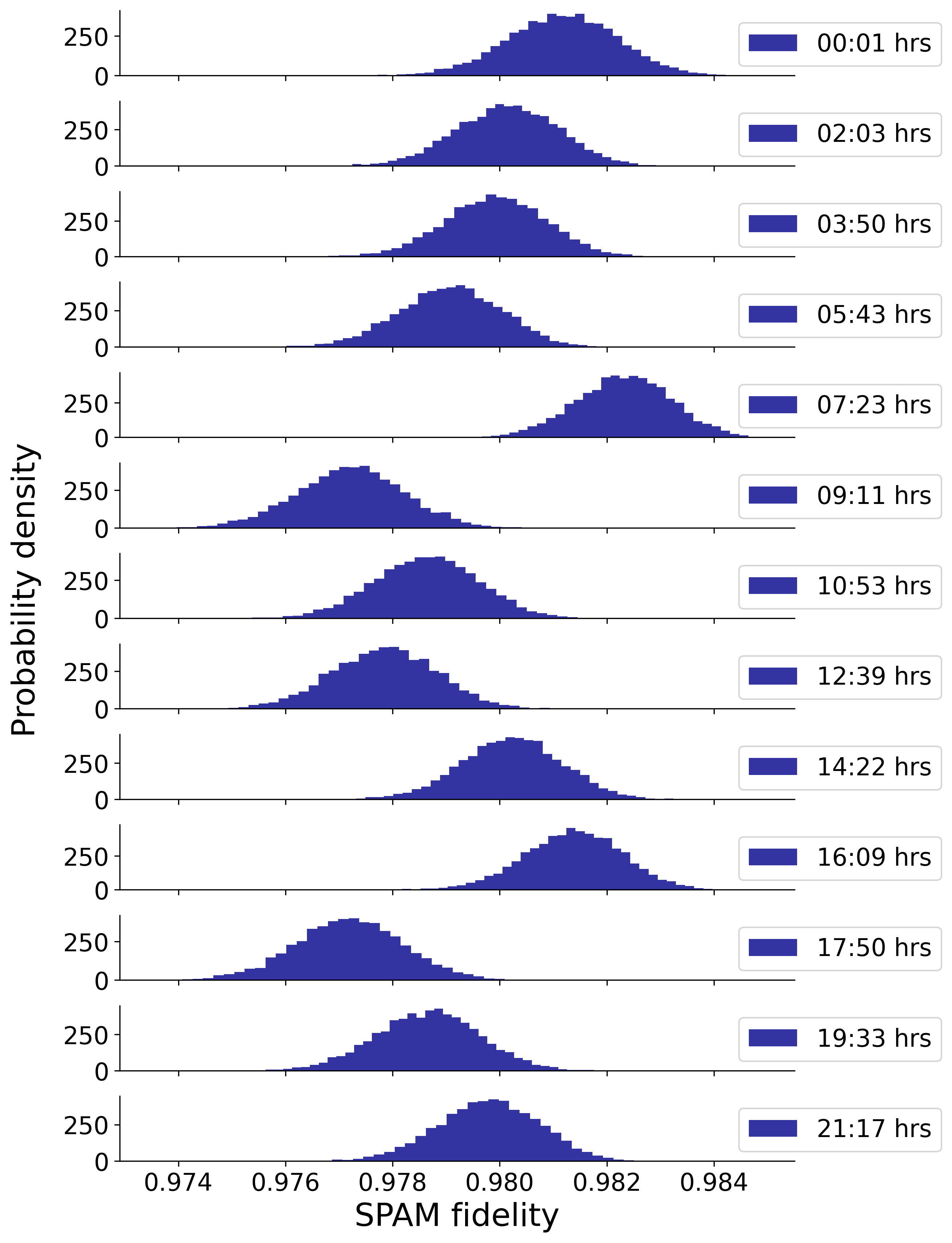}
\caption{Non-stationary distribution function of SPAM fidelity for qubit 0 of \kolkata as observed on Jan 15, 2024.}
\label{fig:tv_noise_f0}
\end{figure}
\vspace{0.5in}
\begin{figure}[htbp]
\centering
\includegraphics[width=\linewidth]{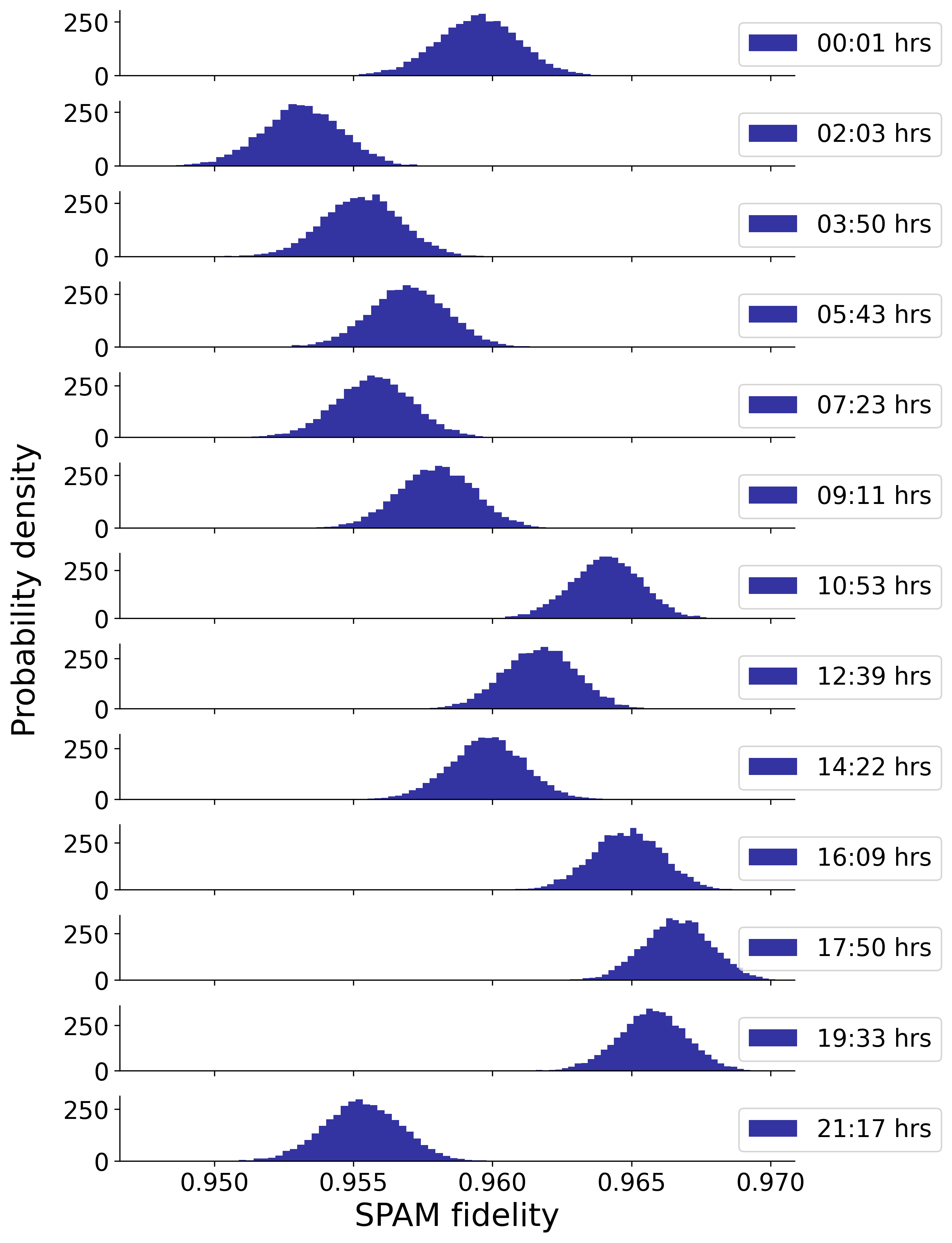}
\caption{Non-stationary distribution function of SPAM fidelity for qubit 1 of \kolkata as observed on Jan 15, 2024.}
\label{fig:tv_noise_f1}
\end{figure}
\vspace{0.5in}
\begin{figure}[htbp]
\centering
\includegraphics[width=\linewidth]{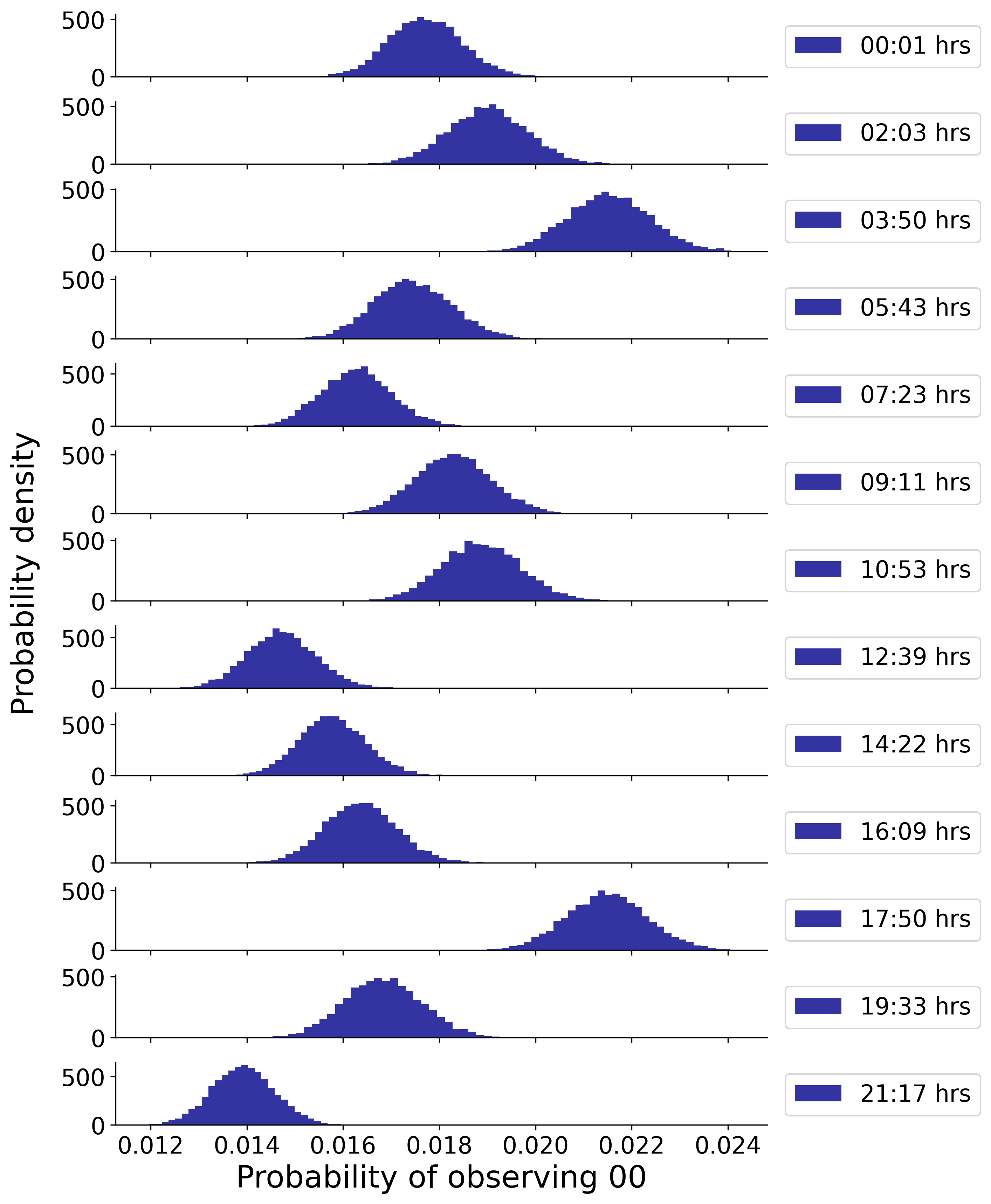}
\caption{Time-varying density for probability of observing `00' on qubits 3 and 4 of \kolkata, for the \BV circuit, as observed on Jan 15, 2024.}
\label{fig:tv_noise_pr00}
\end{figure}
\vspace{0.5in} 
\begin{figure}[htbp]
\centering
\includegraphics[width=\linewidth]{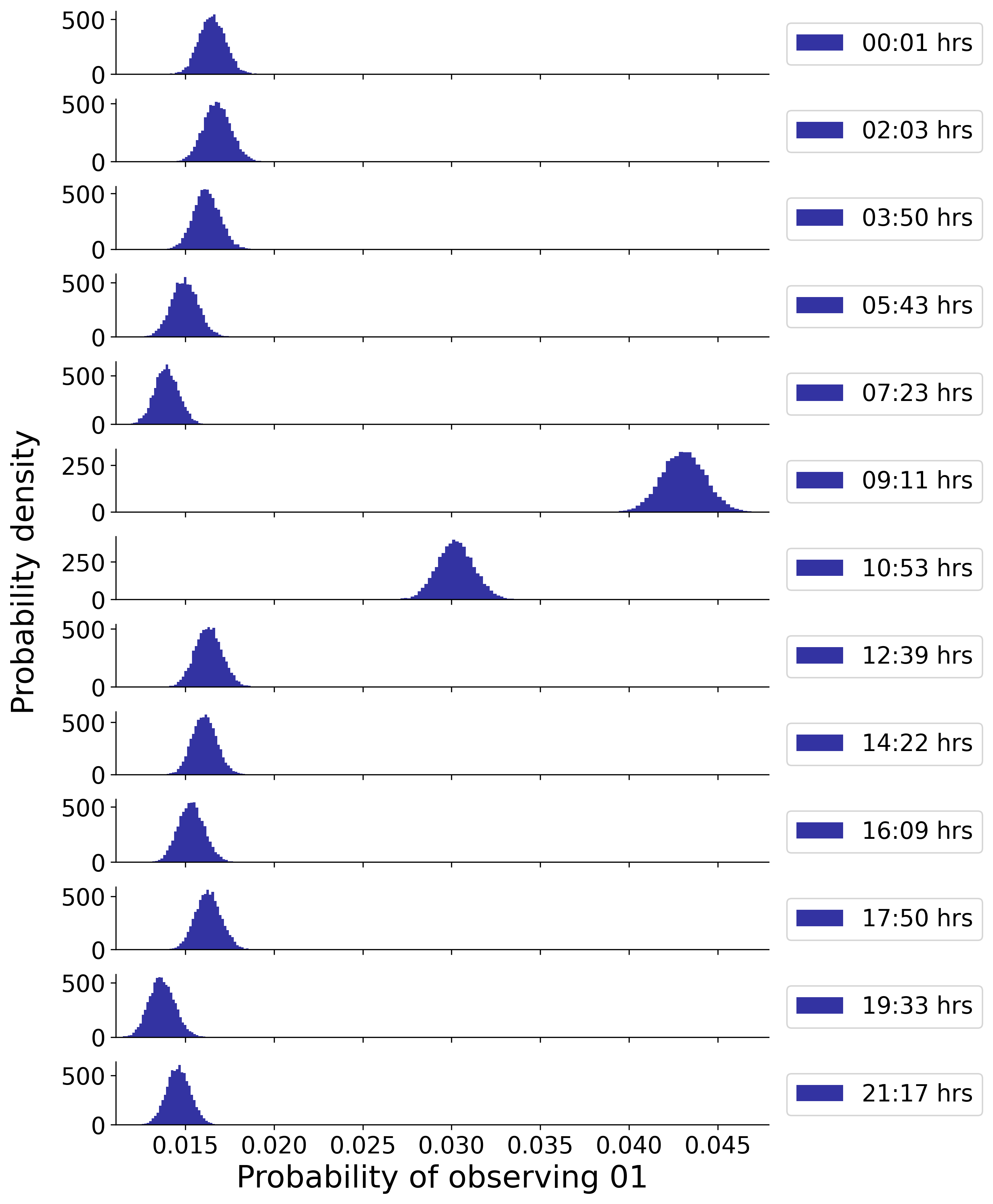}
\caption{Time-varying density for probability of observing `01' on qubits 3 and 4 of \kolkata, for the \BV circuit, as observed on Jan 15, 2024.}
\label{fig:tv_noise_pr01}
\end{figure}
\vspace{0.5in}
\begin{figure}[htbp]
\centering
\includegraphics[width=\linewidth]{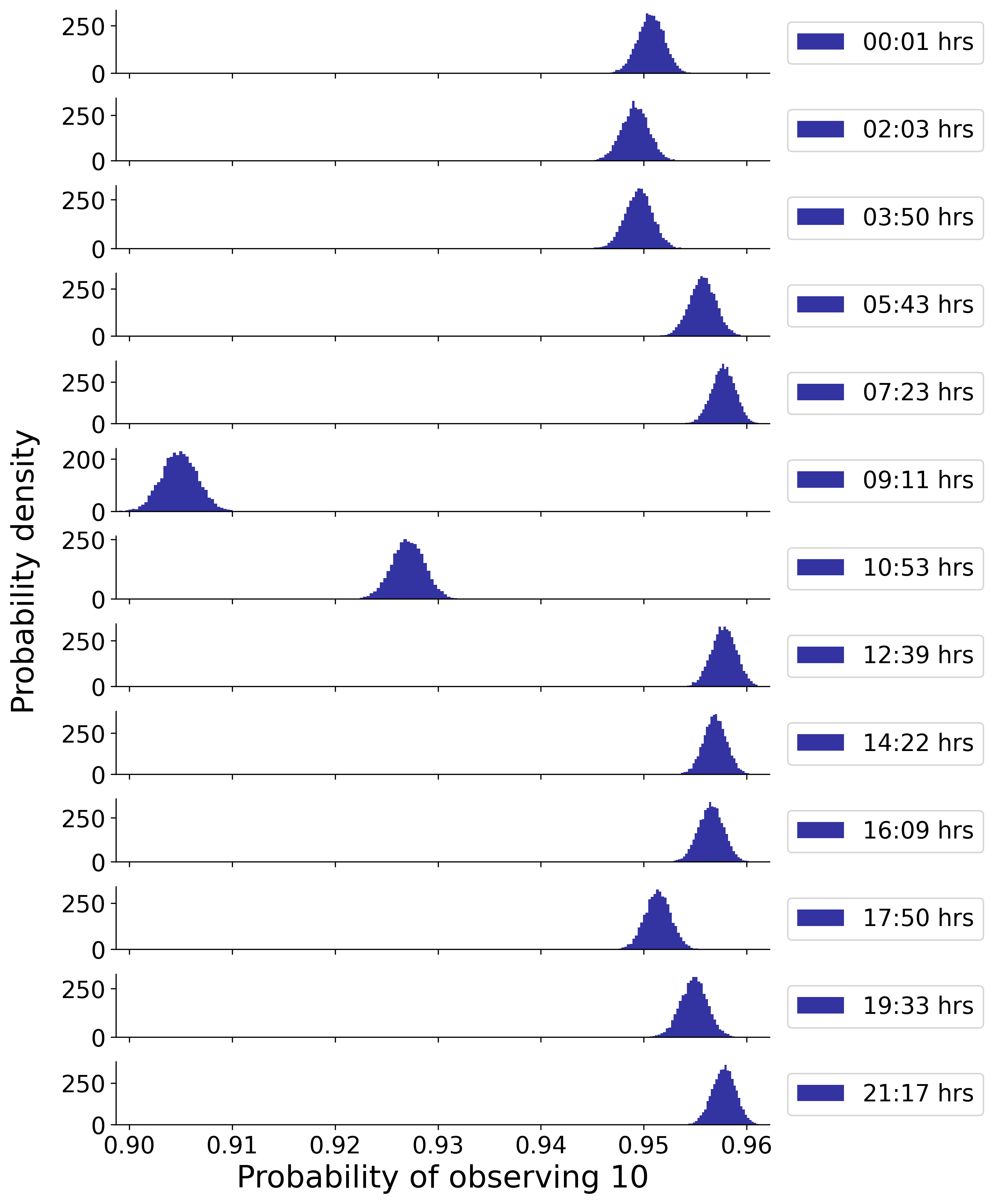}
\caption{Time-varying density for probability of observing `10' on qubits 3 and 4 of \kolkata, for the \BV circuit, as observed on Jan 15, 2024.}
\label{fig:tv_noise_pr10}
\end{figure}
\vspace{0.5in}
\begin{figure}[htbp]
\centering
\includegraphics[width=\linewidth]{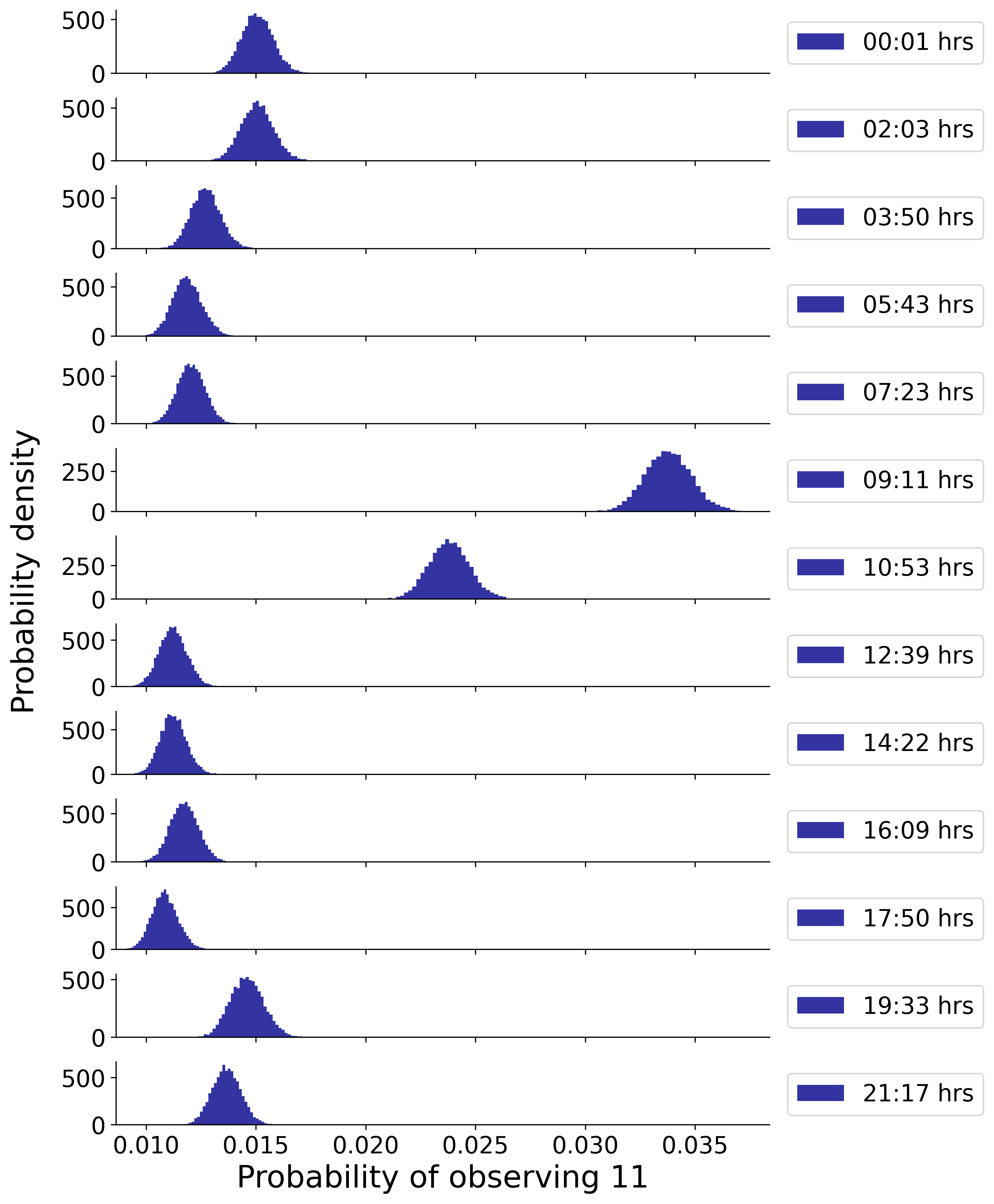}
\caption{Time-varying density for probability of observing `11' on qubits 3 and 4 of \kolkata, for the \BV circuit, as observed on Jan 15, 2024.}
\label{fig:tv_noise_pr11}
\end{figure}

\removefigs
\chapter{Conclusion}\label{sec:conclusion}
Quantum computing's tremendous potential~\cite{humble2019quantum} is curtailed by noise. 
Characterizing noise in contemporary quantum computers remains challenging due to its non-stationary statistics. 
The non-stationarity manifests across time, across different parts of the chip, and across devices. 
It impacts the verification and validation of quantum computing demonstrations and adversely impacts error mitigation strategies~\cite{bharti2022noisy}. 
This hinders the production of trustworthy results. 

This dissertation focused on the non-stationarity~\cite{etxezarreta2021time, muller2015interacting, klimov2018fluctuations}  of noise in \superc~processors~\cite{roth2021introduction}. It established a framework comprising computational accuracy, device reliability, outcome stability, and result reproducibility for assessing noisy outcomes. It determined upper and lower bounds for the performance metrics, in terms of the Hellinger distance between time-varying noise densities.  It demonstrated that if the noise stays within the theoretical bounds, outcome stability can be ensured with high confidence. It developed a Bayesian strategy to improve the stability and accuracy of results obtained from probabilistic error cancellation. Refinement in noise model selection, optimization algorithms, and data collection frequencies can yield further improvements and is a rich space to explore further. 

In particular, chapter~\ref{ch:decoherence_characterization} discussed noise in quantum computing and experimentally analysed the decoherence of \superc~qubits over a 24-hour period on September 12, 2023. 
The time-varying decoherence characterization was connected to the coefficients of a Pauli channel model.

While the majority of quantum computing research has focused on achieving accuracy~\cite{kliesch2021theory, blume2010optimal, ferracin2021experimental}, limited attention has been given to reproducibility, reliability, and stability. Chapter~\ref{ch:statistical_taxonomy} precisely defined these terms (computational accuracy, result reproducibility, device reliability and stability of error mitigated outcomes) for assessing the performance of noisy, quantum computers. Chapter~\ref{ch:exp_characterization} evaluated hardware reliability by analyzing experimental data across various time scales (monthly, daily, hourly, and seconds). It quantified the degree of non-stationarity in SPAM fidelity, CNOT fidelity, duty cycle, and addressability. It developed and validated a method to examine holistic reliability using the method of copulas~\cite{10.1177/1748006x13481928}.

Chapter~\ref{ch:analytical_bounds} developed bounds on a proxy parameter to encapsulate the array of parameters characterizing a noisy device. Such a proxy parameter is valuable for streamlining high-dimensional noise analysis. Experimental validation of the theoretical bound was performed using a 27-qubit \superc~device. The chapter further illustrated how to determine the minimum sample size to achieve reproducibility with $1-\delta$ confidence. It established bounds on device reliability necessary for achieving an $\epsilon$-stable outcome.

Chapter~\ref{ch:adaptive_algorithms} applied adaptive techniques~\cite{lukens2020practical, zheng2020bayesian, gordon1993novel, kotecha2003gaussian} to enhance accuracy of histograms, demonstrating a reduction in the Hellinger distance from 15\% to 3.1\%, while chapter~\ref{ch:adaptPEC} examined the impact of non-stationary noise on probabilistic error cancellation. It introduced a Bayesian approach to improve stability and accuracy of PEC outcomes. The algorithm was tested on the \kolkata device on January 15, 2024. The dataset covered a 24-hour period, consisting of 13 complete PEC datasets for the Bernstein-Vazirani test circuit, with approximately 67 million observations. Results indicated a 42\% increase in accuracy and a 60\% enhancement in stability. Consistent improvement trends across time-stamps and qubits demonstrated the effectiveness of the algorithm. 

The improvements in accuracy and stability of adaptive PEC vs non-adaptive PEC, across time and qubits, presented in this dissertation, indicate that the noise was well-characterized. 
In general, the choice of noise model influences the noise estimation process. 
If the chosen noise model is incorrect or insufficiently granular, the estimated noise parameters will reflect the specific dataset's patterns rather than accurately representing device noise. 
Mitigation using such mis-estimated noise models will be unsuccessful because the mitigation algorithm will introduce additional errors to the already noisy data. 
However, scalability becomes an issue with the Bayesian approach in the presence of correlations. 
While the method can accurately estimate numerous noise parameters in the absence of correlations, scalability diminishes when dealing with highly correlated parameters due to the need to estimate joint distributions and perform Monte Carlo simulations for maximum-a-posterior optimization. 

Before concluding, I offer a perspective on the significance of this dissertation. Firstly, this work delves into the intricate nature of non-stationary noise in contemporary quantum computers. It sheds light on strategies for improving the reproducibility and stability of noisy quantum computations, by incorporating adaptive processes to manage non-stationary noise.

Quantum error correction (QEC) methods represent a formal way to mitigate and manage errors but in practice need to be tailored to the noise. This is similar to 5G networks where low density parity check (LDPC)~\cite{ryan2004introduction, bravyi2024high} codes are employed to meet the high data transfer demands. These codes use a quasi-cyclic structure that adapt their encoding and decoding graph size depending on prevailing error rates. The wireless channel noise is often analyzed using methods that take into account the variability in space, time, and frequency when using correlation functions and power spectrum densities. In parallel, when dealing with quantum noise channels, this work adopted a strategy that incorporates time-varying correlated distributions to reflect the non-stationary nature of quantum noise and utilized quasi-probability distributions (in probabilistic error cancellation) whose weights are influenced by the strength of the prevailing noise.

This dissertation can also help draw parallels to classical cellular communications in handling non-stationary noise. 
Just as non-stationary channel conditions in cellular networks once caused frustrating service quality issues due to inadequate noise characterization, quantum computing faces similar issues with error mitigation fluctuating due to poor noise characterization. Moreover, the concept of temporal decoherence, caused by interactions with the environment, can be likened to the phenomenon of signal fading in classical cellular systems due to atmospheric interactions. Both provide case studies for how non-stationary processes can impact the reliability of information transmission. 
Lastly, just like early cellular communication systems suffered high error-rates from poor multiplexing capabilities due to limited spectrum resources, similarly, contemporary  quantum computing systems are limited in their error mitigation and correction capabilities due to constraints in qubit resources.  

The unpredictable and rapidly changing nature of non-stationary noise presents a challenge for fault-tolerant quantum computing. This type of noise can lead to error patterns that change more quickly than contemporary error correction codes can adjust, making it difficult to maintain fault tolerance. For instance, cosmic rays can cause sudden and sporadic error bursts, temporarily pushing error rates beyond the limits that quantum error correction codes are designed to handle. This can result in uncorrected errors and, potentially, the failure of logical qubits. Additionally, superconducting processors exhibit varying error rates across the chip, which challenges the assumption of a uniform threshold error rate, as often applied in techniques like the surface code. Such issues of non-stationary temporal and spatial error rates affecting error correction have previously been encountered in the field of cellular communications, indicating that the challenges faced by quantum computing are not entirely unprecedented. Note that non-stationary noise does not prevent the application of fault-tolerant quantum error correction but needs a tailored approach.

The results from this study can also help in noise modeling in quantum communication scenarios involving entanglement distribution over long distances as the quality of entanglement can be degraded by non-stationary atmospheric conditions when using free-space optical channels. These conditions can change rapidly due to weather phenomena, causing fluctuations in entanglement fidelity and compromising the integrity of information transfer.

Lastly, this dissertation underscores the need for interdisciplinary collaboration to advance the field of quantum computing. 
For example, it stresses the role of reliability engineering in improving manufacturing processes for quantum chips to ensure high standards of quality. It urges software developers to develop adaptive algorithms to tackle real-time challenges posed by non-stationary noise in quantum systems. It calls for the utilization of advanced statistical models and database management techniques by data scientists to aid quantum physicists using models should embrace the inherent quantum noise model uncertainties as a feature rather than a flaw. And, it challenges information theorists to extend their work beyond simple noise models. While foundational insights are invaluable, there is a need to extend, validate and apply these theories using real-world data to make them useful. By fostering collaboration across such diverse disciplines, we can accelerate the development of quantum computing technologies that are reliable, reproducible, and stable.  

\makeBibliographyPage
\bibliographystyle{unsrtnat}
\bibliography{references.bib}


\end{document}